# Roadmap on Deep Learning for Microscopy

**Giovanni Volpe[1]**, **Carolina Wählby[2]**, **Lei Tian[3,4]**, Michael Hecht[5], Artur Yakimovich[5,6,7,8], Kristina Monakhova[9], Laura Waller[9], Ivo F. Sbalzarini[10,11], Christopher A. Metzler[12], Mingyang Xie[12], Kevin Zhang[12], Isaac C.D. Lenton[13,14], Halina Rubinsztein-Dunlop[14,15], Daniel Brunner[16], Bijie Bai[17], Aydogan Ozcan[17], Daniel Midtvedt[1], Hao Wang[3], Nataša Sladoje[18], Joakim Lindblad[18], Jason T. Smith[19,20], Marien Ochoa[19,21], Margarida Barroso[22], Xavier Intes[19], Tong Qiu[23], Li-Yu Yu[23], Sixian You[23], Yongtao Liu[24], Maxim A. Ziatdinov[24,25], Sergei V. Kalinin[26], Arlo Sheridan[27], Uri Manor[27], Elias Nehme[28,29], Ofri Goldenberg[29], Yoav Shechtman[29], Henrik K. Moberg[30], Christoph Langhammer[30], Barbora Špačková[31], Saga Helgadottir[2,32], Benjamin Midtvedt[1], Aykut Argun[1], Tobias Thalheim[33], Frank Cichos[33], Stefano Bo[34,35], Lars Hubatsch[36], Jesus Pineda[1], Carlo Manzo[37], Harshith Bachimanchi[1], Erik Selander[38], Antoni Homs-Corbera[39], Martin Fränzl[33], Kevin de Haan[17], Yair Rivenson[17], Zofia Korczak[1], Caroline Beck

[1] Department of Physics, University of Gothenburg, 41296 Gothenburg, Sweden
[2] Dept. IT and SciLifeLab, Uppsala University, Sweden
[3] Department of Electrical and Computer Engineering, Boston University, Boston, MA 02215, USA
[4] Department of Biomedical Engineering, Boston University, Boston, MA 02215, USA
[5] Center for Advanced Systems Understanding (CASUS), Helmholtz-Zentrum Dresden-Rossendorf e. V. (HZDR), Görlitz, Germany.
[6] Bladder Infection and Immunity Group (BIIG), Department of Renal Medicine, Division of Medicine, University College London, Royal Free Hospital Campus, London, United Kingdom
[7] Artificial Intelligence for Life Sciences CIC, Dorset, United Kingdom
[8] Roche Pharma International Informatics, Roche Diagnostics GmbH, Mannheim, Germany
[9] Department of Electrical Engineering and Computer Sciences, UC Berkeley, USA
[10] Technische Universität Dresden, Faculty of Computer Science, Dresden, Germany
[11] Max Planck Institute of Molecular Cell Biology and Genetics, Center for Systems Biology Dresden, Dresden, Germany
[12] University of Maryland, College Park, USA
[13] Institute of Science and Technology Austria (ISTA), Klosterneuburg 3400, Austria
[14] School of Mathematics and Physics, The University of Queensland, Brisbane 4072, Australia
[15] Australian Research Council Centre of Excellence for Engineered Quantum Systems, The University of Queensland, Brisbane 4072, Australia
[16] Département d'Optique P. M. Duffieux, Institut FEMTO-ST, Université Franche-Comté CNRS UMR 6174, Besançon, France
[17] Department of Electrical and Computer Engineering, University of California Los Angeles, Los Angeles (CA), USA
[18] Department of Information Technology, Uppsala University, Sweden
[19] Department of Biomedical Engineering, Rensselaer Polytechnic Institute, Troy, NY, 12180, USA.
[20] Elephas, 1 Erdman Pl., Madison WI, 53705, USA
[21] University of Wisconsin–Madison, Department of Medical Physics, Madison WI, 53705, USA
[22] Department of Molecular and Cellular Physiology, Albany Medical College, Albany, NY 12208, USA.
[23] Research Laboratory of Electronics, Electrical Engineering and Computer Science, Massachusetts Institute of Technology, Boston (MA), USA
[24] Center for Nanophase Materials Sciences, Oak Ridge National Laboratory, Oak Ridge, TN 37830, USA
[25] Computational Sciences and Engineering Division, Oak Ridge National Laboratory, Oak Ridge, TN 37830, USA
[26] Department of Materials Science and Engineering, University of Tennessee, Knoxville, TN 37996, USA
[27] Salk Institute for Biological Studies, La Jolla, CA, USA
[28] Department of Electrical and Computer Engineering, Technion – IIT, Israel
[29] Department of Biomedical Engineering and Lorry I. Lokey Interdisciplinary Center for Life Sciences & Engineering, Technion – IIT, Israel
[30] Department of Physics, Chalmers University of Technology; Göteborg, Sweden
[31] Department of Optical and Biophysical Systems, Institute of Physics of the Czech Academy of Sciences; Prague, Czech Republic
[32] Department of Biochemistry and Biophysics, Stockholm University, Sweden
[33] Peter Debye Institute for Soft Matter Physics, Molecular Nanophotonics Group, Leipzig University, Linnéstraße 5, 04103 Leipzig, Germany
[34] Max Planck Institute for the Physics of Complex Systems, Nöthnitzer Straße 38, DE-01187 Dresden, Germany
[35] Department of Physics, King's College London, London WC2R 2LS, U.K.
[36] Max Planck Institute of Molecular Cell Biology and Genetics, Pfotenhauerstraße 108, 01307 Dresden, Germany
[37] Facultat de Ciències, Tecnologia i Enginyeries, Universitat de Vic – Universitat Central de Catalunya (UVic-UCC), Vic, Spain
[38] Department of Marine Sciences, University of Gothenburg, Sweden
[39] Research and Development Department, Cherry Biotech SAS, Paris, France

Adiels[1], Mite Mijalkov[40], Dániel Veréb[40], Yu-Wei Chang[1], Joana B. Pereira[40], Damian Matuszewski[41], Gustaf Kylberg[42], Ida-Maria Sintorn[42,43], Juan C. Caicedo[44], Beth A Cimini[44], Muyinatu A. Lediju Bell[45], Bruno M. Saraiva[46], Guillaume Jacquemet[47,48,49,50], Ricardo Henriques[46, 51], Wei Ouyang[52], Trang Le[53], Estibaliz Gómez-de-Mariscal[46], Daniel Sage[54], Arrate Muñoz-Barrutia[55,56], Ebba Josefson Lindqvist[57], Johanna Bergman[57]

**Guest Editors:**
Giovanni Volpe: giovanni.volpe@physics.gu.se
Carolina Wählby: carolina.wahlby@it.uu.se
Lei Tian: leitian@bu.edu

**Abstract**

Through digital imaging, microscopy has evolved from primarily being a means for visual observation of life at the micro- and nano-scale, to a quantitative tool with ever-increasing resolution and throughput. Artificial intelligence, deep neural networks, and machine learning are all niche terms describing computational methods that have gained a pivotal role in microscopy-based research over the past decade. This Roadmap is written collectively by prominent researchers and encompasses selected aspects of how machine learning is applied to microscopy image data, with the aim of gaining scientific knowledge by improved image quality, automated detection, segmentation, classification and tracking of objects, and efficient merging of information from multiple imaging modalities. We aim to give the reader an overview of the key developments and an understanding of possibilities and limitations of machine learning for microscopy. It will be of interest to a wide cross-disciplinary audience in the physical sciences and life sciences.

[40] Department of Clinical Neuroscience, Karolinska Institutet, Stockholm, Sweden
[41] Department of Information Technology, Uppsala University, Sweden
[42] Vironova AB, Sweden
[43] Department of Information Technology, Uppsala University, Sweden
[44] Broad Institute of MIT and Harvard, Boston (MA), USA
[45] Johns Hopkins University, United States
[46] Instituto Gulbenkian de Ciência, Oeiras, Portugal
[47] Turku Bioscience Centre, University of Turku and Åbo Akademi University, Turku, Finland
[48] Faculty of Science and Engineering, Cell Biology, Åbo Akademi University, Turku, Finland
[49] Turku Bioimaging, University of Turku and Åbo Akademi University, FI- 20520 Turku, Finland
[50] InFLAMES Research Flagship Center, Åbo Akademi University, FI- 20520, Turku, Finland
[51] MRC-Laboratory for Molecular Cell Biology, University College London, London, UK
[52] Science for Life Laboratory, Department of Applied Physics, KTH - Royal Institute of Technology, 171 65 Stockholm, Sweden
[53] Department of Bioengineering, Stanford University, Stanford, CA 94305, USA
[54] Biomedical Imaging Group and EPFL Center for Imaging, Ecole Polytechnique Fédérale de Lausanne (EPFL) Lausanne, Switzerland
[55] Bioengineering Department, Universidad Carlos III de Madrid, Leganés, Spain
[56] Instituto de Investigación Sanitaria Gregorio Marañón, Madrid, Spain
[57] AI Sweden, Gothenburg, Sweden

# 1 — Stability for inverse problems

Michael Hecht[1], Artur Yakimovich[1-4]

1. Center for Advanced Systems Understanding (CASUS), Helmholtz-Zentrum Dresden-Rossendorf e. V. (HZDR), Görlitz, Germany.
2. Bladder Infection and Immunity Group (BIIG), Department of Renal Medicine, Division of Medicine, University College London, Royal Free Hospital Campus, London, United Kingdom
3. Artificial Intelligence for Life Sciences CIC, Dorset, United Kingdom
4. Roche Pharma International Informatics, Roche Diagnostics GmbH, Mannheim, Germany

**Status**

The ability to experience the wonders of the microscopic world with one's own eyes has been fascinating researchers and enthusiasts for hundreds of years. This fascination, as well as the ability to explain the phenomena of the macroscopic scale through the occurrences in the micro-world, has led to the development of a plethora of techniques to visualize, probe, and reconstruct minute objects from the scale of the small animals to the scale of the atom (**Figure 1.1**). These techniques include a variety of functional (labeled by a molecular dye) and label-free light (optical) microscopy modalities, including brightfield, epi- and confocal fluorescence, and lightsheet microscopy (reviewed in Ref. [1]). Attempts to overcome the limitations of the light diffraction limit have led to the development of electron microscopy (EM) techniques like scanning and transmission EM (SEM and TEM, respectively), featuring relatively complex sample preparation steps. The necessity to minimize the sample preparation artifacts and visualize the biological entities in their native state has in turn led to the development of cryo-EM techniques [2]. While the contribution of EM to our understanding of the microworld is difficult to overstate, the sheer complexity of sample preparation and the expensiveness of the equipment has sparked in recent years the development of superresolution microscopy (SRM). Further notable and up-and-coming techniques include X-ray microscopy, live time-lapse microscopy, and holographic microscopy as well as atomic force microscopy (AFM) which uses interaction force to map the microworld.

Remarkably, one common facilitator of this cambric explosion of microscopic techniques, which occurred mostly in the past half a century, is digital microscopy. Originating from micrography, the departure from the necessity to project the microscopic image on the microscopist retina and the ability to capture and describe the images digitally has also has turned microscopy into a quantitative discipline. Beyond facilitating the recording and storage of microscopy data, digitalization allowed improved image processing, denoising, and direct pattern recognition. This, in turn, paved the way for computer vision (CV) algorithms, including machine learning (ML) and deep learning (DL), to facilitate further advances in microscopy.

**Current and Future Challenges**

Given the immense diversity of the microscopic techniques (**Figure 1.1**), it becomes obvious that the microscopy datasets are incredibly domain-specific. This represents a significant challenge for ML and DL efforts, as out-of-domain inference is far from trivial for the vast majority of CV algorithms. This is especially pronounced with quite different modalities, for example, EM and confocal fluorescence microscopy. Furthermore, conventional image augmentation approaches that work well for ImageNet work very poorly for microscopy datasets. Vendor-specific data formats are certainly not facilitating harmonization and transferability of datasets. Very often, models trained on images obtained using hardware of a specific vendor simply don't generalize to other vendors. Finally, while microscopy image data is gradually becoming available, high-quality annotations, especially those with a high level of consensus are still problematic to obtain. All these challenges are positioning ML and DL strategies for microscopy data into a low-data regime, dictating the choice of algorithms available to researchers.

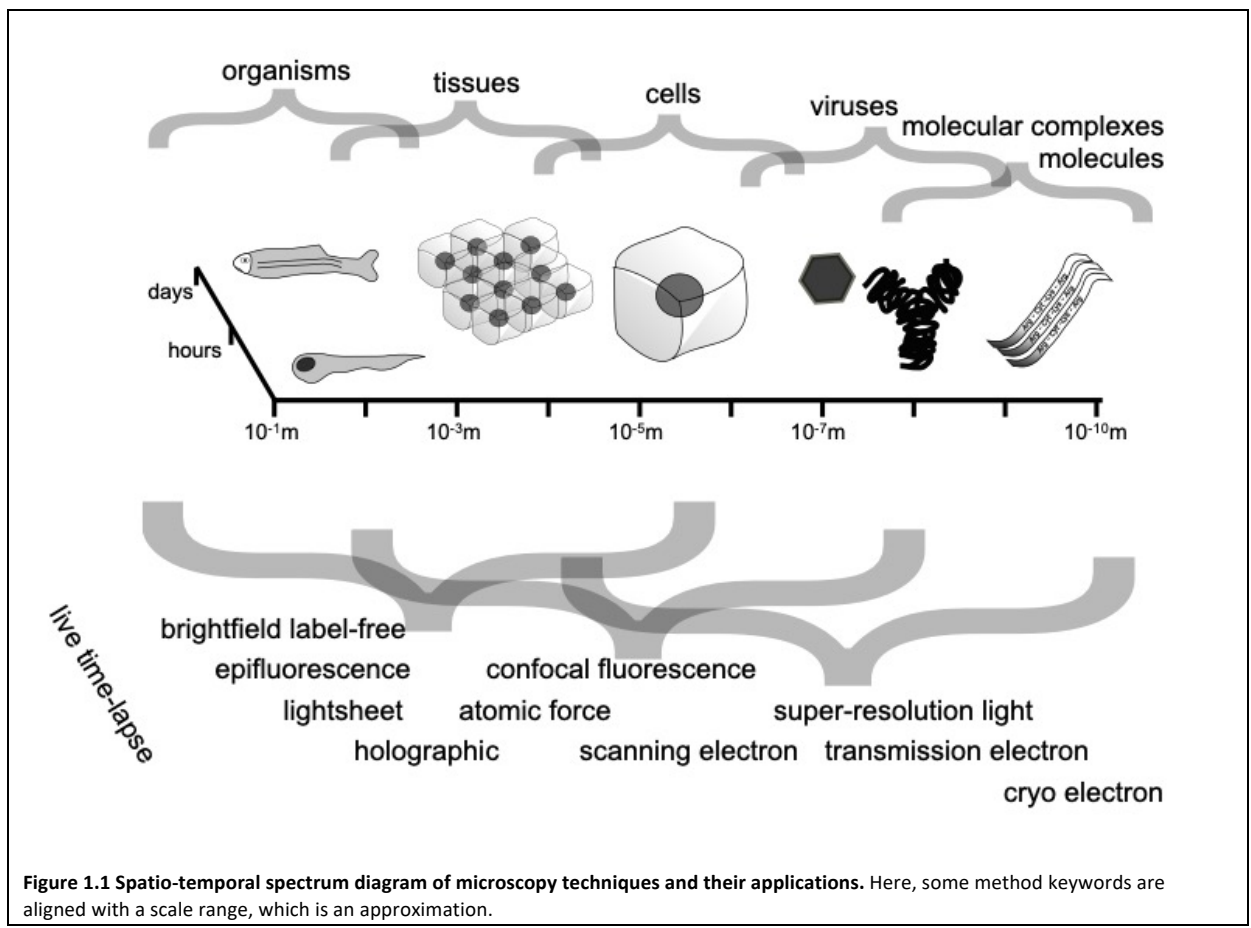

**Figure 1.1 Spatio-temporal spectrum diagram of microscopy techniques and their applications.** Here, some method keywords are aligned with a scale range, which is an approximation.

**Advances in Science and Technology to Meet Challenges**

Recent advances in ML methodology for microscopy have decisively demonstrated the ability of the established ML/DL algorithms to meet the pre-described challenges or their combinations [3], and provide solutions for image processing and analysis. Thus far, this has allowed to successfully address microscopy ML tasks, such as image reconstruction and superresolution, classification and generation, denoising, segmentation, cell tracking, feature selections [3–7]. However, the gravitation of the established algorithms towards the abovementioned low-data regime motivated Hansen et al. [9] to investigate the techniques by asking the questions: "How reliable are such algorithms when applied in the sciences?" and "do AI-based algorithms have an unavoidable Achilles heel: instability?"

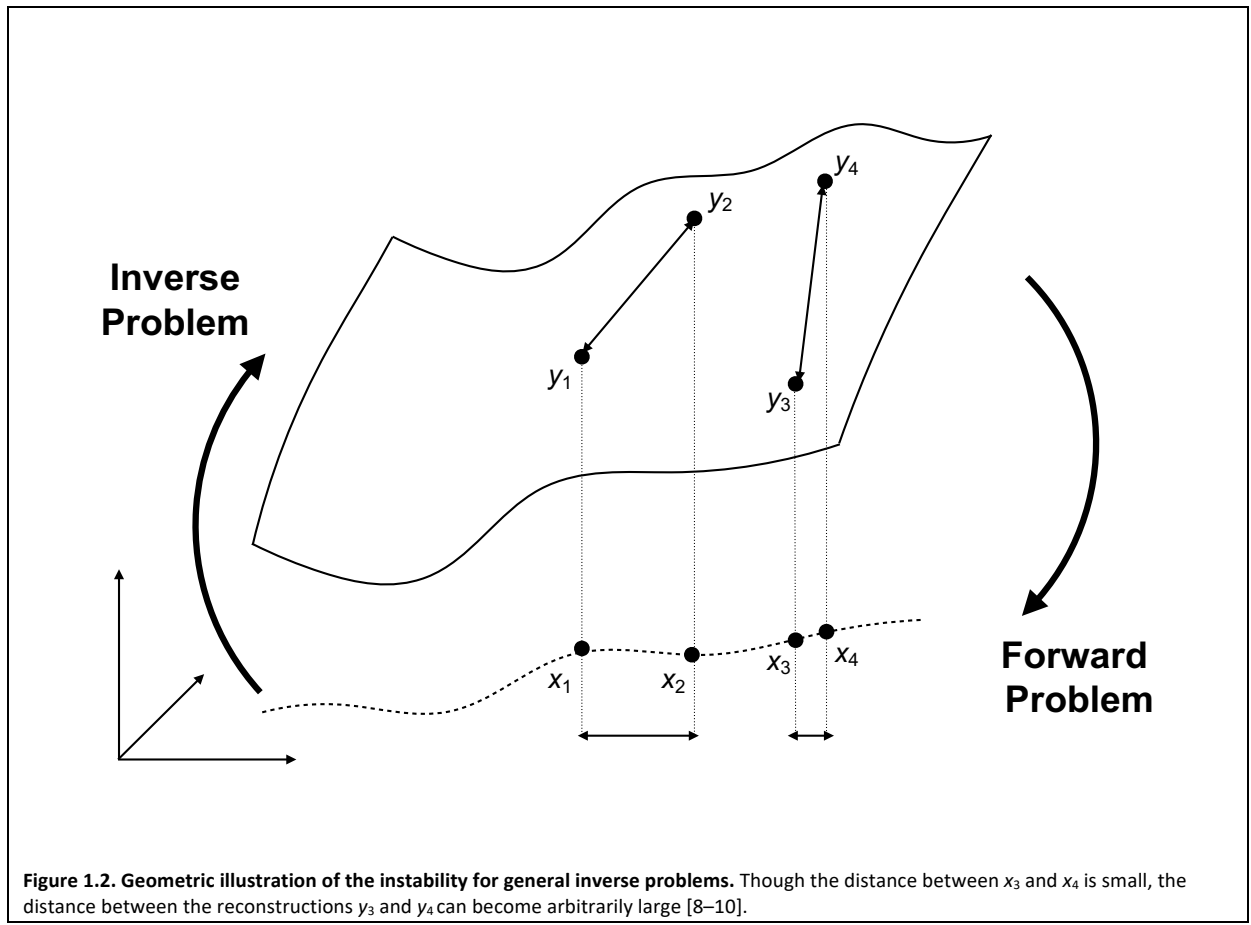

**Figure 1.2. Geometric illustration of the instability for general inverse problems.** Though the distance between $x_3$ and $x_4$ is small, the distance between the reconstructions $y_3$ and $y_4$ can become arbitrarily large [8–10].

In mathematical terms, the mentioned tasks seek solutions of non-linear, inverse problems. However, the *Universal Instability Theorem* [8–10] proves general inverse problems to be inherently unstable. The issue is illustrated in **Figure 1.2**. Here, the forward model is given by x-projection of a 2D-curved plane onto a curved line. The loss of information in the forward direction (null space) can cause reconstruction distances of arbitrary close instances to become arbitrarily large. In other words: small noise perturbations of an observable *x* can cause a huge change in the reconstruction of *y*.

While stable solutions to linear inverse problems can be given by classic principal component analysis (PCA), the non-linearity present here demands the development of novel strategies resisting the instability phenomenon. Recently, approaches addressing the robustness for classification and denoising tasks proposed adversarial re-training, generative-adversarial-network-based denoising, and explicit kernel (null space) control to prevent this issue. These results were discussed and complemented by novel regression techniques in Ref. [10]. Furthermore, the current state of research suggests that, when following and incorporating the mathematical insights and delivered ML techniques, stability can be ensured for a general instance class of inverse problems relevant here. Altogether, this suggests that employing novel more stable ML/DL algorithms may help avoid the inherent instability in ML/DL models for microscopy. Additionally, these approaches may facilitate generative algorithms alleviating low-data regimes and improving generalization.

**Concluding Remarks**

Recent advances in microscopy techniques and ML algorithms go hand in hand in powering a new generation of biomedical imaging methods. However, data scarcity and domain-specificity of microscopic datasets represent a major obstacle to this synergy. Furthermore, we identified a bottleneck for providing the strong reliability requested for scientific reasoning. That is the necessity of ensuring the stability of the ML methods addressing the (inverse) problems occurring for microscopy image processing tasks. While enriching the low-data regime may suppress the occurrence of instabilities, we pointed to the objectives hampering this strategy that are omnipresent in practice. Being aware of the mathematical limitations, however, provides an exit strategy from the dilemma by incorporating techniques that deliver the needed resistance to the instability phenomenon in practice.

## Acknowledgments

*This work was partially funded by the Center of Advanced Systems Understanding (CASUS), financed by Germany's Federal Ministry of Education and Research (BMBF) and by the Saxon Ministry for Science, Culture and Tourism (SMWK) with tax funds on the basis of the budget approved by the Saxon State Parliament.*

## 2 — Physics-based Learning

Kristina Monakhova[1], Laura Waller[1]

1. Department of Electrical Engineering and Computer Sciences, UC Berkeley, USA

**Status**

Computational imaging focuses on the co-design of imaging optics and algorithms to create better and more capable imagers that can see more than just 2D images. In astronomy, this co-design enabled the first pictures of a black hole. In photography, computational imaging allows us to take high-dynamic-range (HDR) and portrait-mode photos with an extremely compact camera. Here, we focus on microscopy, where computational imaging has been used for super-resolution, single-shot 3D, and phase microscopy, with great potential to push new scientific discovery by allowing us to observe smaller, faster, invisible things in more dimensions.

Both the optical design and the algorithm design are critical for computational microscopes. Over the years, optical design has largely been based on heuristics, such as lens sharpness and hand-designed metrics for performance, which were not necessarily optimal given the reconstruction algorithms or imaging task. Similarly, the algorithms have largely been based on optimization approaches consisting of a data-fidelity term and hand-picked priors. These algorithms are often slow, taking many iterations to converge, relied on priors that were not necessarily optimal for the application, and could suffer from model-mismatch given any errors in the optical model. More recently, deep learning-based approaches have been introduced, which can more tightly couple the optical design with the algorithms [1] and improve algorithms through a data-driven approach [2].

Deep-learning-based reconstruction algorithms leverage trainable neural networks and large datasets to learn better ways of solving imaging inverse problems. These methods have shown great promise for speeding up imaging inverse problems by multiple orders of magnitude, improving image quality [3], and providing better priors for underdetermined problems, such as compressive single-shot 3D microscopy [4]. By using differentiable physics-based optical models, the optical design can be represented as the first layer of the neural network, and optimized end-to-end with the reconstruction algorithm [1,5] (**Figure 2.1**). This breaks the paradigm of traditional optical design, and opens the door for a new era of optical design where each element in a microscope is tailored specifically for a given algorithm and higher-level task, rather than for a sharpness metric.

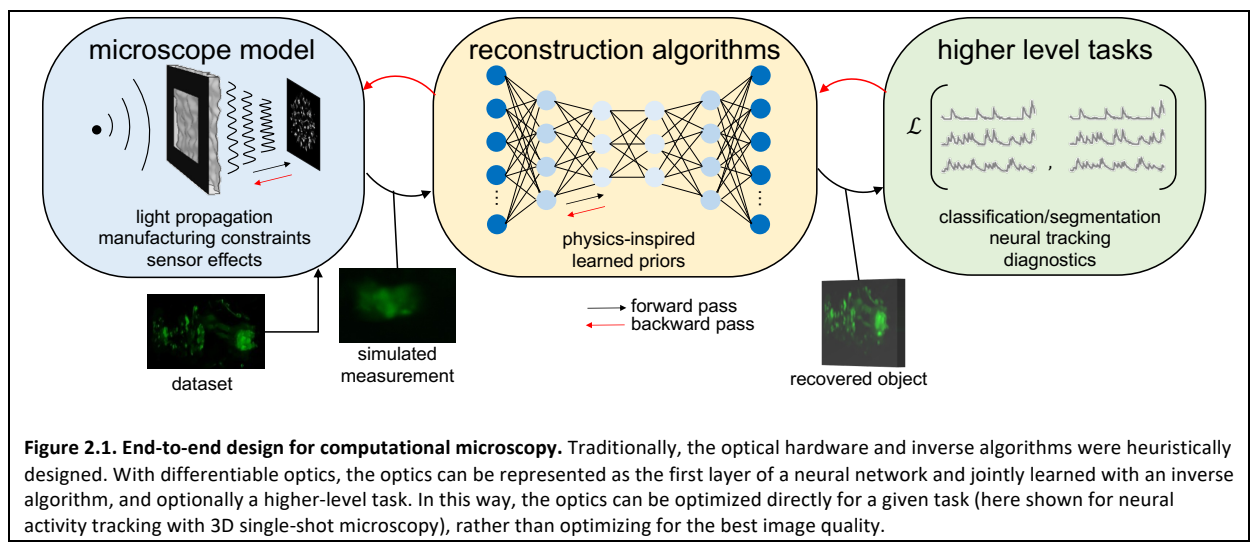

**Figure 2.1. End-to-end design for computational microscopy.** Traditionally, the optical hardware and inverse algorithms were heuristically designed. With differentiable optics, the optics can be represented as the first layer of a neural network and jointly learned with an inverse algorithm, and optionally a higher-level task. In this way, the optics can be optimized directly for a given task (here shown for neural activity tracking with 3D single-shot microscopy), rather than optimizing for the best image quality.

**Current and Future Challenges**

Using deep-learning-based techniques for computational microscopy often necessitates large, custom datasets during the training phase. For example, in order to learn the best optical "encoder" and "decoder" for hyperspectral microscopy, there is a need for a large dataset of high-resolution hyperspectral images to use during training. For certain imaging modalities or tasks, such datasets do not exist, are too small, have insufficient resolution, or may be infeasible to acquire. For other areas of machine learning, such as self-driving vehicle research, high-quality physics-based simulators have shown great success in synthesizing realistic sensor data. Similarly, generating accurate physics-based simulators for microscopy (e.g., modeling complex light interactions, multiple-scatting, and complex biological samples) could enable computational microscopes to be trained using mostly synthetic datasets, eliminating the need for real experimental datasets. Alternatively, unsupervised learning approaches have the potential to leverage the structure of neural networks, but without needing training data [6,7].

Combining domain knowledge of optical physics with deep learning has the potential to improve the interpretability and performance of deep-learning-based reconstructions [3,5,8]. Off-the-shelf networks, such as convolutional neural networks (CNNs), have no knowledge of optical physics and must therefore learn this information from scratch when used to solve an inverse problem, leading to the need for large datasets and lengthy training times. Physics-based networks, in contrast, use differentiable physical models to incorporate optical domain knowledge and create physics-informed networks that are more efficient. These physics-informed networks can also be used to calibrate optical systems, such as by learning how to synthesize and represent realistic camera noise [8], or potentially by learning other effects such as aberrations and non-linearities. This could be useful for synthesizing realistic datasets for a given optical system, or perhaps could be incorporated into the reconstruction pipeline as more accurate, nonlinear forward models.

Finally, computational imaging approaches excel in creating task-specific microscopes that are optimized for a specific higher-level task (e.g., cell counting, disease diagnosis). For many scientific and clinical applications, a high-resolution image is not needed to make a decision — there may be certain features within that image (e.g., polarization, wavelengths, phase) that are more important than others. Through end-to-end design, where the optics are learned together with the reconstruction and higher-level task, there is the potential to create better, faster, smaller, and more capable microscopes that are specially tailored to make decisions, or extract the most useful information from the world.

**Advances in Science and Technology to Meet Challenges**

Despite the promise of deep learning for computational imaging, major challenges include robustness and guarantees. A microscope that you cannot trust is a microscope that is unsuitable for scientific and clinical applications. For machine-learning-based reconstruction algorithms, knowing when the algorithm works and when you can trust the reconstruction is difficult. The structure of the network and the training data used can impact the solution and potentially introduce artifacts that are indistinguishable from signal. Research in deep learning theory on uncertainty quantification [9] and robustness has the potential to resolve some of these challenges. Furthermore, fundamental research in signal processing applied to deep learning could bring some of the mathematical guarantees from classic signal processing, such as compressive sensing, to the realm of deep learning [10]. Building in guarantees and quantifying uncertainty will be pivotal in the broad adoption of machine-learning-based algorithms for scientific and medical computational microscopes.

**Concluding Remarks**

Using deep learning-based methods to learn better optical designs and algorithms is fundamentally changing the way we design microscopes, helping push the limits of what we can observe, while potentially delivering devices that are smaller, cheaper, and more compact. Although this avenue has many exciting possibilities, there exist a number of scientific challenges that need to be addressed. Open problems include building in interpretability into learned optical designs and algorithms, robustness, quantifying uncertainty, and the need for high quality datasets. Building in domain-specific knowledge and known physics into neural networks has promise in addressing some of these challenges. Moving forward, interdisciplinary collaborative research between machine learning theorists, optical physicists, and microscopists has the potential to further address these challenges and bring in a new era of more capable, machine-learning-designed computational microscopes.

**Acknowledgements**

*This work was supported by the David and Lucile Packard Foundation and the Gordon and Betty Moore Foundation Data-Driven Discovery Initiative Grant GBMF4562.*

## 3 — Learning interpretable physical models from large 3D images

Ivo F. Sbalzarini[1,2]

1. Technische Universität Dresden, Faculty of Computer Science, Dresden, Germany
2. Max Planck Institute of Molecular Cell Biology and Genetics, Center for Systems Biology Dresden, Dresden, Germany

**Status**

With the advent of volumetric microscopy modalities in biology, such as light-sheet microscopy, it became possible to image entire developing tissues and embryos at sub-cellular resolution over the full time-course of development [1]. This accelerated developmental biology and our understanding of how cells form tissues by providing us with a direct means of observation. However, it also created a new problem: handling the terabytes of image data these microscopes produce at rates of up to several gigabytes per second. One solution is to visualize the time-lapse images in real time, as they are being acquired, and only record and analyze the "interesting" bits. Since the images are 3D, this is best done using Virtual Reality (VR) displays. A performant open-source platform for that is *Scenery* [2], pioneering the era of immersive volumetric microscopy. Providing rendering performance of several giga-voxels per second, *Scenery* has made VR microscopy real-time on commodity hardware, even for the fastest microscopes. This has enabled developmental biologists and neuroscientists to gain a better intuition of the complex space-time organization of dynamic tissues.

Yet, one would like to formalize this intuition in predictive mathematical models that are, ideally, physically consistent. It has been shown that governing equations for the space-time dynamics of fluorescently labeled molecules can be algorithmically inferred from microscopy videos with sufficient robustness against imaging noise [3] and guaranteed physical consistency [4]. So far, however, it is not feasible to apply these ideas on the large 3D images produced by state-of-the-art volumetric imaging studies. A promising approach is to represent the raw images on some better-suited data structure than a regular grid of pixels. Approaches such as the Adaptive Particle Representation (APR) of images [5], for example, improve the information-to-data ratio of the images by adaptively re-sampling them, storing intensity values only at points where they contribute information to the image (**Figure 3.1**). While the concept of adaptive sampling is classic in signal processing, the APR provides unprecedented accuracy and optimality guarantees. Moreover, it can directly be used for downstream image processing [5,6], enabling end-to-end APR-native pipelines for images up to the petabyte scale at more than a terabyte per second on consumer GPUs [6].

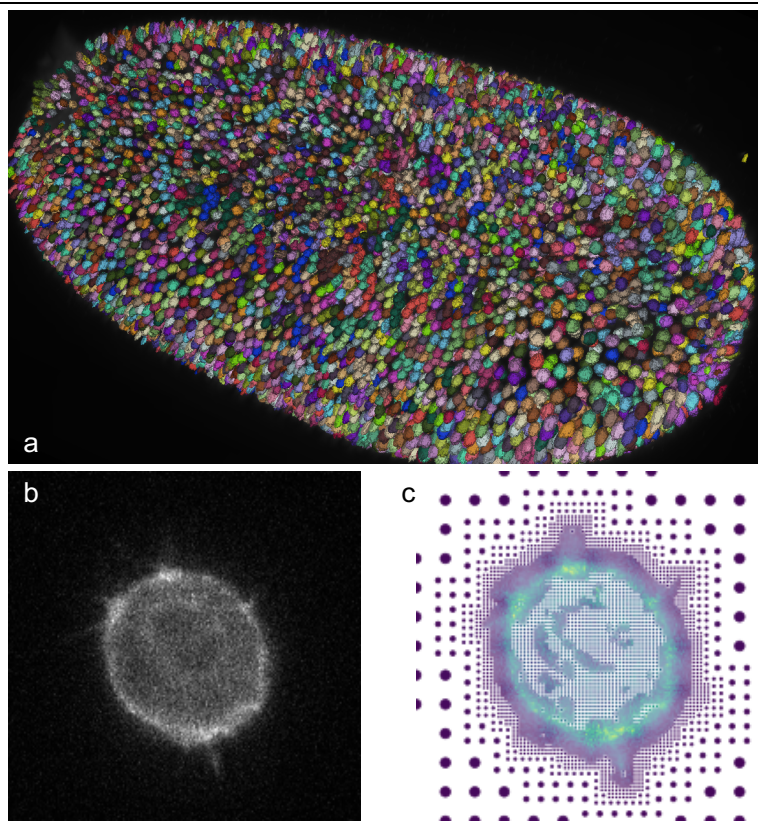

**Figure 3.1. The Adaptive Particle Representation (APR).** The APR of images [5] provides an information-optimal sampling of an image at a fraction of the computational cost. (**a**) An embryo with cells segmented (original image: data set "Fluo-N3DL-TRIF" from the publicly available ISBI cell tracking challenge; segmentation and visualization: Joel Jonsson, Sbalzarini lab). (**b**) A single A549 cell from the cell tracking challenge. (**c**) APR particles (shown as individual dots) of the cell from (b) with color encoding the intensity signal (APR and visualization by Joel Jonsson, Sbalzarini lab).

**Current and Future Challenges**

Current microscopy datasets are chronically under-used in large bio-imaging studies. With sample preparation and image acquisition no longer the bottleneck, analyzing the images and turning them into scientific insight and knowledge became the rate-limiting step. We envision a future where volumetric microscopy data sets can be immersively experienced in VR environments, from within which the user can interact with the sample in real time using hand gestures to, e.g., apply optogenetic manipulations or laser ablation and thereby train machine learning modules running in-the-loop. This brings within reach human-augmented algorithmic inference of the physical laws and processes that characterize the fascinating properties of self-organized living matter. While the bits and pieces for making this vision reality are becoming available, combining them remains a challenge.

It is, for example, challenging to combine the APR [5] with real-time VR rendering, because the performance of rendering algorithms on GPUs deteriorates with irregular adaptive data structures. The APR [5] and APR-based image processing [6] enable large imaging studies at a fraction of the computational cost (both storage and processing time) with guaranteed optimality. *Scenery* [2] enables real-time immersive visualization of voxel images in virtual and augmented reality, but it is not yet able to do so for APRs. Another example is that, while learning interpretable mathematical models of the physical processes that pattern and shape tissues is becoming possible [3,4], this does not currently scale to large data.

Successfully scaling to large data, emerging deep-learning architectures have been adapted to biological imaging, including the use of graph neural networks to accurately estimate physical properties of biological systems directly from microscopy videos [7]. Another example is the Open-Cell project, which developed the self-supervised variational autoencoder network *Cytoself* to infer models of intra-cellular mesoscale organization from microscopy images [8], inspiring thoughts of a sequence-to-image generative model of cells. The physical and biochemical interpretation of these deep-learning models, however, remains a challenge, which could potentially be addressed by combining them with mathematical model inference [3,4]. This could include endowing them with a notion of spatial interaction [9]. Currently, deep neural networks operate on the morphological appearance of objects in the image, but not on their global spatial arrangement with respect to each other and with respect to reference structures. Now-classic concepts from spatial statistics [9] could be recast in a data-driven framework to enable the next leap in understanding.

**Advances in Science and Technology to Meet Challenges**

Both scientific and technological advancements are still needed to realize the vision of integrating immersive visualization, information-adaptive image representation, and machine learning to infer interpretable physical models from large-scale 3D image data. Obvious missing pieces are the ability to immersively render information-adaptive image representations, such as the APR [5], in VR environments, and to develop Convolutional Neural Networks (CNNs) that directly operate on such adaptive image representations without the need of intermediately going back to pixels. Finally, this needs to be linked up with algorithms that extract physical fields, such as velocity fields of flows and deformations, from the images and use those to infer governing mathematical equations that are simple enough to be physically interpreted, but sufficient to explain the dynamics observed in the video.

This link is likely going to come from modern machine-learning architectures, such as transformer networks or graph neural networks. Ideally, again, directly on information-adaptive image representations, such as the APR [5], in order to save time and electric power. Since the APR represents an image in terms of particles that are concentrated around objects in the image, it naturally lends itself to a graph neural network approach. When combined with approaches like MAGIK [7], this could enable direct estimation of macroscopic physical quantities from very large 3D images. Those estimated quantities can then be used to infer mathematical equation models of the observed space-time dynamics [3,4], closing the loop back to the biological mechanism.

This inference loop potentially enables experiments in which scientists use hand gestures in a VR environment to interactively ablate a tissue in the sample, are able to directly observe the recoil and tissue rearrangement this causes, and within seconds get the estimated elastic constants of the tissue as well as a simulation results overlaid with the image (**Figure 3.2**). Further interacting with the sample, or manually correcting estimation mistakes, the machine-learning or simulation model is re-trained in an active learning loop. An interesting question is how to visualize uncertainty of the machine-learning output to the users in order to focus their attention and prompt for additional perturbations that carry significant information. Finally, bringing these exciting advances in machine learning and bio-imaging into everyday laboratory usage also requires user-friendly, easily deployable, and robust software implementations with support for cloud computing.

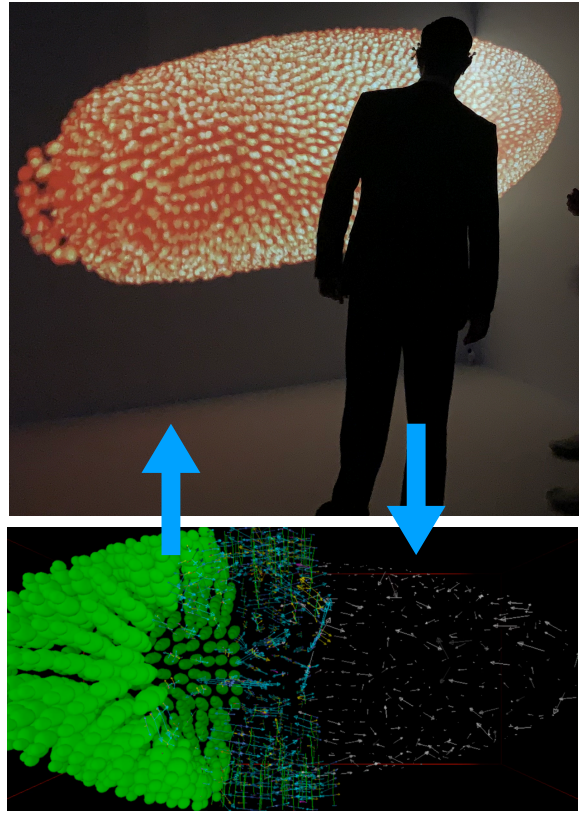

**Figure 3.2. Vision of human-AI co-working in a virtual-reality space.** The top panel shows a scientist interacting with a fruit fly embryo in the microscope in 3D Virtual Reality and in real time (photo credit: Ulrik Günther, CASUS & Sbalzarini lab). The bottom panel shows output (synthetic image) of a simulation model [10] of the cells, the forces acting on them, and their velocity fields (image with permission by Vladimir Ulman, Technical University of Ostrava).

## Concluding Remarks

Exciting times for biological live-cell microscopy! We are witnessing accelerated progress in the molecular markers and sensors available, the optics of the microscopes, and our ability to use light to perform quantitative measurements in living, growing, and deforming organs and organisms. The challenge is to ensure that the computational tools for handling, storing, processing, and visualizing the images progress on par, and that we work towards making better use of the information contained in the images, e.g., by providing machine-learning solutions that increase the throughput at which physically interpretable, mechanistic models can be inferred from the image data. This ideally goes directly from images or videos to equations, models, and knowledge, bypassing the so-far common intermediate steps of instance segmentation and object tracking. This will not only make large bio-imaging projects more insightful, but also potentially more accurate and less resource (energy) demanding. Ultimately, a new ecosystem of biological imaging arises, combining approaches from VR, AI, and signal theory that could become common laboratory methods in a couple of years from now, just like single-particle tracking has become a commodity over the past years.

## Acknowledgements

*We thank Vladimir Ulman (Technical University of Ostrava) for kindly providing the bottom panel of Figure 3.2. Particular thanks also to Bevan Cheeseman, Joel Jonsson, Suryanarayana Maddu, and Ulrik Günther, all from the Sbalzarini lab, whose pioneering works have demonstrated the feasibility of the roadmap outlined here.*

# 4 — Plug-and-Play Learning-Based Computational Imaging

Christopher A. Metzler[1], Mingyang Xie[1], Kevin Zhang[1]

1. University of Maryland, College Park, USA

**Status**

The performance of any computational imaging system is determined, in part, by the performance of its reconstruction algorithm. The role of a computational imaging reconstruction algorithm is to reconstruct a signal/image of interest $x$ from measurements $y$ of the form

$$y = A(x),$$

where $A(x)$ represents the system's forward model. This mathematical model is general and can represent almost any measurement process, from sampling K-space in magnetic resonance imaging (MRI) to capturing band-pass filtered images in Fourier Ptychographic microscopes.

Classically, computational imaging reconstruction has been performed by framing imaging as an optimization problem

$$\hat{x} = \mathrm{argmin}_x f(x) + r(x),$$

where $f(x)$ is a data fidelity term, which ensures $x$ is consistent with the measurements $y$, and $r(x)$ is a regularization penalty, which ensures $x$ is consistent with our prior beliefs on how $x$ should be structured. In the special case that $f(x) = -\ln p(y|x)$ and $r(x) = -\ln p(x)$, solving the above optimization problem provides a maximum a posteriori (MAP) estimate of $x$.

One can minimize the objective, $f(x) + r(x)$, using any number of algorithms, such as proximal gradient descent:

$$v^{t+1} = x^t - \nabla_x f(x^t),$$

$$x^{t+1} = \mathrm{argmin}_x \|x - v^{t+1}\|^2 + r(x).$$

The first line above takes a gradient step to minimize $f(x)$ while the latter, which is known as a proximal mapping, reduces $r(x)$.

Classical iterative algorithms such like proximal gradient descent offer several benefits: They are easy to interpret and, by changing the data fidelity term $f(x)$, can easily incorporate domain expertise about the measurement process $A(\cdot)$. Unfortunately, the performance of classical algorithms falls well behind that of purely deep-learning-based methods, which, given a vast training set of paired examples $\{(x_l, y_l)\}_{l=1}^{L}$, can teach a neural network to effectively map measurements $y$ to images $x$. This performance generally comes at the cost of flexibility — a network is trained and specialized for a specific measurement process $A(\cdot)$, and if the measurement process changes the network is useless.

Plug-and-play (PnP) optimization is a hybrid reconstruction framework which allows one to maintain the interpretability and flexibility of classical algorithms while taking full advantage of powerful data-driven priors [1]. The key idea behind PnP optimization is that one can interpret an off-the-shelf (learning-based) image denoiser $D(z)$ as a solution to

$$D(z) = \mathrm{argmin}_x \|x - z\|^2 + r(x),$$

for some implicit, data-driven prior $r(x)$. This interpretation allows one to "plug-in" the denoiser into an existing iterative algorithm. For example, PnP proximal gradient descent can be written

$$v^{t+1} = x^t - \nabla_x f(x^t),$$

$$x^{t+1} = D(v^{t+1}).$$

This algorithm follows the same form as standard proximal gradient descent, but with the proximal mapping step replaced by a powerful image denoising algorithm.

PnP algorithms stand apart from purely data-driven and purely classical computational imaging reconstruction techniques in their unique ability to combine (1) domain expertise, in the form of the forward measurement operator $A(\cdot)$, and (2) data-driven-priors on the distribution of the dataset, in the form of an image denoiser trained using a vast set of training examples $\{x_l\}_{l=1}^{L}$ (**Figure 4.1**). PnP algorithms represent the current state-of-the-art in computational imaging reconstruction [2].

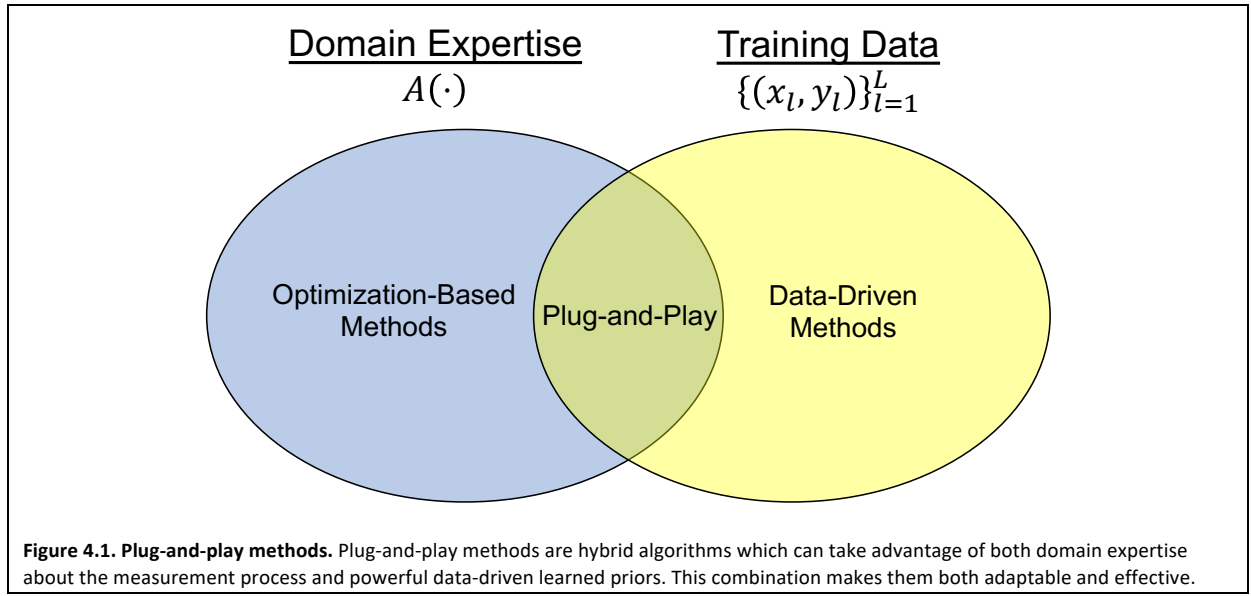

**Figure 4.1. Plug-and-play methods.** Plug-and-play methods are hybrid algorithms which can take advantage of both domain expertise about the measurement process and powerful data-driven learned priors. This combination makes them both adaptable and effective.

**Current and Future Challenges**
PnP algorithms face several hurdles which can make them challenging to apply in practice.

*Forward model mismatch*: Like almost any other reconstruction method, PnP algorithms rely on accurate knowledge of the forward measurement model $A(\cdot)$. If this forward model is mis-specified severe artifacts can appear in the reconstructions.

*Denoiser model mismatch*: PnP algorithms typically use off-the-shelf denoising algorithms, which often assume the noise they are removing follows an additive white Gaussian distribution. In general, however, the "effective noise", i.e., the difference between the true signal and the intermediate solution fed into the denoiser, is neither Gaussian nor white, making existing denoising algorithms suboptimal. In addition, denoisers are trained on a specific class of data (e.g., natural images) and their performance can suffer when applied to other distributions (e.g., MRI).

*Parameter tuning*: Most PnP algorithms have multiple hyperparameters (step sizes, regularization strengths, etc.). These parameters can vary iteration to iteration and setting them correctly is critical to getting the best performance. One solution to the parameter tuning problem is to "unroll" an iterative PnP algorithm into a feedforward neural network (**Figure 4.2**), whose parameters can be automatically tuned with training data and backpropagation [3]. However, unrolling gives up flexibility, the network is now specific to a particular forward operator $A(\cdot)$, and comes with its own set of limitations.

*Memory usage*: To train an unrolled algorithm, one needs to back-propagate errors through multiple copies of the denoiser, which requires storing many intermediate variables. When dealing with high-dimensional data, e.g., time-varying volumetric data, the memory costs associated with storing all these intermediate variables can become prohibitive.

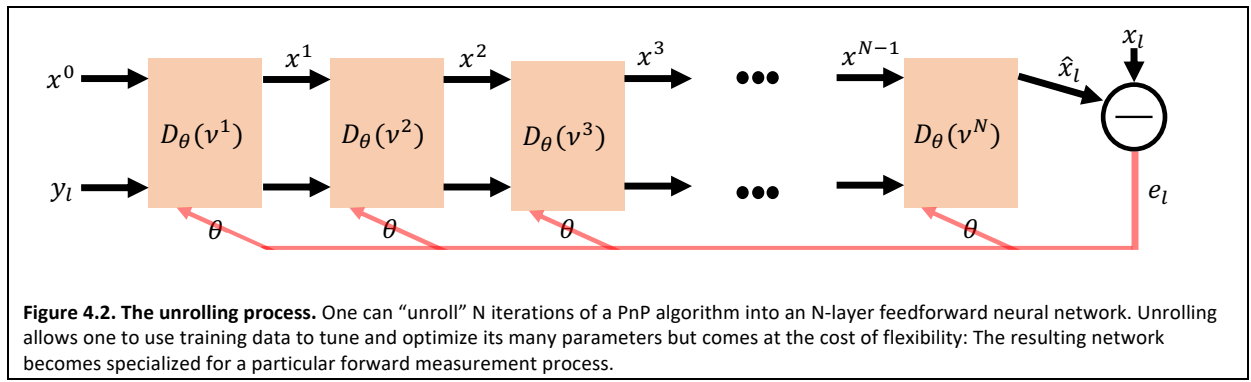

**Figure 4.2. The unrolling process.** One can "unroll" N iterations of a PnP algorithm into an N-layer feedforward neural network. Unrolling allows one to use training data to tune and optimize its many parameters but comes at the cost of flexibility: The resulting network becomes specialized for a particular forward measurement process.

**Advances in Science and Technology to Meet Challenges**

A host of solutions have been put forward to address the above challenges.

*Forward model mismatch*: If one is dealing with a parametric forward model $A_\gamma(\cdot)$, one can jointly recover the target image $x$ and the forward model parameters $\gamma$ through alternating minimization. Specifically, in addition to the usual PnP steps, one can also minimize the objective function with respect to the forward model parameters. Variations on this idea have been applied successfully to X-ray computed tomography [4], MRI [5], and holography [6].

*Denoiser model mismatch*: Denoiser model mismatch comes in the form of mismatched noise distributions and mismatched data distributions. The former problem can be addressed through unrolling [3] or by careful characterization of the per iteration noise distribution [7]. The latter problem is largely still open, though denoisers tend to be relatively robust to this mismatch in practice.

*Parameter tuning*: While unrolling is the most straightforward way to perform parameter tuning [3], unrolled algorithms largely (though not entirely [5]) give up the ability to adapt to new measurement operators. Alternative parameter tuning methods, such as reinforcement learning [8], have recently been put forward to avoid this trade-off.

*Memory usage*: One can reduce the massive memory costs associated with training an unrolled algorithm by utilizing invertible network architectures, which allow one to recompute intermediate activation from a network's output [9]. Alternatively, one can improve memory usage by relying upon stochastic approximations of the data fidelity update step [10].

**Concluding Remarks**

PnP algorithms provide a convenient framework to combine the adaptability and interpretability of classical optimization-based computational imaging reconstruction methods with the powerful priors and efficiency of deep learning. Here, we have highlighted some of the recent developments in PnP reconstruction and have described some of the pitfalls to avoid when applying learning-based PnP algorithms.

**Acknowledgements**

*This work was supported by the AFOSR Young Investigator Program Award FA9550-22-1-0208.*

# 5 — Accelerating simulations of optical forces

Isaac CD Lenton[1,2], Halina Rubinsztein-Dunlop[2,3]

1. Institute of Science and Technology Austria (ISTA), Klosterneuburg 3400, Austria
2. School of Mathematics and Physics, The University of Queensland, Brisbane 4072, Australia
3. Australian Research Council Centre of Excellence for Engineered Quantum Systems, The University of Queensland, Brisbane 4072, Australia

**Status**

Optical tweezers (OT) are a powerful tool for manipulating and handling microscopic samples, whether in liquids, in gases, or in vacuum [1]. Since their invention by Arthur Ashkin [2], OT have developed as a robust tool that is routinely employed in multiple disciplines [3]. OT use the optical force of light to trap and manipulate objects; in their simplest form OT can be thought of as an *optical harmonic spring*, which, under the right circumstances, can allow a particle to be held in all three dimensions and moved around by simply moving the optical beam focus. The strength of the spring is proportional to the beam's optical power. When the optical trap is well characterised, i.e., the stiffness of the spring is well known, optical tweezers become a powerful method for measuring piconewton scale forces, nanometre displacements, and torques by directly monitoring how the particle moves while held in the trap.

Simulations and models of OT have often aided their development, and can be useful for verifying and understanding experimental observations [4]. For instance, when trying to understand the observed dynamics of an optically trapped motile bacterium, a thorough understanding of the optical forces can help to shed light on the non-optical forces involved in the bacterium's motility. Depending on the size and shape of the particle, different descriptions of the trap are often needed, ranging from simple methods analogous to one or more harmonic springs, through to methods solving Maxwell's equations. The main limitation has often been the ability to use these models to make predictions about experiments. This can be either due to computational constraints, when advances in computing power are needed to evaluate models [5] or due to how easy-to-use and well-documented available codes are. Complex models often have many parameters and verifying that these models accurately describe a given experiment can be difficult.

Machine learning (ML), which encompasses a range of techniques including deep learning and computer-enabled statistical inference, has seen a huge growth in popularity in recent years. Aided by a rise in computational power and simultaneously an increase in the number of easily accessible algorithms, ML has been adopted into many fields. In the past few years, we have seen ML techniques applied to optical tweezers related problems, including modelling optical forces [6, 7] and extracting relevant information from experiments [8, 9]. These early works are very promising and suggest ML could significantly accelerate OT simulations, while simultaneously making accurate models more accessible for researchers wanting to model and understand their experiments.

**Current and Future Challenges**

Multiple challenges limit the application of accurate OT models including available computational power, verifying models accurately represent reality, and accessibility of models to researchers.

**Accurately modelling large, complex particles, and structured light fields** can be difficult. Although computers have significantly improved in speed and memory capacity (extending the range of cases that can be modelled), for many researchers it is still difficult to model even large spherical particles using available hardware[1], let alone the more complex geometries of bacteria or deformable cells often studied using OT. To simulate the dynamics of these particles in OT, the simulations must be sufficiently fast. One solution to modelling large and complex particles/beams is to pre-compute the forces using a more powerful computer or computer cluster. Interpolation can then be used to approximate the forces at intermediate points not included in the data set. However, this simply transfers the problem from being a processor-limited problem, to a memory/storage problem. An alternative option is to use ML to encode all this information in a much smaller representation. One of the main challenges in using interpolation or this kind of ML is generating sufficient training data spanning a representative range of parameters (beam shape, particle size, etc.).

One of the main difficulties with developing computational models for OT is **determining whether these models accurately depict reality**. Not all models work well on different size scales: for instance, the dipole approximation can be very effective for dipole-sized particles but give unreliable results for larger particles; similarly, geometric optics can be great for large particles, but produce erroneous results for small particles. Sometimes it can be necessary to combine multiple models, and care must be taken for when a particular model should be used/is valid. Often it is up to the user to assess if the model is suitable/accurate for the particular experiment they are trying to model. Care must be taken when using both traditional models and ML enabled models that the model predictions accurately depict reality.

The final challenge is **usability and accessibility**. There are many excellent models for optically trapped particles, however many of them are not the easiest to use or access. In some cases, users need specific operating systems or familiarity with certain programming languages to use/run codes (some of which may be proprietary). Once the user has a running model, they need to learn how to use it and assess its validity. For example, the discrete dipole approximation is extremely useful in modelling a range of particles but the choice of voxel placement, size, and spacing can have huge effects on the simulation results and care must be taken to choose the appropriate parameters and verify results. This requires good documentation including good release notes highlighting differences between versions of the code. This extra work sometimes creates a barrier between researchers who develop codes and users wanting to apply these codes to their experiments. Further still, in some cases researchers are hesitant to release code in the fear that it might be used incorrectly.

---

[1] We were only able to simulate particles up to about 100 μm and 1 mm with implementations of generalized Lorentz-Mie theory and geometric optics via the optical tweezers toolbox and optical tweezers simulations software packages. We only tested if these methods would run in a feasible amount of time for these sizes. No checks were made to see if the results for these sizes were accurate.

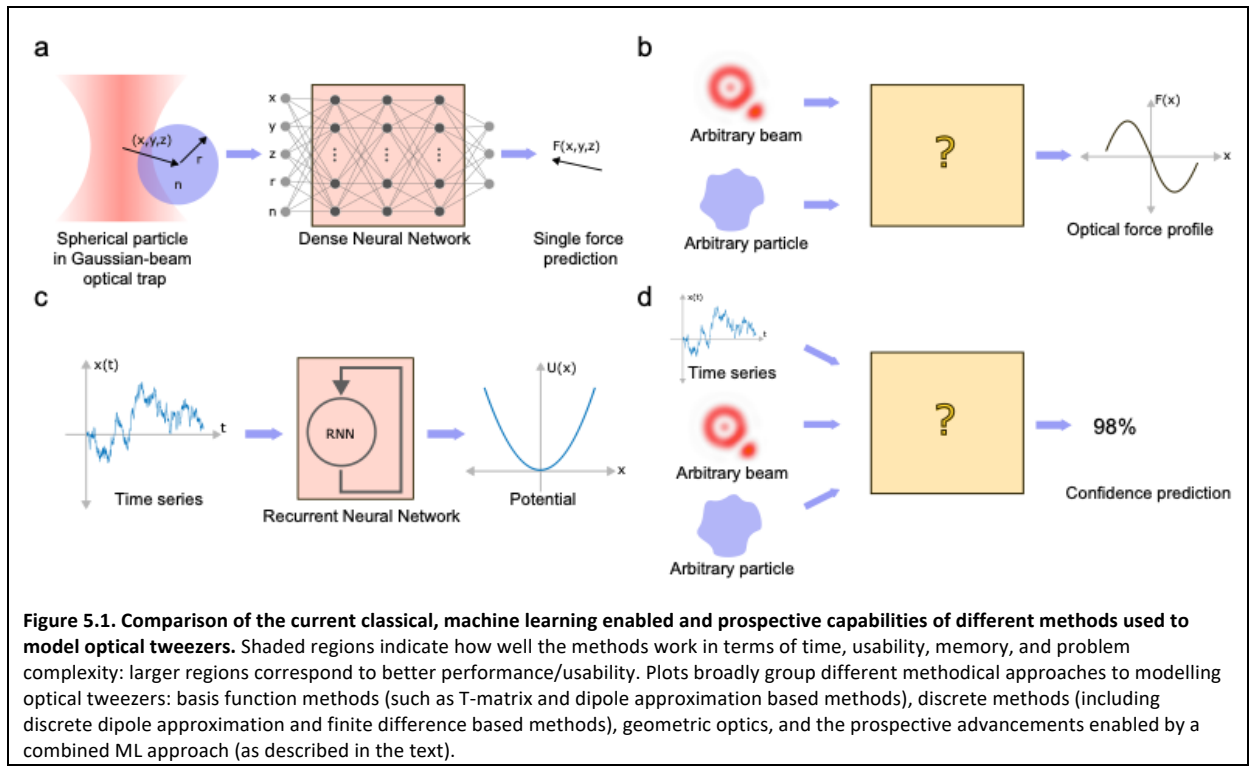

Figure 5.1. Comparison of the current classical, machine learning enabled and prospective capabilities of different methods used to model optical tweezers. Shaded regions indicate how well the methods work in terms of time, usability, memory, and problem complexity: larger regions correspond to better performance/usability. Plots broadly group different methodical approaches to modelling optical tweezers: basis function methods (such as T-matrix and dipole approximation based methods), discrete methods (including discrete dipole approximation and finite difference based methods), geometric optics, and the prospective advancements enabled by a combined ML approach (as described in the text).

**Advances in Science and Technology to Meet Challenges**

There is no shortage of good algorithms for modelling OT. While there are still plenty of active challenges (for example, see sections 7-11 of [10]), for many researchers working with OT the problems are rather in accessibility/usability. In some cases this is due to computational limits, uncertainties about applicability to a particular problem, or simply how easy the method is to download/install. To summarise this problem, **Figure 5.1** shows a current version of the methods' comparison in Ref. [4] with the addition of a usability axis. We have grouped methods into three broad categories. Tractable problem complexity is represented by a single axis (somewhat representing minimum/maximum size and beam/particle complexity).

ML has already pushed the boundaries of what traditional techniques can achieve. For example, in our work [6], we showed how an artificial neural network (i.e., a specific ML technique) depicted in **Figure 5.2a** can be used to significantly reduce simulation memory and time requirements. This network is trained on representative force values for a particular beam/particle combination and can rapidly predict values at other locations while using significantly less memory compared to interpolation. We demonstrated the technique for the T-matrix method and a similar approach has since been demonstrated for accelerating geometric optics simulations [7]. The technique has the potential to increase accessibility/usability since the resulting networks are often small and fast enough to be embedded in an interactive web-browser based simulation. The main shortcoming of these kinds of models is generalisability: i.e., a model trained on spherical particles is not well suited to predicting forces and torques on ellipsoidal particles. Ideally, we would want a model that looks more like Figure 2b: takes a beam and particle as input and produces a force field[2]. For this to be achieved, sufficient training data has to be generated. This will likely involve choosing a network architecture which generalises well to learn features with much less training data. For instance, a convolutional network might perform better when given images of the optical field and particle as compared to the network in **Figure 5.2b**. Other techniques such as transfer learning or model re-training might also be key to reducing the required volume of training data.

---

[2] For non-spherical particles, this would be a 6-dimensional space spanning possible positions and orientations of the particle.

ML also offers a potential solution to validating models correctly describe experiments. ML techniques have been used to analyse experimental measurements, and can be used, for instance, to extract potentials from particle trajectories [8]. **Figure 5.2c** shows a depiction of a recurrent neural network (another ML technique) which converts a time series measurement into a potential. By comparing these potentials with model predicted optical forces, we can verify the accuracy of our models. **Figure 5.2d** depicts a possible extension involving training a network that can discriminate between if a model (beam/particle description) is sufficient to describe a particular experimental observation with some confidence level. One advantage of this problem formulation is we can apply the same optimisation functions we used to train the network also for optimising the model parameters (beam/particle shape).

As a final note, the ML boom has coincided with many other advances in available tools for sharing and collaborating on code and ML model training. Particular tools of note include:

1. Version control software (such as Git and SVN) and related hosting sites (such as GitHub and GitLab) which provide tools for tracking both code and known issues/features/developments.
2. Papers with Code: a website where anyone can share implementations of algorithms described in papers.
3. Containers (such as Docker and Singularity) offer a way to package not just the code but the entire working environment (i.e., specific software versions and dependencies).
4. Read the Docs: an online documentation hosting site which can be setup to automatically hook into popular source distribution tools and automatically generate and publish documentation when a new version of the code is released.

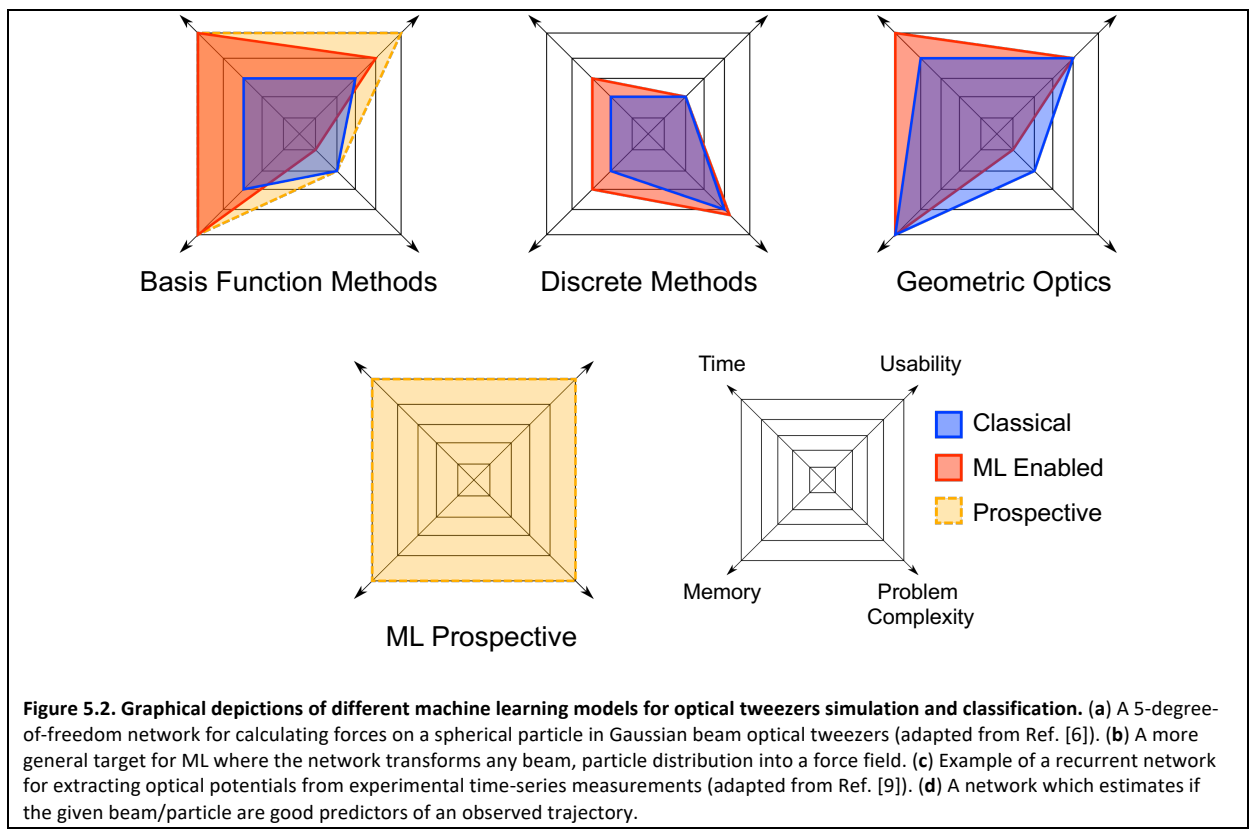

**Figure 5.2. Graphical depictions of different machine learning models for optical tweezers simulation and classification.** (**a**) A 5-degree-of-freedom network for calculating forces on a spherical particle in Gaussian beam optical tweezers (adapted from Ref. [6]). (**b**) A more general target for ML where the network transforms any beam, particle distribution into a force field. (**c**) Example of a recurrent network for extracting optical potentials from experimental time-series measurements (adapted from Ref. [9]). (**d**) A network which estimates if the given beam/particle are good predictors of an observed trajectory.

**Concluding Remarks**

We have already seen that ML can improve access to powerful OT models without users needing super-powerful computers. We have also seen knock-on effects of the ML boom improving software tools for better documentation and collaboration. Improvements in particle tracking and OT calibration, enabled by ML, have the capabilities to ensure the models we generate are accurate. The advances described above are summarised in **Figure 5.1**. We currently envision these being extended further, one of the main limitations is resources including time to generate large data sets, train models, and explore alternative model architectures. We should expect to see new ML models pushing the bounds of what is possible in OT simulation. We might even envisage a universal OT simulator that spans and unifies the capabilities of our present numerical models for a range of cases that would be useful to the OT community. This would require accurate datasets spanning a wide range of beams and particles. In order to achieve this goal, we need to make the existing models more accessible. While we have focussed on OT here, many of these techniques could be applied to simulation of other systems such as acoustic tweezers, electrodynamic traps and optical imaging.

**Acknowledgements**

*H.R-D received funding from the Australian Research Council Discovery Project DP180101002; and the Australian Research Council Centre of Excellence for Engineered Quantum Systems (EQUS, CE170100009).*

# 6 — Optical networks for all-optical computing and imaging

Daniel Brunner[1], Bijie Bai[2], Aydogan Ozcan[2]

1. Département d'Optique P. M. Duffieux, Institut FEMTO-ST, Université Franche-Comté CNRS UMR 6174, Besançon, France
2. Department of Electrical and Computer Engineering, University of California Los Angeles, Los Angeles (CA), USA

**Status**

Optical neural networks (NNs) have a long history. It started in 1985 in seminal experiments that implemented an optical version of Hopfield networks [1]. The prime motivation, which remains unchanged, was to exploit attributes inherently linked to optics: massive parallelism and the resulting advantage for the *connectionist* NN computing concepts. And this promise actually materialized in 1993, as real-time face recognition was demonstrated in an experiment where holographically stored weights were optimized during a training procedure [2].

Since then, artificial intelligence has gone through numerous hype and bust cycles, until in recent years it has become an indispensable tool. Today's high-performance electronics can implement NN topologies with ever-increasing complexity and capabilities. Yet, the fundamental mismatch between computational hardware and NN topology creates enormous challenges in terms of scalability, energy consumption and speed. The result is that adding NN-functionality to other devices comes associated with a substantial power budget, and that real-time processing of high-volume data is unattainable. Employing NNs for either real-time high-resolution wide-field microscopy or small, non-lab-based systems remains challenging. Optical NNs, therefore, remain of interest, and recently, attention towards such unconventional hardware exploded.

**Current and Future Challenges**

Currently, efforts toward next-generation photonic hardware for NNs include integrated photonic solutions [3] and free-space solutions [4-6], as well as methods for addressing the fundamental challenge when integrating NNs using 3D photonic integration [7] or unlocking the ultra-fast potential of fully parallel and autonomous photonic NNs [8].

As a free-space optical computing platform, diffractive deep neural networks ($D^2$NNs) have attracted growing interest as they compute a given task by engineering light diffraction through a series of complex, spatially structured surfaces, one layer following another. Given a targeted task, these layers are optimized using deep learning concepts to minimize a task-specific loss function. The resulting diffractive layers collectively form an all-optical computer, performing the desired transformation between the input and the output fields-of-view. Once the deep learning-based training is complete, the diffractive layers can be fabricated through, e.g., 3D printing or lithography, creating a physical network that performs the designed computational task at the speed of light propagation without requiring any external power except for the illumination light.

Since its first demonstration in 2018 [4], diffractive NNs have been successfully employed for various applications, for example, holography [10], quantitative phase imaging (QPI) [9] and class-specific imaging [13]. Compared to using electronic digital processors, $D^2$NN inference in computational imaging offers some unique advantages since the optical information of a scene is directly accessible to a diffractive network. Therefore, no digitization or complicated pre-processing steps (such as phase retrieval) are needed, enabling computer-free "computational imaging". Using a trained $D^2$NN, the spatial information of any unknown objects can be instantly retrieved from raw holograms at the speed of light propagation without requiring any digital processors [10] . A $D^2$NN can also be trained to directly convert the phase information of an arbitrary input scene into an intensity distribution at its output plane, creating an all-optical QPI system [9] (**Figure 1a**). It was demonstrated that this QPI diffractive network could generalize to unseen, entirely new phase objects.

D$^2$NNs have also been applied to imaging through diffusive media, which is crucial across various fields such as biomedical optics, atmospheric physics, astronomy, oceanography, and autonomous systems. A trained D$^2$NN was demonstrated to all-optically recover the images of arbitrary objects completely distorted by unknown, random phase diffusers, presenting a real-time, computer-free, and power-efficient solution to imaging through random and unknown diffusers [14] (**Figure 1b**).

As another example, D$^2$NNs also present unique opportunities for designing novel computational camera systems with customized functions that cannot be implemented through standard optical designs. As an example, a new type of privacy-preserving imager was designed based on D$^2$NNs, which images only certain desired types of objects, while all-optically and instantaneously erasing other undesired types of objects from its output images [13] (**Figure 1c**).

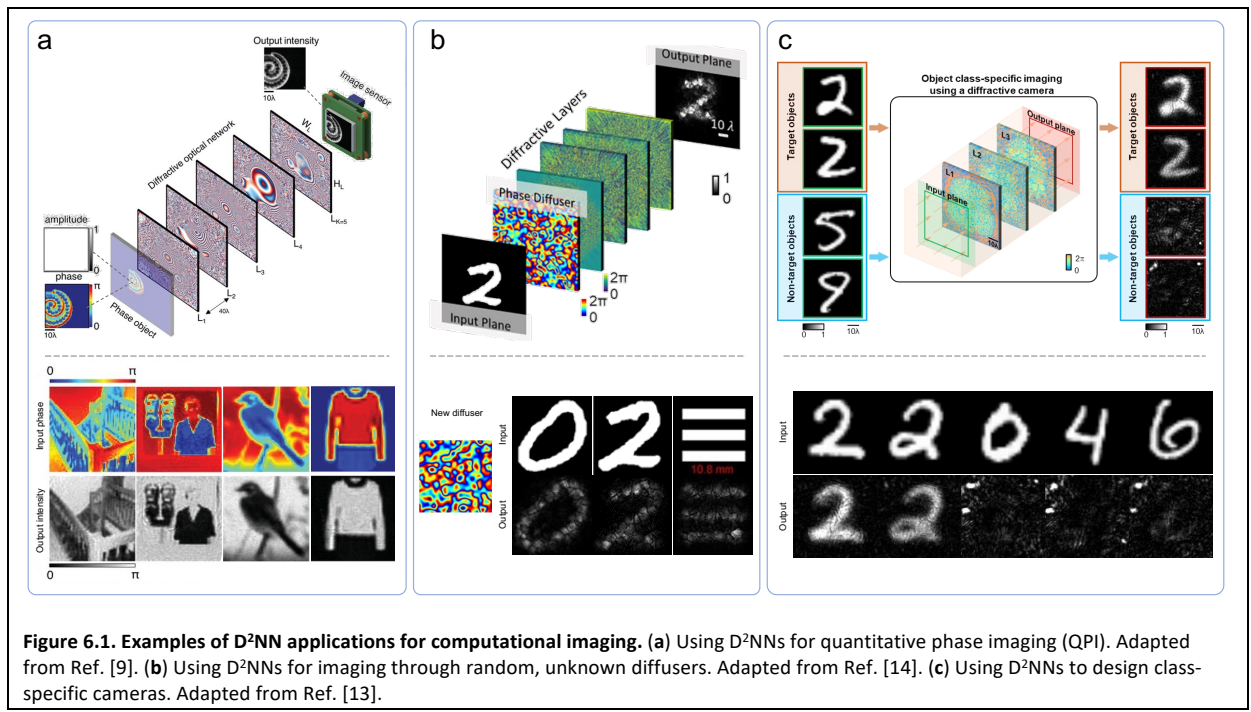

**Figure 6.1. Examples of D$^2$NN applications for computational imaging.** (**a**) Using D$^2$NNs for quantitative phase imaging (QPI). Adapted from Ref. [9]. (**b**) Using D$^2$NNs for imaging through random, unknown diffusers. Adapted from Ref. [14]. (**c**) Using D$^2$NNs to design class-specific cameras. Adapted from Ref. [13].

**Advances in Science and Technology to Meet Challenges**

There exist, therefore, a wide range of high-potential applications. An important challenge remains to physically integrate the many values that define a NN's topology, i.e., the weights to its connections. In two dimensions, such physical implementations do not scale, and recent attention has shifted towards full-scale integration in 3D. Moughames et al. [7] demonstrated optical convolutional filters integrated via 3D photonic waveguides (**Figure 2a**), while some of the advances in the associated fabrication technology now allow for low-loss and large-scale 3D photonic integration of optical couplers [11, 12].

Finally, substantially improving the speed of NN computation requires implementing all the involved processes in parallel hardware, abolishing the slow-down induced through the large-scale serial communication used in schemes that do not leverage full parallelism on each computation stage. This was recently demonstrated by implementing an optical NN in the high-dimensional state space of a multi-mode semiconductor laser diode, while trainable network weights were realized via a spatial light modulator. As a consequence, both the linear operations through optimized weights as well as the nonlinear transformations happened in parallel and with approximately 20 GHz bandwidth [8], (**Figure 2b**).

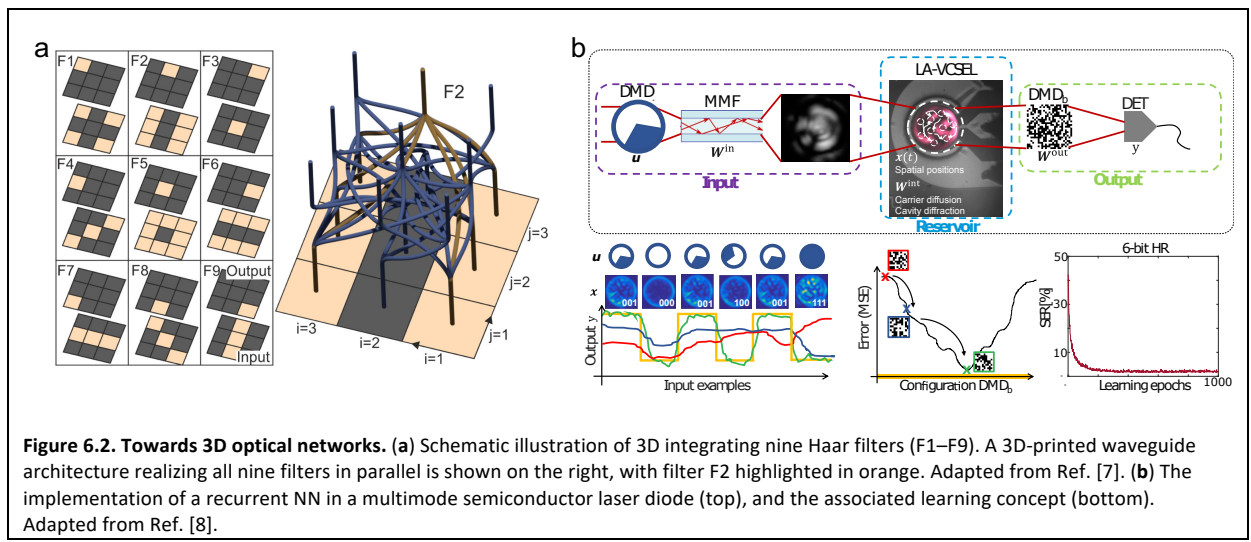

**Figure 6.2. Towards 3D optical networks.** (**a**) Schematic illustration of 3D integrating nine Haar filters (F1–F9). A 3D-printed waveguide architecture realizing all nine filters in parallel is shown on the right, with filter F2 highlighted in orange. Adapted from Ref. [7]. (**b**) The implementation of a recurrent NN in a multimode semiconductor laser diode (top), and the associated learning concept (bottom). Adapted from Ref. [8].

**Concluding Remarks**

In summary, using fully parallel optical NNs, computational imaging tasks can be executed at the speed of light propagation through compact, and potentially integrated photonic systems, establishing low-power and fast all-optical computing platforms beyond what existing electronic systems can offer. The free-space diffractive computing framework is scalable to different parts of the electromagnetic spectrum, including the visible and infrared wavelengths, which could find wide-ranging applications for high-throughput computing tasks and inspire the design of intelligent optical front-ends for advanced machine vision and communication systems. 3D photonic integration, on the other hand, provides the tools and techniques to enable scalable photonic integration of such concepts, while the use of high-dimensional nonlinear optical media such as semiconductor lasers unlocks another potential of photonics for computing: ultra high speed.

**Acknowledgments**

*The authors acknowledge the support of the Volkswagen Foundation, the region of Bourgogne Franche-Comte, the European Commission, the US Office of Naval Research and the US Air Force Office of Scientific Research (AFOSR).*

# 7 – Quantitative microscopy

Daniel Midtvedt[1]

1. Department of Physics, University of Gothenburg, Sweden

**Status**

Optical microscopy has had a tremendous impact on our understanding of the microscopic world and is widely utilized across many disciplines. The traditional workflow consists of (1) data acquisition, relying on manual adjustments of condenser lenses, illumination intensity and camera exposure time, (2) ocular inspection of the acquired microscopy images followed by (3) a qualitative assessment of the sample. The lack of standardization in this workflow makes optical microscopy a qualitative, as opposed to quantitative, technique.

To resolve this limitation, we first note that quantitative microscopy requires that uncontrollable and/or unmeasurable parameters of the optical system do not influence the measurement result. This restricts the measurable quantities to those that are agnostic to the parameters of the optical system. The set of such measurable quantities depends on the choice of microscopy method. The development of microscopy techniques and data analysis techniques go hand in hand: new microscopy techniques can extend the set of parameter-agnostic quantities, calling for new analysis techniques capable of quantifying them. In this context, deep-learning-powered analysis has emerged as an alternative for fast, accurate, and automated analysis and quantification of microscopy data. As an example, the positions, and motion, of objects within the sample do not depend on the parameters of the illumination or the optical system. Particle tracking is, as such, a prime example of a parameter-agnostic technique for most microscopy techniques. It is therefore not surprising that large effort has been put into developing accurate and efficient tracking techniques. Here, deep-learning-powered analysis has been demonstrated to surpass algorithmic techniques in terms of accuracy and speed while being fully automated and objective [1].

Beyond this, interferometric techniques such as holography, interferometric scattering microscopy, and optical coherence tomography, have the potential to provide a quantification of the amount, and in some cases the angular distribution and/or the optical phase shift, of scattered light from objects. This can in turn be related to physical parameters of the objects in the sample such as their mass [2], size and refractive index [3]. Again, deep learning powered techniques have demonstrated superior accuracy and speed in quantifying such parameters, in particular at low signal-to-noise ratios [4].

Finally, multi- or hyperspectral imaging modalities, such as Raman imaging or Brillouin imaging, record a spectrum of optical responses in each pixel, generating enormous datasets. In this context, deep learning has provided a powerful method for automated extraction of relevant data from such datasets [5].

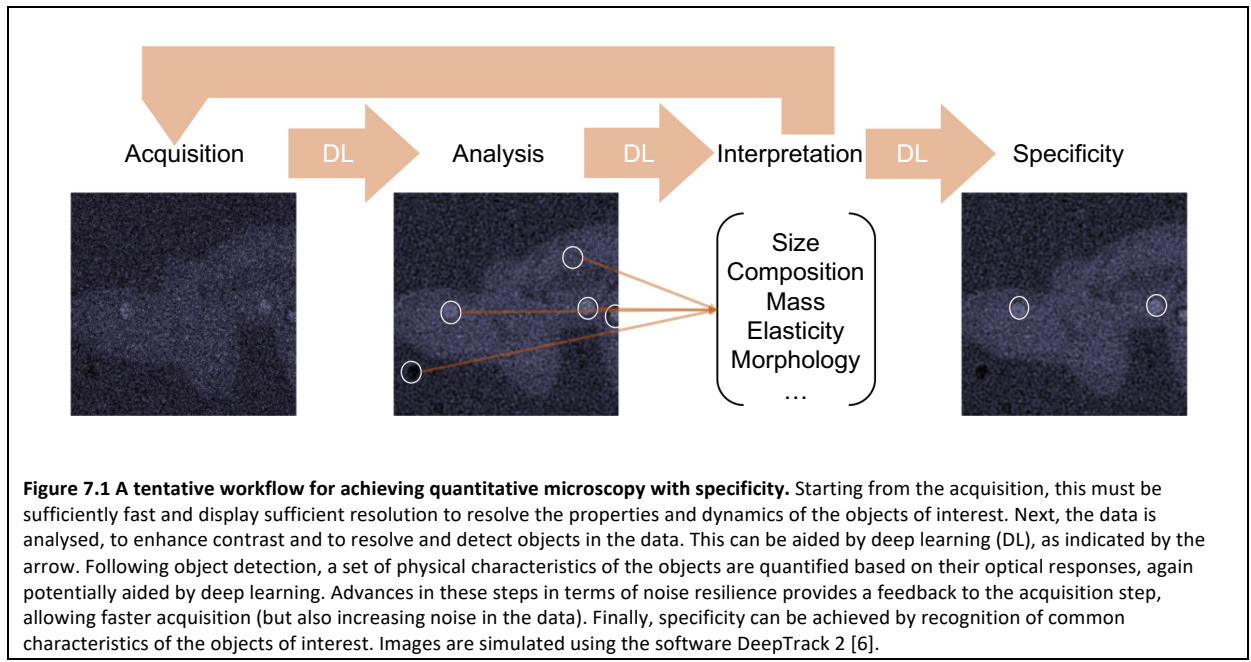

**Figure 7.1 A tentative workflow for achieving quantitative microscopy with specificity.** Starting from the acquisition, this must be sufficiently fast and display sufficient resolution to resolve the properties and dynamics of the objects of interest. Next, the data is analysed, to enhance contrast and to resolve and detect objects in the data. This can be aided by deep learning (DL), as indicated by the arrow. Following object detection, a set of physical characteristics of the objects are quantified based on their optical responses, again potentially aided by deep learning. Advances in these steps in terms of noise resilience provides a feedback to the acquisition step, allowing faster acquisition (but also increasing noise in the data). Finally, specificity can be achieved by recognition of common characteristics of the objects of interest. Images are simulated using the software DeepTrack 2 [6].

**Current and Future Challenges**

Arguably, the holy grail within the field of quantitative microscopy is the accurate quantification of physical properties of specific components in a sample. Generally speaking, achieving this goal requires distinguishing the signal of the components of interest from that of other components (see **Figure 7.1** for a possible work flow). Doing this in a quantitative manner requires, in turn, that (**1**) sufficiently many physical parameters can be quantified, on an individual object level, to classify objects in the sample. In complex environments, as is often the case when considering biological samples, multiple components are often in close proximity and as a result the signals from such components are intertwined and not readily distinguishable. Therefore, in order to be useful, such specific imaging needs (**2**) to be robust to noise and (**3**) to feature a high spatial resolution (preferably sub-diffraction limited). Further, since components are often dynamic, it is also crucial (**4**) to have a high temporal resolution (1-10 milliseconds).

More specifically, interferometric techniques typically display a high temporal resolution, meeting the requirement (**4**) above. It has also been demonstrated that deep-learning-powered analysis techniques have the potential to quantify physical parameters of individual objects at high noise levels (**2**) and short length scales (**3**) (comparable to the diffraction limit) [4], much faster than traditional methods [3]. The set of parameters that can be quantified from individual nanoscale objects depends on the object size and imaging modality, and is currently limited to mass (for Rayleigh scatterers) [2], and size and refractive index (for particles larger than the diffraction limit) [4]. While these parameters arguably are key physical parameters in many cases, they may not be sufficient to do a robust classification and to perform quantitative imaging with high specificity. Extending this set of parameters without sacrificing spatiotemporal resolution is therefore a key challenge for these techniques (**Figure 7.1**, "Interpretation"). A particular challenge will be the quantification of objects larger than the Rayleigh limit, but smaller than the diffraction limit, as the relation between mass and light scattering is ambiguous in this regime and depends on the collection angle of the optical system and the internal mass distribution within the object [8].

For Raman and Brillouin imaging, the information content in each pixel consists of a frequency spectrum quantifying the local photon/phonon interaction. In this case the signal is in fact material specific, and can be used to quantify the abundance and distribution of different materials across a sample [7], resolving requirement (**1**) above. The remaining challenges in this field relate to the detection and quantification of individual objects in a complex sample (**2**) at high spatial (**3**) and temporal (**4**) resolution (**Figure 7.1**, "Analysis").

The key challenges are therefore:

1. extending the set of quantifiable parameters for interferometric imaging techniques, without sacrificing spatiotemporal resolution, and
2. improving spatiotemporal resolution and noise resilience of multispectral quantitative imaging techniques without sacrificing material specificity.

**Advances in Science and Technology to Meet Challenges**
The necessary technological and scientific steps in order to address these challenges are the following.

First, the amount of information in interferometric images can be enhanced through wavelength and/or modality multiplexing, which will ease the quantification of auxiliary physical parameters.

Second, the full information content of interferometric scattering patterns need to be utilized in the analysis. Specifically, the angular distribution of the optical field scattered from an object depends not only on its size, but also on its shape and internal mass distribution. This information is encoded in interferometric scattering patterns, but extracting it would generally require inversely solving Maxwell's equations. This is computationally expensive and extremely sensitive to measurement noise, in particular when approaching the Rayleigh limit. It has been demonstrated that deep-learning-powered approaches enable much faster determination of object properties [3], while retaining accuracy even at low signal-to-noise ratios at subwavelength length scales [4]. Extending such quantification approaches to quantify also the morphology and internal mass distribution of objects covering a range of sizes from < 100 nm to several micrometers will be a necessary step to achieve specific interferometric imaging.

For multispectral quantitative imaging, the key challenge is related to increasing spatiotemporal resolution. Recent works have demonstrated that stimulated Raman [7] and Brillouin microscopy [9] do have the potential for fast multispectral imaging, and work along those lines is expected to push the resolution of those techniques in the coming years. Deep learning powered analysis techniques can assist in this development by reconstructing physical parameters from incomplete spectral information [7]. In this way, less data may be needed to reach the same level of specificity, thereby increasing temporal resolution further.

Combining such developments with advances in virtual staining, which has been demonstrated to recognize, classify and quantify structures in non-specific imaging modalities [10] is predicted to enable specific quantitative imaging through deep learning assisted analysis.

**Concluding Remarks**
In conclusion, the big challenge in quantitative microscopy for the foreseeable future relates to combining quantitative measurements with having high specificity, thereby transforming such techniques into viable alternatives to fluorescence imaging. In this roadmap, I have discussed how deep learning powered data analysis may aid this development, highlighting some key steps that need to be realized in order to achieve this goal.

**Acknowledgements**

*The author acknowledges funding from the Swedish Research Council, grant number 2019-05071*

## 8 — Computational phase microscopy

Hao Wang[1], Lei Tian[1,2]

1. Department of Electrical and Computer Engineering, Boston University, Boston, MA 02215, USA
2. Department of Biomedical Engineering, Boston University, Boston, MA 02215, USA

**Status**

Quantitative phase imaging enables label-free imaging of transparent biological samples such as unstained cells and tissues. Many holographic-based phase imaging techniques have been developed to extract the phase information based on the principle of interferometry. However, these techniques generally suffer from complex instrumentation. In this perspective, we focus on computational phase microscopy techniques which recover phase information from a non-holographic setup using intensity-only measurements. The main difficulty of reconstructing phase from intensity-only measurements is that it is generally an ill-posed inverse problem. Classical "phase retrieval" algorithms are developed based on the Gerchberg-Saxton-Fienup algorithm, which uses an iterative reconstruction procedure and incorporates additional physical constraints to find the solution. In recent years, deep learning (DL) has shown tremendous success in solving such computational imaging problems. Here, we highlight several major advances on how DL pushes the imaging performance of computational phase microscopy, as summarized in **Figure 8.1**.

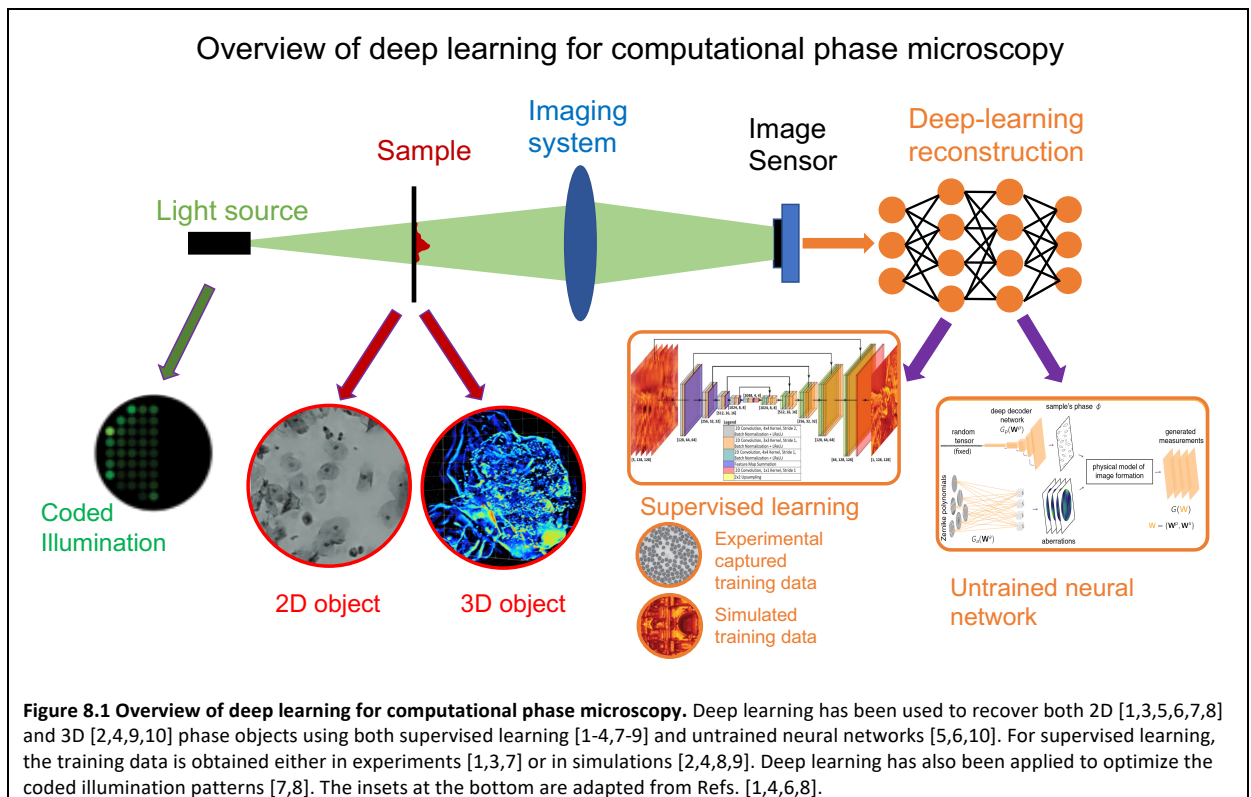

**Figure 8.1 Overview of deep learning for computational phase microscopy.** Deep learning has been used to recover both 2D [1,3,5,6,7,8] and 3D [2,4,9,10] phase objects using both supervised learning [1-4,7-9] and untrained neural networks [5,6,10]. For supervised learning, the training data is obtained either in experiments [1,3,7] or in simulations [2,4,8,9]. Deep learning has also been applied to optimize the coded illumination patterns [7,8]. The insets at the bottom are adapted from Refs. [1,4,6,8].

In general, DL techniques have been developed to solve both the 2D phase recovery and 3D tomographic reconstruction problems, as summarized in **Figure 8.1**. For 2D computational phase imaging, the first supervised learning-based reconstruction algorithm proposed in Ref. [1] showed that the twin-image and self-interference related diffraction artifacts can be eliminated by a deep neural network. To push the performance at low-light conditions for 2D phase imaging, a DL technique has been developed to provide high-fidelity recovery when the photon numbers are low [3]. For 3D computational phase imaging, DL frameworks have been developed to solve the challenging 3D tomographic phase reconstruction from intensity-only measurements [2,4]. In addition, frameworks have been developed to mitigate multiple-scattering effects and overcome the limitations in commonly used linear single-scattering approximation models [2,4].

Although these supervised DL methods have made significant progress, they require large-scale training datasets containing paired measurements and ground-truth images. While early attempts focus on generating experimentally captured dataset [1, 3], in practice the accessibility to large-scale training data in experimental settings can be limited. To overcome this limitation, several innovative approaches have been developed. For example, computationally efficient and accurate simulations based on multiple-scattering models have been developed to generate synthetic training datasets for 3D problems [2, 4]. Instead of using the supervised DL framework, the unsupervised "untrained neural network" framework based on Deep Image Prior [5] and Deep Decoder [6] have been adapted for 2D phase retrieval without needing any network pre-training nor training data. In this approach, the phase map is parameterized by an untrained network, whose parameters are optimized by matching the predicted measurement to the actual measurement using the known physical model via a standard iterative "network training" procedure.

In addition to only performing the post-hoc 2D or 3D reconstruction tasks in computational phase imaging, DL methods have also been developed to co-optimize the physical design of measurement process along with the reconstruction. Pioneering work demonstrates an "unrolled neural network" framework to optimize the coded-Illumination pattern and high-resolution 2D phase reconstruction using much reduced measurements than model-based method [8].

**Current and future challenges**
Despite achieving state-of-the-art performances in many tasks, multiple challenges remain to be solved both in terms of physical principles and computational frameworks, including:

1. **Multiple scattering effects and missing-cone problems**
   The multiple scattering effects become significant with the increase of refractive index contrast, structure complexity, and size of the biological samples. The measurement is often confounded by the "missing-cone" problem, which does not provide access to a large amount of axial spatial frequency information. Together, these make the 3D reconstruction suffer from poor axial resolution and degraded reconstruction accuracy. Although DL frameworks have been developed to reconstruct highly scattering objects [2, 4, 9], the effectiveness is still limited.
2. **Reliable DL prediction**
   Another major limitation is that the trained DL model is often not robust to experimental variations. To overcome this issue, an uncertainty quantification framework based on the Bayesian learning framework has been proposed [7], which allows evaluating the confidence of the prediction result without knowing the ground truth. However, this type of Bayesian learning framework is still at its nascent stage and requires significant process to provide more reliable DL predictions.
3. **Large-scale computation**
   In DL-based reconstruction, the size of the neural network generally increases as the size of the input images. Thus, the practically achievable space-bandwidth product (SBP) of the reconstructed image without excessive image stitching is limited by the memory of the computer. The associated computational cost becomes a bottleneck for emerging large-SBP imaging techniques, such as Fourier ptychographic microscopy for gigapixel 2D phase imaging [7] and diffraction tomography on large-scale 3D objects [4].

**Advances in science and technology to meet challenges**

Here, we outline a few promising directions to pursue to overcome the above challenges and further push the fundamental limit of deep learning for computational phase imaging.

1. **Prior knowledge incorporation**

   An overall strategy to overcome the multiple scattering and missing-cone problems as well as the reliability of DL prediction is to incorporate additional physical knowledge into the neural network designs. In simulation-based training [2, 4], the multiple-scattering physics is incorporated into the simulator. In untrained neural networks [5, 6], the physical model of the imaging system is incorporated during network training. In Bayesian learning [7], the priors are learned by a neural network, which is then used to quantify the uncertainty at the test stage. A few other network designs implicitly take the physical insights into consideration. For example, Ref. [3] separately processes low and high spatial frequency information using a synthesis network. Ref. [9] treats the sequential measurement as a dynamical system and solves the tomographic reconstruction problem by a recurrent neural network. We anticipate that hybrid strategies that incorporate multiple types of priors, e.g., multiple-scattering physics, imaging model, and experimental data distribution, are promising directions to investigate in the future.

2. **Computationally efficient framework**

   Future development is needed to enable memory-efficient computational frameworks for large-SBP imaging applications. A promising solution is the neural fields (a.k.a. implicit neural representations) that parameterize the object by a coordinate-based deep network. The object is represented by the parameters of a small-scale neural network, instead of the dense voxel/pixel grids. Promising results using this framework have been recently demonstrated for large-scale 3D phase recovery in Ref. [10]. We anticipate novel neural representations for efficient information embedding may be a promising direction to pursue for large-SBP reconstructions.

# 9 — Multimodal microscopy image registration

Nataša Sladoje[1], Joakim Lindblad[1]

1. Department of Information Technology, Uppsala University, Sweden

**Status**

The technological advancement enables highly informative imaging, allowing scientists to see what previously was hard to even imagine. The vast collection of imaging techniques used in the biological and preclinical areas indicates the breadth of possibilities [1]. Powerful and complex devices reveal a variety of properties of a specimen, such as morphological structure, chemical composition, dynamics, function; however, most often they reveal only one such property at a time. We are witnessing a growing popularity, particularly within life sciences, of *correlative multimodal imaging* (CMI) approaches, which combine complementary information from different imaging modalities to create a holistic, composite view of the sample, maximising the extracted information about an object of interest [2]. The most well-established CMI technique is Correlative Light and Electron Microscopy (CLEM), combining spatial and functional information within the subcellular context.

To enable joint analysis and fusion of the heterogeneous information captured by different devices, the first requirement is to establish precise geometric correspondence between the acquired images, i.e., to find a spatial transformation which best aligns the data in the same coordinate system. This process, known as *image registration*, is traditionally performed as an iterative optimization process, aiming to find the transformation which maximises a similarity measure between the images to align. Depending on the application scenario, different types of transformations (e.g., rigid, affine, deformable) are used, making the task more or less constrained. The optimization is typically highly non-convex, which makes the process complicated and time consuming, while still often only delivering suboptimal solutions. This in particular holds for multimodal registration, where the images that need to be aligned, aiming to capture complementary information, often look very different from each other.

To solve or circumvent the observed problems, researchers have turned to approaches based on deep learning (DL). Three main directions have been followed: (i) learning a suitable similarity measure, (ii) speeding up the registration process by directly predicting the transformation, and (iii) reducing multimodal alignment to a monomodal task, by applying image-to-image (I2I) translation techniques. These approaches have been evaluated on medical data [3-5] and the most successful ones are steadily growing in popularity. However, their use in biomedical/microscopy image registration is still very limited. In particular, DL-based multimodal microscopy image registration has been attempted in only a few works [6,7]. The extraordinary boost, observed for other image analysis tasks, such as image segmentation and classification, is still lacking — the DL-revolution is yet to come to multimodal microscopy image registration.

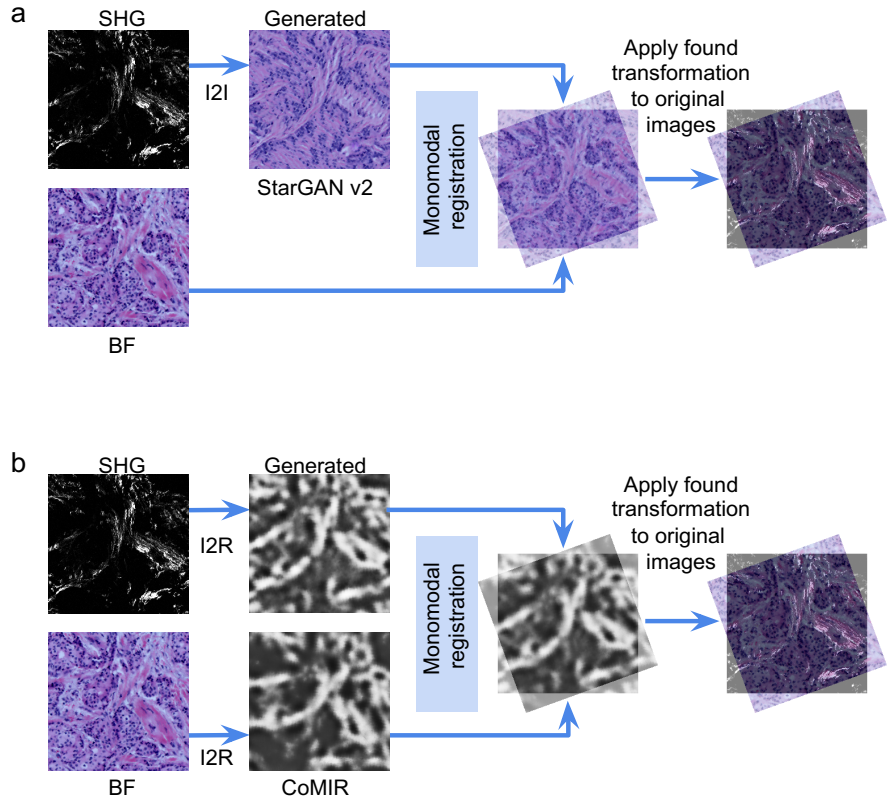

**Figure 9.1 Multimodal image registration.** Multimodal image registration can be addressed by converting the multimodal problem into a, presumably less challenging, monomodal one. Here, this is illustrated by the difficult task of aligning Second Harmonic Generation (SHG) and BrightField (BF) microscopy images. (**a**) One approach is to synthesise images acquired by one modality to appear as if acquired by the other; this type of Image-to-Image (I2I) translation is typically done by Generative Adversarial Networks (here, using StarGAN v2). However, a comparative performance analysis [7] indicates that the synthesised images, even when convincing in appearance, seem to lack sufficiently reliable features to ensure successful subsequent monomodal alignment. (**b**) Another approach is to, instead of learning the appearances, learn representations (I2R) of the joint content of multimodal images, transforming them into a common "virtual" modality. This coordinated representation learning approach, in combination with a suitable monomodal registration method, reaches excellent performance on alignment of SHG and BF images [6].

**Current and Future Challenges**

To reach, and benefit from, the full potential of DL in multimodal microscopy image registration, researchers need to respond to several challenges (**Figure 9.2**):

*Challenge 1 — DL methods still do not meet the expectations in image registration*

Considering that different combinations of modalities may require different measures of image similarity, the idea to learn such a measure from data comes naturally. Given sufficient annotated training data, proposed learned similarity measures perform well. However, they are most often used to replace conventional measures within a slow iterative registration procedure, which reduces the benefits.

Once trained, DL models are typically very fast and therefore appealing to use as regressors, to directly predict the transformation parameters to reach image alignment. This approach is taken by currently most popular methods which use DL for registration. While considerably reducing runtime, they still fall behind conventional iterative methods in terms of accuracy.

I2I translation approaches, typically based on Generative Adversarial Networks, perform impressively in learning to mimic, and combine, content and styles captured by image data, e.g., enabling "virtual staining" of label-free tissue. However, generated images, even if convincing in appearance, often do not preserve well the information important for image alignment [7]. Relying on aligned pairs, the coordinated representation learning approach [6] delivers superior performance (**Figure 9.1**).

*Challenge 2 — DL methods require a lot of data, but not any data*

Correlative multimodal microscopy typically results in few very large images. This is far from the millions of relatively small annotated images collected in publicly available datasets which are used for development of state-of-the art DL models in Computer Vision. The complexity of the acquisition process, typically requiring significant manual labour, combined with exploratory aims of the performed research, implies that images acquired are often counted in single-digit numbers. Transfer learning, aimed at reusing trained models on new datasets, often delivers diminishing returns on the diverse and heterogeneous biomedical image data. At the same time, very high-resolution imaging provides terabyte-sized (or even larger) image volumes, necessitating advanced algorithmic solutions for their processing.

*Challenge 3 – Annotated multimodal microscopy data are critically lacking*

To ensure reliable performance, DL approaches generally require large amounts of annotated training data. Annotation of biomedical data requires not only extensive time, but also considerable expertise, making it costly to collect. DL models which require aligned image pairs for training find limited use, since the alignment is, for many modality combinations, simply too difficult to be performed manually. Few-shot and un/self-supervised methods may reduce the need for training data, but this often comes at a cost reduced performance.

**Advances in Science and Technology to Meet Challenges**

Availability of high quality training data is identified as a *sine qua non* condition for successful DL approaches. The challenges of DL-based multimodal microscopy registration all relate to the need for, and difficulty to provide in sufficient amounts, accurately aligned multimodal microscopy image pairs for diverse combinations of modalities. This indicates two directions forward:

i. develop methods which reach good performance while requiring only few, or no aligned image pairs;
ii. collect large annotated multimodal microscopy datasets and make them broadly available.

*Advances in methodological development*

Iterative maximisation of mutual information (MI) is still the most popular multimodal registration method; it is generally applicable, performs reasonably well, and does not require any training data. A recently proposed FFT-based algorithm [8] advances MI-driven rigid multimodal registration, both in terms of speed and accuracy, outperforming DL-based competitors on two multimodal microscopy datasets. Novel DL-based registration methods need to deliver more.

Requirements for extensive training data of aligned multimodal image pairs need to be reduced, or removed. We need to turn to novel and innovative learning strategies which deliver highly performing, yet scalable solutions. General improvement of data-efficient DL strategies, such as few-shot learning, domain transfer, and self-supervised learning, will advance multimodal microscopy image registration as well.

Unsupervised methods, not requiring any aligned image pairs for training, are already available. Examples include I2I translation approaches based on unsupervised GANs [7]. However their performance needs to be further improved.

*Advances in data collection, open science, and standardised benchmarks*

Annotated (ground truth) data is needed not only for DL model training, but also for evaluation of novel methods. It is therefore of critical importance to assemble and publish high quality datasets, enabling quality control and reproducibility. Automated approaches may reduce need for manual annotation, e.g., by generating landmarks through segmentation of identified common structures in multimodal data.

Integrated imaging, where different modalities are simultaneously acquired and therefore aligned, may be a rewarding path for generating large amounts of high-quality ground-truth registration data, for both training and performance evaluation. This option is, however, only available for a few combinations of modalities. Advances in integrated imaging will at the same time reduce the need for multimodal registration methods.

Considering the specificities and large size of multimodal microscopy datasets, availability of suitable data sharing platforms is a necessity. Standardisation of acquisition processes will contribute to improved quality of data and increased homogeneity (e.g., within a particular modality), leading to increasingly successful training of DL models and improved registration quality. Open platforms for standardised method evaluation will boost the quality of novel approaches. A recent contribution is an open evaluation framework for rigid multimodal registration methods, [7].

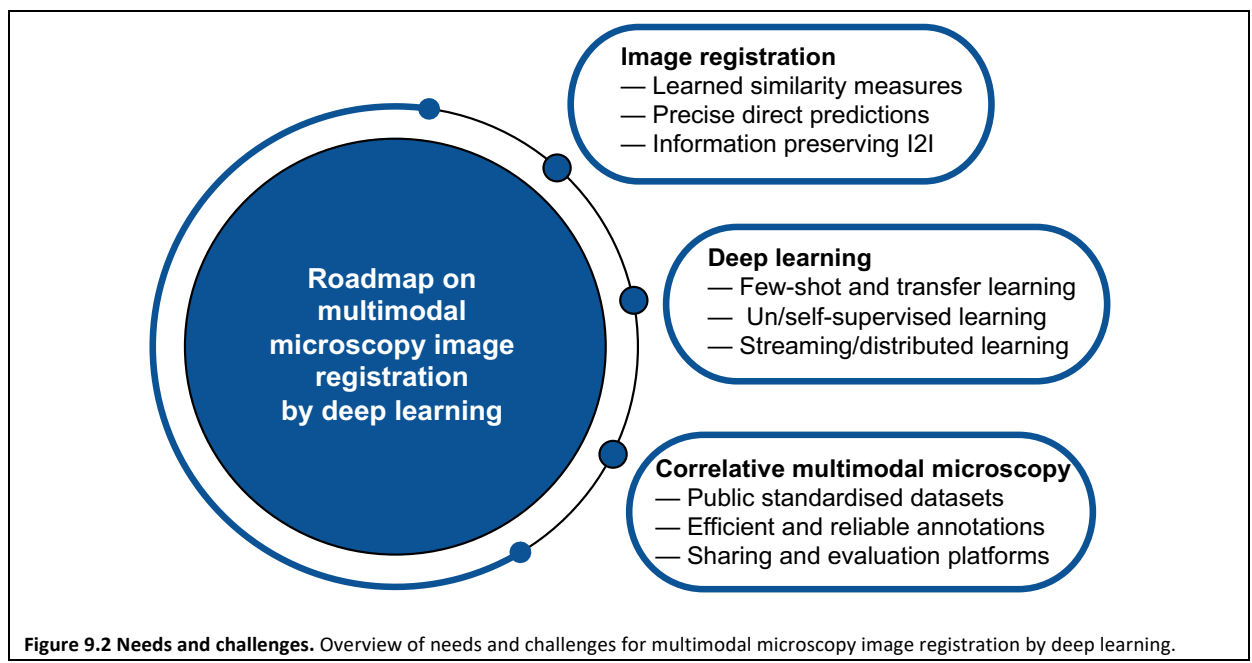

**Figure 9.2 Needs and challenges.** Overview of needs and challenges for multimodal microscopy image registration by deep learning.

**Concluding Remarks**
In the era of outstanding performance of deep learning methods, which continuously advance image data analysis, most popular methods used in practice for multimodal microscopy image registration still rely on semi-automatic approaches. A successful example is eC-CLEM [9] — relying on user-interaction, it has demonstrated applicability to registration tasks involving a wide range of modality combinations. However, the increasing diversity of multimodal microscopy image registration problems and the growing scale and dimensionality of acquired data urgently call for the power and flexibility of efficient data-driven approaches, to bring the quality of the methods and analysis results to the next level. By addressing the identified challenges of the field (overviewed in **Figure 9.2**), DL-based techniques have the potential to deliver modality-agnostic, fast, and generally applicable image registration solutions, to advance correlative multimodal imaging, and ultimately – life sciences.

**Acknowledgements**

*The authors are inspired by the numerous activities and initiatives of the COST consortium 'Correlated Multimodal Imaging in Life Sciences' (COMULIS) — a collaborative European network aiming to foster interdisciplinary communication and collaboration on all aspects of Correlative Multimodal Imaging in Life Sciences. The authors are financially supported by VINNOVA (MedTech4Health project 2017-02447).*

# 10 — Fluorescence Lifetime Imaging

Jason T. Smith[1,2], Marien Ochoa[1,3], Margarida Barroso[4], Xavier Intes[1]

1. Department of Biomedical Engineering, Rensselaer Polytechnic Institute, Troy, NY, 12180, USA.
2. Elephas, 1 Erdman Pl., Madison WI, 53705, USA
3. University of Wisconsin–Madison, Department of Medical Physics, Madison WI, 53705, USA
4. Department of Molecular and Cellular Physiology, Albany Medical College, Albany, NY 12208, USA.

**Status**

Fluorescence lifetime imaging (FLI) provides distinctive contrast mechanisms for the interrogation of biological samples. A key strength of FLI is that it can uniquely provide absolute measurements directly related to the fluorescent molecule(s) state and its/their interaction with the surrounding molecular micro-environment: such as temperature, pH, viscosity, polarity and mechanical forces. The principles and various technical implementations of FLI have been established over the last three decades but, until recently, have typically remained an expert field. With the advent of turn-key commercial FLI capable imaging platforms, FLI is currently being increasingly embraced by end-user communities (e.g., molecular biologists, drug development experts) [1,2] with demonstrated increased utility in microscopic [1] and macroscopic preclinical and clinical applications [3]. Still, a main challenge in FLI resides in the estimation of the lifetime(s) or associated parameters. This is a complex computational task, in which accuracy can be highly dependent on the model selected, the set of parameters used, and the signal-to-noise ratio (SNR) of the acquired measurements (typically a photon starved application, hence, low SNR). Hence, there are still large efforts focusing on providing robust, user-friendly, and accurate methodologies, including deep learning models, for FLI image formation. The use of machine learning approaches for FLI image formation was first using an Artificial Neural Network (ANN) approach with promising results [4]. This work was followed by the first Deep Learning (DL) model, FLI-NET, designed to simultaneously produce 2D images of all the lifetime-based parameters associated with a double exponential model [5]. FLI-NET was validated both with microscopic and preclinical data sets and for two main instrumental detection technologies, Time-Correlated Single Photon Counting (TCSPC) and gated ICCDs. Since then, an increasing number of reports have demonstrated the potential of DL models to estimate accurately and without any user input Lifetime parameters, over large FOVs, with extremely fast inference times (at, or close to real-time), and with better performances at low photon counts. There is also an increased interest in developing end-to-end DL models for pixel-wise classification based on spatio-temporal inputs. An example of an ANN based FLI image formation over a large FOV is provided in **Figure 10.1**.

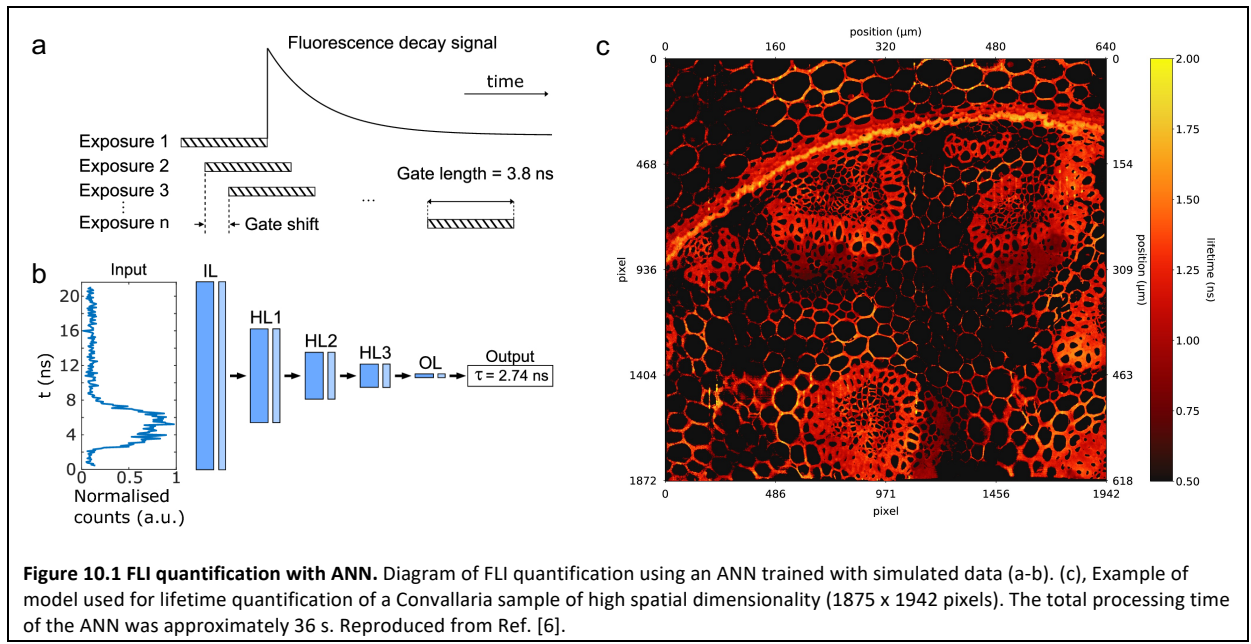

**Figure 10.1 FLI quantification with ANN.** Diagram of FLI quantification using an ANN trained with simulated data (a-b). (c), Example of model used for lifetime quantification of a Convallaria sample of high spatial dimensionality (1875 x 1942 pixels). The total processing time of the ANN was approximately 36 s. Reproduced from Ref. [6].

**Current and future challenges**

As the field of deep learning (DL) for FLI continues to mature, it is facing the same challenges currently encountered in the development and validation of DL models for biomedical imaging at large. These can be summarized as: the acquisition/availability of large representative data sets, the generalization of DL models, and instilling explainability and/or trustworthiness of the DL model output. In numerous fields, DL successes at large can be attributed to the confluence of increased computational power with accessibility to large data sets. However, for biomedical applications, large data sets that have been thoroughly curated are rarely available to the community. This is particularly true for FLI. To date, most of the work reported has implemented experimentally representative data simulation routines to generate large data sets for efficient DL model training and validation. This approach is cost effective and has been demonstrated to perform well for processing experimental data not used during training with high accuracy [15]. This mitigates the requirements of large experimental data sets for training, but still requires expertise in developing the simulation environment that closely replicates the specificity of the application – such as instrumental characteristics (especially the Instrument Response Function, IRF, which characterizes the temporal behavior of the system), lifetime-based parameters range and noise distributions.

Though, there is a large set of factors that can affect FLI quantification from experimental data and that can be challenging to represent during the training phase. These include laser jitter and instrumental drift, changes in sample to detector distance, pixel-dependent variation of the IRF, photobleaching of the fluorophore, saturation of the detector. Other factors include simulation model bias such as imaging a sample with fluorescence lifetime outside of the range of that used during model training, more complex signal signatures such as multi-exponential features beyond bi-exponential behaviors. All and each of these factors can negatively affect the model output independently. This also highlights the difficulty to generalize the developed DL models beyond the imaging system and application at hand. In the case of FLI-NET, the only DL model applied to different FLI applications and technologies, the DL model needed to be trained specifically for each case. This comes with an added computational burden comparatively to classical iterative fitting approaches. Additionally, this leads to assessing the trustworthiness of the DL model output. In this regard, classical iterative fitting-based approaches can assess the fidelity of the quantification, as well as the quality of the FLI data itself, through residual error between the data and the approximated fits. This is an important means for quality assessment that is commonly employed during FLI analysis, as FLI data can be comprised of pixels with decays that are not suitable for quantification due to a range of factors (e.g., low SNR, motion artifacts, improper parameter settings). When using DL for either image formation or for classification, these poor-quality decays can be either removed or artificially enhanced via rudimentary pre-processing steps prior to network inference. However, this preprocessing step precludes the application of DL models towards real-time and/or can lead to large bias as it enforces expected features in the data that could not always be valid (for instance bi-exponential behavior while more complex biological distribution is present in the sample). To date, no DL work applied to FLI has tackle this issue of providing a measure of the output trustworthiness.

**Advances in science and technology to meet challenges**

The use of DL methodologies for FLIM is a nascent field that promises to greatly increase the widespread utility of lifetime-based contrast in biology, as well as its use in translational medicine. To date, a pioneering set of reports have laid the foundation for the development of efficient DL models that are dedicated to specific applications and technologies. One important aspect is that, in many cases, experimental validation of DL models has been primarily performed with relatively bright and long lifetime samples that are "best case" scenarios. Still, it can be argued that the biggest challenge in the field of lifetime imaging at present is that of low photon count detection with high background noise levels. Indeed, proper FLI data acquisition requires many photons to provide high quality decays in all pixels. However, low counts often force the use of high exposure times or illumination powers that can lead to fluorophore photobleaching or cell/tissue damage. Low counts also lead to increased binning resulting in reduced resolution. However, collection of suitably high counts is oftentimes unfeasible when dealing with sensitive samples or with applications requiring even modest (i.e., sub-second) framerates. Only a few studies have reported DL validation in the case of challenging samples, such as in metabolic imaging (low photon counts) or short lifetime (such as in NIR emitting fluorophores). Recently, next-generation time-resolved detectors, such as single-photon avalanche SPAD diode arrays, are poised to enable faster and more photon-efficient FLI data acquisition with improved SNRs [7,8]. Furthermore, next generation microscopy systems that leverage computational imaging approaches like single-pixel detection will allow for improved collection efficiency especially when coupled with DL models [9,16]. Also, the implementation of detectors that leverage additional information content (e.g., hyperspectral detection arrangements) will allow for increased specificity, especially for conditions where spectral emissions from target fluorophores are hard to isolate due to spectral bleedthrough [10]. Additionally, there are still large efforts focusing on developing more stable exogenous fluorophores with high quantum efficiencies and low cytotoxicity for improved FLI signal detection [11].

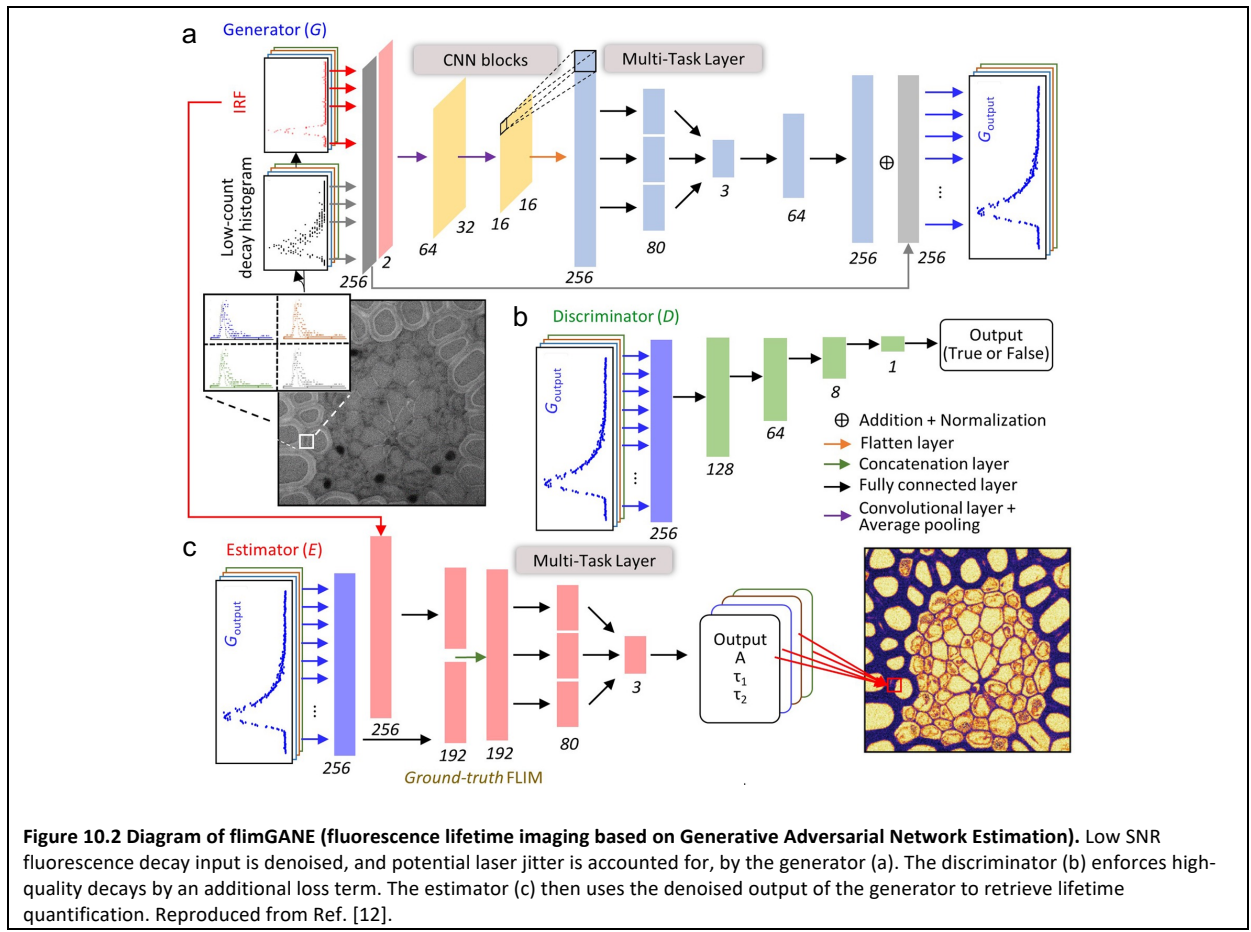

**Figure 10.2 Diagram of flimGANE (fluorescence lifetime imaging based on Generative Adversarial Network Estimation).** Low SNR fluorescence decay input is denoised, and potential laser jitter is accounted for, by the generator (a). The discriminator (b) enforces high-quality decays by an additional loss term. The estimator (c) then uses the denoised output of the generator to retrieve lifetime quantification. Reproduced from Ref. [12].

On the account of output trustworthiness, there has been an increased number of tools available for eXplainable AI (XAI). Though, if current methods can be leveraged efficiently in FLI classification tasks, there is still a lack of appropriate tools in DL image formation. To date, DL model outputs must be assessed by an expert and unexpected results, forensically investigated. From experience, these unexpected DL results are typically attributed to some variations in the experimental acquisition parameters. Such variations can be mitigated by employing generative models. For instance, flimGANE takes into account the noise distribution as well as laser jitter in FLI data collected by TCSPC (**Figure 10.2** [12]). In turn, this enables the development of DL models that are more generalizable. Still, they typically come with an additional computational complexity that doesn't make them competitive for fast inference, which is required in certain clinical scenarios. In this regard, simpler DL models have been proposed that process data for individual pixels with potential to be implemented directly on the acquisition hardware. Coupled with efficient training methodologies, they herald embedded hardware implementation, coupling with sensors and readout circuits to achieve fast on-chip training and inference [13].

Last, to maximize reproducibility and accessibility across the FLI community, the development of user-friendly open-sourced software (such as FLIMJ [14] — a plugin for the widely used ImageJ) will have a large and significant impact. Providing a workflow that allows for deconvolution of the end-users IRF (i.e., the detector-specific response) along with subsequent data simulation and model training across wide parameter bounds could provide a cross-reference platform for benchmarking. Combined with the mandate from institutional funding agencies to support the release of publicly available data, it will elicit even further developments in the field by opening it to the computer science community.

**Concluding remarks**

FLIM has firmly established itself as a pillar of cellular imaging from microscopic to macroscopic platforms. However, FLI utility is associated with parameter(s) quantification obtained via computational approaches, which has reduced its impact and wide-spread use due to the expertise requirements. The implementation of DL methodologies is expected to enable real-time FLI quantification free of user-related bias. Beyond simplifying and enhancing the data processing pipeline, DL is poised to greatly impact FLI imaging protocols and imaging platforms. Due to its enhanced robustness at dim signals, DL is expected to relax the need of long integration time, spatial binning, and increased illumination power. Altogether, DL-enhanced FLI will be better-positioned to impact applications ranging from fundamental biology to clinical practice.

## 11 — Multi-modal nonlinear microscopy

Tong Qiu[1], Li-Yu Yu[1], Sixian You[1]

1. Research Laboratory of Electronics, Electrical Engineering and Computer Science, Massachusetts Institute of Technology, Boston (MA), USA

**Status**

The development of nonlinear optical theory and microscopy has enabled unprecedented opportunities to look at living cells, tissues, and animals with submicron resolution in real-time. Over the past two decades, a variety of biological phenomena have been investigated using images based on the fluorescence excited by multiphoton processes, harmonic generation from specially structured molecules, and chemical profiling based on coherent Raman scattering. Integration of these modalities has been shown to provide metabolic, chemical, and structural profiling of cells in the context of living tissues. Despite its strong promise, multimodal nonlinear microscopy has not yet reached its full potential. Improvements in hardware and software are needed for further translation to biomedicine. This perspective focuses on the important gaps that can be potentially addressed by deep learning from two angles: 1) how deep learning can help overcome the technical limits of multimodal nonlinear microscopy through DL-based reconstruction and augmentation, and 2) how deep learning can help increase the clinical and biological relevance of multimodal nonlinear microscopy through DL-based image and video understanding.

**Current and Future Challenges**

One challenge lies in the longstanding technical limitations of nonlinear microscopy, such as the penetration depth, imaging speed, signal strength, and resolution associated with in vivo imaging. Various hardware improvements have been proposed in recent decades, including adaptive optics, adaptive laser sources, and fiber-based endoscopy for deeper tissue imaging, multi-foci and wide-field temporal focusing for higher speed, and pulse shaping for stronger signals. However, optimization of one of these parameters based on hardware is usually associated with the degradation of performance in the other parameters. For example, pulse shaping boosts signal generation efficiency by compensating for dispersion at the expense of laser power and cost. Computational reconstruction, such as optimization and learning-based algorithms, provides a promising alternative to overcome such inherent tradeoffs by regularization based on principle-based or learned data priors.

The other challenge is the automated image and video understanding of multimodal nonlinear microscopy. The information captured by multimodal nonlinear microscopy is a pixel-coregistered multimodal quantitative measurement of fluorophores (two-photon, three-photon absorption fluorescence), molecular structures (second, third-harmonic generation), and chemical bonds (coherent Raman anti-Stokes scattering, stimulated Raman scattering). Despite its rich information, the translation of the multimodal information to biological and pathological analysis is not yet readily accessible to biologists and clinicians who are experts in immunohistopathology. Algorithms that can address the challenges of image and video understanding of multimodal nonlinear microscopy are in great demand. One direction is to directly transform multimodal nonlinear microscopy to H&E-like images, which facilitates the biomedical relevance of the new image dataset. Cahill et al. have shown a color metric-based method that reliably translates nonlinear microscopy images to H&E images [1]. The other direction is the direct extraction of quantitative information for specific applications. For example, Walsh et al. performed classification and single-cell analysis of quiescent and activated T cells using quantitative features of nonlinear autofluorescence microscopy, which revealed the correlation between the autofluorescence features and the metabolism of T cells [2]. These insights establish robust protocols and propose new mechanisms but rely on users' mastery of both nonlinear microscopy and the specific biomedical applications. Deep learning methods promise to alleviate the burden of domain expertise and further streamline the process via data-driven and end-to-end learning.

**Advances in Science and Technology to Meet Challenges**

As mentioned in the previous section, computational reconstruction is uniquely positioned towards pushing the physical limits of nonlinear microscopy via proper data priors. For example, to faithfully reconstruct wide-field multiplexed measurements in deeper layers, Zheng et al. [3] leveraged temporal focusing for randomly patterned wide-field illumination and physics-based inversion for computational reconstruction of the multiplexed signals (**Figure 1a**). To enable video-rate multiphoton endomicroscopy, Guan et al. [4] devised a two-stage learning transfer strategy to generate augmented training datasets that would be otherwise challenging to obtain experimentally in vivo, which effectively recovered the loss of signal-to-noise ratio and resolution associated with a high frame rate (**Figure 1b**). Towards high-speed high-fidelity hyperspectral nonlinear microscopy, Lin et al. [5] achieved ultrafast tuning based on a polygon scanner and enhanced the weak Raman signals via a spatial-spectral residual learning network (**Figure 1c**). Weigert et al. [6] and McAleer et al. [7] showed promising results in high-quality image reconstruction with reduced light dosage or acquisition time via learning-based algorithms. Fan et al. demonstrated an enhanced second harmonic imaging using a deep learning decipher for efficient and resilient phase retrieval [8]. Although these computational strategies have the potential to bypass the physical limitations of nonlinear microscopy, a few challenges remain open to more investigations, including how to obtain ground truth data in demanding applications, how to avoid hallucination in data-driven reconstruction, and how to avoid data-driven bias in learning-enabled biological discoveries and clinical diagnosis.

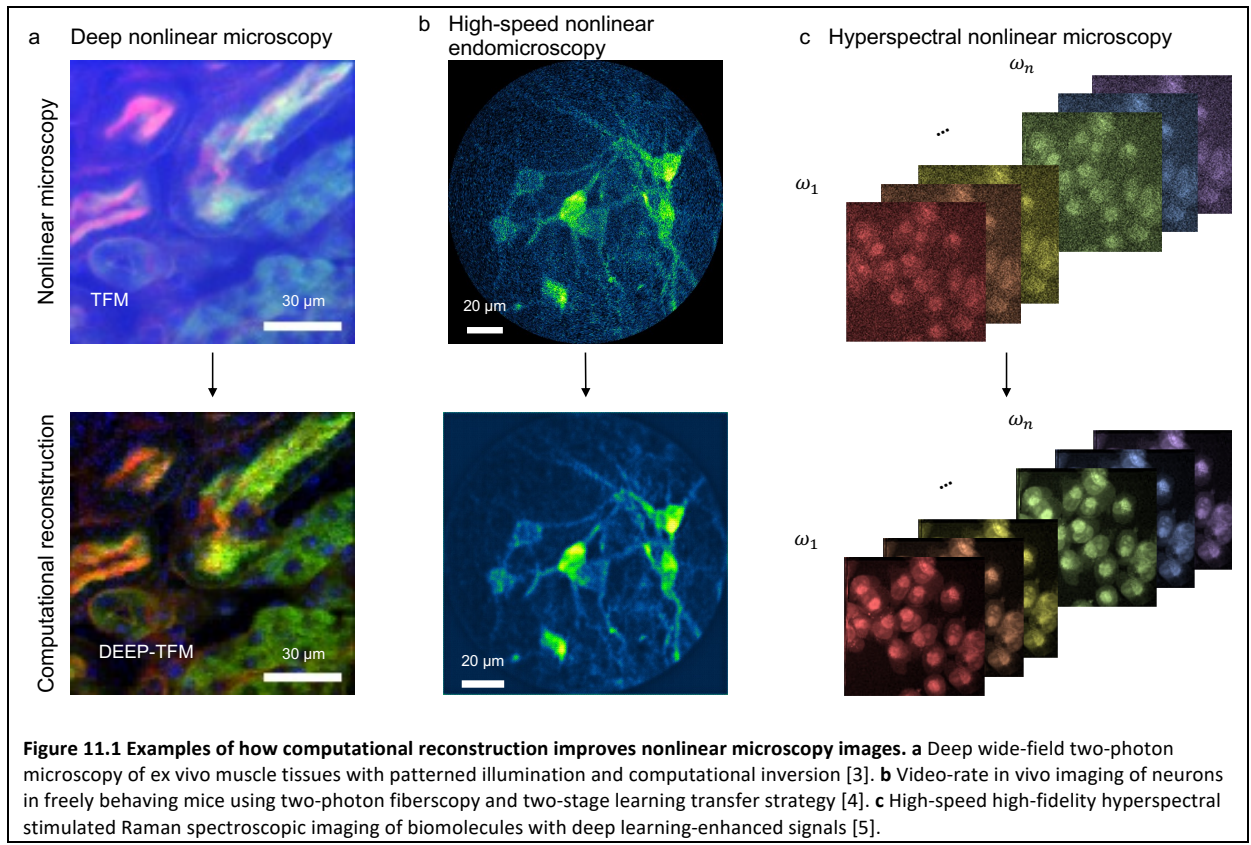

**Figure 11.1 Examples of how computational reconstruction improves nonlinear microscopy images. a** Deep wide-field two-photon microscopy of ex vivo muscle tissues with patterned illumination and computational inversion [3]. **b** Video-rate in vivo imaging of neurons in freely behaving mice using two-photon fiberscopy and two-stage learning transfer strategy [4]. **c** High-speed high-fidelity hyperspectral stimulated Raman spectroscopic imaging of biomolecules with deep learning-enhanced signals [5].

Towards better image and video understanding of multimodal nonlinear microscopy, deep learning is a promising tool due to its end-to-end data-driven principle, i.e., not relying on hand-crafted features, and its demonstrated capability of generalization to unseen data. Rapid advances in computer vision pave the way for numerous applications in optical microscopy, and many methods can be adapted to nonlinear microscopy analysis such as virtual staining [9,10], cell segmentation [11], and cancer classification [12,13]. For example, to generate histology-like images, Sun et al. [10] have demonstrated virtual H&E staining based on SLAM images, segmentation neural network, and color translation metrics (**Figure 2a**). To carry out single-cell analysis of the tumor microenvironment, You et al. [11] performed multiclass pixel-wise segmentation using a supervised U-net (**Figure 2b**). To enable real-time intraoperative assessment, Hollon et al. [12] demonstrated real-time tumor detection based on stimulated Raman histology images using a CNN-based architecture (**Figure 2c**). Despite these encouraging results, many real-world biomedical applications face the scarcity of adequate, curated training datasets, which poses a significant problem for supervised learning. To tackle this challenge, Shi et al. [14] have demonstrated weakly supervised learning to extract conventional and unconventional cancer biomarkers from the optical signatures obtained by multimodal nonlinear microscopy. While these deep-learning-based approaches open a wide array of applications, several challenges need to be addressed in the future. First, to alleviate the requirement of extensive training datasets, weakly supervised or self-supervised methods are in great demand. Generalizability is another issue. It remains a challenge to deploy well-developed methods trained on one dataset to other different biomedical applications. You et al. [15] attempted to address the issue of quality discrepancy between lab-based data and intraoperative data through physics-based data augmentation. More efforts are needed to develop algorithms that can easily adapt to microscopy systems and user differences.

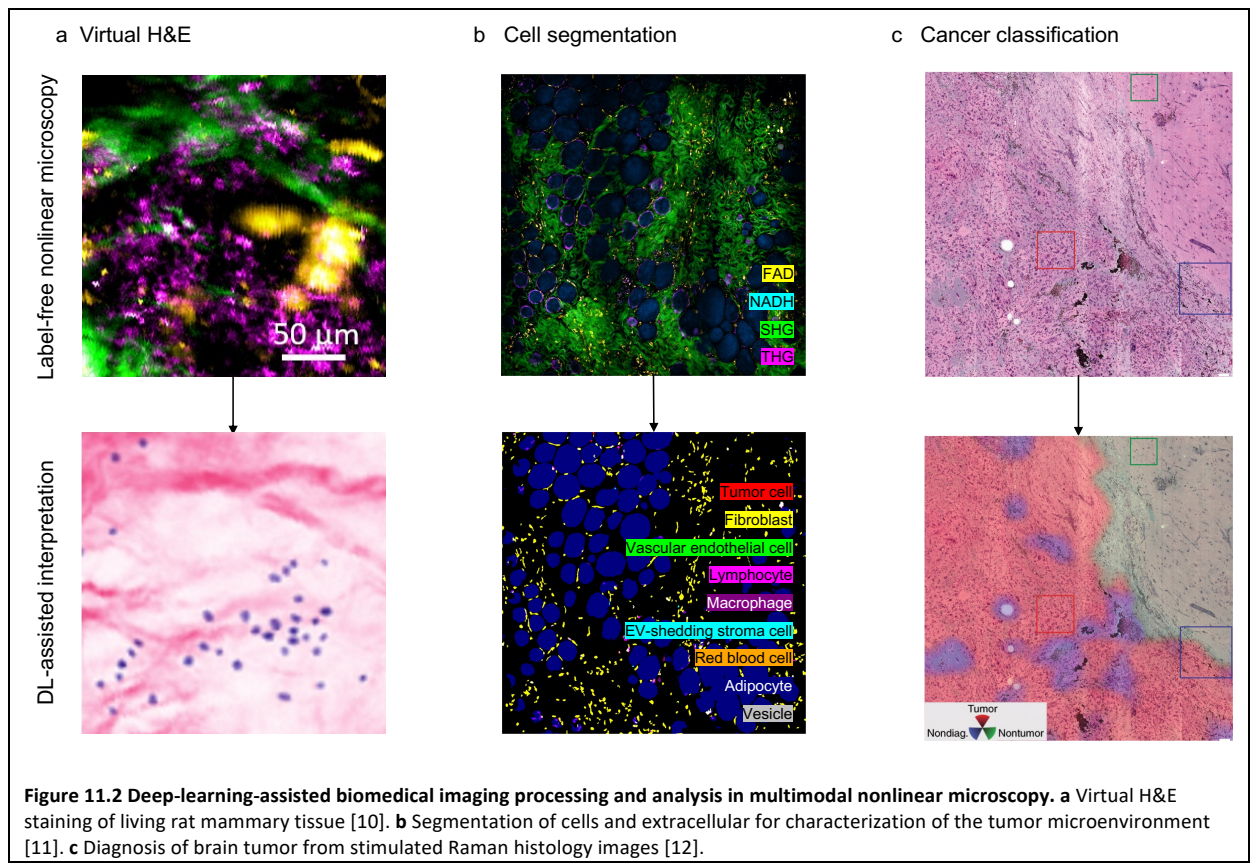

**Figure 11.2 Deep-learning-assisted biomedical imaging processing and analysis in multimodal nonlinear microscopy. a** Virtual H&E staining of living rat mammary tissue [10]. **b** Segmentation of cells and extracellular for characterization of the tumor microenvironment [11]. **c** Diagnosis of brain tumor from stimulated Raman histology images [12].

**Concluding Remarks**

Multimodal nonlinear microscopy has become an indispensable tool for high-resolution imaging of living biological systems. Rapid advances in deep learning are and will keep being leveraged to push the limits of multimodal nonlinear microscopy. This perspective has focused on how deep learning helps push the technical limits of multimodal nonlinear microscopy as well as enhancing its biomedical translation and impact. Looking forward, we expect these efforts will accelerate the clinical translation and biological discovery based on nonlinear microscopy with caution and innovation. There will also be other exciting work showing how deep learning can further advance the design of nonlinear microscopy, such as learning-based imaging system design, as demonstrated in the prospering community of computational photography and microscopy.

**Acknowledgments**

*T. Q., L. Y., and S. Y acknowledge the support of Jameel Clinic, Scialog RCSA Award, Amazon Research Award, and MIT.*

## 12 — Machine Learning Driven Automated Scanning Probe Microscopy

Yongtao Liu[1], Maxim A. Ziatdinov[1,2], Sergei V. Kalinin[3]

1. Center for Nanophase Materials Sciences, Oak Ridge National Laboratory, Oak Ridge, TN 37830, USA
2. Computational Sciences and Engineering Division, Oak Ridge National Laboratory, Oak Ridge, TN 37830, USA
3. Department of Materials Science and Engineering, University of Tennessee, Knoxville, TN 37996, USA

**Status**

35 years after the development of atomic force microscopy (AFM) by Binnig and Rohrer [1, 2], Scanning Probe Microscopy (SPM) has become the mainstay technique in areas ranging from condensed matter physics and materials science to biology and medicine [3]. The highly robust nature of SPM has spawned multiple types of machines operating in the ambient environment, controlled atmosphere, liquid, and ultrahigh vacuum. Combined with the electrical, chemical, magnetic functionalization of the probes, the gamut of SPM imaging modes is truly broad. Complementing imaging, many SPM techniques allow a broad variety of spectroscopic measurements, ranging from the force-distance curves in conventional AFM, current-voltage curve in Scanning Tunnelling Microscopy (STM), and a broad variety of time- and voltage-spectroscopies in techniques such as Piezoresponse Force Microscopy (PFM) [4]. The development of the SPM modes is seamlessly tied to the priorities of the R&D community, with the continuous growth wave of magnetic force microscopy in the late 1990s being driven by the magnetic hard drive industry, development of PFM stimulated by the ferroelectric non-volatile memories and tunnelling barriers and recently discovery 2D ferroelectricity, and force-distance measurements providing insight into statistical physics of biomolecules. It can be argued that exponential growth of experimental effort in materials for energy storage and conversion will guide the SPM progress over the next decade. Here, techniques such as Electrochemical Strain Microscopy (ESM)[5] can become invaluable for probing electrochemical reactivity in nanoscale volumes of batteries and fuel cells, whereas light-assisted electrical SPMs are likely to emerge as techniques of choice for probing photovoltaic materials and devices. Similarly, STM and associated spectroscopies will grow as techniques of choice to explore quantum behaviour of materials on atomic level, as well as enabling tool for single-atom manipulation and assembly of atomic scale devices.

**Current and Future Challenges**

Despite the tremendous progress in SPM instrumentation and continuously growing number of imaging modalities and experimental platforms worldwide, the basic principles of SPM remained unchanged from the early days of Binnig, Quate, and Rohrer [2]. The SPM is based on continuous raster scan of the probe, effectively sampling the probe-surface interactions over the uniform rectangular grid of points. The spectroscopic measurements are enabled either as point and click approach by operator, or via hyperspectral imaging modes in which the spectroscopic data is acquired over the uniform grid. This acquisition mode is convenient from the instrumental implementation, human perception, and mathematical analysis perspectives. For hyperspectral measurements, a number of physics- or data-driven approaches have been developed to convert the high dimensional spectroscopic data sets to the set of 2D representations [6]. However, for realistic materials the information of interest is often concentrated in a small amount of locations. For example, in biological systems the molecules deposited on the surface are often of a higher interest then the substrate between them. In ferroelectric materials, functional responses of structural defects such as grain boundaries or topological defects such as domain walls are often of higher interest then the responses in the uniform-domain regions. This consideration is particularly important for the spectroscopic measurements. Here, the grid measurements are often very time consuming and can be associated with the tip damage. Perhaps even more importantly, many spectroscopic measurements can affect the state of material due to reversible or irreversible processes, for example shift ferroelectric domain walls, induce local electrochemical reactions, or plastic deformation of material. Hence, it is of interest the development of the instrumental workflows with the varying density of imaging points, and particularly methods for active experiment in SPM. In these, the algorithm updates the locations for image or spectroscopic measurements based on the measurements results within the same experiment. Here, we disambiguate three types of automated experiment, namely (a) adapting sampling for the scalar or multimodal measurements (**Figure 12.1a**), (b) forward spectroscopic experiment in which the a priori known objects of interest are discovered in real time and spectroscopic measurements are taken (**Figure 12.1b**), and (c) the inverse experiments in which the spatial structures that correspond to functional behaviours of interest are discovered (**Figure 12.1c**). We note that while (b) roughly corresponds to the operation of a human microscopist, the tasks (a) and (c) are purely amenable to human operation.

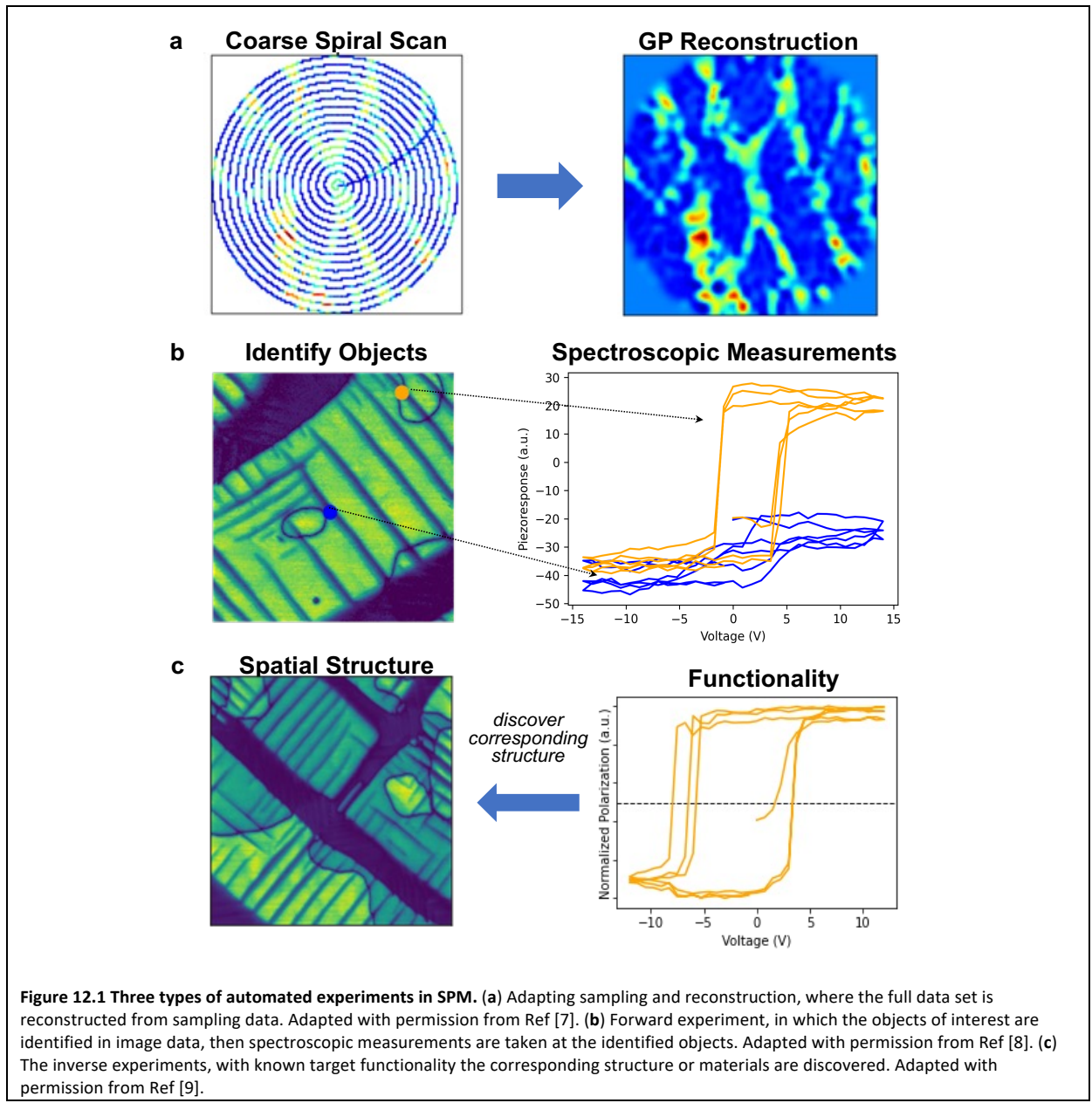

**Figure 12.1 Three types of automated experiments in SPM.** (**a**) Adapting sampling and reconstruction, where the full data set is reconstructed from sampling data. Adapted with permission from Ref [7]. (**b**) Forward experiment, in which the objects of interest are identified in image data, then spectroscopic measurements are taken at the identified objects. Adapted with permission from Ref [8]. (**c**) The inverse experiments, with known target functionality the corresponding structure or materials are discovered. Adapted with permission from Ref [9].

**Advances in Science and Technology to Meet Challenges**

Developing the automated experiment workflows requires the synergy of three components, namely engineering controls (i.e., the algorithm should be able to issue the control commands to the microscope), machine learning algorithms, and definition of the reward function. While the former two components are obvious, the third component (reward) is traditionally less recognized. However, it is clear that even for applications such as automated driving, the chosen pathway will be very different if reward is safety (favouring very slow driving) vs. time. For physical experiments, the reward is considerably more complex and has to be defined in the context of the specific physical experiment, prior hypotheses, etc. Here, for the reconstruction problems (a), the enabling algorithm can be variants of adaptive sparse sampling, e.g., based on Gaussian Processes. The reward function in this case is defined by the balance between the quality of reconstruction and minimization of samplings. However, while initially perceived to be promising, these algorithms often lead only to the insignificant (factor of 2-3) reduction of sampling points at the cost of more complex scanning protocols. This behaviour can be traced to the presence of the multiple length scales of the images that preclude effective construction of GP kernels. From a more general perspective, the classical uniform scanning of the SPM corresponds to the fully open prior, and hence is often optimal. The experiments (b) rely on the a priori defined objects of interest that can be recognized in real time using deep convolutional networks. Here, the emergence of the ensemble and iterative training methods allowed to address the inevitable out of distribution effects (i.e., capability of the trained network to recognize objects of interest if microscope parameters have changed). Recently, a deep residual learning framework with holistically-nested edge detection (ResHedNet) has been ensembled to minimize the out-of-distribution drift effects in real-time SPM measurement [8]. The ensembled ResHedNet is implemented in operating SPM, and converts the real-time PFM data stream to segmented ferroelastic domain wall images. Then, a pre-defined workflow uses the discovered domain walls as the coordinates for spectroscopic measurements. In doing so, the approach allows a thorough investigation of domain walls (virtually all locations at domain walls) in an automated manner, in contrast, traditional manual operation only allows us to investigate a limited amount of locations at domain walls. Using this approach, alternating high- and low-polarization dynamic ferroelastic domain walls in a $PbTiO_3$ thin film is observed. Finally, the emergence of the deep kernel learning (DKL) methods allows to implement the inverse spectroscopic workflow. Here, the operator defines the characteristics that make the spectrum of interesting, e.g., intensity of a specific feature, specific aspect of spectrum shape, or even maximal variability of spectra within the image. In other words, each collected spectrum can be associated with a single number defining how "interesting" it is, in absolute sense for compared to previously acquired spectrum. The DKL algorithm learns what elements of the materials structure maximize this reward, and guides the exploration of materials surface accordingly. This DKL algorithm is recently implemented in SPM to investigate the relationship between ferroelectric domain structure and polarization dynamics [9]. Both the DKL exploration process and results are interesting. As show in **Figure 12.2**, the DKL exploration process demonstrates the domain walls to be interesting and the DKL results indicate the high polarization dynamic of 180º domain walls. Although these are expected by ferroelectric experts, DKL itself does not have any physical knowledge, all information is actively learned during the experiments.

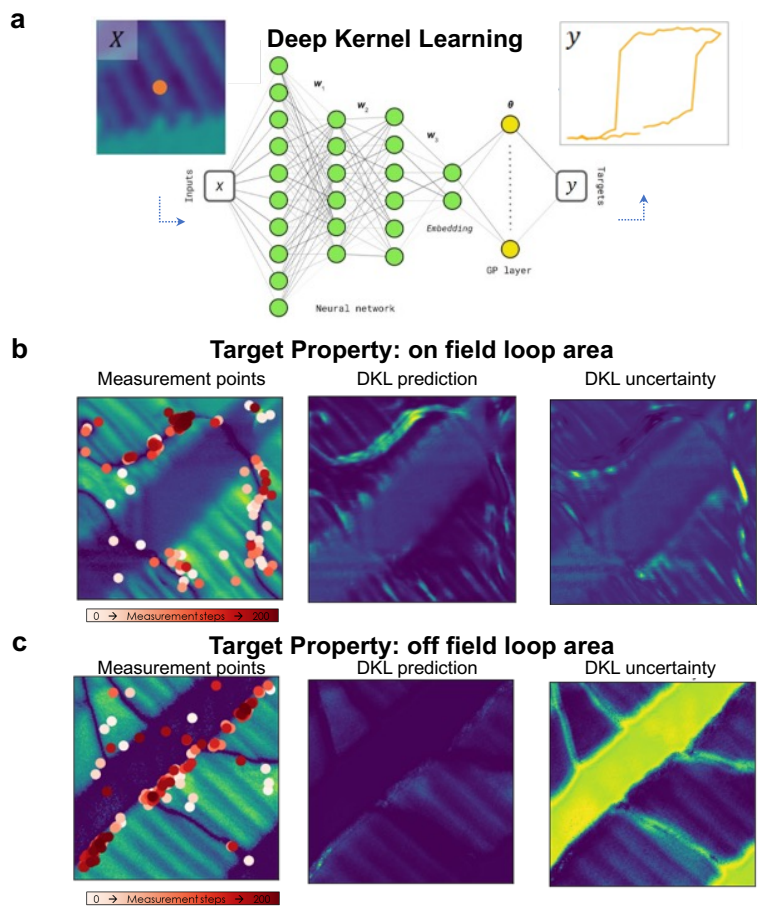

**Figure 12.1 Deep kernel learning piezoresponse force microscopy (DKL-PFM).** (**a**) Deep kernel learning structure: it contains a neural network to encode the properties of image data into a low-dimensional latent space and a Gaussian process layer operating over the latent space to analyze the relationship between image data and target property encoded in spectroscopic data. (**b-c**) DKL-PFM results on a PbTiO$_3$ thin film, which show that the DKL decided measurement points are around domain walls (indicating domain walls to be interesting) and predict that domain walls have large polarization dynamics. Adapted with permission from Ref. [9].

**Concluding Remarks**

Machine learning (ML) methods have revolutionized many aspects of modern science including computer vision and image generation, medical and biological imaging, planning and prediction. The growth of the open code and data culture results in rapid propagation of ML algorithms between domains. Combined with the introduction of the Python control interfaces in modern microscopes, this poses the field of SPM for a transformative change. However, taking full advantage of this opportunity requires developing the connection between the domain areas and ML, including defining the domain-specific rewards that will steer the experiment, introducing the invariances and defining the biases that should be ignored, and adoption of the required ML tools required to build them. With these, automated experiments provide opportunities to perform experiments that are challenging in traditional methods because of experimental parameters are too complex and numerous (e.g., too many locations/conditions are required to investigate). In addition, automated experiments offer the opportunities for solving problems that are challenging to human beings, e.g., inverse problem. The active learning approach can learn the relationship between target functionality and material behavior quicker by several orders of magnitude than traditional method. As the development of ML and its connection to domain science, automated experiments are expected to accelerate the material and physics discovery.

**Acknowledgements**

*This effort was primarily supported by the Center for 3D Ferroelectric Microelectronics (3DFeM), an Energy Frontier Research Center funded by the U.S. Department of Energy (DOE), Office of Science, Basic Energy Sciences under Award Number DE-SC0021118. The effort was partially supported (piezoresponse force microscopy) at Oak Ridge National Laboratory's Center for Nanophase Materials Sciences (CNMS), a U.S. Department of Energy, Office of Science User Facility.*

# 13 — Image restoration for scanning microscopy

Arlo Sheridan[1], Uri Manor[1]

1. Salk Institute for Biological Studies, La Jolla, CA, USA

**Status**

Point-scanning imaging systems are among the most commonly used microscopy imaging modalities due to their versatility and accessibility [1,2]. While these tools have had a profound impact on advances in the life sciences, they come with a cost. Since it is difficult to simultaneously optimize imaging speed, resolution and sample preservation, one parameter is always compromised at the benefit of the others (aptly deemed the "eternal triangle of compromise") [2]. For example, imaging speed and sample damage can be enhanced by ensuring a fast pixel dwell time, but at the expense of a lower resolution and signal-to-noise ratio (SNR). It is vitally important to correct for these shortcomings to facilitate future scientific breakthroughs. To address these challenges, deep learning-based methods have been developed to subsequently improve the negatively affected component(s) following image acquisition. Over the past several years, deep-learning breakthroughs have advanced the field in both denoising and super-resolution of point-scanning microscopy modalities including fluorescence and electron microscopy. While initial approaches [1-3] centered around fully supervised learning via pairs of images of the same samples acquired at low versus high quality, promising new semi-supervised [4] and fully unsupervised methods [5-7] are beginning to demonstrate competitive results that allow investigators to avoid the technically difficult and costly process of acquiring paired data. Optimization of image acquisition is a costly process. Deep-learning-based restoration and super-resolution not only provide a solution to the "eternal triangle of compromise", but substantially decrease the amount of imaging time (and subsequent costs) to conduct necessary experiments for downstream analysis. In order to further minimize these costs, future work should aim to improve fully unsupervised approaches and, ideally, integrate these into imaging hardware. Additionally, it is imperative to ensure that these methods are openly accessible to the broad community via intuitive, easy-to-use software that does not depend on a resident computer scientist. These advances will alleviate the burdens on researchers and allow them to focus on their scientific questions.

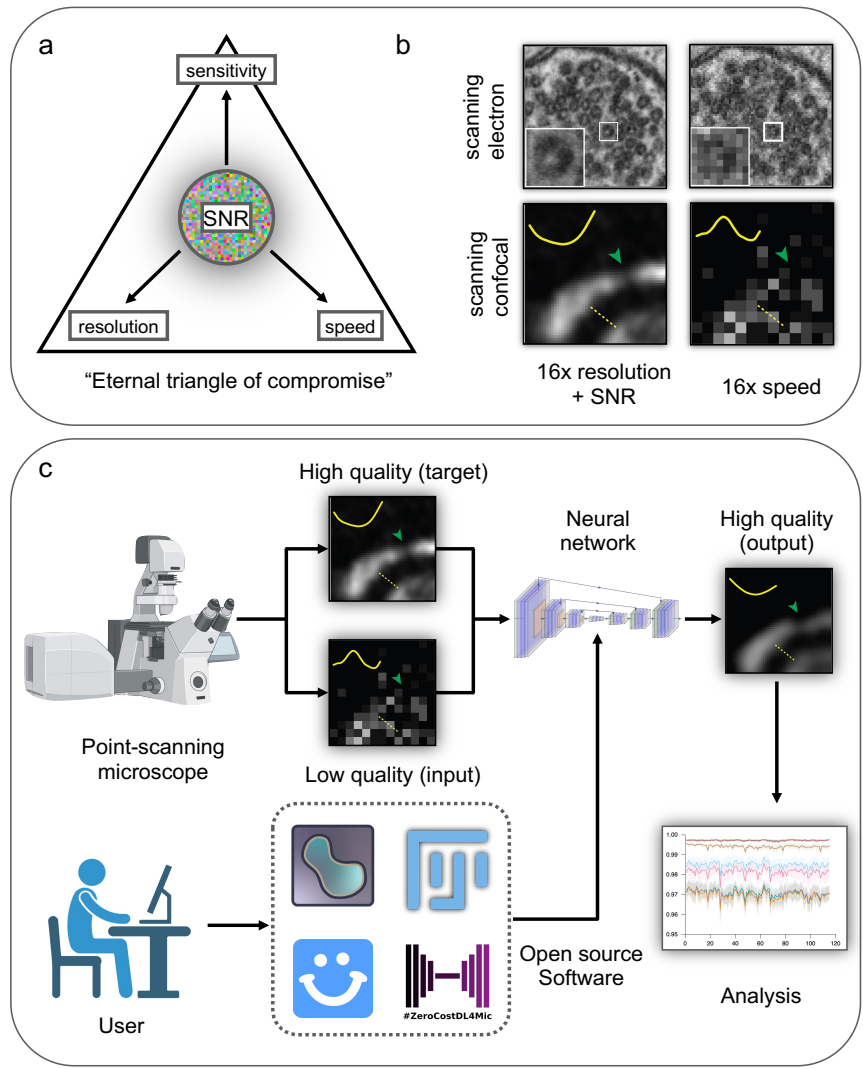

**Figure 13.1 Image restoration for scanning microscopy.** (**a**) The "eternal triangle of compromise" illustrates the necessary compromises for optimizing each imaging component. (**b**) Examples illustrating the tradeoffs between imaging speed and resolution/SNR for scanning electron microscopy (SEM) and confocal microscopy data (top and bottom rows, respectively). (**c**) Overview of an example workflow for integrating deep learning with open-source software to accelerate scientific discovery. Microscope and user schematics created with BioRender.com.

**Current and Future Challenges**

While deep-learning approaches offer a practical solution for overcoming common pitfalls with scanning microscopy, there are several limitations that pose a challenge for current methods. Fully supervised methods still require the acquisition or generation of pairs of high- and low-quality images of the same samples to create training data [1,2,3]. Here we are using the term "quality" to include both signal-to-noise ratio (SNR) and/or resolution. The high-quality data typically must exceed the desired quality of the output of the trained network. Acquiring image pairs of the same sample can be technically difficult, costly, and sometimes impossible; often, samples can only be imaged once [2]. Generating and validating semi-synthetic data (e.g., a "crappifier" that generates low-quality data from high-quality data, [2]) also requires significant effort, and validation still depends on real-world image pairs of the same sample. These data requirements are also hampered by the high cost of imaging and storing high quality data, which is proportional to image resolution [2]. Unsupervised approaches aim to solve this by enabling image restoration from noisy or low-resolution input images alone. While leading methods [5] generate impressive reconstructions, it is still challenging to match the accuracy of fully or semi-supervised results. For example, successfully denoising pixel-wise independent noise does not necessarily remove spatially correlated (structured) noise [5]. The extra effort to generate target data for fully supervised approaches is therefore sometimes unavoidable. Additionally, image restoration is an inherently ill-posed problem, i.e., multiple different high-quality images could be used to generate a single low-quality image, [4,7]. While this is less of a concern for popular tasks such as restoring photographs, restoring scientific imaging data can be challenging because of a greater need for accuracy. Any inconsistency in the fidelity of reconstructions can have harmful effects on downstream analyses. For example, super-resolution can be used to quantify objects in microscopy data imaged with a lower resolution than is normally required to resolve these structures. For this approach to be useful, the model must be validated to show it did not introduce errors resulting in inaccurate quantification of these objects [2]. Furthermore, for image restoration methods to realize their full potential they must be accessible to the community via easy-to-use graphical user interfaces (GUIs) [8,9]. Most scientists do not have the expertise necessary to re-implement deep learning approaches, and running established code often poses a steep learning curve. However, it is not always trivial to implement robust solutions that satisfy all requirements. Ideally, software should be easily packaged, distributed, and hosted on a well-maintained platform. Critically, having turn-key methods for generating model outputs will greatly facilitate validation of model outputs by domain experts. Validation should never rely solely on pixel-based metrics such as PSNR or SSIM. Instead, validation workflows should integrate visual inspection by human experts, multiple pixel-based metrics, and, most importantly, comparing the final measurement of the experiment with ground truth, validated data. For example, the number and size of objects of interest (e.g., presynaptic vesicles) generated in the model output should match the ground truth data as closely as possible [2].

**Advances in Science and Technology to Meet Challenges**

Even though fully supervised methods require the acquisition of high-quality data, the standardization of imaging techniques and advances in data storage have been transformative over recent years. Many microscopy imaging systems now allow for automated collection and alignment of images. It is also considerably cheaper nowadays to store large amounts of data (for example, storing a terabyte of data on AWS on average costs a relatively modest $20 per month[1]). As imaging datasets become more widely shared, it may be feasible to generate learned crappifiers that can be used to quickly train new supervised models. Nevertheless, in an ideal world, unsupervised approaches would be preferred. Despite the aforementioned challenges, denoising methods can now handle structured noise in addition to pixel-wise noise and recent results indicate that accuracy on par with fully supervised methods is achievable. Self-supervised and unsupervised methods for super-resolution are also rapidly advancing [10]. Furthermore, some methods now provide a distribution of restorations instead of a single output [7]. This is of great practical relevance for solving inverse problems in which multiple solutions may be proposed. Theoretically, it should allow a user to choose different solutions across regions based on additional context. For example, if one corner of an image is much noisier than the center, a single solution might provide reasonable results in the center and poor results in the corner. In these cases, having a range of solutions that a user can choose from is preferable. The full impact of these advances can only be achieved if they are accessible to a wide range of users with various backgrounds. In order to cater to the broad community, platforms with intuitive user interfaces and continued support must exist. Over the

---

[1] https://aws.amazon.com/s3/pricing/

years, FIJI/ImageJ[2] (the *gold standard*) has allowed scientists to implement a wide range of image analysis techniques on their data. Next generation tools such as Napari[3], ImJoy[4], and ZerocostDL4mic [8] aim to extend this accessibility and enhance it through the integration of deep learning approaches. These tools and many others [9] will allow users to analyze their data in ways that were previously unattainable without sufficient expertise.

**Concluding Remarks**

Advances in biological imaging are moving at an exponentially increasing pace. As it becomes easier to acquire large amounts of images via point-scanning microscopy systems, there is an increasing need to accurately and efficiently analyze the resulting data. Since image acquisition poses challenges for the optimization of the "eternal triangle of compromise", deep learning methods are exciting solutions for image restoration. Since these are difficult inverse problems to solve, ideal solutions need to be robust to data ambiguities and should limit the burden on researchers to collect and store high quality target data. Additionally, efforts should ensure the accessibility of methods to a broad user base across scientific domains. With the development of recent deep learning approaches and the growing communities dedicated to developing open-source software and datasets, it is clear that the field is moving in the right direction.

**Acknowledgements**

*Arlo Sheridan and Uri Manor are supported by the Waitt Foundation, Core Grant application NCI CCSG (CA014195), NIH (R21 DC018237), NSF NeuroNex Award (2014862) and the Chan-Zuckerberg Initiative Imaging Scientist Award.*

---

[2] https://imagej.net/software/fiji/
[3] https://napari.org
[4] https://imjoy.io

# 14 — Single Molecule Localization

Elias Nehme[1,2], Ofri Goldenberg[2], Yoav Shechtman[2]

1. Department of Electrical and Computer Engineering, Technion – IIT, Israel
2. Department of Biomedical Engineering and Lorry I. Lokey Interdisciplinary Center for Life Sciences & Engineering, Technion – IIT, Israel

**Status**

The spatial resolution of conventional microscopes is fundamentally bounded by the diffraction limit at approximately half the wavelength of the light, which, in the visible range, practically corresponds to 200-300 nm. To overcome this limitation, a variety of super-resolution microscopy techniques have been developed including structured illumination microscopy, stimulated emission depletion (STED), as well as single molecule localization microscopy (SMLM) [1], which is the focus of this perspective. The main working principle of SMLM relies on a space–time trade-off: instead of capturing a single image of a fluorescent sample, a movie consisting of many frames (typically thousands) of temporally blinking fluorophores is acquired. In each frame, only a sparse, random set of fluorophores is activated, and their positions are determined computationally. There are multiple chemical and physical mechanisms to obtain blinking, and a myriad of associated acronyms, most notably PALM, STORM, and PAINT [1]; however, subsequent analysis is similar between the different variations. After data acquisition, the resulting localizations are combined numerically to render a single, computationally super-resolved image, typically, with an order of magnitude improvement in resolution (**Figure 14.1**).

SMLM revolutionized biological research, enabling *nanoscale* imaging of biological structures and tracking of single-particles [2], thereby earning its pioneers the Nobel Prize in Chemistry in 2014. Compared to other high-resolution imaging modalities like electron microscopy, SMLM offers the high specificity and SNR of fluorescence microscopy, as well as the possibility to image living cells.

With the great advancement in resolution introduced by SMLM, came unique experimental and algorithmic challenges. The main issue with trading-off space over time, manifested in the per-frame emitter *sparsity* constraint, is the sacrifice of temporal resolution, which imposes strict limitations on the ability to image dynamic processes. This problem becomes more acute as we are interested in sensing more and more physical properties of the imaged sample, such as depth (3D), color information, and molecular orientation.

In recent years, Deep Learning (DL) has found tremendous success in handling some of these challenges [3]. In particular, DL-based end-to-end optimization (E2E) for joint design of sensors and algorithms [4,5,6] has led to powerful task-driven experimental designs, significantly pushing the barriers of the spatiotemporal trade-off in localization microscopy (**Figure 14.2**). Such designs are starting to find their applications in scientific research [7,8]; however, work remains to be done in making these methods fully-mature and widely adoptable by the biological research community. Specifically, efforts should be invested in robustifying SMLM algorithms for a wide range of experimental conditions and providing accessible software packages to end users [9].

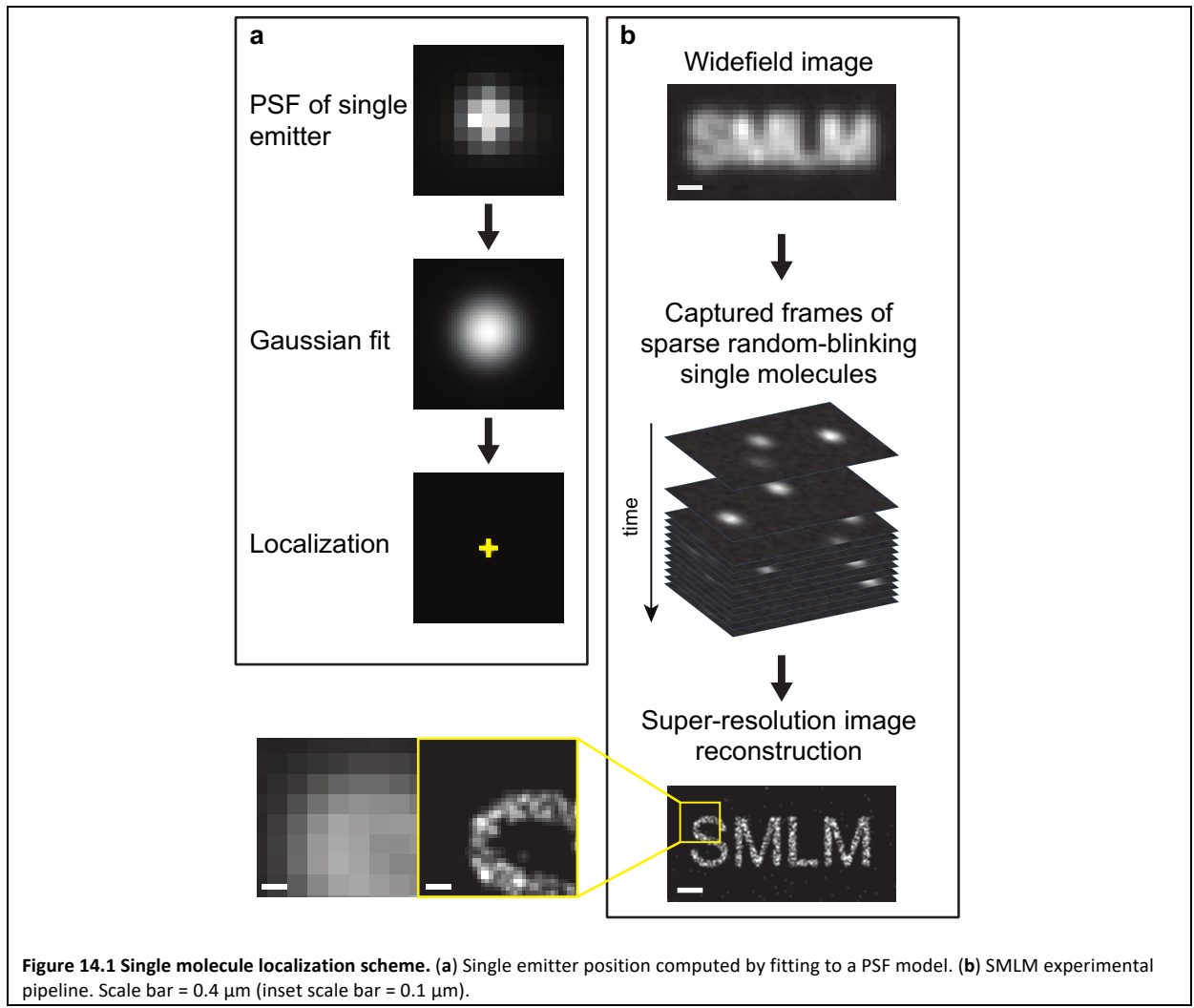

**Figure 14.1 Single molecule localization scheme.** (**a**) Single emitter position computed by fitting to a PSF model. (**b**) SMLM experimental pipeline. Scale bar = 0.4 µm (inset scale bar = 0.1 µm).

**Current and Future Challenges**

Here, we outline four main challenges that are key to address in order to improve the capabilities and accessibility of SMLM:

1. **Improving spatiotemporal resolution**. For imaging fast dynamical processes at high resolution using SMLM, its temporal resolution needs to be improved. The trick to enhance its spatio-temporal resolution is to trade off experimental time with information-rich image data containing complex patterns. The latter needs to be then analyzed by powerful algorithms to extract the underlying physical information. For example, relaxing the emitter sparsity constraint to increase temporal resolution in SMLM requires image processing algorithms that can handle nearby emitters with overlapping Point Spread Functions (PSFs) [10,11].
2. **Enhancing information extraction**. There is often a need to look beyond the 2D positions of emitters. For example, by tagging different targets (e.g., proteins) with distinct fluorescent emitters, we can capture correlative information that reports on inter-specie relations. One exemplary challenge, in this case, is to classify the color of each emitter based on the acquired PSFs, which are typically captured on highly photon sensitive, yet spectrally insensitive (grayscale) detectors. Another physical property of interest is emitter depth; since biological structures are three-dimensional, we can gain significant insight by looking into their 3D organization (3D SMLM).

3. **Optimal information acquisition**. Imaging systems in SMLM are typically optimized through physical intuition-based heuristics or through mathematical measures quantifying information content, e.g., Fisher Information. However, such methods are still limited in their ability to adapt to challenging experimental conditions and are difficult to generalize to systems acquiring multiple physical properties jointly. Hence, optimal designs of acquisition schemes are key to unlocking the full potential of SMLM.
4. **Increasing algorithm reliability**. DL algorithms are extremely powerful and have led to a performance revolution in solving inverse problems. However, little is understood about their inner-workings and their failure modes. Specifically, for SMLM, there is need for sample-adaptive self-tuning algorithms, that can handle arbitrary experimental conditions without need for extensive calibration. More importantly, in a field such as biological research where algorithms are expected to drive scientific discovery, we need to be able to quantify uncertainty and bias in the reconstructions [11]. At the moment, DL algorithms have a hard time "knowing when they don't know", and we anticipate extensive work to be done on this front to improve their reliability.

**Advances in Science and Technology to Meet Challenges**

Deep Learning has proven to be highly successful in handling some of the fundamental challenges in super resolution microscopy [3]. DL is particularly suited for SMLM, because large, paired training sets, which are the main bottleneck in supervised learning, can be generated easily using accurate simulators based on physical models. The first application of DL to 2D SMLM was presented in Ref. [10]. The authors showed that the acquisition time could be shortened significantly down to a few hundred frames by proposing a DL algorithm that is able to reconstruct dense emitters with overlapping PSFs. A similar strategy was used later in Ref. [7] for tracking chromatin dynamics in living cells. Concurrent work proposed content-aware strategies to increase the temporal resolution [12], and even decrease the number of necessary frames to a single image [13]. Similarly, DL was also applied for sensing additional physical properties from grayscale 2D measurements such as color classification [4,14] and depth estimation from high-density data with engineered PSFs [5,11,15].

A particularly promising application of DL in SMLM is the ability to tailor the acquisition scheme to the task at hand. Specifically, End-to-End (E2E) optimization of the physical acquisition system jointly with the data processing algorithm, holds great promise [4,5,6]. E2E optimization taps into the full potential of existing hardware as well as making use of new hardware in a task-driven manner. This synergy between acquisition and reconstruction enables us to design significantly more complex sensing paradigms that make full use of the photons, as well as powerful DL-based reconstruction algorithms. DL was originally invented to approximate functions which were hard to mathematically define, and optical instrumentation design is no different. Recently, E2E optimization was shown to be extendible to multiple sensor designs [6,16], offering even greater temporal resolution at the expense of a more complex optical setup.

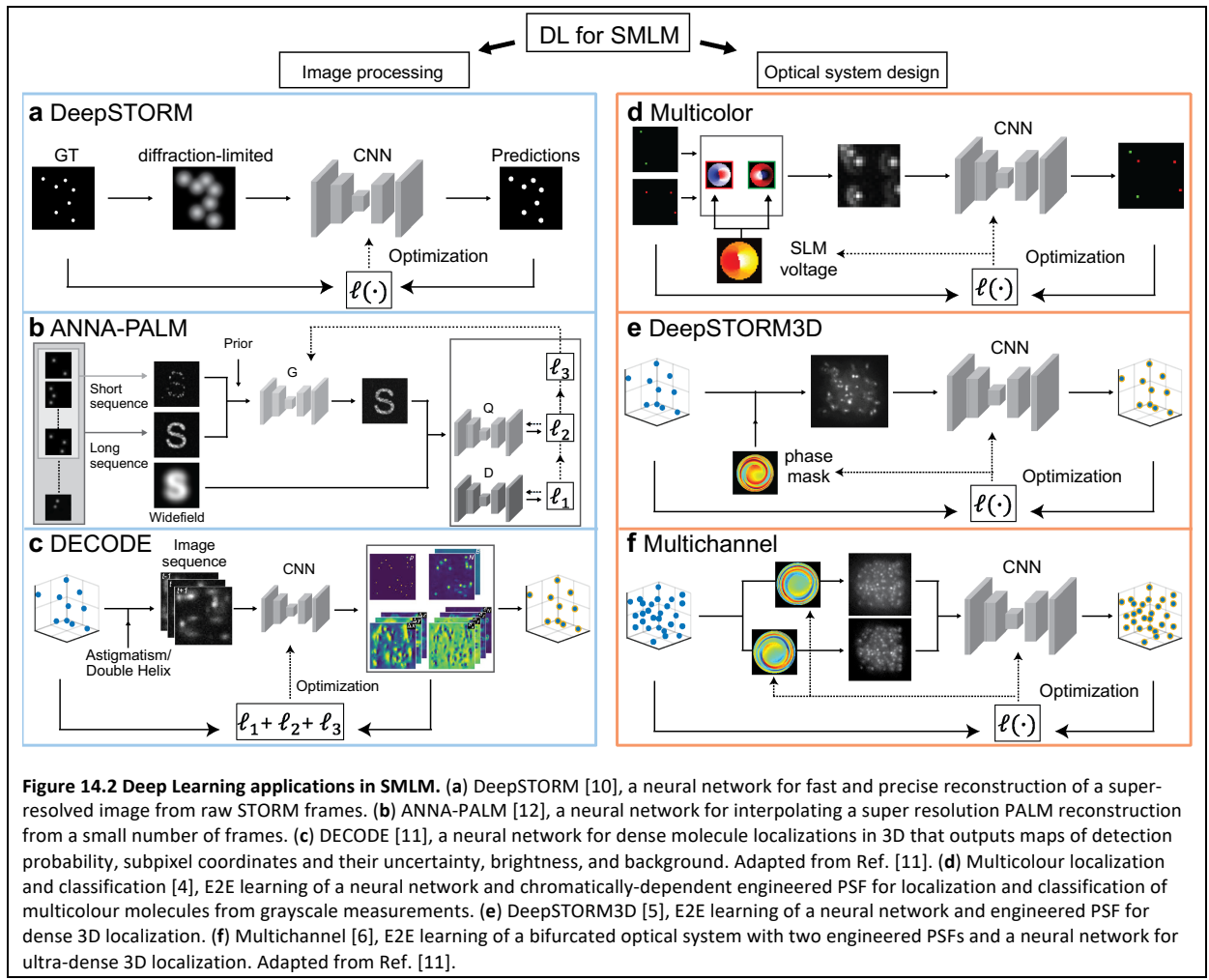

**Figure 14.2 Deep Learning applications in SMLM.** (**a**) DeepSTORM [10], a neural network for fast and precise reconstruction of a super-resolved image from raw STORM frames. (**b**) ANNA-PALM [12], a neural network for interpolating a super resolution PALM reconstruction from a small number of frames. (**c**) DECODE [11], a neural network for dense molecule localizations in 3D that outputs maps of detection probability, subpixel coordinates and their uncertainty, brightness, and background. Adapted from Ref. [11]. (**d**) Multicolour localization and classification [4], E2E learning of a neural network and chromatically-dependent engineered PSF for localization and classification of multicolour molecules from grayscale measurements. (**e**) DeepSTORM3D [5], E2E learning of a neural network and engineered PSF for dense 3D localization. (**f**) Multichannel [6], E2E learning of a bifurcated optical system with two engineered PSFs and a neural network for ultra-dense 3D localization. Adapted from Ref. [11].

## Concluding Remarks

SMLM is a powerful tool in bio-imaging, with new applications emerging quickly. The combination of SMLM with DL enables efficient analysis and system-design for obtaining information-rich imaging data. Due to the nature of E2E designs, this interdisciplinary field requires tight collaboration between experimentalists and data scientists to fully exploit existing sensors and compute power. Focus thus far has been mainly in improving reconstruction algorithms [17]; however, we anticipate that more and more applications will emerge in which acquisition systems and algorithms are designed jointly. Furthermore, the development of new sensors and hardware will drive DL-based design of new optimal acquisition systems. In any case, there will be need for standardized metrics to assess performance, reproducibility, and reliability of DL-based SMLM algorithms. Finally, caution needs to be taken and uncertainty quantification should be addressed before existing tools can be used for scientific discoveries; this is especially important for methods that assume strong prior information on the object to be recovered.

## Acknowledgements

*This work was supported by H2020 European Research Council Horizon 2020 (802567) and by the Israel Science Foundation (grant No. 450/18).*

## 15 — Nanofluidic scattering microscopy

Henrik K. Moberg[1], Christoph Langhammer[1], Barbora Špačková[2]

1. Department of Physics, Chalmers University of Technology; Göteborg, Sweden
2. Department of Optical and Biophysical Systems, Institute of Physics of the Czech Academy of Sciences; Prague, Czech Republic

**Status**

Recent years have witnessed the rapid development of new fields of science, such as nanobiology, single-molecule biochemistry, and biophysics. A myriad of information has been obtained which had not been anticipated based on conventional ensemble-averaged measurement and microscopy tools. At the same time, single-molecule imaging methods have long relied on fluorescent labeling [1]. This labeling involves the chemical attachment of light-emitting moieties to the biomolecules of interest, which in turn may alter their natural state and, thus, their biological activity and function. This drawback has been the driving force for the development of label-free optical microscopy methods. Among these techniques, Interferometric Scattering Microscopy (iSCAT) has been successfully employed in many applications of imaging and analysis of individual biomolecules [2]. These studies are however limited to biomolecules attached to a surface and do not reveal their natural diffusive motion in solution.

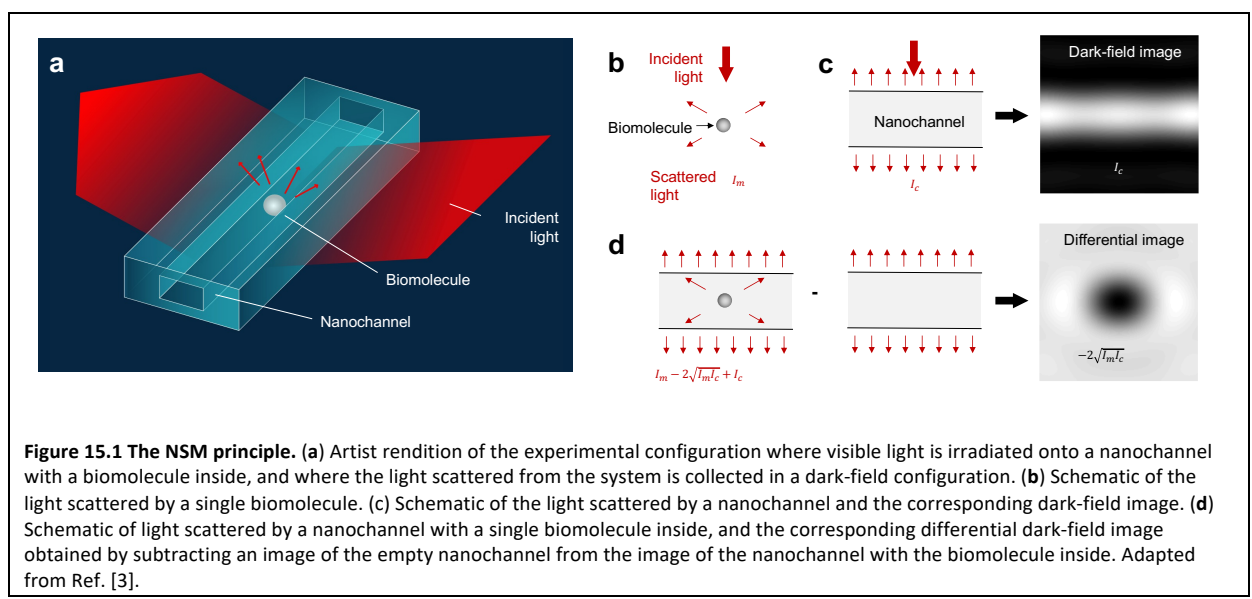

**Figure 15.1 The NSM principle.** (**a**) Artist rendition of the experimental configuration where visible light is irradiated onto a nanochannel with a biomolecule inside, and where the light scattered from the system is collected in a dark-field configuration. (**b**) Schematic of the light scattered by a single biomolecule. (**c**) Schematic of the light scattered by a nanochannel and the corresponding dark-field image. (**d**) Schematic of light scattered by a nanochannel with a single biomolecule inside, and the corresponding differential dark-field image obtained by subtracting an image of the empty nanochannel from the image of the nanochannel with the biomolecule inside. Adapted from Ref. [3].

To this end, we have recently introduced Nanofluidic Scattering Microscopy (NSM) [3], whose unique underlying principle enabled to bypass those limitations: NSM is capable of imaging individual biomolecules in free motion without any label. This tool relies on the interference of light scattered from a nanofluidic channel and a nano-object inside it — such as a biomolecule (**Figure 15.1**). To enable quick and accurate analysis of the recorded images, we employ a deep learning-based computer vision (CV) pipeline, consisting of three convolutional-based neural network (CNN) architectures (**Figure 15.2**). In Ref. [3], this pipeline is employed to identify and characterize a diverse array of biological nanoparticles ranging from extracellular vesicles down to DNA molecules and proteins of molecular weight down to the tens of kDa regime. Specifically, given an image containing the scattered light of several biomolecules inside a nanochannel over time (known as a "kymograph"), the CV pipeline outputs the trajectory of each separate biomolecule along with its two key characteristics — its molecular weight ($MW$) and its hydrodynamic radius ($R_S$).

Their determination is enabled by the fact that the *MW* is proportional to integrated optical contrast (*iOC*) of a molecule and that $R_S$ correlates with the diffusivity (*D*), i.e., the characteristic of its movement.

With further advances in this novel technology, concurrently on the experimental and deep learning side, we expect NSM to have a significant impact in the field of single-molecule science.

**Current and Future Challenges**

NSM has proven its unprecedented performance in visualizing and analysing of individual biomolecules diffusing in a solution inside a nanofluidic channel. However, numerous challenges have still to be addressed to allow the widespread adoption of the technique in research and industry.

The most significant drawback is its **low throughput,** which makes the analysis of heterogenous samples very time-consuming. This is the consequence of the current limitation of the experiment to measure one nanochannel at a time. Parallel analysis of hundreds of nanochannels filling the whole field of view of a microscope is possible, in principle, but it is hampered by the lower performance of cameras when operated in the full frame regime. **Clogging of nanochannels** is to be expected for complex biofluids and has to-date only partially been mitigated. Specifically, the pre-filtering of the sample is time-consuming and the passivation of the nanochannel walls needs to be more robust. In terms of data analysis, there is a significant limitation related to the current CV pipeline, as the performance **degrades with high-concentration samples**. Finally, the reported performance of NSM allows to analyze biomolecules down to tens of kDa in molecular weight. There is however a whole realm of biomatter below this limit that is central in many applications of biomolecular research. Therefore, strategies to **push the detection limit** await their implementation.

Besides the weighting and sizing of individual biomolecules, NSM has potential in several other applications and future challenges will likely relate to their development. For instance, label-free and tether-free **investigation of biomolecular interactions** is central for progress in many important areas (e.g., medical and pharmaceutical research). To provide for quantitative analysis of interaction kinetics and to record a statistically relevant number of binding and unbinding events of particular species, tools that enclose the biomolecules in a reaction volume need to be utilized. To provide information regarding the **specificity of the investigated biomolecules**, affinity-based single molecule detection can be utilized. This will require immobilization of analyte-specific receptors on the nanochannel walls.

The fundamental principle of NSM is also applicable beyond the visible range of the light spectrum, providing the performance of available infrared detectors and light sources is sufficient. The **combination of single-molecule microscopy and spectroscopy** holds enormous promise in the analysis of nanomatter and can provide a rich pool of information about the chemical and structural composition.

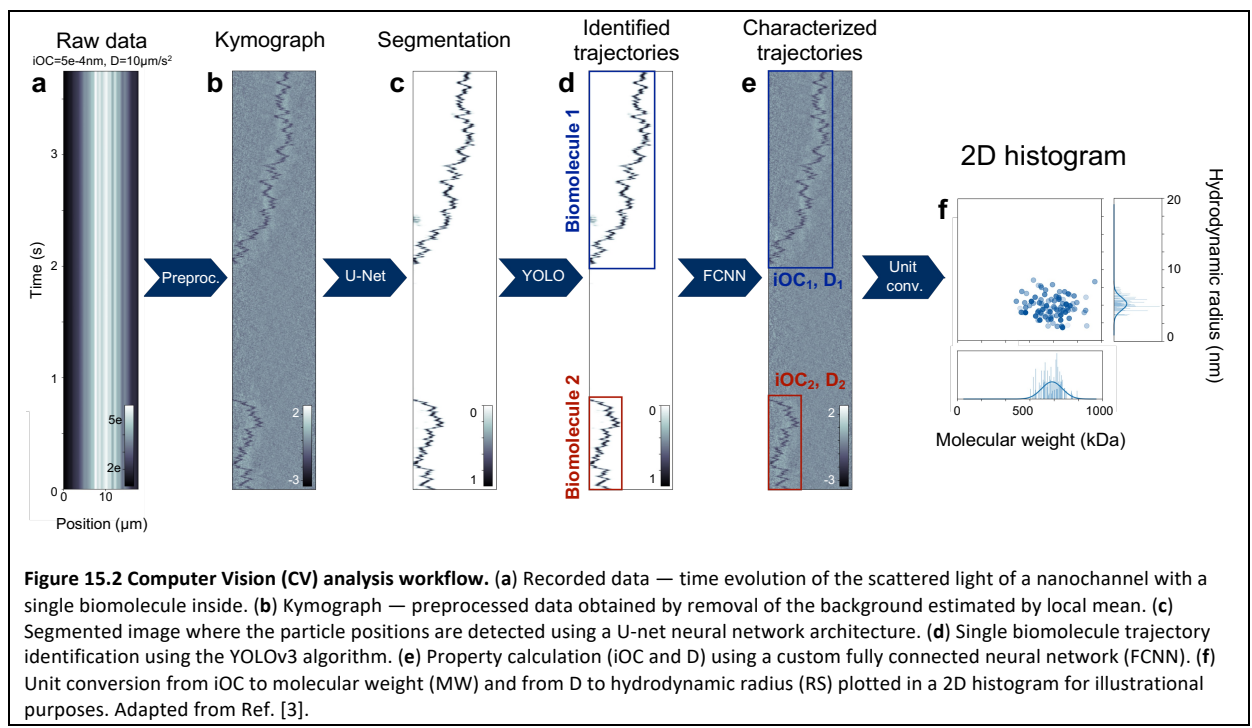

**Figure 15.2 Computer Vision (CV) analysis workflow.** (**a**) Recorded data — time evolution of the scattered light of a nanochannel with a single biomolecule inside. (**b**) Kymograph — preprocessed data obtained by removal of the background estimated by local mean. (**c**) Segmented image where the particle positions are detected using a U-net neural network architecture. (**d**) Single biomolecule trajectory identification using the YOLOv3 algorithm. (**e**) Property calculation (iOC and D) using a custom fully connected neural network (FCNN). (**f**) Unit conversion from iOC to molecular weight (MW) and from D to hydrodynamic radius (RS) plotted in a 2D histogram for illustrational purposes. Adapted from Ref. [3].

**Advances in Science and Technology to Meet Challenges**

Further development of NSM is an effort involving many fields of science and technology and will depend on the advancements of modern machine learning analysis tools, micro- and nanofluidic technology, nanofabrication, and optical detectors.

In terms of AI advancement, recent progress in geometric deep learning for particle tracking and characterization represents promising complementary methods for use in high-fidelity high-sample-concentration studies [4, 5], where the current CV pipeline struggles most. Since the NSM AI models are trained entirely on simulated data, their performance is limited by available computational power. Thus, the performance of the NSM analysis will improve in tandem with the breakneck speeds at which newer generations of GPU and TPU operates. Most recent theoretical analysis predicts that the access to already-available higher computational power combined with further model optimization leads to an order of magnitude improvement in terms of detection limits. Thus, it can be expected that the analysis of biomolecules in the single-digit kDa regime will become available soon.

Regarding the advances in **detector technology**, the CMOS cameras available on the market have quickly increasing frame rate, improved sensitivity, and larger dynamic range. The NSM method will thus reach higher throughput via parallel analysis of multiple channels and achieve more accurate measurements of even smaller molecules.

**Nano- and microfluidic technology** is a swiftly evolving field which keeps bringing new possibilities for further development and application of the NSM method. For example, microfluidic sorting systems [6] allow on-chip filtering to effectively eliminate clogging issues. New tools (such as nanofluidic valves [7] and traps [8]) are introduced to provide means for efficient confinement and release of ultra-low volumes. This ability is crucial for development of future applications, such as sorting at single-molecule level or investigation of biomolecular interactions inside a nanofluidic reactor.

New materials available for **nanofabrication** will potentially open up new possibilities for increasing the sensitivity of the method [9]. In addition, advances in large-scale nanofabrication will pave the way for NSM becoming a reliable and cost-effective bioanalytical method.

**Concluding Remarks**

In its current form, NSM already enables highly accurate label-free and tether-free characterization of individual biomolecules and biological nanoparticles in a wide range of biofluids. The expected advances of the instrumentation and deep-learning tools will push the performance even further. In particular, high-throughput and resolution in single-digit kDa regime will find numerous bioanalytical applications requiring analysis of highly heterogenous samples such exosome characterization or direct detection of small-molecule binding [10]. Long-term monitoring of individual biomolecules diffusing in solution represents a yet unexplored opportunity for studies of conformational changes, aggregation processes and interactions between individual biomolecules. In addition, due to minimized dilution of the sample by the nanofluidic platform, NSM is highly efficient for transporting ultra-low volumes, such as intracellular content or secreted metabolites of a single cell, thereby paving the way to real-time label-free single-cell studies. Moreover, it can be expected that NSM will find applications outside the field of bioanalysis, such as characterization of inorganic nanoparticles, particle counting, or single particle analysis.

**Acknowledgements**

*The authors acknowledge financial support from the Swedish Foundation for Strategic Research project FFL15-0087 (C.L.) and the Czech Science Foundation project 22-18203S (B.S.). Part of this research has been executed at the Chalmers Nanofabrication Laboratory MC2 and under the umbrella of the Chalmers Excellence Initiative Nano.*

## 16 — Particle tracking

Saga Helgadottir[1,2]

1. SciLifeLab BioImage Informatics Unit, Dept. of Information Technology, Uppsala University
2. Department of Biochemistry and Biophysics, Stockholm University, Sweden

**Status**

Tracking the motion of single particles has become a critical tool for probing the microscopic world. Particle tracking has come a long way from manually locating particles' positions over a century ago to using standard algorithmic approaches to the use of deep learning in microscopy. A pioneering example of particle tracking is when Jean Perrin proved the physical existence of atoms in 1910. Perrin projected the image of microscopic colloidal particles in a solution on a sheet of paper and manually tracked their positions, managing to quantify their Brownian motion despite a time resolution of just 30 seconds [1].

Modern particle tracking has largely been dominated by a technique generally referred to as "digital video microscopy", introduced over 20 years ago [2], in which a video of microscopic particles is acquired and the particles' positions in each frame are determined using computer algorithms. Until a few years ago, the standard algorithms have often been based on the measures of the centroid of the particles in a black-and-white thresholded version of the image or the radial symmetry center of the particles [3], and can successfully achieve subpixel resolution when their underlying assumptions for ideal experimental conditions are satisfied — generally that the particle is spherically symmetric and imaged with a homogenous and constant illumination. Their performance decreases drastically with less-than-ideal experimental conditions and significant user intervention is required, which in turn is time-consuming and can introduce user bias.

In the last few years deep learning, a kind of machine learning built on artificial neural networks, has started to be employed for digital video microscopy. In contrast to standard algorithms, deep learning algorithms autonomously learn to determine rules to perform specific tasks using a series of input data and corresponding desired outputs — which, in the case of particle tracking, would be images of particles and their coordinates in the image. Early success of deep learning for particle tracking has already shown that deep learning outperforms standard algorithms in accurately localizing particles in challenging experimental conditions and in many cases is able to eliminate user bias by simulating training data [4] (**Figures 16.1a-d**). However, there is still a lot of room for improvement and deep learning for particle tracking has yet to reach its peak potential.

**Current and Future Challenges**

One major challenge is the availability of training data, both in quantity and quality [5]. Even though it is possible to train deep-learning algorithms using experimentally-acquired data, this becomes increasingly difficult for the application of particle tracking. Considerable amount of data needs to be acquired for each experimental setting, for which it is also extremely difficult to determine the ground-truth particle positions with sufficient accuracy as they often need to be manually annotated. This process is time-consuming and limits the algorithms to human-level accuracy as well as introducing bias. This has been partly solved by training the algorithms using simulated particle images for which the ground-truth positions can be known exactly. It is especially useful to be able to physically simulate particle images replicating each user's experimental setup, from properties of the particle to optical properties of the instrument used to capture the data. However, simulating training data is often not feasible and this is in fact true for non-symmetric and biological objects acquired in most imaging modalities. When it is not possible to accurately recreate the experimental setups numerically, the original experimental dataset can be artificially expanded using augmentation [6]. A special type of deep learning algorithm has even been developed that is able to generate additional synthetic training images as a form of data augmentation. Augmentation can aid in the training process but cannot replace high-quality training data.

A second major challenge is, as for standard algorithms, that the trained algorithms are usually tuned for a specific problem and are not easily generalizable, meaning that they most often cannot be used to analyze data containing different particle types or obtained with another experimental setup. This, in addition to the steep learning curve for developing custom deep-learning solutions, makes deep learning underutilized in particle tracking and digital video microscopy as a whole. Pre-trained algorithms have been made available and are in practice easily applied to new data [7]. The results are however often acceptable only in the absence of a better alternative customized for the problem at hand.

**Advances in Science and Technology to Meet Challenges**
The most obvious way forward for deep learning for particle tracking is to focus on easier access to high-quality training data in order to custom train algorithms for each set of experiments or even to be able to train a generalized particle tracking algorithm. In this case, the continuous development of accurate ways to synthetically recreate more complex experimental setups imaging non-symmetrical particles with various imaging modalities is of upmost importance. The collaborative culture of the field provides a prime environment for a fast growth with open-source software packages emerging with the possibility of other researchers to contribute with additional functionalities related to their areas of expertise [6]. These software packages also attempt on being user friendly to allow for the development of custom deep-learning solutions without the need for special coding skills, making them available for a broader audience.

Another interesting approach is to develop clever and simple deep-learning algorithms that do not require large amounts of data with known ground-truth labels. Novel deep-learning approaches are being introduced that can use a single label-free particle image for training, reaching sub-pixel accuracy for arbitrary particle morphologies [8] (**Figures 16.1e-f**). This type of solution bypasses the need to simulate particle images in order to have accurate and unbiased ground-truth positions and has the potential to work for most particle types and imaging modalities, especially in homogeneous samples. Thanks to its simplicity it could also allow for easily customizable solutions for each set of problems, requiring little-to-no user expertise and can be trained on ordinary computers in the matter of few minutes.

Deep-learning approaches have so far followed the conventional way of particle tracking, providing only a data-driven version of standard algorithms focusing on finding the coordinates of single particles in an image. For further analysis, the coordinates are then linked into single particle trajectories using other standard approaches. New approaches are able to take more advantages of the possibilities of deep learning to analyze the particles' dynamics to directly produce linked trajectories of particle coordinates as well as inferring local and global dynamic properties of the sample [9]. This has the potential to improve the analysis of samples where particles are lost, new particles introduced or change shape during the experiment, which is highly relevant in biological samples. When trained with manually annotated experimental data, the algorithm can even learn how to override imperfect annotation of the training dataset.

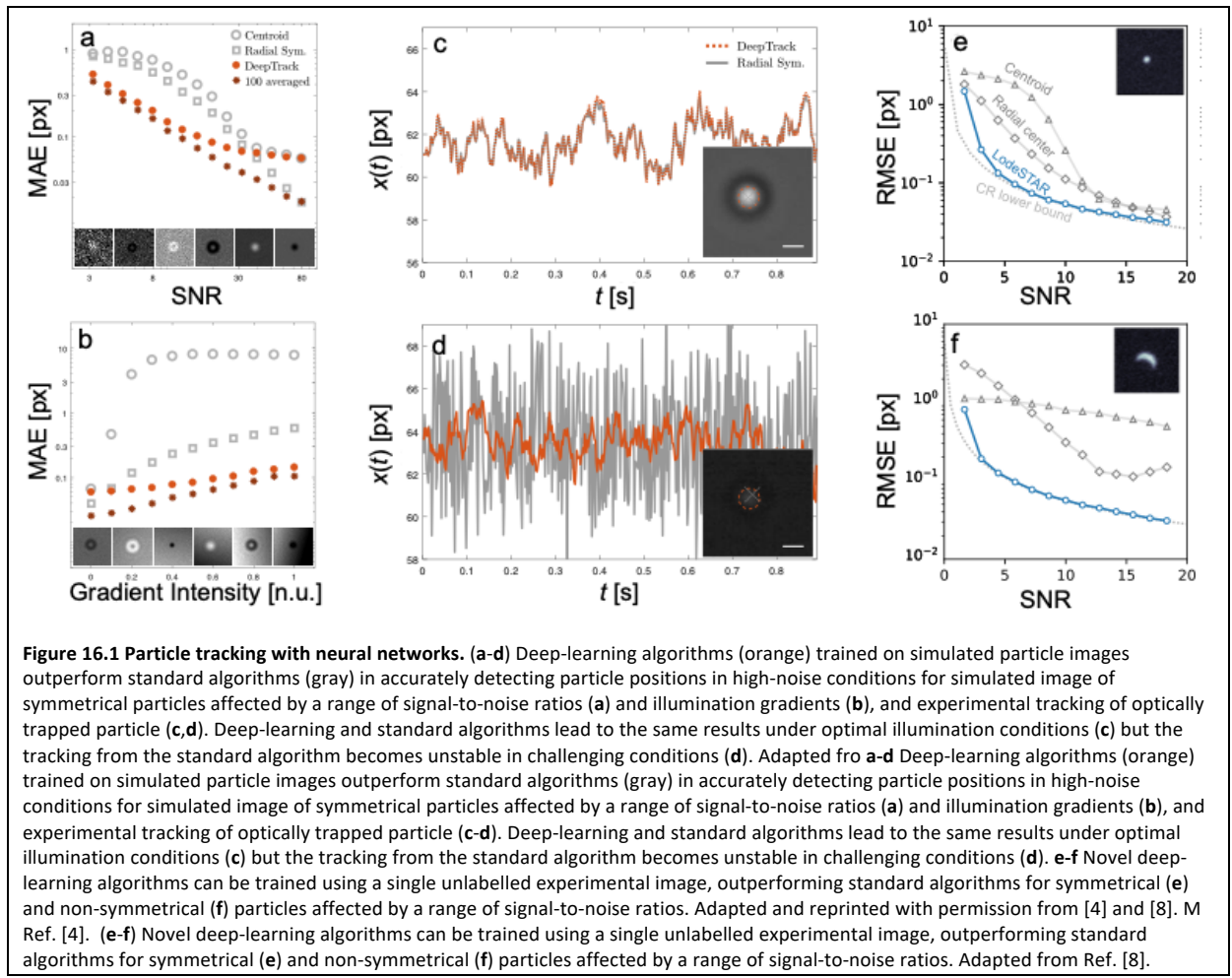

**Figure 16.1 Particle tracking with neural networks.** (**a**-**d**) Deep-learning algorithms (orange) trained on simulated particle images outperform standard algorithms (gray) in accurately detecting particle positions in high-noise conditions for simulated image of symmetrical particles affected by a range of signal-to-noise ratios (**a**) and illumination gradients (**b**), and experimental tracking of optically trapped particle (**c**,**d**). Deep-learning and standard algorithms lead to the same results under optimal illumination conditions (**c**) but the tracking from the standard algorithm becomes unstable in challenging conditions (**d**). Adapted fro **a**-**d** Deep-learning algorithms (orange) trained on simulated particle images outperform standard algorithms (gray) in accurately detecting particle positions in high-noise conditions for simulated image of symmetrical particles affected by a range of signal-to-noise ratios (**a**) and illumination gradients (**b**), and experimental tracking of optically trapped particle (**c**-**d**). Deep-learning and standard algorithms lead to the same results under optimal illumination conditions (**c**) but the tracking from the standard algorithm becomes unstable in challenging conditions (**d**). **e**-**f** Novel deep-learning algorithms can be trained using a single unlabelled experimental image, outperforming standard algorithms for symmetrical (**e**) and non-symmetrical (**f**) particles affected by a range of signal-to-noise ratios. Adapted and reprinted with permission from [4] and [8]. M Ref. [4]. (**e**-**f**) Novel deep-learning algorithms can be trained using a single unlabelled experimental image, outperforming standard algorithms for symmetrical (**e**) and non-symmetrical (**f**) particles affected by a range of signal-to-noise ratios. Adapted from Ref. [8].

## Concluding Remarks

Despite its success in recent years, deep learning for particle tracking still has huge potential. Improved physical simulations of particle images will allow for more accurate tracking of single particles, bypassing the time-consuming and bias-inducing manual ground-truth annotations. Thanks to increasingly available inference speed, trained algorithms could also be implemented in the experimental setup for real-time particle tracking and decision making. Going one step further, new, fast and easily trained algorithms that only require a single unlabeled training image even give rise to the possibility to train and track particles on the spot during experiments.

Tracking particles is usually only the first necessary step in order to further analyze the dynamics of a system. Deep learning can be used to go beyond only acquiring particle coordinates and simultaneously calculate other particle characteristics, as well as directly returning single particle trajectories and the underlying dynamics of the whole system.

## Acknowledgements

*The work for this project has received funding from the H2020 European Research Council (ERC) Starting Grant ComplexSwimmers (677511) and the SciLifeLab BioImage Informatics Unit.*## References

## 17 — Single-shot self-supervised object detection

Benjamin Midtvedt[1], Giovanni Volpe[1]

1. Department of Physics, University of Gothenburg, Gothenburg, Sweden

**Status**

Deep learning is a rapidly growing field in microscopy that aims to improve image analysis and automate the process of identifying and extracting information from images, enabling more efficient and accurate data analysis. Object detection is one such task where deep learning excels over traditional methods. However, most deep learning methods for object detection in microscopy require either large manually annotated datasets or highly realistic physical simulations of the experiment [1]. For many applications, this is a prohibitively expensive barrier to overcome. Recently, we introduced LodeSTAR, which is a self-supervised, low-shot deep learning method that addresses this challenge by exploiting the inherent roto-translational symmetries of the task of object detection [2].

LodeSTAR's key advantage over other deep learning methods is its ability to achieve high performance with minimal training data. Unlike traditional methods, which require large, manually annotated datasets or highly realistic physical simulations, LodeSTAR can be trained on a single or a small number of unannotated experimental images. In fact, LodeSTAR has been shown to match the performance of state-of-the-art cell detection methods in the Cell Tracking Challenge [3] using 1000 times less training data (**Figures 1a-b**). This makes it an ideal solution for researchers working in fields where annotated datasets are not available, or where the annotation process is time-consuming and costly.

Another notable feature of LodeSTAR is its ability to find the sub-pixel position of objects with high accuracy. In fact, LodeSTAR reaches the theoretical Cramér-Rao bound on accuracy for a wide range of objects, using only a single image for training. This is particularly useful in applications where precise object localization is critical, such as in particle tracking, single-molecule localization, or super-resolution microscopy. This level of precision is difficult to achieve with traditional methods, and opens up new possibilities for researchers working in fields where sub-pixel precision is required.

Beyond detection, LodeSTAR can also perform certain regression tasks if there is a corresponding relationship to exploit. In many interferometric modalities, such as holographic microscopes, the focal plane can be numerically changed after-the-fact. This creates a relationship between the z-position of the object and a numerical repropagation transformation. Similarly, in many quantitative modalities, the strength of the signal is directly proportional to some physical property of the object. LodeSTAR has used these relations to position sub-wavelength polystyrene beads in 3D (**Figure 1c**), and to measure the optical mass of intracellular aggregates (**Figure 1d**).

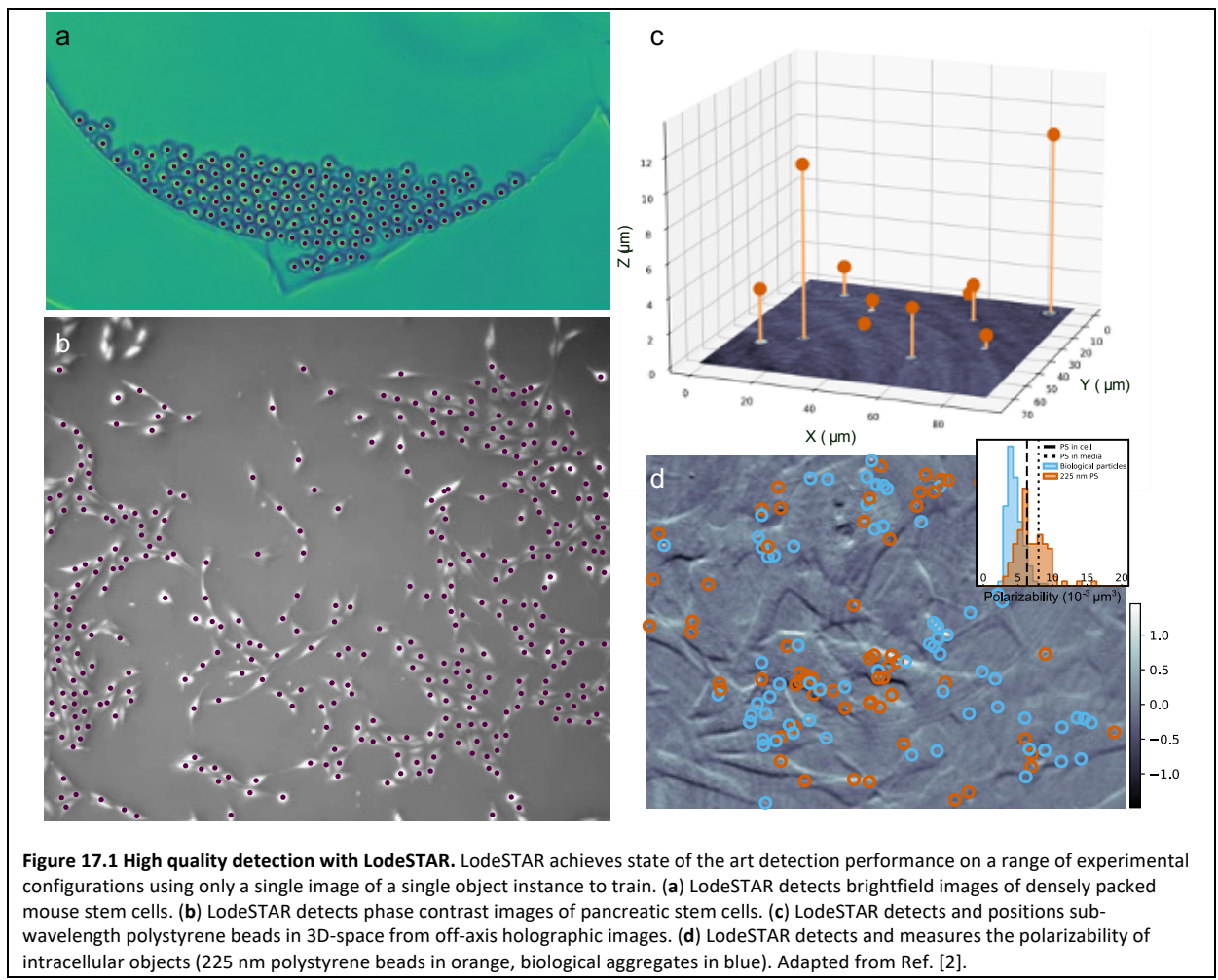

**Figure 17.1 High quality detection with LodeSTAR.** LodeSTAR achieves state of the art detection performance on a range of experimental configurations using only a single image of a single object instance to train. (**a**) LodeSTAR detects brightfield images of densely packed mouse stem cells. (**b**) LodeSTAR detects phase contrast images of pancreatic stem cells. (**c**) LodeSTAR detects and positions sub-wavelength polystyrene beads in 3D-space from off-axis holographic images. (**d**) LodeSTAR detects and measures the polarizability of intracellular objects (225 nm polystyrene beads in orange, biological aggregates in blue). Adapted from Ref. [2].

**Current and Future Challenges**

While LodeSTAR achieves high performance with minimal training data, it should be noted that it currently only provides the position of objects, and not their size or morphological information. This may be a limitation for certain applications that require more detailed information about the objects being detected. Nonetheless, in applications such as cell counting and particle tracking, this is not a limiting factor.

Another limitation of LodeSTAR is that it cannot efficiently utilize large amounts of data to gain more specificity. While it outperforms other methods in the low-shot regime, it may fall behind when unlimited data is available. This is a general problem with self-supervised learning, which often is less directed due to the lack of supervision.

Another limitation of LodeSTAR is that it is optimized for small to medium size objects, typically up to 60-100 pixels in size. While it has been shown to be effective for detecting objects of this size, it may not perform as well on larger objects. This is because the method relies on exploiting the inherent roto-translational symmetries of the task of object detection, which may not be as prevalent in larger objects. This limitation may be less important for certain applications such as single-molecule localization, where the objects of interest are relatively small. However, it could be a limitation for other applications such as detecting large cells or structures in tissue imaging. In these cases, the issue can be resolved by sampling the image at a lower resolution. This improves detection at the cost of positional accuracy. However, for large structures, the notion of a single position is less well-defined and often less relevant to the applications.

**Advances in Science and Technology to Meet Challenges**

An important avenue for future research is to expand the set of transformations and symmetries that LodeSTAR utilizes in its learning process. The goal would be to train LodeSTAR to produce morphological information. For example, by incorporating scale transformations, LodeSTAR could be taught to predict the relative sizes of objects. However, this alone would not be sufficient to provide absolute sizes without additional transformations.

Another approach could be to use a more interpretable network architecture, such as vision transformers [4]. These architectures have been shown to naturally produce segmentations of objects when solving consistency tasks. However, it's important to note that replacing the current convolutional backbone with a vision transformer would likely lead to a loss of translational-equivariance, which may negatively impact LodeSTAR's positioning performance.

Currently, LodeSTAR is trained on positive samples only, which are images containing the object to be detected. One potential area of research is to expand the training procedure to include negative samples, which are images with no instances of the object. This could help improve specificity in samples with multiple visually similar classes, where only a subset of the classes is of interest.

Lastly, LodeSTAR is currently designed to train on images with exactly one instance of the object per view. However, it could be extended to allow multiple instances of the object per view during training, which would enable it to learn from larger datasets where the number of objects in view may vary over time. However, this would also decrease the specificity of the training, as all objects in view would be detected regardless of interest. Finding ways to mitigate this would be an important area of research.

**Concluding Remarks**

LodeSTAR is a deep learning method for object detection in microscopy that is particularly well-suited for low-data regimes. LodeSTAR exploits the inherent roto-translational symmetries of object detection to achieve high performance with minimal training data. Its unique self-supervised training process allows it to learn from a single or a small number of unannotated experimental images, making it a cost-effective solution for researchers working in fields where annotated datasets are not available. Furthermore, LodeSTAR's ability to find the sub-pixel position of objects with high accuracy is a key feature that sets it apart from traditional methods. This level of precision is particularly useful in applications where precise object localization is critical, such as in particle tracking, single-molecule localization, or super-resolution microscopy. This, combined LodeSTAR's ability to perform utilize additional relationships and symmetries to measure additional physical quantities makes it a versatile and powerful method in the field of microscopy.

**Acknowledgements**

*The authors would like to acknowledge funweding from the H2020 European Research Council (ERC) Starting Grant ComplexSwimmers (Grant No. 677511), the Horizon Europe ERC Consolidator Grant MAPEI (Grant No. 101001267), and the Knut and Alice Wallenberg Foundation (Grant No. 2019.0079).*

## 19 — Diffusion characterization

Stefano Bo[1,2], Lars Hubatsch[3]

1. Max Planck Institute for the Physics of Complex Systems, Nöthnitzer Straße 38, DE-01187 Dresden, Germany
2. Department of Physics, King's College London, London WC2R 2LS, U.K.
3. Max Planck Institute of Molecular Cell Biology and Genetics, Pfotenhauerstraße 108, 01307 Dresden, Germany

**Status**

Live imaging has revolutionised our understanding of biological processes. Resolution and signal-to-noise ratio have steadily improved over the past decades. As a consequence, the trajectories of single particles can be recorded, enabling detailed descriptions of microscopic dynamics while avoiding some artefacts of bulk labelling. Single-particle tracking (SPT) unveils features of the surrounding environment and of processes that would be lost by ensemble averaging, thus opening an observational window that provides a deeper understanding of many biological processes.

To probe the molecular properties in heterogeneous environments and to enable quantitative modelling of biological processes, it is important to tell whether particles undergo normal or anomalous diffusion (defined by the power-law scaling of the Mean Squared Displacement: MSD $\sim t^\alpha$) and to measure their diffusion coefficients or anomalous scaling exponent α. Motion at these scales is inherently stochastic and can involve processes characterised by significant variability in time and space. This makes the study of diffusion from particle trajectories difficult.

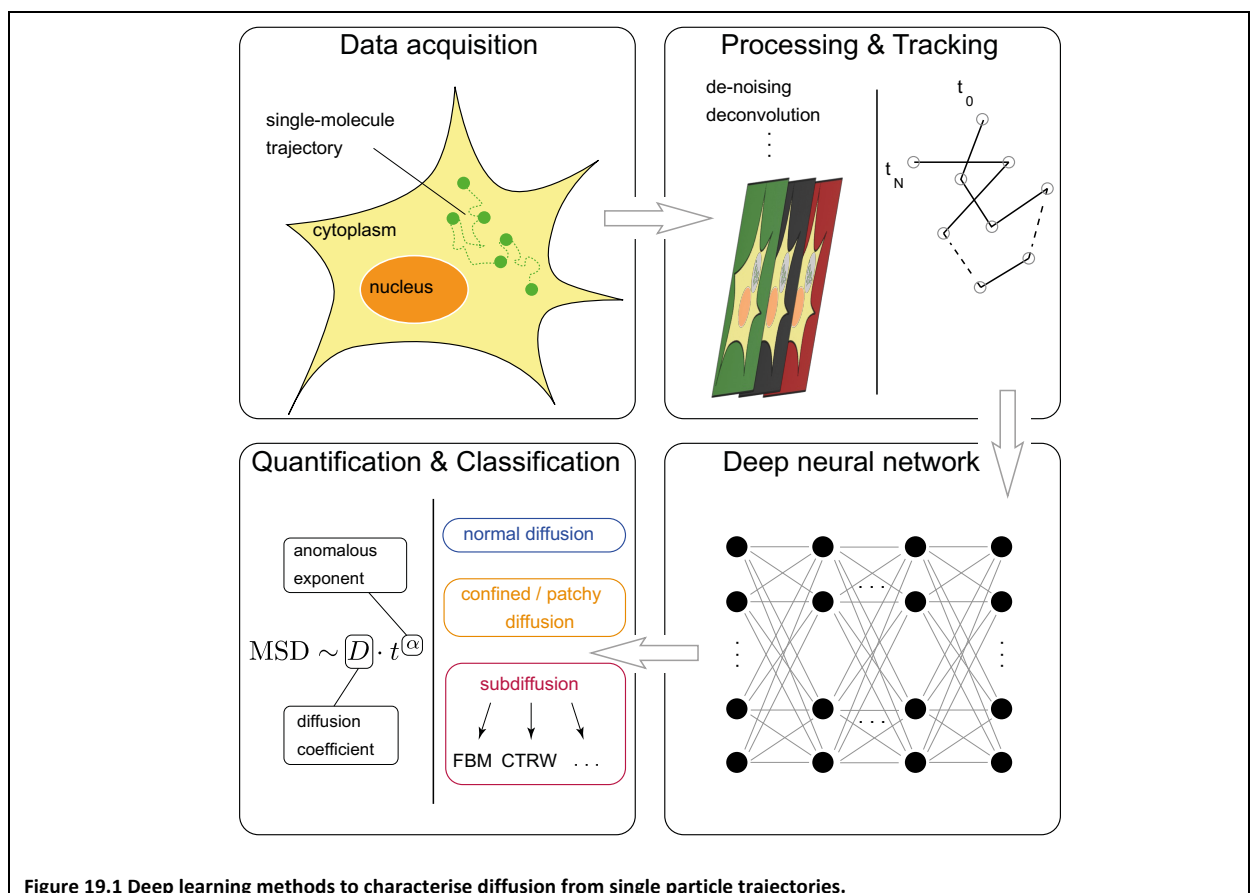

Figure 19.1 Deep learning methods to characterise diffusion from single particle trajectories.

A first challenge consists in identifying the particles, localising them and reconstructing their trajectories. Various methods have been developed for these purposes in the past decade [1], and deep learning techniques are gaining considerable attention [2].

The second main challenge is the extraction of information from the reconstructed trajectories. This is complicated by the fact that diffusion is a phenomenon related to particle fluctuations, in other words, the signal is in the noise. Even for simple diffusion, the direct application of Einstein's 1905 result MSD ~ Dt can result in severe biases, for example depending on the localization precision, track length, number of available tracks, or fluctuations due to movement of the surrounding media. To tackle these issues, methods accounting for experimental noise and optimal fitting algorithms for different use cases were developed [3]. Anomalous diffusion dynamics, which can be modelled, for example, by *continuous time random walks* (CTRW), *fractional Brownian motion* (FBM) and *Lévy walks* (LW), are even more challenging to analyse, due to long-term correlations in the dynamics and possible ergodicity breaking [4]. Most of the methods to analyse such dynamics rely on asymptotic behaviour, requiring many long trajectories, which are typically difficult to obtain experimentally. This has motivated the development of deep learning techniques to measure the diffusion coefficient, the anomalous exponent and to identify the model underlying the anomalous behaviour [4-7]. These techniques typically involve artificial convolutional neural networks (CNN) or recurrent neural networks named Long Short-Term Memory (LSTM). They have remarkably improved the analysis of short individual trajectories, also including cases where individual trajectories switch between different diffusive dynamics.

**Current and Future Challenges**

A promising feature of deep-learning techniques is that they have the potential to generalise, i.e., correctly analyse scenarios that differ from the training data. For applications to single-particle trajectories this suggests that algorithms trained on synthetic data, which are cheaply generated by simulation, can be applied to experiments. The techniques developed in Refs. [4-7] are networks trained on simulated data, using supervised methods, i.e., the parameters used to simulate the input data, such as diffusion coefficient, anomalous exponent or the type of model, are known and used to label the input. The networks are then asked to return a prediction that is as close as possible to the label. These methods have proven their versatility, correctly predicting experimental trajectories that they had not seen during training, and the ability to deal with measurement noise. These promising examples, however, do not generally guarantee that networks trained on simulated data can be carried over to experiments. To assess the issue of generalisation, more work is required to understand what leads to a network's decision. This would contribute to making the predictions more interpretable and transparent. Interpretability can significantly extend the scientific applicability of deep learning models to diffusion, linking network behaviour to the underlying physical principles and structures. Much like developing a physical theory provides a mechanistic understanding of a physical phenomenon and goes beyond measurement and inference, improving the interpretability of deep learning contributes to deeper physical understanding. Furthermore, understanding how deep learning methods estimate diffusion from particle trajectories might contribute to the development of new algorithmic and statistical techniques, which are inspired by the solution found by the deep learning methods. Indeed, the progress in interpretability might pave the way towards learning from the machines. However, due to their high dimensionality and many parameters, it is notoriously difficult to unveil the inner workings of deep learning models.

An additional challenge is related to supervised methods. These methods, by their very construction, can only predict the labels they were taught during training. For instance, a method trained to only infer the diffusion coefficient will not be able to detect anomalous diffusion or its scaling exponent. Similarly, for the classification of the dynamics, typical networks are bound to return an answer within the classes they have been designed for. A related challenge concerns the use of deep learning to capture the transient nature of anomalous diffusion, where different scaling regimes are observed at different scales.

**Advances in Science and Technology to Meet Challenges**
The field of deep learning is evolving rapidly, providing insights that can be applied to various sub-domains. For instance, several tools for the interpretability of deep learning techniques are currently being developed (see, *e.g.*, Ref. [8]). The most promising results in this direction concern deep neural networks, such as CNN, employed for the classification of images. These novel tools identify the most significant parts, or features of the images that led to their respective classification. For SPT, one could apply these algorithms (which should be extended also to LSTM) to determine which parts of a trajectory most heavily influence its classification.

In parallel, autoencoder and transformer architectures have been developed [9]. Transformers are currently among the best techniques to process natural language, for example for translation purposes. They feature attention layers in their architectures, whose role is to identify how much different parts of an input are related to each other (*e.g.*, learning that the subject in a sentence is strongly related to the verb).

The combination of the development of tools to interpret existing deep-learning architectures and of more interpretable architectures offers the possibility of identifying the parts of a diffusive trajectory that are the most informative ones to characterise it. One can then perform statistical analysis on these results to extract information about what features or statistical properties of the trajectories are the most relevant to characterise their diffusive nature. This sheds light on how the deep learning methods make their predictions, contributing to their interpretability.

Additional advances in deep unsupervised learning can provide methods that are more insightful when applied to trajectories belonging to none of the models they have been trained on (see [10] for a first step in this direction using anomaly detection). Indeed, these models are not trained using specific labels but focus on reconstructing the input trajectory, thereby learning its salient features. Unsupervised methods might also successfully address the challenges posed by trajectories featuring different diffusive scalings at different scales.

The advances discussed so far concern general improvements for the interpretability of deep-learning methods. Additional progress is likely to come from the development of physics-informed methods for the study of diffusion. Such methods are designed to take into account the physical knowledge about the diffusive dynamics under consideration, for example considering key physical and statistical properties, such as the self-similarity of certain anomalous diffusion dynamics.

**Concluding Remarks**
Deep learning methods have proven to be invaluable tools for the study of diffusive dynamics, displaying remarkable robustness and showing promising potential to generalise results to unseen data and experiments. They excellently complement traditional algorithmic and statistical techniques, providing better performance. This makes it possible to characterise short individual trajectories, thereby expanding the range of systems for which diffusion can be quantitatively investigated, including many biological systems, for which track lengths are typically short. At the current stage, this performance improvement comes at the cost of a lower interpretability of the results compared to traditional algorithmic and statistical techniques. However, recent and ongoing progress in deep learning techniques with the development of novel architectures and methods to interpret their prediction together with physics-informed approaches suggest that more transparent deep learning techniques might be within close reach. These techniques have the potential to reliably unveil the physical properties of diffusive particles and the environments in which they move. These advances provide an important building block of future quantitative models of complex biological processes.

### Acknowledgements

*SB gratefully acknowledges inspiring discussions with Aykut Argun, Giovanni Volpe, Ralf Eichhorn, Alessia Gentili, Giorgio Volpe and the participants and organisers of the AnDi workshop hosted at ICFO in December 2021.*

# 20 — Microscopic motion characterization with MAGIK

Jesus Pineda[1], Giovanni Volpe[1], Carlo Manzo[2]

1. Department of Physics, University of Gothenburg, Gothenburg, Sweden
2. Facultat de Ciències, Tecnologia i Enginyeries, Universitat de Vic – Universitat Central de Catalunya (UVic-UCC), Vic, Spain

**Status**

The characterization of dynamic processes in living systems provides essential information for advancing our understanding of life processes in health and diseases, as well as for developing new technologies and treatments [1]. In the past two decades, optical microscopy has undergone significant developments, enabling us to study the motion of cells, organelles, and individual molecules with unprecedented detail at various scales in space and time. However, analysing the dynamic processes that occur in complex and crowded environments remains a challenge [2,3,4]. The diversity of biological motion is a significant contributor to the complexity of dynamic analysis. Cells, for example, exhibit a high degree of variability in their motion depending on factors such as cell type and environmental cues. This variability poses analytical challenges for current methods, which are typically designed for specific experiments or motion models. Thus, the analysis of experiments often requires manual adjustments of parameters to adapt to different dynamics, limiting their utilization and applicability.

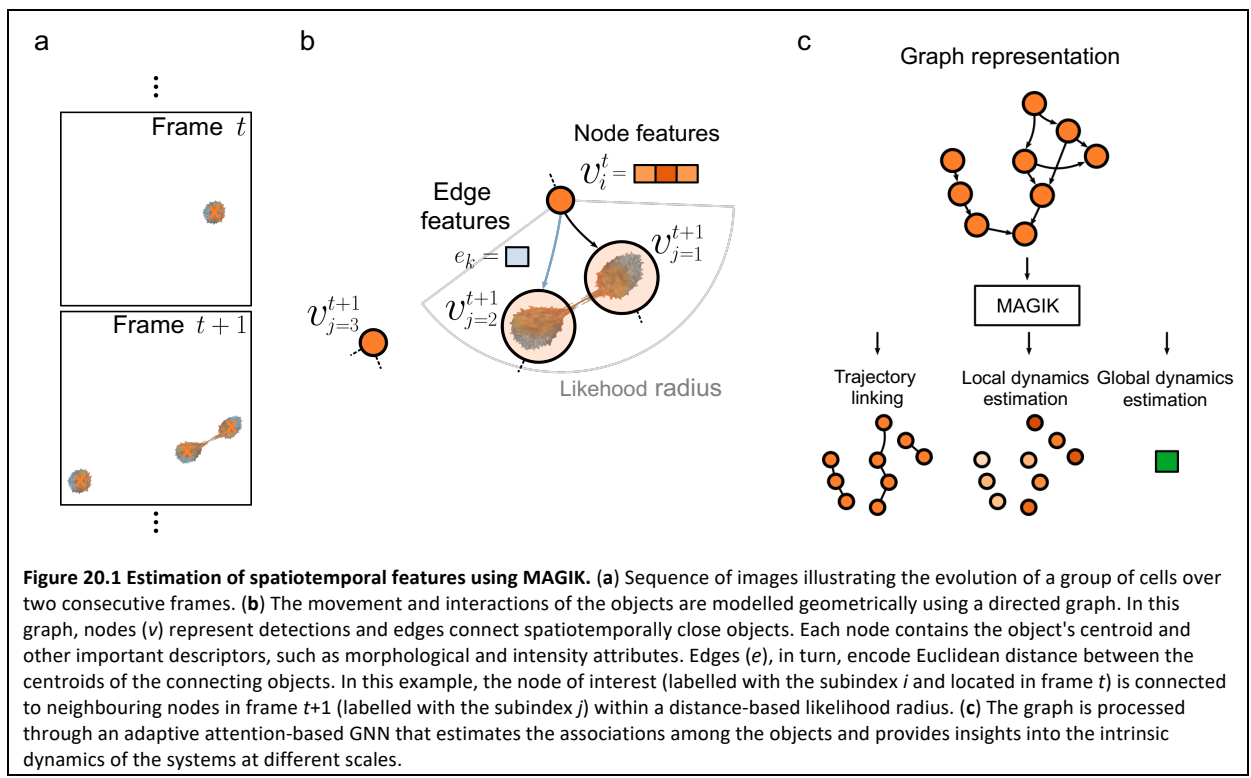

**Figure 20.1 Estimation of spatiotemporal features using MAGIK.** (**a**) Sequence of images illustrating the evolution of a group of cells over two consecutive frames. (**b**) The movement and interactions of the objects are modelled geometrically using a directed graph. In this graph, nodes ($v$) represent detections and edges connect spatiotemporally close objects. Each node contains the object's centroid and other important descriptors, such as morphological and intensity attributes. Edges ($e$), in turn, encode Euclidean distance between the centroids of the connecting objects. In this example, the node of interest (labelled with the subindex $i$ and located in frame $t$) is connected to neighbouring nodes in frame $t$+1 (labelled with the subindex $j$) within a distance-based likelihood radius. (**c**) The graph is processed through an adaptive attention-based GNN that estimates the associations among the objects and provides insights into the intrinsic dynamics of the systems at different scales.

Recently, we have proposed MAGIK [5], a deep-learning framework for the analysis of biological system dynamics from time-lapse microscopy. MAGIK models the movement and interactions of particles through a directed graph where nodes represent detections and edges connect nodes that are spatiotemporally close (**Figures 20.1a-b**). The framework utilizes an attention-based graph neural network (GNN) to process the graph and modulate the strength of associations between its elements through two mechanisms. The first mechanism is a learnable local receptive field [6] that captures the complexity of local particle interactions. The second is a gated self-attention mechanism [7], enabling MAGIK to derive insights into the dynamics of each particle from regions within the graph that are not directly connected but provide valuable information about the overall dynamics.

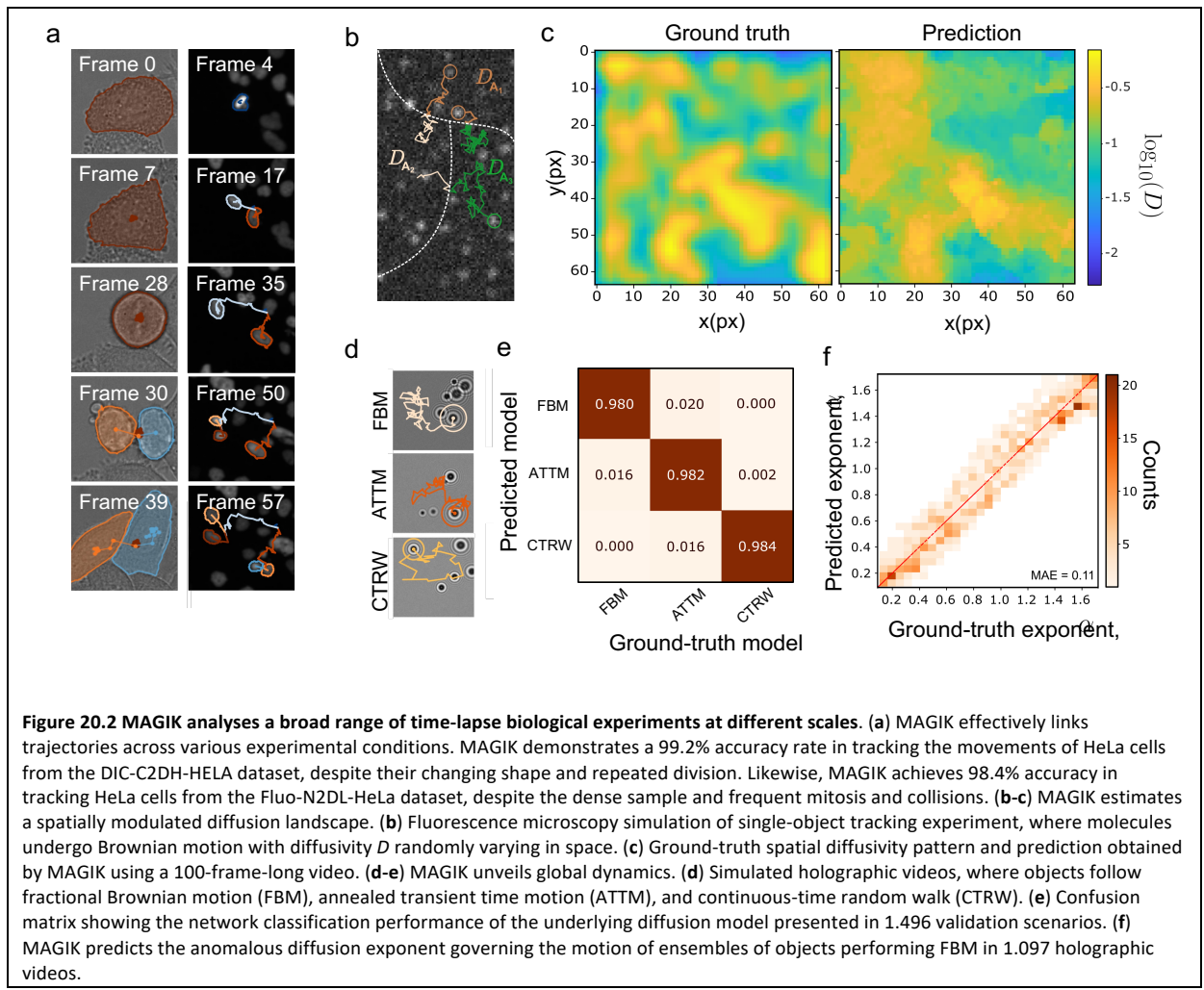

**Figure 20.2 MAGIK analyses a broad range of time-lapse biological experiments at different scales**. (**a**) MAGIK effectively links trajectories across various experimental conditions. MAGIK demonstrates a 99.2% accuracy rate in tracking the movements of HeLa cells from the DIC-C2DH-HELA dataset, despite their changing shape and repeated division. Likewise, MAGIK achieves 98.4% accuracy in tracking HeLa cells from the Fluo-N2DL-HeLa dataset, despite the dense sample and frequent mitosis and collisions. (**b-c**) MAGIK estimates a spatially modulated diffusion landscape. (**b**) Fluorescence microscopy simulation of single-object tracking experiment, where molecules undergo Brownian motion with diffusivity $D$ randomly varying in space. (**c**) Ground-truth spatial diffusivity pattern and prediction obtained by MAGIK using a 100-frame-long video. (**d-e**) MAGIK unveils global dynamics. (**d**) Simulated holographic videos, where objects follow fractional Brownian motion (FBM), annealed transient time motion (ATTM), and continuous-time random walk (CTRW). (**e**) Confusion matrix showing the network classification performance of the underlying diffusion model presented in 1.496 validation scenarios. (**f**) MAGIK predicts the anomalous diffusion exponent governing the motion of ensembles of objects performing FBM in 1.097 holographic videos.

MAGIK is a versatile tool capable of performing various tasks, from linking coordinates into trajectories to determining local and global dynamics (**Figure 20.1c**). As shown in **Figure 20.2a**, MAGIK offers reliable performance in tracking HeLa cells, despite the challenges posed by the heterogeneity in cell shape and dynamics. MAGIK accurately identifies cell divisions and estimates trajectories in edge regions where cells are partially observed and move out of the field of view. Furthermore, MAGIK delivers outstanding results on several datasets from the 6th Cell Tracking Challenge across a range of microscopy techniques and cell types [5].

A unique feature of MAGIK is its ability to characterize dynamic aspects without the need for detection linking. In this way, MAGIK can provide information from high-density experiments, e.g., by resolving a spatially modulated diffusion landscape solely from particle localizations (**Figures 20.2b-c**). Notably, most spatial features are accurately estimated from a 100-frame-long video. By skipping the linking step, MAGIK inherently reduces the propagation of linking errors to the quantification of relevant parameters and can thus unveils global dynamics, as shown for two relevant examples: the classification of the mode of motion of diffusing particles (**Figures 20.2d-e**), and the quantification of anomalous diffusion [4] from ensembles of Brownian particles (**Figure 20.2f**).

**Current and Future Challenges**
The rapid progress in deep learning has led to the development of various techniques for characterizing motion in biological systems [2-4]. However, several technical and interpretative challenges hinder the widespread use of these algorithms.

The scarcity of high-quality annotated microscopy datasets hampers the effective training and validation of deep learning models. Available datasets are often small and unrepresentative, making it difficult to create large and dependable models. In addition, the diversity of microscopy data and limited access pose further challenges to their utilization and integration.

Deep learning models typically learn complex representations of the input data in an abstract, high-dimensional space, making interpretation of these abstractions difficult for even experts. This lack of interpretability is a barrier to the widespread adoption of deep learning in medical and biological fields, particularly in contexts outside of research where interpretability is crucial [8]. Moreover, the proliferation of multiple methods for the same task without a clear evaluation of their performance can confuse non-experts and limit their usage.

The scalability of deep-learning graph models is a major technical challenge. Processing large graph representations with numerous interactions is computationally demanding and requires significant memory and computing resources. This makes it challenging to directly use graph neural networks (GNNs) for analysing dense and lengthy dynamic experiments.

**Advances in Science and Technology to Meet Challenges**
The MAGIK framework is continuously improved to address the limitations in the current deep learning approaches for motion characterization in biological systems. The focus of MAGIK's development is to create a framework that is both general and easily convergent. It has shown the capability to train using just one labelled video by utilizing tools that maximize information extraction from limited data, ensuring proper representation and stability during neural network training [5]. MAGIK also uses transfer learning for migration experiments, enabling the trained network to be applied to other cell data without any reduction in performance, as shown in the MAGIK GitHub repository [9]. Further advancements in this regard are desirable through the implementation of self-supervised algorithms.

MAGIK is equipped with attention mechanisms that provide users with interpretability into the specific aspects of the data structure that the framework focuses on when making predictions. This offers a reliable and efficient method for analyzing dynamics and provides opportunities for discovering new features in the movement of living systems.

MAGIK is included in the deep learning package for microscopy, DeepTrack 2.1 [10], and is, therefore, undergoing continuous development and optimization with a focus on scalability and deployment improvements. Future efforts will focus on developing self-destillation-guided graph subsampling techniques [11] and resource-efficient GNN architectures [12].

**Concluding Remarks**
Deep-learning frameworks for the analysis of biological system, such as MAGIK, can successfully handle the complexities of dynamic analysis in complex and crowded environments utilizing an attention-based graph neural network. MAGIK can perform various tasks, including tracking cells, determining local and global dynamics, and characterizing dynamic aspects without the need for detection linking. Despite these progresses, technical and interpretive challenges that hinder widespread use of these tools. With the continuous development of the MAGIK framework within the DeepTrack 2.1 package, we aim to address these limitations and offer a generalizable tool able to further provide interpretability into the data structure.

## 21 — Quantification of biological dynamics from single molecules

Carlo Manzo[1]

1. Facultat de Ciències, Tecnologia i Enginyeries, Universitat de Vic – Universitat Central de Catalunya (UVic-UCC), Vic, Spain

**Status**

Time-lapse microscopy images of biological processes are widely used to observe the dynamics and behaviour of live cells and unicellular microorganisms, with applications ranging from fundamental aspects of molecular and cell biology to medical practice. The development of single-molecule imaging and super-resolution microscopy has further extended the capability to resolve the dynamics of biological processes, reaching the subcellular and molecular scales [1]. The current technology thus enables the visualization of the motion of organelles, proteins, and lipids in their native environment. The observations provided by these experiments are valuable to decipher the interactions between cellular components and to disclose their role in fundamental processes such as signalling and function regulation. They are also helpful for biomedical applications related to pathogen infection and drug design. Nowadays, microscopes capable to perform live-cell single-molecule imaging are accessible in many research laboratories and, therefore, experiments are routinely performed. However, mining quantitative information from these experiments still poses several challenges.

Typically, the analysis pipeline is divided into two steps: single-particle tracking and trajectory analysis (**Figure 21.1**). In the first step, the information contained in the image stream is converted into trajectories, i.e., time series of features associated with imaged particles (such as position, intensity, and size). Considerable efforts have been dedicated to developing automated algorithms for this task [2]. Relying on the advances of the research community in computer vision and multi-object tracking, the tracking-by-detection paradigm has gained increasing prevalence for single-particle tracking. Thus, images are first processed to detect features (detection), then features obtained at different times are connected using assignment algorithms to obtain trajectories (linking) (**Figure 21.1**).

Once the trajectories are obtained, they are analysed using statistical methods to extract information about the underlying dynamics of the particle, using estimators such as the mean-squared displacement (**Figure 21.1**). These analyses aim at providing details about the type of transport being observed (Brownian, directed, confined, or anomalous), interactions with other particles and/or with the surrounding medium. Trajectories are also used to estimate biophysical parameters (e.g., the diffusion coefficient) or to determine whether the motion is compatible with a given theoretical model.

Life scientists dispose of a variety of algorithms to precisely track individual particles in living biological systems as well as many methods to interrogate trajectories. Recently, approaches based on deep learning have also been proposed, claiming remarkable improvements. The objective assessment of the performance of these methods is thus required to help end-users to pick the suitable tool.

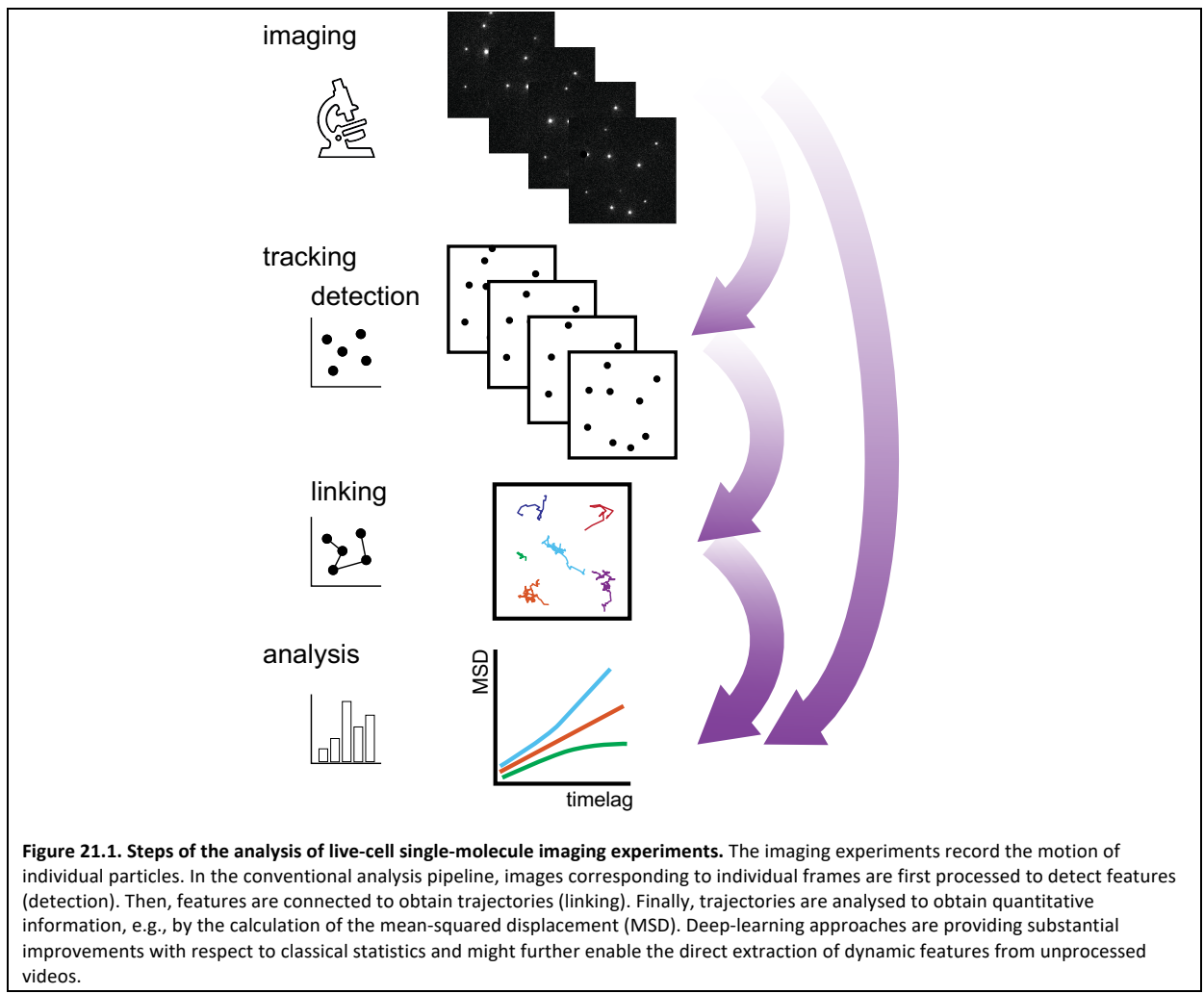

**Figure 21.1. Steps of the analysis of live-cell single-molecule imaging experiments.** The imaging experiments record the motion of individual particles. In the conventional analysis pipeline, images corresponding to individual frames are first processed to detect features (detection). Then, features are connected to obtain trajectories (linking). Finally, trajectories are analysed to obtain quantitative information, e.g., by the calculation of the mean-squared displacement (MSD). Deep-learning approaches are providing substantial improvements with respect to classical statistics and might further enable the direct extraction of dynamic features from unprocessed videos.

**Current and Future Challenges**

Live-cell single-molecule imaging experiments typically record the motion of a subpopulation of individual particles (molecules, viruses, organelles) taking place in heterogeneous environments with the objective of detailing the molecular mechanism of transport and interactions with the environment. Technical and instrumental drawbacks impose limitations on the experimental conditions (e.g., the density of imaged particles and the temporal resolution) and affect quantitative parameters (e.g., the localization precision and the trajectory length) that eventually impact the precise characterization of the system [2]. A current challenge entails deploying approaches that can improve the performance of the methods that carry out the individual steps of the traditional analysis pipeline. In the last years, machine learning and single-molecule localization microscopy have produced a surge of methods for single-molecule detection and localization, mainly based on convolutional neural networks [3,4]. More recently, deep learning approaches have also been proposed for the trajectory linking task [5]. These methods aim to provide an automated, unbiased, and reliable analysis of the image stream. The improvement of their performances enables experiments to be performed at faster image acquisition rates and higher labelling densities, increasing the temporal resolution and the spatial sampling.

Because of the variety of phenomena taking place inside living cells, numerous approaches focusing on different aspects of particles' motion have been proposed to analyse trajectories. A challenging aspect is the characterization of individual trajectories, in particular when experimental conditions limit their length or localization precision. Very recently, pioneering works using machine learning have shown substantial improvements with respect to classical statistics and have demonstrated the ability of several architectures (random forest, convolutional, and recurrent neural networks) to provide the precise estimation of parameters (such as the diffusion coefficient for Brownian motion and the anomalous diffusion exponent for anomalous diffusion) as well as trajectory classification, either with respect to the diffusion mode (described as immobile/confined/Brownian/directed or sub-/Brownian/super-diffusive) or to the underlying physical model [6-9]. These results have led to the organization of the first Anomalous Diffusion (AnDi) challenge [10], a competition to objectively assess these methods, which fostered the development of novel approaches with outstanding performance [10]. The challenge also featured a task on trajectory segmentation for the detection of changes in dynamic behaviour associated, e.g., with interactions with the environment. Due to its implications for the characterization of biological systems, the development of trajectory segmentation methods for the detection of transient and short-lived events is likely to gain further momentum in the immediate future.

**Advances in Science and Technology to Meet Challenges**

Deep-learning methods developed for both single-particle tracking and trajectory analysis outperform classical statistics counterparts in a wide range of conditions and promise to relax experimental constraints, providing more information at a faster speed from live-cell single-molecule imaging. However, several of the proposed methods remain stuck at the stage of proof-of-principle and do not reach a widespread application in actual experiments. Various reasons might be contributing to this process. Most of the methods introduced so far involve supervised learning, but the lack of annotated data for this kind of experiment forces the training and validation over simulated datasets. Despite the realism of the simulations, the transfer learning to actual data might generate concerns from end-users, in addition to the black-box model concern. As done in other fields, progresses in this sense might be obtained by creating community efforts aimed at i) producing public datasets to evaluate novel methods; ii) periodically benchmarking existing methods using objective metrics to determine the state-of-the-art. It must be also considered that to work optimally, deep-learning architectures must be trained on simulations reproducing the specific experimental conditions. It is thus recommendable to implement user-friendly interfaces to help non-experts to train and fine-tune the model.

The deep-learning approaches implemented for single-particle tracking have incrementally improved on existing methodologies but have so far been bound to follow the standard analysis pipeline, providing data-driven versions of conventional approaches. A leap forward might be taken by developing "tracking-free" methods capable of directly extracting dynamic features from unprocessed videos. Such approaches might be based on geometric deep learning or physics-informed machine learning architectures that include informative priors, i.e., physical constraints and inductive biases, on top of the observational data. In fact, besides producing faster training and more accurate predictions, these architectures will also increase the interpretability of the model. The direct use of raw data would also prevent the propagation of errors generated at the different steps of the pipeline that finally impact the extraction of dynamic information.

The development of techniques not requiring labelled datasets might further accelerate the application of deep learning to real data. Both unsupervised and self-supervised learning methods are advancing at a very rapid pace. In combination with innovative network architectures such as transformers, self-supervised techniques have demonstrated the ability to learn representations from unlabelled data, achieving outstanding results for image-based analysis. As such, they represent a promising approach for the development of the next generation of tools for single-particle tracking and analysis.

**Concluding Remarks**

The deep learning revolution is yielding exciting perspectives for the quantitative analysis of live-cell single-molecule imaging. Yet, while recently developed tools are demonstrating their gain in performance, they still have important challenges to overcome to reach a widespread use of data-driven methods over classical tools. In this sense, it is advisable to promote community-driven actions to benchmark and validate methods [2,10]. Stepping beyond the tracking-by-detection paradigm might lead to a boost in performance by digging information directly from raw data. The integration of physical constraints and inductive biases into deep learning models can improve performance and interpretability. It is foreseeable that novel techniques not requiring labelled datasets will further boost this field and enable the study of molecular dynamics in living cells beyond current capabilities.

**Acknowledgements**

*CM acknowledges support through grant RYC-2015-17896 funded by MCIN/AEI/10.13039/501100011033 and "ESF Investing in your future", grants BFU2017-85693-R and PID2021-125386NB-I00 funded by MCIN/AEI/10.13039/501100011033/ and FEDER "ERDF A way of making Europe", and grant AGAUR 2017SGR940 funded by the Generalitat de Catalunya*

## 22 – Plankton life trajectories

Harshith Bachimanchi[1], Erik Selander[2]

1. Department of Physics, University of Gothenburg, Sweden
2. Department of Marine Sciences, University of Gothenburg, Sweden

**Status**

Life likely began in the oceans 3.8 billion years ago [1]. Early cyanobacteria spread across the oceans and oxygenated the atmosphere [2]. Since then, the phytoplankton — the microscopic equivalents or terrestrial plants — have played a pivotal role in the Earth's ecosystem. Phytoplankton generate approximately half of the oxygen produced on Earth and fix 50 Peta-grams (Pg) of carbon from carbon dioxide, around five times the total emissions from fossil fuels [3], every year. Yet, because of their smaller size, our understanding of the lower aquatic food-web in many aspects is limited.

In order to effectively interrogate these organisms at a closer level, we depend on microscopy-based methods. Some of the application scenarios in this direction are plankton detection and counting, plankton segmentation and species classification, and long-term tracking of plankton cells. However, manual identification of single cells is a labour-intensive process considering the extraordinary diversity among plankton taxa that require trained taxonomists [4]. This also limits the possibilities to monitor and follow plankton communities with adequate resolution in time and space. Over the past decade, a multitude of computer vision-based imaging methods have been developed for object detection from microscopy images. These techniques range from traditional approaches that follow segmentation and binary thresholding methods, to more advanced techniques based on deep learning [5].

Particularly, deep-learning-based methods for plankton analysis have seen considerable success in recent years. Emerging as an alternative approach to established methods, deep learning offers objective schemes for investigating microorganisms in different environments. For example, a family of convolutional neural networks (CNNs) such as Faster R-CNN, U-Net, and YOLO are widely used for object detection, classification and segmentation problems [5]. The convolutional layers in these networks help to identify the high-level and low-level morphological features of cells in the image in contrast to traditional approaches. A family of generative models known as generative adversarial networks (GANs) are being used for generating new plankton imagery data from the existing data, to better evaluate the existing models and to further improve the accuracy in measurements [6]. Efforts are also being directed towards new problems such as quantitative tracking of planktons, and biomass estimation through deep learning [7].

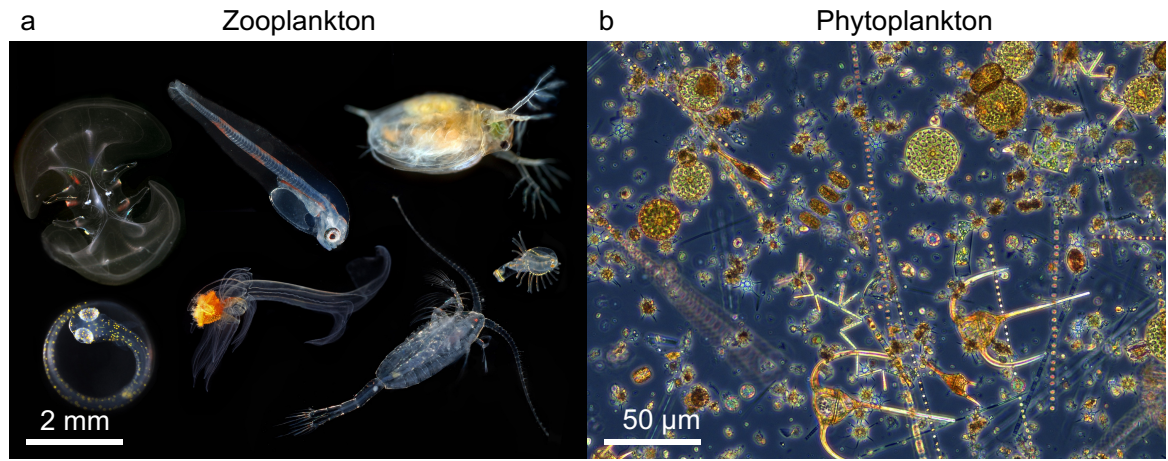

**Figure 22.1 Plankton.** Example of zooplankton (left) and phytoplankton (right) organisms. Plankton, as a broad term includes morphologically diverse organisms with complex geometries and deviating life stages that challenge AI driven classification and segmentation of images. Right panel photo by Malin Mohlin.

**Current and Future Challenges**

The rapid development of AI inspires applications in plankton analysis, such as automatic identification and high volume and throughput *in-situ* monitoring efforts. In parallel there is a more demand-driven development to fine-tune and improve existing methods and identify techniques that work with the limited amount of labelled data, which is typically the bottleneck with deep learning based methods. Below we discuss the most pressing challenges.

*Manual annotation of data:* To train a deep learning network, one needs labelled data. Which means if an image contains an organism, we need experts identifying the organism to species level based on taxonomic knowledge from the literature and experience. This is a laborious task and often prone to errors. The problems with manual annotation of data can be tackled with unsupervised and self-supervised deep learning algorithms. Unsupervised classification of organisms, for example, can be performed by investigation the latent space distribution of variational autoencoders, followed by a subsequent evaluation of the predicted clustering by taxonomic experts.

*Practical applications:* Deep learning applied research in plankton ecology is mostly restricted to detection, counting, and segmentation problems. This needs to be leveraged for more practical application scenarios to better understand the plankton dynamics. For instance, there is need to develop AI based tracking algorithms, that can track individual plankters over extended time periods. Mechanistic understanding of individual interactions such as predation, and resource exploitation are unexplored areas where boundaries are being pushed by deep learning algorithms.

*Practical constraints*: In field experiments and ship-based monitoring efforts, plankton communities are sampled at too low temporal and spatial resolution. Which means, plankton communities in ocean are rapidly evolving, and short-lived features such as blooms can easily form and disappear between two discrete sampling events. Apart from infrastructural difficulties, there is also a bottleneck in image analysis speeds. With the development of marine profiling instruments with AI based real time classification without manual annotations, the lower marine food web can be sampled at the appropriate spatial and temporal resolution. Furthermore, successful classification is a challenge given the overlapping morphologies in plankton cells (**Figure 22.1**). While some can be accurately classified, there are other where overlapping morphology will prevent accurate classification even with large training sets.

*Usability*: Though deep learning methods are becoming increasingly accessible, most often they are not production-ready and are not packaged in a ready-to-use interface. Efforts in this direction would benefit technicians and users with minimal programming knowledge.

**Advances in Science and Technology to Meet Challenges**

Below we discuss the recent developments in deep learning that can be used to tackle some of the challenges in plankton ecology and likely open the doors for new applications.

*Labelled data*: Obtaining the labelled ground truth data is an undisputed problem in many deep learning applications. Lately, unsupervised and self-supervised deep learning algorithms have shown some promising results in this direction. Cycle-GANs, which belong to the family of generative adversarial networks are widely used in style-transferring the images from one domain to different domain with *unpaired* images and ground truth data. From an image segmentation viewpoint of planktons (**Figure 22.2**), this indicates that Cycle-GANs can be used for segmentation tasks when there is a limited amount of manually annotated data. Since the plankton images and the corresponding segmentation masks needed not be paired in order to train a cycle-GAN, synthetically generated masks with comparable morphologies can be used as the ground truth data for real plankton images.

Additionally, microscopy imagery data of planktons can itself be synthetically generate either by GANs [6], or by simulating plankton-like objects. By employing the state-of-the-art computational optics and replicating the optical properties of the experimental devices that are used to record the data, a representative set of microscopy images can be generated on a large scale [8]. The advantage of synthetically generated data is that the ground truth is known beforehand and can be easily controlled. This has the potential to overcome the challenges that arise with manual annotation of data, specifically for the segmentation tasks where careful labelling of cell borders is crucial. As an example, the segmentations of plankton species shown in **Figure 22.2** are obtained by a U-Net model trained on simulated plankton data. This methodology also offers cell classification based on morphological properties, apart from the segmentation.

Recently, self-supervised deep learning algorithms have also shown some promising results in object segmentation and detection tasks. Unlike supervised algorithms where learning is based on labelled examples, self-supervised algorithms are provided with labels which are transformed versions of input images themselves. Particularly, Vision transformers (ViTs), which employ an attention based mechanism to emphasize different regions of input image, have outperformed CNNs in many computer-vision tasks [9]. Belonging to the family of ViTs, DINO (distillation with no labels), a self-supervised vision transformer [10], was able to successfully segment objects from images without any labelled data with a higher accuracy. Considering the diversity in plankton morphologies, self-supervised vision transformers can used for segmentation and classification of complex taxa.

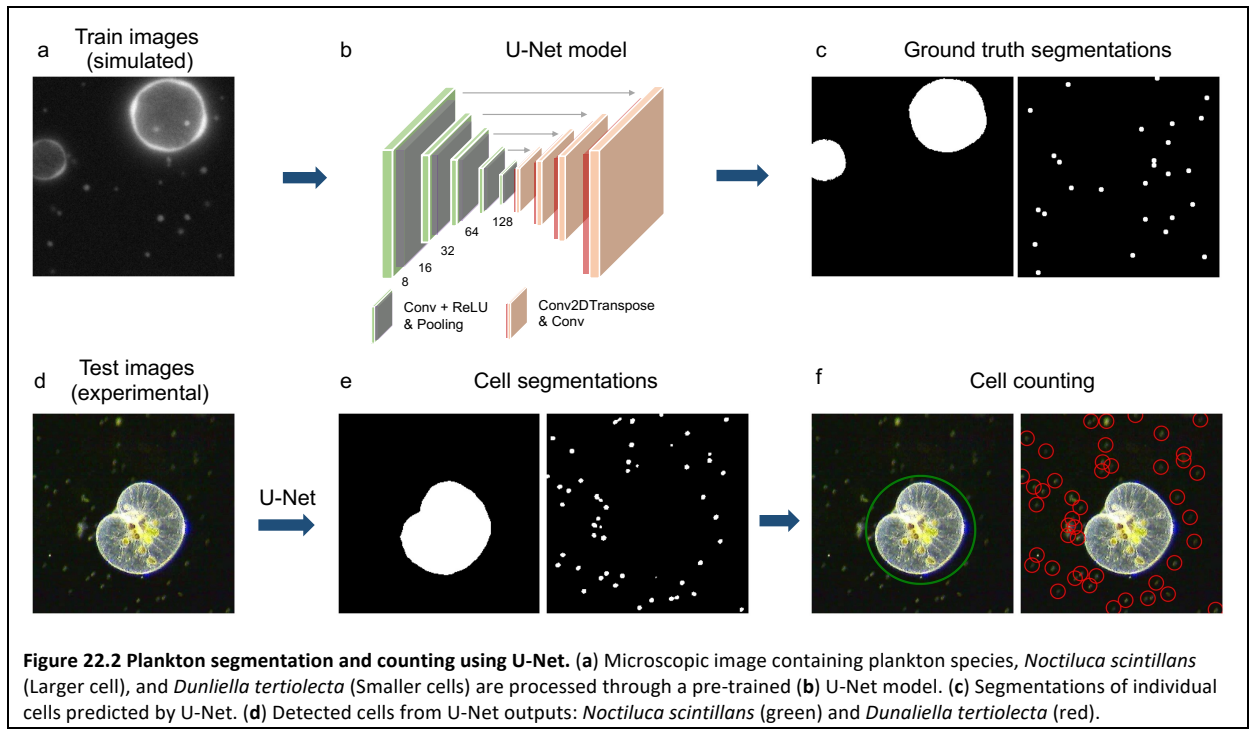

**Figure 22.2 Plankton segmentation and counting using U-Net.** (**a**) Microscopic image containing plankton species, *Noctiluca scintillans* (Larger cell), and *Dunliella tertiolecta* (Smaller cells) are processed through a pre-trained (**b**) U-Net model. (**c**) Segmentations of individual cells predicted by U-Net. (**d**) Detected cells from U-Net outputs: *Noctiluca scintillans* (green) and *Dunaliella tertiolecta* (red).

**Concluding Remarks**

AI driven microscopy is a rapid developing field of research. The automatic identification, segmentation, and tracking of individual plankton organisms provides mechanistic insights beyond the current state and will revolutionize our understanding of the lower aquatic food web and allow observations of high numbers of organisms at the relevant spatial and temporal scales. Moreover, recent combinations of well-established techniques such as digital holography with deep learning algorithms will facilitate individual resolution of plankton organisms and interactions in a way that will catalyse plankton ecology in coming years.

## 23 —Micro-physiological systems

Antoni Homs-Corbera[1]

1. Research and Development Department, Cherry Biotech SAS, Paris, France

**Status**

Micro Physiological Systems (MPS) lying at the interface of microfluidics and biology are a promising tool for better understanding human biology, physiology and physiopathology. Otherwise termed organ on chip (OoC) devices, these systems exploiting physical and chemical phenomena at the microscale, are able to replicate biological cellular and biomaterial ensembles as well as their conditions in the human body [1]. They provide functional units, emulating organs, that can be interconnected together [2], as is done by the blood and lymphatic streams in the naturally occurring systems, to provide simplified versions of the human body leading to a better understanding of complex multifactorial behaviours [3].

MPS are hybrid devices in which cells and biomaterials status during the experiments play a fundamental role on the accuracy and the relevance of results. Guaranteeing time-dependent spatial conditions and the monitoring of experimental events require gathering, managing and processing complex data in a non-invasive way while preserving the validity of the biological in vitro model (phenotype) at least in functional terms. The capability to provide large amounts of time-dependent data is inherent of the MPS which are potentially ready to integrate multiple sensors [4,5] and to be imaged with sophisticated microscopy techniques [6]. This provides the means of controlling and analysing biological complexity of several interconnected units, working synergistically, to gain insights into their systemic response.

The usage of machine learning techniques [4,7] to control and to evaluate specific behaviours of MPS when disturbed by stimuli, such as drug compounds, could provide a reliable tool to predict the behaviour of new potential therapies, to stir therapies discovery, to enable personalized medicine, or to spot processes that remain hidden to the human eye due to the complexity and volume of data that should be simultaneously evaluated. This could open a paradigm shift in the pharma and biotech industries with important socioeconomic positive impacts and the substitution, at least partially, of the current *in vivo* animal models [3,8,9]. The current parallel advancement of machine learning, MPS, sensing and imaging technologies provides a fertile field to achieve autonomous MPS platforms providing the necessary high-throughput and robustness to be used as true human physiology and physiopathology emulators. Despite current limitations, the combination of these technologies is already providing encouraging scientific results [7]. An example of how such a system could work is presented in **Figure 23.1**.

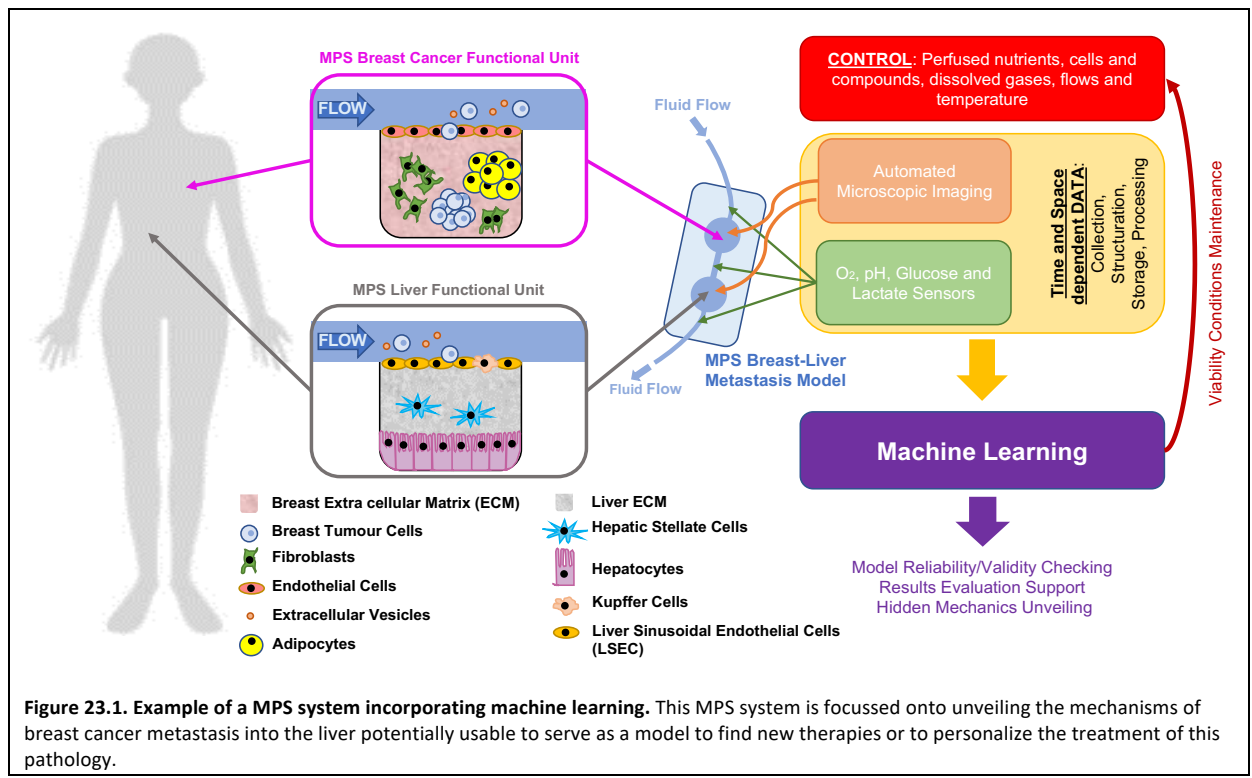

Figure 23.1. **Example of a MPS system incorporating machine learning.** This MPS system is focussed onto unveiling the mechanisms of breast cancer metastasis into the liver potentially usable to serve as a model to find new therapies or to personalize the treatment of this pathology.

**Current and Future Challenges**

Human physiology has a strong dependence on local physicochemical parameters. For long-term studies on MPS accurate control is critical to provide reliable and meaningful data. This control, when aimed massively, particularly for potential applications related to therapeutic screenings, cannot be done by human operators. Automated control systems able to take decisions in real time to fine-tune and maintain MPS parameters, such as dissolved oxygen contain, pH, temperature, fluidic profiles or induced mechanical forces, will be critical (**Figure 23.1**). Furthermore, they could assess how the system is responding to specific stimulus and also provide an indicator of validity of the MPS model after each experiment based on monitored media components and metabolites as well as other events occurred.

The combination of sensing and imaging technologies with machine-learning approaches provides a promising tool to achieve such a degree of control and evaluation in these complex systems [5-7]. However, sensing and imaging techniques need to be chosen wisely not to alter the MPS biology itself. Chemical and physical conditions that introduce bias (e.g., phototoxicity, genetic alterations, molecule absorption) have to be avoided to keep the models viable and realistic. Some of these undesirable effects can be also eliminated by machine-learning techniques using advances coming from microscopic image analysis such as virtual staining [10] that could avoid the use of chemical staining and eliminate phototoxicity in some applications.

Nevertheless, acquiring and labelling massive input data to train future autonomous decision-making systems is a tedious and user-intensive task that is subject to certain bias. Also, continuous monitoring of the development and the function of these biological models implies real-time comparison and integration of multiple data sources from heterogeneous conditions which is a challenge by itself. The machine learning evaluation system should be able to classify time-dependent events expressing in different dimensional scopes. This is another considerable challenge linked to evaluate simultaneously cellular level characteristics, such as phenotypes, migration (e.g., immune and tumoral) or distant communication (e.g., extra cellular vesicles), as well as overall larger multicellular functional units and systemic responses.

The model proposed in **Figure 23.1** illustrates the degree of complexity that can be attained in an MPS. Recently, cell-derived vesicles secreted by breast cancer cells have shown to travel and activate liver sinusoidal endothelial cells (LSEC) in a liver MPS promoting the destruction of vessel barriers and unveiling metastatic mechanics [11]. At imaging level, higher resolution microscopy could be also required on top of conventional one to follow the morphology and density of fenestrations, nanostructured apertures present in the liver LSEC.

Multisource automated study of multidimensional time-dependent data is still in its youth for MPS. So far, machine learning for automation of single-sensor measurements, imaging systems or identification of key candidates in a drug screening procedure have been implemented, but a long road has to be paved to achieve reliable, usable and feasible multiplexed and complex MPS.

**Advances in Science and Technology to Meet Challenges**

Machine-learning techniques have already demonstrated their value and feasibility in simple MPS to generate virtual staining, to study cells phenotypes or to track migration events [7] both in supervised and non-supervised approaches. However, the usage of machine-learning techniques to control the multiple parameters of MPS microenvironments to guarantee their viability and phenotype has so far not been demonstrated. Furthermore, limited data sources have been used to evaluate the MPS, such as a unique time-lapsed imaging approach or a limited number of combined analytical sensors, reducing the true potential capabilities of the approach and dismissing important physiological and physio pathological data.

MPS systems improvements in terms of usability, repeatability, high-throughput and robustness have to go hand in hand with imaging, sensing and machine-learning advances. The provision of completely automated MPS systems with these characteristics for continuous, standardized, condition-controlled and reliable data provision for researchers should be the main aim of current developments. This will allow test and optimization of different machine learning strategies and mechanisms leading to exploit the full potential of these technologies. It will also lead to facilitate high-throughput analysis which is critical to generate the envisaged healthcare disrupting system that are aimed to be. Eventually, other data obtained by sampling the MPS fluids and tissues, such as genetic information, or by medical evaluation of human sources could be added to improve the results when therapeutic evaluation outcomes are targeted.

From the pure machine-learning algorithms point of view, data augmentation, semi-supervised learning and transfer learning can be explored as means to reduce the amount of data needed for the training. Similarly, automatic data annotation can be used to reduce the cost of manual labelling. Furthermore, specific requirements for deep learning algorithms applied to the MPS field involve the development of novel customized architecture and layout designs. Different approaches to automate design and gain efficiency have to be explored and catastrophic interference has to be prevented in any explored iterative upgrading. Also, the large amount of real-time data gathered by these systems coupled to sensors and imaging devices implies an accurate study of how to improve inference speeds by compressing the model volume while ensuring accuracy. Equivalently, the hardware utilized for training the deep learning algorithm and evaluating MPS systems needs to be chosen wisely and performance of the overall system optimized. Finally, interpretable deep learning technologies, mitigating the "black box" effect, should be developed to facilitate the generation of meaningful physical explanations from a biological point of view.

**Concluding Remarks**

Combination of MPS and machine learning techniques is an emerging field holding a great potential to impact human biology understanding, medical therapies development and personalized medicine. Ideally, the application of machine-learning techniques to those multiparametric complex systems should in the future provide currently unattainable biological insights by using unsupervised learning approaches. On the other hand, by comparing results obtained with well-known and characterized therapeutic compounds libraries to the ones resulting with novel compounds and to the real *in vivo* outcomes these algorithms could also achieve good levels of prediction in drug discovery and therapy personalization. Eventually, medical and biological data coming from the human subjects providing tissue or cells in the studies, and obtained by other means, could be also included in the machine learning models providing extra information in therapeutic discovery and personalized medicine applications.

In the years to come, efforts into automatizing and standardizing the MPS data sources, providing experimental robustness, high-throughput and scalability, should improve the machine learning outcomes in this domain. The need of condition-controlled, massive and reliable data is paramount. Besides this, the specificities of MPS devices, their data heterogeneity and its multiples sources, have to be considering when designing and exploring strategies for any machine or deep learning approaches.

**Acknowledgements**

*The author would like to thank the European Commission for continuous research and innovation financial support to Cherry Biotech SAS in the subjects treated in this article through different initiatives and programmes (MSCA-ITN-2018-812780, HORIZON-EIC-2021-PATHFINDEROPEN-01-01-101046928, HORIZON-EIC-2021-PATHFINDEROPEN-01-01-101046620, HORIZON-EIC-2022-TRANSITIONOPEN-01-101099775).*

## 24 — Self-learning thermofluidics

Martin Fränzl[1], Tobias Thalheim[1], Frank Cichos[1]

1. Peter Debye Institute for Soft Matter Physics, Leipzig University

**Status**

In microfluidics, liquids and suspended objects are manipulated in tiny channels to trigger chemical reactions, assemble new materials, analyse single cells at high throughput, or to imitate organs on a chip [1]. Its innovation potential for fields like biology, chemistry, drug design, and medicine is driven by the superior mass and heat transport at these small length scales combined with an easy experimental accessibility and control. Most fluid manipulation approaches are pressure-based and require external pressure differentials applied at the inlet/outlet of complex channel systems. Such methods allow for complex flow designs, but need to move the fluid throughout the channel to affect the local composition, though other local approaches based on electro-kinetic (electrophoretic) actuation exist. In recent years new techniques for the manipulation of nano-objects and flows in a fluidic environment have been developed that are remotely controlled by light and go beyond the force generation of optical tweezers. These approaches employ local temperature gradients to drive the migration of species suspended in liquids by thermophoresis [2], thermo-osmosis [3], or thermo-viscous [4] or other secondary effects [3] and may be summarized as thermofluidics (**Figure 24.1**). In this way, schemes for manipulating colloids [5] and single molecules [6] have been developed to even provide new access to the study of protein aggregation [7]. Temperature gradients at liquid–solid boundaries in simple fluidic slit pores allow the generation of local flow patterns to guide, manipulate, and separate objects suspended in liquids without any external pressure [3]. Thermo-viscous effects allow dynamic flow generation by dynamic local heating of the liquid. Due to the small dimensions of microfluidic systems, heat-transfer is fast paving the way for feedback-controlled techniques in fluid manipulation [8]. A combination of all these effects would allow to stir, mix, separate, compress, and even heat/cool in microfluidic systems based on the simple application of light. Yet, large scale integration and application in microfluidics is still missing.

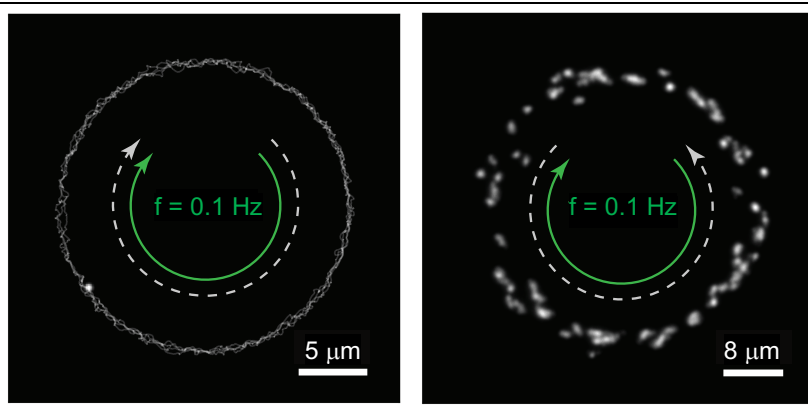

**Figure 24.1 Colloidal motion through thermo-osmotic flows.** Colloidal motion induced by controlled thermo-osmotic flows (left) and a combination of thermo-viscous and thermo-osmotic flows (right). Adapted from Ref. [3].

**Current and Future Challenges**

The increasing complexity in the phase composition of liquids, in the objects suspended, in multi-modal sensing techniques used together with the specific goal of microfluidic applications raises the need for new machine-learning-based analysis and control schemes that harvest the advantages of microfluidics [1] (**Figure 24.2**). One of the current challenges that, for example, is often met in various applications is the appearance of heterogeneities in samples, which hampers the data analysis of molecular species. Such heterogeneous samples are, for example, highly relevant in the study of protein aggregation in assays to understand the origins of neurodegenerative diseases [7].

The experimental investigation of heterogeneous ensembles leads to ensemble averages that are dominated by the most abundant species, which, however, might not be the most important ones for the specific disease. The new local approaches of thermofluidics can be size- or even species-selective in their action allowing the spatial dispersion of the different species. Yet, their interaction with the temperature fields is often unknown. Machine learning approaches, for example, for real-time visual classification and localization of species in a sample [9] combined with reinforcement learning [10] seem well-suited to meet this challenge of a self-learning variant of thermofluidics. Together with local spectroscopic information in real-time to improve the local homogeneity while increasing the spatial heterogeneity will readily lead to new reconfigurable self-learning tools to tackle chemical, medical, or physical questions with high flexibility.

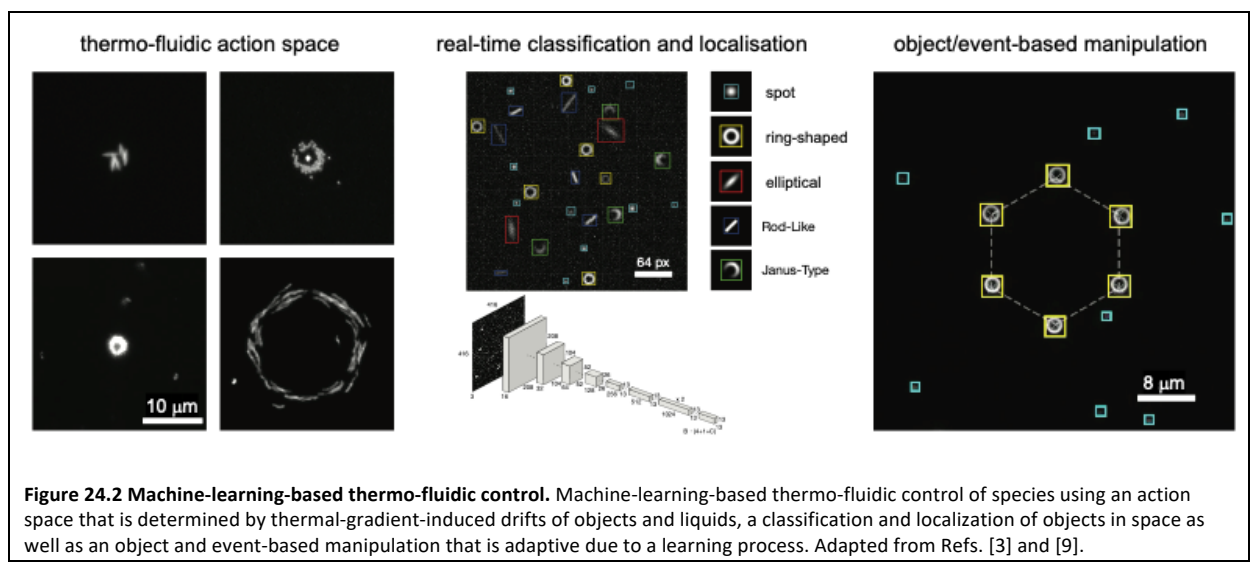

**Figure 24.2 Machine-learning-based thermo-fluidic control.** Machine-learning-based thermo-fluidic control of species using an action space that is determined by thermal-gradient-induced drifts of objects and liquids, a classification and localization of objects in space as well as an object and event-based manipulation that is adaptive due to a learning process. Adapted from Refs. [3] and [9].

**Advances in Science and Technology to Meet Challenges**

The main advances to meet these challenges constitute on one side the large-scale integration and testing of thermofluidic approaches into conventional microfluidic systems. This includes simple absorptive layers to induce laser-controlled local temperature increments to yield strong boundary flows that can be combined with additional fields such as electric fields to deliver even stronger effects of thermo-electrohydrodynamics. These approaches shall be studied with pressure-driven flows as they will provide many new variants, such as temperature-driven flow field fractionation. On the other side, the key approach that is suggested here is, however, the use of the freely configurable dynamic temperature fields with newly developed machine learning approaches (Recurrent Networks, CNNs, Reinforcement Learning) that pick up local signals with chemical resolution at a high sensitivity and speed. The machine learning techniques make use of the local character of the temperature perturbations to yield a goal-driven microfluidics. This further requires highly sensitive detection techniques, that deliver at best chemical resolution in real-time to allow the adaptive improvement suggested by machine learning such as deep reinforcement learning. Such highly sensitive experimental techniques may, for example, involve new variants of photothermal infrared microscopy, which recently revolutionized infrared microscopy but still have to be adopted for microfluidic applications.

## Concluding Remarks

Local thermofluidic effects have the potential to drive new approaches of machine-learning-driven microfluidics as they provide a rich action set that can be applied to fluids of almost any type to manipulate suspended objects, but also to induce local flows that separate or mix different constituents. While the individual interactions of species with temperature fields might not be known in detail, machine learning of actions to yield specific outcomes directly during the experiment, will provide new approaches to manipulate species for applications in physics, chemistry, biology, and medicine.

## Acknowledgements

*We acknowledge support from the SFB TRR102 "Polymers under multiple constraints: restricted and controlled molecular order and mobility" (DFG project number 189853844).*

## 25 — Deep Learning in Digital pathology

Carolina Wählby[1]

1. Dept. IT and SciLifeLab, Uppsala University, Sweden

**Status**

Microscopy-aided visual assessment of tissue samples, such as biopsies and cytological smears, has been essential for diagnosing cancer and other disease for more than 100 years. Grading disease severity is critical to the clinical management of patients. It is however a difficult task, and the variability between pathologists poses clinical challenges, leading to both under- and overtreatment, which impacts patient morbidity, mortality, and healthcare costs. The first approaches to machine-aided diagnostics in pathology were made already in the 1930s, and more recently, efficient slide scanners and automated sample handling has boosted digital pathology and the development of deep learning-based image analysis systems for assisting pathologists in sample evaluation [1]. Such systems have the potential to increase both accuracy and cost efficiency in cancer screening, and are a prospective solution to the problem of high inter-pathologist variability [2]. Some of the most impressive results have been reached by global challenges such as the PANDA challenge [3], including training- and test-data from multiple hospitals and countries, leading to deep-learning-based solutions based on ensembles of diverse models, featuring, for example, different data preprocessing approaches and different neural network architectures.

At the same time, automated systems based on deep learning are often very sensitive to sample-to-sample variation and artifacts stemming from processes during sample collection, sample handling, staining, and scanning. While a human is very efficient in adapting to such variability, very subtle variations, sometimes not even possible to notice by the human eye, can have catastrophic effects on automated detection and grading. Large multi-site efforts such as the PANDA challenge can overcome these issues simply by ensuring that the massive amounts of data included during training cover as much as possible of the sample variation that can be expected during model deployment in the clinic. However, such large-scale efforts are costly, and may be difficult to organize for rare disease where sample availability is limited. Another bottleneck is the reliability of the training data, depending heavily on the inter-pathologist variability in visual annotation of samples. With robust approaches for image normalization, augmentation, and novel learning-regimen that ignore non-essential variability, deep learning has the potential to become a broadly applicable and reliable tool in the clinic.

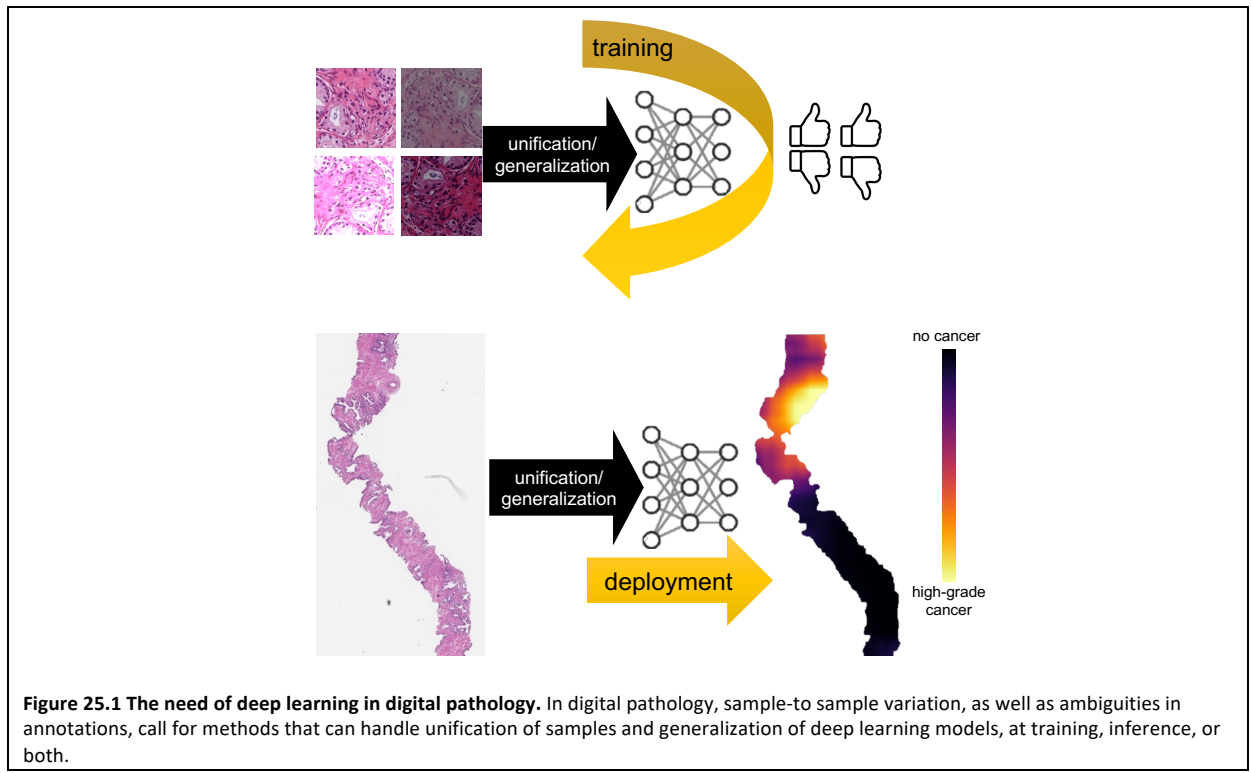

**Figure 25.1 The need of deep learning in digital pathology.** In digital pathology, sample-to-sample variation, as well as ambiguities in annotations, call for methods that can handle unification of samples and generalization of deep learning models, at training, inference, or both.

**Current and Future Challenges**

The two largest challenges in deployment of deep learning for digital pathology is the availability of reliable training data, and sample-to-sample variability.

Another word for training data is 'ground truth', an expression that comes from remote sensing, where data is collected from imaging devices attached to satellites or aircrafts, and automated analysis results, such as mapping of roads or classification of tree species, are compared to 'the truth' collected from observations made on the ground. Such observations of 'the truth' are not straight forward in digital pathology. Typically, the 'ground truth' is manual visual annotations by pathologist. Many times, people ask the question 'What precision does a learning-based decision system have to reach to be useful?'. This number must always be answered in relation to the 'truth' to which it is compared, and since pathologists often disagree in their visual assessments, both training and evaluation has to be done with care. One approach is to compare automated result from a learning-based decision system to multiple manual annotations, and in the same way also compare each of the manual annotators to one another [4]. One can also use other metrics as a means of evaluating method's performance. Patient survival is such a metric. This is however a very noisy metric, especially since a mm-sized tissue sample may not at all be representative of the cause of death of a patient.

Sample-to-sample variability and limited generalization performance is a fundamental problem when using deep learning applied to digital pathology, and lack of generalization may even introduce bias. For example, digitized tissue samples may be collected from a number of different hospitals. If one of these hospitals is specialized in, e.g., a severe type of breast cancer, there is a risk that the system learns to associate irrelevant hospital-specific image effects with sever breast cancer, rather than learning the actual features of the tissue morphology. This is typically approached by some form of *unification of data* from different sites to each other, for example by normalization [5], or by *generalization of the model* during training by creating artificial data so that samples from different sites span the same parameter space, see **Figure 25.1**. The simplest approach to stain normalization is to separate the RGB-image into its stain components and scale each channel to a fixed intensity interval. More advanced methods, such as sparse non-negative matrix factorization have been successfully used to normalize individual stains [6].

**Advances in Science and Technology to Meet Challenges**

When training a deep learning model, the input data influences the model's ability to learn relevant features of the data and generalize to new data. A standard technique is to use color and texture augmentation of the training data, artificially generating more variations for the network to learn. However, it is typically difficult to produce a dataset without some bias toward any specific feature. Deep learning models used in digital pathology have a tendency to overfit to the stain appearance of the training data. If a model is trained on data from one lab only, it will usually not be able to generalize to data from other labs.

Recent advances in Generative Adversarial Networks (GANs) can reduce the effects of sample variation by being trained to mimic what an observed image would have looked if it was captured in a different batch or at a different site [7]. This type of neural network-based sample unification can transfer images from one 'mode of variation' to another while preserving the phenotype of the tissue morphology. In this way, the training data can be extensively expanded to represent a feature space spanning both the non-important variation due to sample handling, and the disease-related variation that is to be learned. It is however important to note that false structures that could influence grading may be added.

Another promising approach is to use so-called domain-adversarial neural networks, which are designed to prevent the model from being biased towards features that in reality are irrelevant, such as the origin of an image. Ultimately, such a system would adapt in a similar fashion as a human, resulting in no need for normalization or augmentation, as indicated by promising results on prostate cancer grading on datasets from different hospitals [8].

New molecular methods have the potential to bring deep learning in digital pathology beyond mimicking ambiguous manual annotations. Recently, new transcriptome-wide analysis technologies have enabled the study of RNA molecules directly in tissue samples, thus maintaining spatial resolution and complementing histological information with molecular information important for the understanding of many biological processes and potentially relevant for the clinical management of cancer patients [9]. Parallel application of standard clinical staining techniques and novel molecular methods introduces a novel type independent sample annotation that can be used for model training, with the potential to discover of previously unseen but medically relevant tissue patterns.

**Concluding Remarks**

Limited generalization across diverse multinational cohorts is one of the central barriers to implementation of deep learning in clinical practice. Strict protocols and quality control, all the way from sample collection to staining and scanning has the potential to reduce variability. At the same time, deep learning approaches that can adapt to shifted domains much like a human, learning to discriminate between image features relevant or irrelevant for decision making, are starting to make their way into digital pathology.

As spatially resolved novel multiplexed and target-specific molecular methods can be directly correlated with prognosis and strategies for treatment, they may function as useful tools by themselves. They may however also function as a means to molecularly 'annotate' parallel slices of tissue samples exposed to standard clinical stains, thus providing input for deep learning systems that have the potential to go beyond mimicking what a human observer could do.

Despite many challenges still remaining before deep learning will be widely used in clinical practice, the attitudes are generally positive: A survey with 487 pathologists practicing in 54 countries [10] showed that nearly 75% reported interest or excitement in AI as a diagnostic tool in pathology.

**Acknowledgements**
*This research was funded by the European Research Council via ERC Consolidator grant CoG 682810 to C.W. and the SciLifeLab Bioimage Informatics Facility.*

# 26 — Virtual staining of histological tissue sections using deep learning

Kevin de Haan[1], Yair Rivenson[1], Aydogan Ozcan[1]

1. Department of Electrical and Computer Engineering, University of California Los Angeles

**Status**

The histochemical analysis of tissue samples is used as the gold standard for diagnosing diseases. It is a hundred-year-old practice based on microscopic analysis of 5- to 10-micron-thin tissue sections stained to provide contrast optimized for the human visual system. While the most commonly used stain is hematoxylin and eosin (H&E), a wide variety of stains are used to give different contrast to different tissue constituents. This staining process can be time-consuming and expensive and can create a significant amount of chemical waste, some of which is toxic. The staining process is also destructive to the specimen, so typically only a single stain can be performed on each tissue section. Two different deep-learning-based methods have been recently developed to computationally generate an accurate artificial/virtual stain: 1) virtually staining label-free tissue sections, and 2) transforming one stain into another.

Virtual staining of an unlabelled tissue section can be used to eliminate the need for chemical staining altogether. It involves using a deep neural network to computationally transform images of label-free tissue into various stains (**Figure 26.1i**). This technique has been developed using different imaging modalities as the input to the neural network, with many different stains being developed. For example, autofluorescence microscopy [1], quantitative phase imaging (QPI) [2], and reflectance confocal microscopy [3] have all been used to image the label-free tissue and achieve virtual staining.

Stain transformations — the second class of methods that can be used as an alternative to histochemical staining — use deep neural networks to perform transformations from one chemically labelled stain to another[4,5] (**Figure 26.1ii**). While these methods rely on images of chemically stained tissue as input to the neural networks, they can be used to avoid the need for as many stains to be performed, without changing existing clinical pathology workflows. Furthermore, H&E, which many of the transformations are based on, can be performed consistently and is already used in nearly every clinical case. Therefore, the virtual stain transformation technique can be targeted to replace more difficult and costly special stains with their virtually stained counterparts.

Both the label-free virtual staining and stain transformation techniques have been proven to create highly accurate computationally generated stains, which are equivalent in quality to histochemical stains, and allow for accurate diagnoses to be performed [1,4,6]. In addition to avoiding the need for chemical labelling (reducing costs and eliminating chemical waste), these techniques have several advantages over standard histochemical staining. For example, virtual stains are standardized, as the same transformation is performed by the neural network every time. This results in stain-to-stain variations being minimized. Another major benefit is that multiple stains can be performed on a single tissue cross-section, allowing pathologists to view the same area (and therefore the exact same cells) with multiple stains rather than relying upon staining of serial tissue sections. Furthermore, when performing virtual staining of an unlabelled tissue, the tissue is also preserved for future use if more advanced (e.g., molecular) analysis is needed.

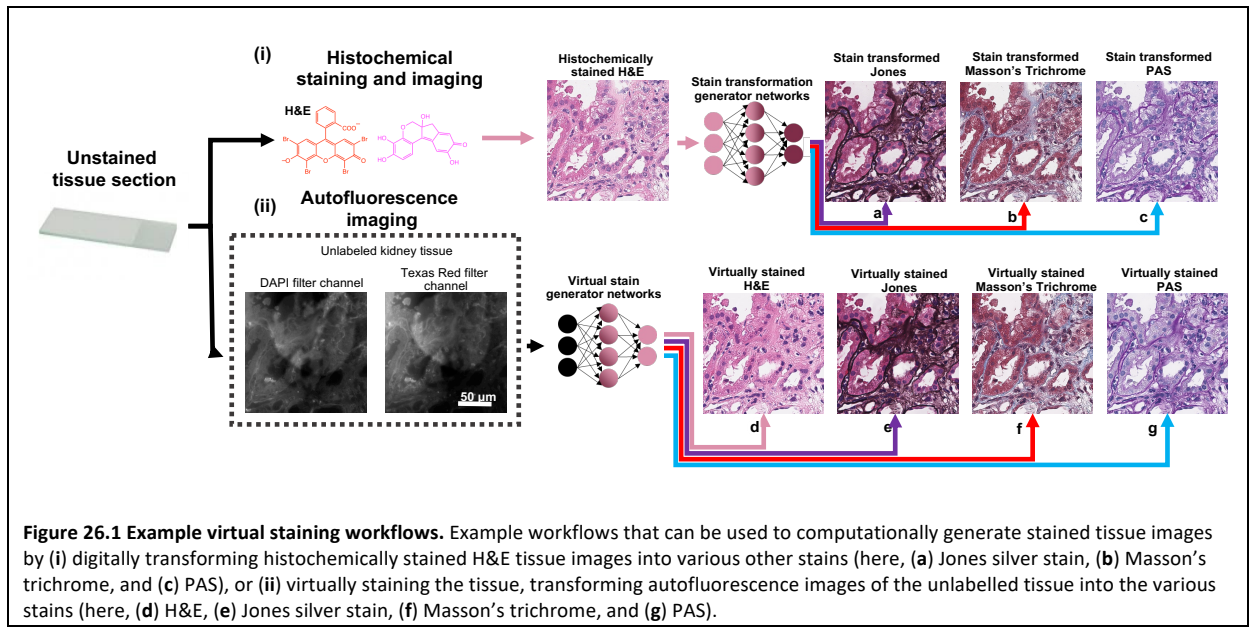

**Figure 26.1 Example virtual staining workflows.** Example workflows that can be used to computationally generate stained tissue images by (**i**) digitally transforming histochemically stained H&E tissue images into various other stains (here, (**a**) Jones silver stain, (**b**) Masson's trichrome, and (**c**) PAS), or (**ii**) virtually staining the tissue, transforming autofluorescence images of the unlabelled tissue into the various stains (here, (**d**) H&E, (**e**) Jones silver stain, (**f**) Masson's trichrome, and (**g**) PAS).

**Current and Future Challenges**

One of the challenges posed during the creation of virtual staining models is the generation of the image data required to train the neural networks. These data often need to go through an extensive pre-processing workflow, which can be time-consuming, particularly if manual steps such as data cleaning are necessary. Furthermore, a significant amount of data from various sources/labs is needed to allow the models to generalize to new patients or different sample processing procedures. This is a particular pain point for stain transformations, as there can be significant variations between the stains performed at different labs or even within a single lab. For these models to be useful, they must be able to generalize to any given stain (that is correctly performed).

Image-to-image translation techniques such as virtual staining are often performed by deep neural networks trained using supervised learning, taking advantage of loss functions that directly teach the network to perform a mapping between pixels. These structural loss functions ensure that an accurate transformation is learned. However, supervised learning techniques rely upon images of the same tissue section captured before and after the staining being matched at the pixel level. This image registration task can be difficult, particularly when the tissue is damaged or structurally altered during the staining process. Furthermore, for the stain transformation technique, the input to the neural network is a stained tissue image. This further complicates the image registration process, as destaining and restaining of tissue is very difficult. While there has been some success with using unsupervised learning (e.g., with distribution matching losses) to perform transformations between stains [5], they may be limited as networks trained only using distribution matching losses are prone to potential hallucinations [7].

**Advances in Science and Technology to Meet Challenges**

These limitations have left room for technological advancements to improve upon previously developed virtual staining techniques. For example, techniques such as data augmentation through stain style transfer are effective at allowing stain transformation networks to generalize across a large sample distribution [4]. By developing and incorporating data augmentation techniques for other modalities such as fluorescence, virtual staining techniques may become more effective and generalizable.

Significant improvements to the quality of virtual staining and ease of dataset development can be achieved by advancing the multimodal image registration techniques used to match tissue images before and after labelling. Synthetic data can be used to avoid the image matching process, and have effectively been used to train stain transformation networks in a supervised manner[4]. More advanced loss functions can also be developed to make the networks less reliant on perfect image registration. Finally, by modifying the chemical labelling process (i.e., the ground truth generation step), the stains may be optimized to reduce the physical damage to the tissue, making any image registration pipeline more accurate and easier to perform.

There is also room to expand upon the unique opportunities afforded by computationally generated stains rather than using histochemical staining. For example, one transformative aspect of virtual staining is the ability to multiplex stains by performing multiple stains all from a single scan (as shown in **Figure 26.1**, where four different stains are generated from autofluorescence images of a single tissue section). When a single network is trained to perform multiple stains, the stains can also be blended to digitally generate new artificial stains, giving different levels of contrast to different tissue constituents [8]. Multiplexed staining also enables micro-structured staining, where different tissue areas in a single image of a tissue sample appear as different stains [8]. In contrast, histochemical staining is a destructive process, so each tissue section can only be stained a single time and by a single type of stain. Researchers may exploit these and other unique aspects of computational stains enabled by deep learning to improve the speed and accuracy of diagnoses, with further research being able to determine the best opportunities for virtual staining to improve existing clinical workflows.

There is also significant room for this technology to advance by taking advantage of the symbiosis between virtual staining and image analysis algorithms. Some preliminary studies have begun to show this potential, for example, by demonstrating that the information extracted by a stain transformation network can be used to improve image segmentation algorithms [9]. This same technology can be adapted to improve other downstream image analyses, such as automated diagnostic algorithms, by incorporating virtual staining into machine learning-based disease detection workflows.

**Concluding Remarks**

The use of virtual staining of label-free tissue, and stain transformations between histochemical stains is a rapidly developing field with significant potential to improve the field of pathology. These technologies can improve the speed of diagnoses while reducing costs and chemical waste. While this is still an emerging field of research, with continued advancements and regulatory approvals, there are many opportunities for it to assist both human and computer-based diagnoses. Furthermore, as the field of pathology continues to move away from manual inspection of glass slides and toward digitization, there will be more and more opportunities for alternative imaging modalities and digital visualization technologies to make their impact and improve upon standard pathology workflows.

**Acknowledgments**
The authors acknowledge the support of NSF Biophotonics Program.

## 27 — Cell phenotype determination using virtual staining

Zofia Korczak[1], Caroline Beck Adiels[1]

1. Department of Physics, University of Gothenburg, Sweden

**Status**

Identifying cell types and following their dynamic behavior is crucial when delineating biological mechanisms. Whether *in vitro* or *in vivo*, cell–environment interactions result in stimuli-dependent cell responses. These can constitute, e.g., stem cell differentiation during embryogenesis, cellular defense responses due to altered extracellular environment, or malfunctioning behaviors due to pathological circumstances. Typically, intracellular gene and protein expression profiles are obtained only after cell isolation, whereby the spatial cell–environment interaction is lost. Instead, phenotypic characteristics can rely on visual traits or cell morphology captured by imaging. The classical methodology uses a probe to target the cell, protein, or even particle of interest, which is detected and visualized by microscopy. The probe can constitute a dye that binds to a specific protein or subcellular organelle, a fusion protein expressed in genetically modified cells, or another type of particle or substance that provides information about the cell's phenotype or behavior. Any structural or behavioral cell differences are then analyzed via image processing, clustered, and commonly confirmed with additional biochemical analyses. The temporal and spatial resolution can be modulated depending on the imaging setup and the probe used. Some probes can be applied to and are compatible with living cells with minimal impact during a limited period. Others are harmful and will affect the cells and possibly the response, therefore are better used on terminated cell cultures or "dead" tissue. Regardless, the specificity and sensitivity of the methodology rely on many different parameters, some hard to control. Further, analyzing the microscopy image outcome of sometimes subtle changes or variations is far from straightforward that may be misleading, especially if using rigid, more traditional image analysis tools, not to mention the risk of being biased.

With the advancement of imaging techniques and artificial intelligence approaches, a new era of biological screening has emerged. Deep learning mimics humans' learning process and includes statistics and predictive modeling. The technology has been successfully used in numerous biological applications such as various clinical diagnostics (reviewed in Ref. [2]), quantitative extraction of information on the single cell level [3], and subcellular localization of organelle-specific proteins [4]. It is also central in developing probe-free cell identification and tracking analysis tools. The raw input data can stem from various microscopy modalities: bright-field [5], fluorescence [3], structured illumination [6], scattering [7], or phase contrast [1] (see **Figure 27.1**), to mention a few. The abovementioned models are trained using supervised learning, which is suitable when the input and the corresponding output are known. The network develops a strategy to transform an input image into the correct output from labeled training data. Another possibility is to train models using unsupervised learning. Unsupervised learning aims to find hidden patterns from the unlabeled dataset. It carries significant uncertainty about the features the model considers crucial but is supposed to be less biased compared to supervised learning models and is foreseen to be the next generation of software models for complex image analysis.

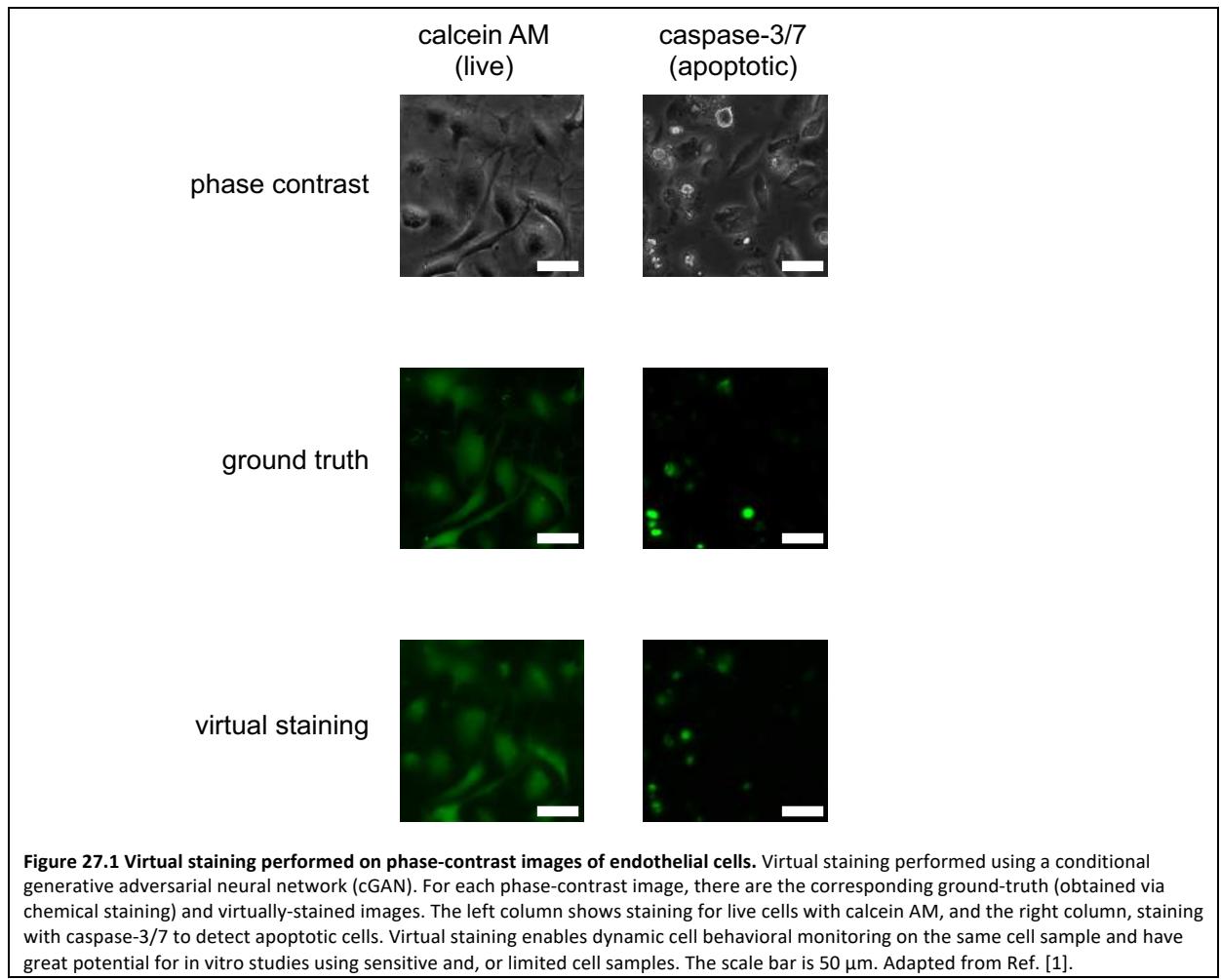

**Figure 27.1 Virtual staining performed on phase-contrast images of endothelial cells.** Virtual staining performed using a conditional generative adversarial neural network (cGAN). For each phase-contrast image, there are the corresponding ground-truth (obtained via chemical staining) and virtually-stained images. The left column shows staining for live cells with calcein AM, and the right column, staining with caspase-3/7 to detect apoptotic cells. Virtual staining enables dynamic cell behavioral monitoring on the same cell sample and have great potential for in vitro studies using sensitive and, or limited cell samples. The scale bar is 50 μm. Adapted from Ref. [1].

**Current and Future Challenges**

Despite the usefulness and many available applications of deep learning in cellular phenotyping, there are still challenges to overcome to improve the efficacy of the technology in complex biological settings. One aspect regards attaining ground truth data; the other regards choosing or developing the optimal imaging modalities for distinct purposes.

A deep-learning-based method successfully predicted progenitor cells' future differentiation direction from bright-field microscopy images, using fluorescent probes to generate ground truth data [8]. In this example, when targeting specific proteins, fixation of the cell culture is required before adding the immunofluorescent antibodies, which eradicates the possibility of dynamic monitoring of the same cell sample. For live cell data to be acquired, live cell-compatible dyes, endogenously expressed fusion proteins or alternative staining methods are necessary. Such an approach is not always applicable, e.g., when focusing on unique protein expressions or using cells that are difficult to manipulate genetically.

The next hurdle is to image and monitor cells in more complex and physiologically relevant settings, e.g., 3D cell cultures. Such volumetric cultures are often very dense with complex cell-cell interaction patterns, resulting in problematic staining and imaging processes. First, dyes targeting specific cell structures or proteins might not even reach their targets, a common scenario seen for organoid staining where dyes have difficulties penetrating the sample. Second, the dense structure scatters the illumination light, limiting the resolution and focus, which masks the readout. For instance, this is the case when using fluorescent probing in a classical setup of light-sheet-based microscopy, where aberrations, and low-intensity illumination, as a result of scattering, limit the sample thickness range.

Another relevant challenge is the simultaneous monitoring of dynamic morphological changes on both population and subcellular-scale levels. Such information is relevant for, e.g., understanding the importance of heterogeneous versus synchronized cell responses in cell communication studies or cancer cell screening. In the latter scenario, discriminating cells' behavior based on migration speed and direction, differentiation, and replication frequency over time will be necessary.

Moreover, although new model architectures require only a few or even just one image for training, many currently best-working networks require high computational power for training and data analysis. Unfortunately, not all users interested in switching to AI-based software have access to such computational clusters or powerful computers with GPUs.

**Advances in Science and Technology to Meet Challenges**

Meeting these challenges requires imaging techniques that provide information-rich data as input to the deep learning models, but also models that can interpret correctly the often non-grid ordered and interconnected nature of the biological information entities (see **Figure 27.2**).

Powerful imaging techniques of multiple modalities, e.g., high-speed light-sheet microscopy [9], 3D-structured illumination microscopy [10], or 3D-pRESOLFT (3D, parallelized, reversible, saturable/switchable optical fluorescence transition) microscopy [11], might overcome the current limitations of volumetric imaging (to monitor time-resolved dynamic cell behavior). 3D-pRESOLFT, for example, relies on reversibly switchable fluorescent proteins and endogenously expressed labels and has been used to monitor dynamic structural alteration upon stimulation from living neurons combining high levels of spatial resolution (sub 80 nm) and a wide field of view (>40 × 40 µm2). The methodology could benefit from combining with AI, such as graph neural networks (GNNs), to classify and predict the outcome. Especially in the biomedical field, GNNs have recently attained much attention, due to their ability to, by signal processing, model unstructured and structured relational data.

GNNs have been utilized in an alternative approach where photoactivatable fluorescent probes target annotated cells, and live cell spatial transcriptomics data is attained [12]. The approach provides spatial structure and high-throughput gene expression profiles for individual cells in parallel, successfully clustered by implementing GNNs.

Instead of using expensive and sophisticated microscopy setups for 3D imaging, there is also the possibility of applying different reconstruction approaches that augment, e.g., light-field microscopy images, to circumvent this technology's limitations of nonuniform resolution and slow reconstruction speeds. By using global voxel transformer networks (GVTNs), subcellular structures were identified with high accuracy from unlabeled transmission microscopy images [13], and even in real-time, volumetric reconstructions of a beating zebrafish heart from light-field images was achieved using a so-called view-channel-depth (VCD) neural network [14]

Regarding the efficacy with which different networks run, improved network architectures will reduce the computational speed and, overall, the computational load on local computers. Further, relying on semi-supervised machine learning networks, where a small number of labeled data is mixed with unlabeled data and used for training, will be especially useful when analyzing large data sets that, in turn, will require less computational power.

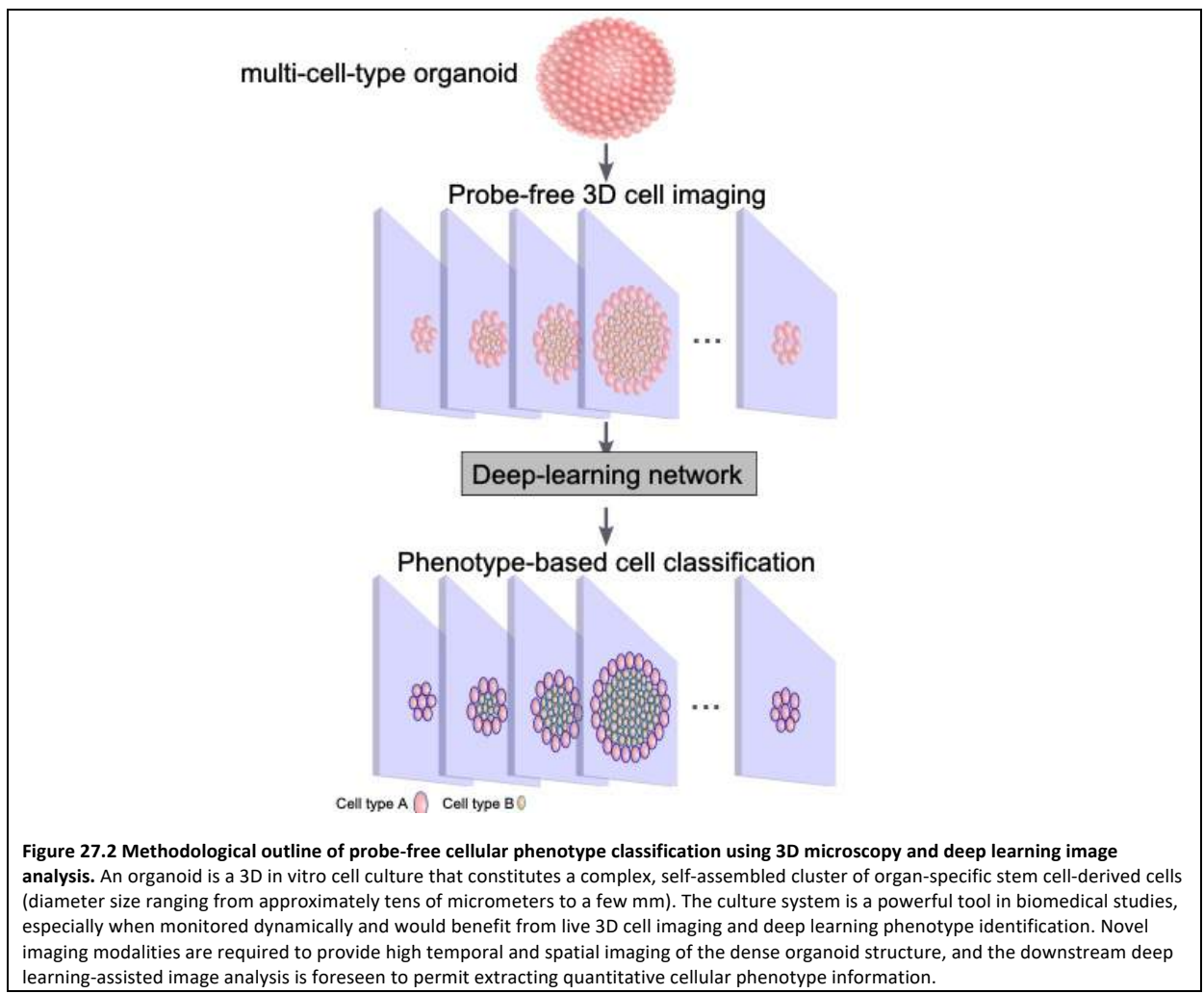

**Figure 27.2 Methodological outline of probe-free cellular phenotype classification using 3D microscopy and deep learning image analysis.** An organoid is a 3D in vitro cell culture that constitutes a complex, self-assembled cluster of organ-specific stem cell-derived cells (diameter size ranging from approximately tens of micrometers to a few mm). The culture system is a powerful tool in biomedical studies, especially when monitored dynamically and would benefit from live 3D cell imaging and deep learning phenotype identification. Novel imaging modalities are required to provide high temporal and spatial imaging of the dense organoid structure, and the downstream deep learning-assisted image analysis is foreseen to permit extracting quantitative cellular phenotype information.

**Concluding Remarks**

AI for cell biology and biomedicine is a rapidly developing interdisciplinary field that urges tight collaborations across disciplines. The field should focus on developing new tools and providing them in a form that users can modify and retrain the networks for custom image processing with minimal effort. Hence, there is an urgent need for user-friendly software interfaces. Without them, it is very challenging to incorporate AI-aided software in biological laboratories where there are often no programming experts. Reliable, probe-free analysis methods are here to stay and will probably alter many biological laboratories' standard routines. By novel imaging techniques combined with AI, not only single cells in 3D will be possible to distinguish and classify, but complex self-assembled patterns where multiple cell phenotypes organize themselves due to environmental cues are foreseen to be tracked over time.

**Acknowledgements**

## 28 — Neuroimaging analysis

Mite Mijalkov[1], Dániel Veréb[1], Yu-Wei Chang[2], Joana B. Pereira[1]

1.  Department of Clinical Neuroscience, Karolinska Institutet, Stockholm, Sweden
2.  Department of Physics, University of Gothenburg, Gothenburg, Sweden

**Status**

Neuroimaging techniques such as magnetic resonance imaging (MRI), computed tomography (CT) and positron emission tomography (PET) have been widely used to understand the structure and function of the brain in healthy individuals as well as of patients with neurodegenerative and psychiatric disorders. The recent technological advancements in these techniques have led to the collection of larger amounts of higher quality brain imaging data, which has been challenging to analyze using traditional machine learning approaches due to the complexity of brain anatomy as well as the inherent variability in the images, e.g., due to different scanners or scanning protocols. As a result, it is imperative to create novel, efficient and generalizable methods capable of processing this increased volume of available data [1].

Deep-learning models are a promising tool to integrate, assess and make predictions from brain imaging data by using artificial neural networks with multiple layers that extract more meaningful information compared to conventional methods. While their application in clinical practice has not been extensive, there are multiple examples showing their benefits [2,3], e.g., in the case of assisted reporting, where deep-learning methods are used to pinpoint or quantify pathological changes in medical images or to extract more features from brain imaging data. In neuroimaging research, deep-learning tools have also been applied to image acquisition and preprocessing by improving the speed of image reconstruction [4], creating higher-resolution images from the originally obtained low-resolution ones or detecting artefacts in the images [1,5]. Furthermore, one of the most important applications of deep learning has been in image segmentation, i.e., dividing the image into several regions with comparable properties, allowing for their subsequent quantification [6]. Finally, deep learning has been used in disease diagnosis and prediction, where subjects are classified into different disease groups based on shared behaviors or biomarkers [2].

Despite the enormous potential of deep learning tools, the growing availability of high-quality data, facilitated in part by the creation of many large-scale open-access databases, suggests that it may play a more significant role in future research. Since these tools are not based on any prior assumptions about the data, deep learning methods can identify novel and unique traits in the data that could provide new knowledge, and therefore allow changing currently established practices, e.g., in studying diseases or categorizing patients. One example is the recent implementation of deep neural architectures to analyze connectivity matrices derived from functional MRI, diffusion weighted imaging or other modalities, for image classification and regression (http://braph.org/ [7]).

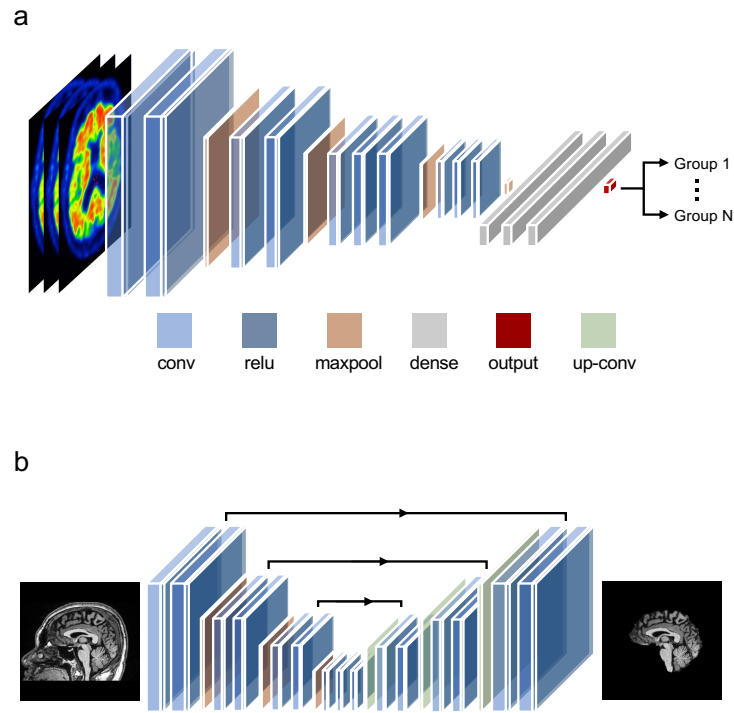

**Figure 28.1. Depiction of commonly used deep learning architectures in neuroimaging studies.** (**a**) Structure of a typical convolutional network for accurate binary or multi-class classification using neuroimaging data. (**b**) U-Net architecture commonly applied in skull-stripping.

**Current and Future Challenges**

While the benefits of using deep-learning methods in data analysis are clear, currently several limitations hinder their widespread application. Some of these limitations are technical, e.g., the higher dimensional methods needed to analyze the three- or four-dimensional medical images are computationally expensive and computer memory intensive. Furthermore, the lack of understanding of the rigorous mathematical framework and underlying theory by end-users can lead to problems with the interpretability of the models, which could limit their applications and lead to wrong conclusions [1].

One of the most frequently encountered obstacle for deep learning is obtaining the data needed to test and validate the model [5]. These models need a large number of training data to achieve a good prediction and to prevent overfitting to the training sample. However, especially in the context of supervised learning, producing labelled data is difficult and expensive [2]. Furthermore, the training data needs to be general and representative of the population data that the model is expected to assess. Most implemented models are tested on local datasets, which could have a distribution that is different from real-world data, leading to decreased generalizability and reproducibility of the particular model and difficulties in the comparison of its performance against alternative models. Finally, even if the model's application is limited to a certain imaging modality or disease, there is a variability in the quality of the image, e.g., due to the use of different hardware or imaging protocols, which could again lower the generalizability of the specific model [1,5].

However, the main issue of deep-learning models is their lack of validation in clinical settings. Model performance has traditionally been evaluated using statistical measures that are frequently obtained from synthetic data. Several visual explanation approaches (e.g., occlusion analysis, gradient-weighted class activation mapping) have been proposed to obtain the heatmap from deep learning models, and therefore, further the understanding of the biomarker representation from neuroimaging data via deep learning. However, in some cases, such metrics can be difficult to interpret by clinical experts and may not be in agreement with experts' predictions, leading to some doubts in the models' predictions. Furthermore, analyses in research are commonly performed on more standardized, high-quality images or even on synthetic data, which greatly differ from the real-world data acquired from patients in the clinic. Hence, this could present an important obstacle for the generalization and practical application of deep learning models in clinical settings [2,4].

**Advances in Science and Technology to Meet Challenges**

To overcome these technical challenges, several methods to reduce the dimensionality of the data and, in turn, decrease the computer load have been proposed, e.g., training on only a part of the image, or representing 3D images as a stack of 2D images and use lower dimensionality models to analyze them. The computational time can be further decreased by employing transfer learning [8], a widely used method that entails pre-training the model on a large dataset, recording the obtained weights and finally, applying them to the neural network assigned to the current task. The recording of the pre-trained weights allows researchers to train only a subset of layers for the current task, which requires less data; also performing the pre-training on a large data set can lead to more robust results. Another benefit of transfer learning is the ability to share the well-trained deep learning models between researchers, which can be an important step to improve the generalizability and robustness of these models.

Another broadly used technique to increase the training set and prevent overfitting is training data augmentation [9]. Using this technique, the size of the dataset is effectively increased by introducing some random variations in the data, e.g., random transformation of the image by translation, rotation or deformation, introducing intensity shifts or scaling factors. In addition to augmentation, other techniques have also been developed with the aim of preventing overfitting, including dropout (i.e., randomly removing nodes from the network at different layers during training) as well as regularization of the network that ascribes weight penalties to different nodes.

To increase the level of use of deep learning methods in clinical settings, it would be necessary to include the end-user in the process of design and testing the model, enabling the user's understanding of the model and allowing an easier interpretation of the results. Steps have already been taken in that direction, including attempts to integrate the different deep learning frameworks into the commonly used analysis pipelines and the scanner itself [4]. If successfully integrated, such approaches would allow for multi-site testing and validation for a given method, improving its generalizability.

Finally, there are several ongoing studies aimed at obtaining and providing large, publicly available data sets, e.g., UK Biobank [10], Human Connectome Project [11] or OASIS-3 [12]. The availability of these data sets is extremely important for the design of novel deep learning methods as they can provide a single setting in which different methods could be compared and benchmarked against each other.

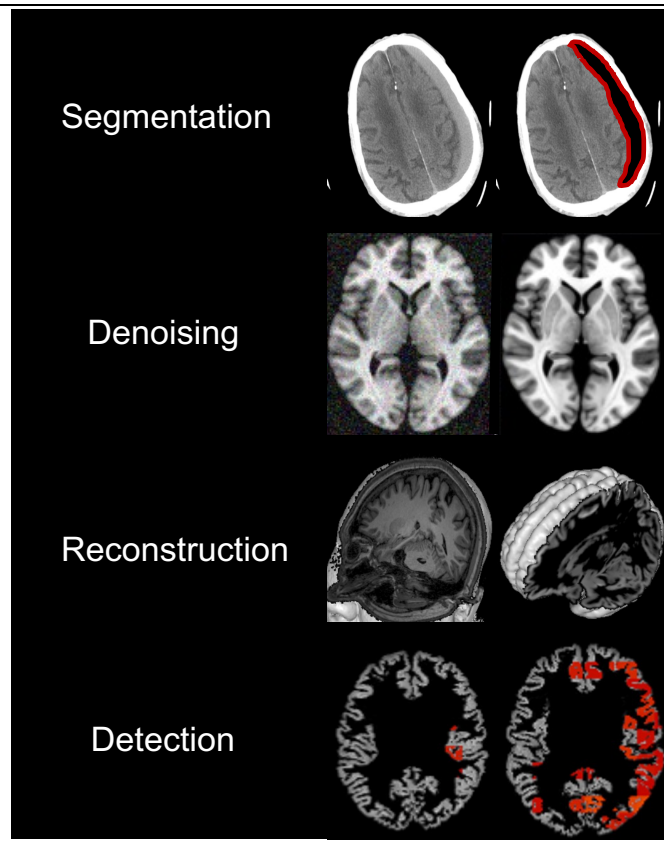

**Figure 28.2. Examples of deep learning applications in neuroimaging pipelines.** Deep learning has been utilized in neuroimaging studies extensively due to its ability to detect abstract and complex patterns. This figure provides several common examples of applications, the most common of which are segmentation, denoising, image or 3D reconstruction, and detection of abnormal intensity patterns. Segmentation and detection examples are from the authors' previous work and depict the automatic segmentation of a subdural haematoma in a CT scan and the identification of abnormal tau deposition patterns in a PET scan.

**Concluding Remarks**

Despite the number of challenges that need to be overcome by deep learning algorithms so that they can be more widely applied to neuroimaging analyses both in research and clinical practice, there is already overwhelming evidence showing their potential in producing very valuable results. With the current advancement in image acquisition technology, increase in the knowledge of the theoretical underpinnings of the models and understanding of the environment where they would be applied, it is feasible to expect that the current limitations will be addressed in the near future. This will allow researchers and clinicians to agree on the exact protocols of applying these techniques to clinical data, e.g., whether the deep learning algorithms should be applied as a sole method to tackle a given analysis or if they should be combined with alternative methods or human expertise. Therefore, deep learning applications in neuroimaging are likely to achieve even more impressive results in the next few years.

**Acknowledgements**

*The authors are funded by the Swedish Research Council, Alzheimerfonden, Brain Foundation, Dementia Foundation, Karolinska Institute, Stratneuro, Center for Medical Innovation, Gamla Tjänarinnor, Stohnes, Stiftelsen Lars Hiertas Minne and Foundation for Geriatric Diseases at Karolinska Institutet.*

## 29 — Bio-analytical and diagnostic transmission electron microscopy

Damian Matuszewski[1], Gustaf Kylberg[2], Ida-Maria Sintorn[2,3]

1. Department of Information Technology, Uppsala University, Sweden
2. Vironova AB, Sweden
3. Department of Information Technology, Uppsala University, Sweden

**Status**

Transmission electron microscopy (TEM) offers the only means of directly seeing and analyzing biological nanoparticles and structures at the nano-level. The recent focus on gene therapy and revival of vaccine development has contributed to an increased interest in TEM as an analytical tool of biological nanoparticles in drug development and formulation, in addition to its established use in disease understanding and clinical diagnosis. The advent of automation capabilities in image acquisition in combination with the expectations of deep learning provide possibilities and promises for making TEM more accessible. An overview of deep learning used in various electron microscopy applications (biology as well as material science) can be found in Ref. [1].

As repeatedly demonstrated in computer vision applications and many medical and microscopy proof-of-concept studies, deep learning offers the great possibility to learn relevant information from examples. Thereby, the human's limited capability to instruct the computer on what information to extract can be excluded. It has the general drawback that a priori expertise and information not represented in the examples will not be incorporated in the model unless explicitly added. It is also difficult to decipher and validate what information is used to reach a decision. This leads to interpretability and reliability issues which practically limit deep learning deployment in real-world TEM bio-processing and clinical applications. These applications often face a scenario with a limited amount of training images, lack of or unreliable ground truth, non-representative or too narrow training sets, as well as regulatory requirements. In addition, for image data in clinical situations, there is also a fear of missing "other" and unexpected information obvious to the experienced eye.

Commonly used AI networks are very self-confident in their predictions, also when the evidence for a certain decision is dubious. This results in so-called silent failing i.e., misclassifications and unreliability due to too narrow training sets, as well as missing structures in the sample not directly asked for. For certain applications, this is a showstopper. However, in other applications, a model doesn't need to be extremely accurate or generalize to be useful. They can solve or contribute to a particular step in the analysis pipeline, be interactively fine-tuned by the user for each dataset to be analyzed, or serve as a detection or segmentation step further processed or validated by the user. Thus, in many practical scenarios, tools that reduce rather than remove human input are in demand.

**Current and Future Challenges**

Common to all types of microscopy is the need to adapt and modify deep learning methods to the prerequisites and characteristics of the different imaging techniques. Some general issues that need TEM domain-specific attention are:

1) **Quality and amount of annotated image data**. The training dataset has to be carefully prepared to represent all variations that can be observed in the application. It is, however, easy to miss rare albeit important instances. In a real setting, the models will not only have to face all varieties of data appearances but also anomalies unlike anything in the training set. Typical supervised models fail when presented with a sample from a different class or appearance than those in the training set. This is exemplified in Figure 1. What might be obvious to a human expert requires careful consideration when training and verifying deep learning models.

2) **Design of performance and evaluation criteria** that are relevant for the imagery and application. A TEM is a complex scientific instrument used to gain information about nano-scale features. In contrast to many computer vision applications, looking pleasant is not interesting while being correct is crucial.

3) **Adaptation of network architectures to focus on the characteristics of TEM**. For example, features at the single-pixel level in combination with large-scale information are usually key for what is studied. The images can in general not be down sampled as is done as a pre-processing step to many standard deep learning models without severe loss of information.

4) **Development of TEM-specific augmentation techniques** that incorporate variations in instrument settings and properties, as well as modality-specific artifacts. This is especially important for complex imaging techniques such as TEM since generating real data representing a lot of variation (instrument makes and models, instrument settings, sample preparation, etc.) is difficult due to the low abundance of microscopes and high cost. The value of appropriate augmentation has e.g., been shown in stain normalization in pathology [3].

5) **Lower the expectations to gain real value**. Shift the focus from trying to apply and "market" deep learning as the magic answer to all problems to seeing it as a tool in the toolbox to aid the imaging and analysis process. For example, to automate instrument control as exemplified in [4] to simplify imaging and increase robustness, or to assist the sample analysis and thus reducing the manual input for corrections or verification provided by a human expert.

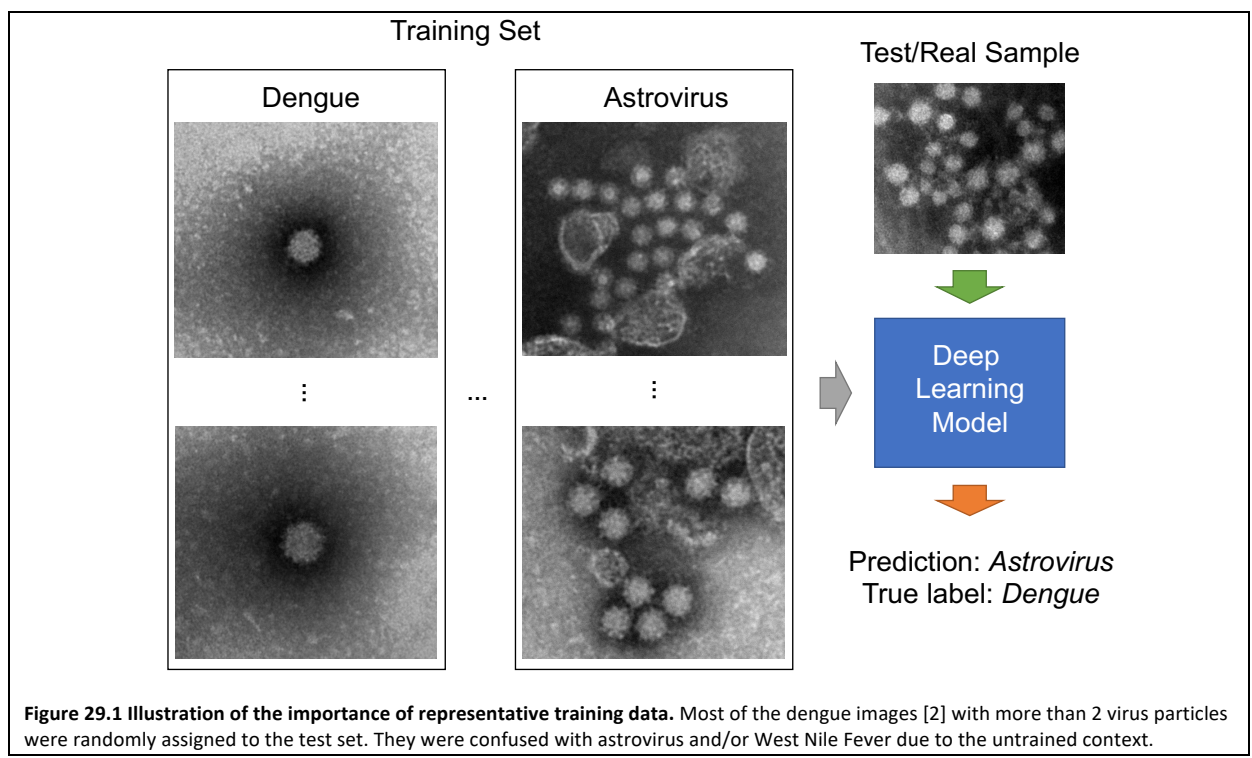

**Figure 29.1 Illustration of the importance of representative training data.** Most of the dengue images [2] with more than 2 virus particles were randomly assigned to the test set. They were confused with astrovirus and/or West Nile Fever due to the untrained context.

**Advances in Science and Technology to Meet Challenges**

As described above, many of the challenges are common and general for microscopy and medical imaging techniques but require a specific modality or instrument focus. Therefore, theoretical deep learning advances, as well as insights gained in other imaging domains and applications, are also of interest to TEM. Incorporating physics and constraints in the deep learning models and/or loss function is one such technological advancement that has proven useful in other domains [5], but is yet to be widely explored in TEM instrument control and applications.

Another example is the Human-in-the-loop trend [6], meaning that the user or expert guides the model training, evaluation, or use (sometimes referred to as human-in-the-loop data analytics – HILDA). This is particularly useful in applications with limited (annotated) data, or when fine-tuning or human validation is required for each new sample/dataset. One example of such a human-in-the-loop approach also showcased on EM data is [7]. The user interactively marks the type of objects to search for in a dataset and then validates suggested objects in iterative steps.

The lack of training data and EM benchmarking datasets is being addressed to some extent via competitions or "challenges" such as the ISBI 2021 challenge "large-scale Mitochondria 3D Instance Segmentation from Electron Microscopy Images" and the 2012 challenge "Segmentation of neuronal structures in EM stacks". In addition, the recent requirements by journals and conferences to publicly share image data will naturally lead to increased availability of also TEM images. Recently authors Matuszewski and Sintorn published a virus classification data set and deep learning benchmark performance [2], focusing on discussing the importance of training and test sets representativeness and its impact on performance and reliability in real-world applications.

In recent years, following a large interest in understanding deep learning decisions, more attention has been put on designing models producing confidence scores [8], anomaly detectors [9], and interpretable models, all to strengthen deep learning reliability and trustworthiness. A sign of technology maturation of deep learning in medical imaging along those lines is to apply deep learning to smaller interpretable parts (cf. as a tool in the toolbox) contributing to the decision support with additional sources of information rather than constructing a deep learning black box total solution [10].

**Concluding Remarks**

Many of the issues with deep learning deployment in real situations are common to all imaging modalities, e.g., lack of sufficient amounts and variations of realistic and annotated training data, as well as reliability and interpretability of the results. General advances in deep learning methodology and making more TEM image data publicly available will of course transfer to and benefit users also in the TEM field. In addition, incorporating electron imaging physics and instrumentation properties into deep learning models and data augmentation properties is a foreseeable adaptation from other imaging fields.

Perhaps somewhat more specific to TEM, we foresee that deep learning will play a big role in instrument control and automation and hence make this complex imaging technique more accessible and available. We also believe that theoretical deep learning development focused on the prerequisites in TEM will lead to better and more efficient architectures for handling the large-scale spanned in TEM imaging and analysis. Often the interplay and correlation of image features at multiple scales including information at the finest pixel level is needed to gain insight, and multiple scales need to be traversed to find objects or regions of interest in the huge search space a sample in TEM constitutes.

**Acknowledgments**

*This work has been supported by a post-doc grant from the Swedish e-science collaboration (eSSENCE) to author DM. Author IS has been supported by Uppsala University's AI4Research initiative and the Swedish Foundation for Strategic Research (grant BD150008).*

## 30 — High-content high-throughput screening

Juan C. Caicedo[1], Beth A Cimini[1]

1. Broad Institute of MIT and Harvard, Boston (MA), USA

**Status**

With modern optical, electronic and chemical technology, imaging experiments can be run efficiently in *high throughput* to address pressing research questions in drug discovery and functional genomics. Given the massive amounts of imaging data generated by such experiments, accurate software platforms are also necessary to transform images into *high content* quantitative data for downstream research. Phenotypic high content imaging assays have been used to determine mechanisms of action of drugs, predict toxicity, and even predict the outcomes of other, non-imaging assays (reviewed in Ref.[1]). While deep learning approaches are not currently universal within high content/high throughput screening(HCS/HTS), work in recent years has illustrated several ways in which deep learning can add value to these assays (**Figure 30.1**).

The computational workflow to transform images into biological insights typically begins with *segmentation* — that is, providing an exact boundary for each desired object within the image. This area has progressed the furthest in recent years, thanks to models that generalize across imaging experiments [2], and successful tools becoming widely available to the community [3].

When the objects of interest — typically single cells — have been identified by segmentation, the next challenge is *feature extraction.* This involves quantifying the unique properties of cell state, cell structure, and phenotype observed in the images using multi-dimensional representations [4]. Recent advances in representation learning may automate data-driven optimization of feature representations, instead of requiring analysts to manually create feature sets.

The quality and content of images themselves can now be enhanced and expanded using *image generation and modification strategies.* Generative models (such as some deployed in Ref. [5]) can now improve the resolution and signal-to-noise ratio of images, which improves the quantitative data obtained. Similarly, networks can predict the staining pattern of many biological stains from unstained images, allowing one to generate measurements of biological structures that were never directly stained (reviewed in Ref. [4])*.* Generative models can also reduce the effects of technical variation (e.g. well or plate effects) by modeling and then removing them; this can allow the researcher to calculate how an observed image would have looked if it was captured in a different batch [6].

Finally, deep learning has the potential to change *the scale at which assays must be run*: training models to predict the outcome of one assay based on already-collected data from an orthogonal assay has been shown to enhance secondary screening hit rates up to 200X [7].

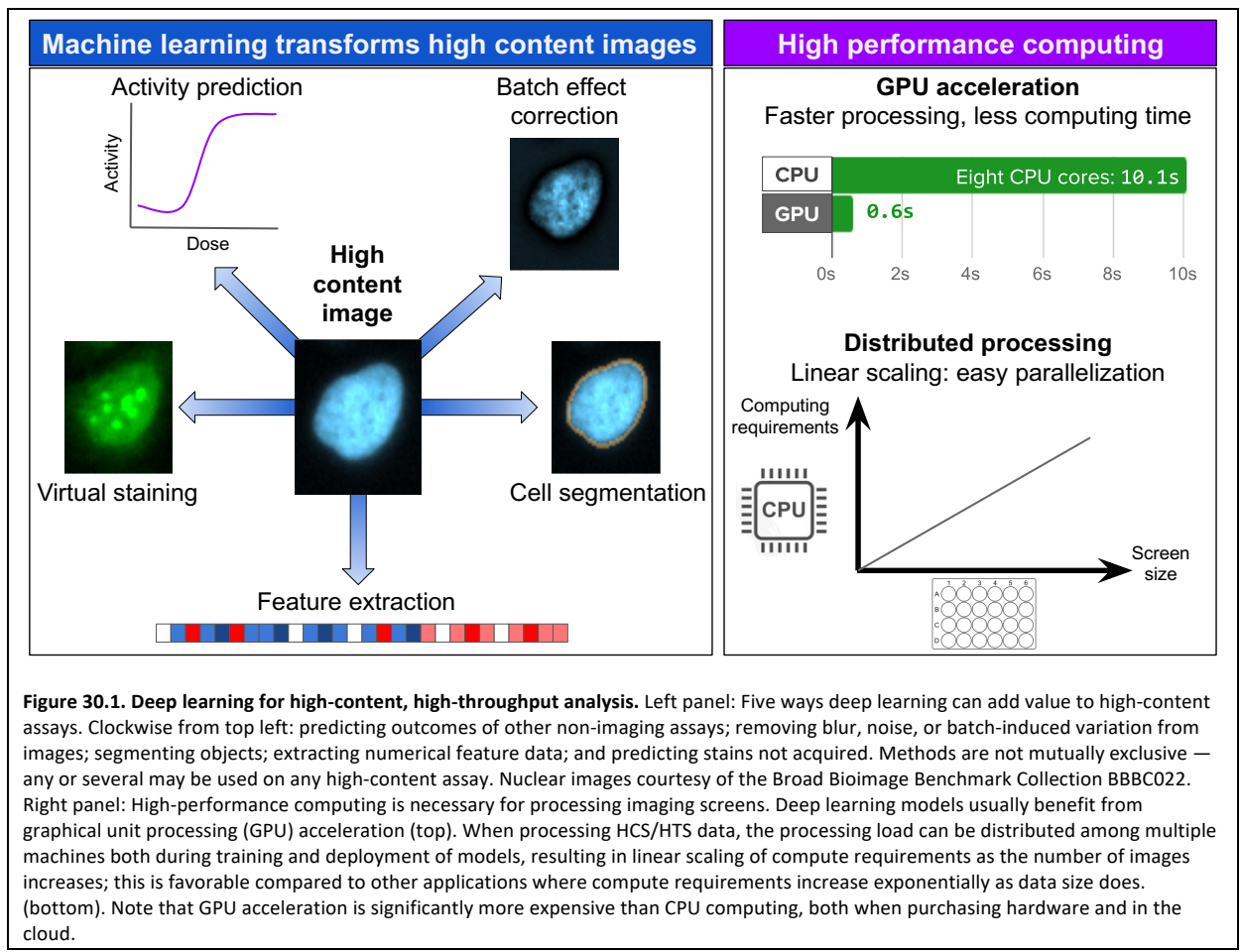

Figure 30.1. **Deep learning for high-content, high-throughput analysis.** Left panel: Five ways deep learning can add value to high-content assays. Clockwise from top left: predicting outcomes of other non-imaging assays; removing blur, noise, or batch-induced variation from images; segmenting objects; extracting numerical feature data; and predicting stains not acquired. Methods are not mutually exclusive — any or several may be used on any high-content assay. Nuclear images courtesy of the Broad Bioimage Benchmark Collection BBBC022. Right panel: High-performance computing is necessary for processing imaging screens. Deep learning models usually benefit from graphical unit processing (GPU) acceleration (top). When processing HCS/HTS data, the processing load can be distributed among multiple machines both during training and deployment of models, resulting in linear scaling of compute requirements as the number of images increases; this is favorable compared to other applications where compute requirements increase exponentially as data size does. (bottom). Note that GPU acceleration is significantly more expensive than CPU computing, both when purchasing hardware and in the cloud.

**Current and Future Challenges**

With recent advances in deep learning research on bioimage, the problem has shifted from assessing whether deep learning could help to *choosing which aspects of your workflow benefit most from deep learning.* As deep learning can potentially improve many aspects of the HTS/HCS process (see above), at each step one must assess which options are available, how much they stand to improve the final data, and what the compute, person, and opportunity costs are to implement them.

*Compute* considerations are critical in HTS/HCS: assays typically involve dozens or hundreds of plates with thousands of fields of view per plate [1]. At this scale, speed matters: each second of analysis time equals ~1 CPU hour per plate. While deep learning techniques can be orders of magnitude faster than conventional approaches [8], these increased speeds often rely on access to GPUs, which are at a premium compared to CPUs in most on-premises and cloud environments.

*Person and opportunity* costs are harder to predict or measure, since one must investigate options (and therefore partially incur these costs) in order to assess the potential benefit. For tasks where tools already have been made for a "substantially similar" task, care must be taken to assess how similar the external training data is to one's own data; differences invisible to the human eye can still require retraining. If the external training data is sufficiently large and varied, publicly available models may be appropriate without further training; this assessment is often easier if the developers have released performance statistics on public benchmarks. Even when existing tools require no retraining, many require advanced computational skills to use and/or scale; a recent conference poll on "Barriers to entering HCS" identified "access to knowledge" as the largest barrier and "Python" as the thing they most wished new employees knew (Cimini, in preparation).

If one decides that no existing tools are appropriate, many more decisions emerge — should one adapt an existing bioimaging tool, or seek out new computational architectures? If using an existing tool, should one train from scratch or fine-tune an existing model? Are there existing biological data sets that you can use for training, or must you generate your own? How should the ground truth be defined, how can it be made efficiently, and what are appropriate metrics of success? These questions are difficult for experts to answer and will take even longer for new users.

**Advances in Science and Technology to Meet Challenges**
The success of deep learning for natural images is partly explained by the careful curation of datasets with ground truth annotations. Ground truth collection has been slower for microscopy analysis tasks, but the field is making progress for problems such as cell segmentation [2,3]**;** more coordinated efforts are necessary to expand the availability of well annotated data for training reusable models under different experimental conditions. Since creation of these sets is costly and time consuming, training strategies that do not require supervision or explicit annotations for learning should be deeply explored. For instance, a family of self-supervised learning techniques, including contrastive learning, use image matching under different transformations (such as cropping or color adjustment) to capture machine-useful (though not human-interpretable) features that make an image unique. These features have shown to be highly informative to recognize objects in natural images and could be used for phenotypic analysis in images of cells or to accelerate training of segmentation and/or classification models by requiring significantly less ground truth to fine-tune.

While more powerful networks are needed, these tools will not gain wide adoption if users find them too challenging to use. Educational materials, tutorials, guides and documentation for non-experts on how to make decisions (about models, hyperparameters, and more) and use these tools in practice are required to reach the scientists that will ultimately use them to drive biological discovery. This will require tools to create easy to use user interfaces [9] or plug-and-play API libraries that can be loaded in Jupyter Notebooks [5] or similar environments to support the development of quick and reusable processing pipelines.

Finally, once an approach is finalized by the scientist, deep learning architectures require significant computing resources; while these needs are greatest during training, they may also be significant for processing new datasets in inference mode, which provides an especial challenge for the scale of HTS. New advances in model compression and efficient architectures must be developed and adopted for bioimage analysis. Energy efficient architectures can run on mobile devices and could be useful to accelerate processing where GPUs are available or to allow deployment where they are not. To serve HTS-scale data, tools must also be developed to allow easier parallelization of models across remote servers or cloud computing services [5,9,10]. The community also needs guidelines to navigate these computational choices and to make architecture and platform decisions in practice.

**Concluding Remarks**
While still in early days, deep learning has the potential to revolutionize the entire HCS/HTS process, from assay design to compound selection to segmentation and feature extraction. As new tools are created and/or refined to tackle these and other as-yet-unimagined possibilities, we believe that emphasizing *reusability*, *interpretability,* and *scalability* will help the entire community. Training on diverse image sets, while more time consuming, leads to reusable tools which require minimal hand configuration [3], and comprehensive documentation increases practical ability to reuse. Tools should maximize interpretability: for some tools, this may involve disclosure of failure modes, for others it may involve adding attention maps so users understand what led the network to a particular conclusion, and for others creation of ground truth benchmarks to better align results across experiments. Finally, reusable and easy ways to generically scale new approaches must be developed to ensure adoption at HTS scale. If such tools are created and made widely available and user-friendly, it is easy to imagine that in a very few years there will be no steps in HCS/HTS that do not routinely incorporate deep learning models.

**Acknowledgements**

*This work was supported by the Broad Institute Schmidt Fellowship program to JCC, NSF grant number 2134695 to JCC, and National Institutes of Health grant P41 GM135019 to BAC. This project has been made possible in part by grant number 2020-225720 to BAC from the Chan Zuckerberg Initiative DAF, an advised fund of the Silicon Valley Community Foundation.*

# 31 — Ultrasound and photoacoustic image formation

Muyinatu A. Lediju Bell[1]

1. Johns Hopkins University, United States

**Status**

The success of diagnostic and interventional medical procedures is deeply rooted in the ability of modern imaging systems to deliver clear and interpretable information. One of the most widespread imaging systems available in hospitals and clinics around the world is the ultrasound imaging system. With its 60+ year history, ultrasound imaging has four major benefits in comparison to other medical imaging and microscopy systems: (1) safety, (2) portability, (3) cost-effectiveness, and (4) real-time delivery of images.

While sound is transmitted and received in ultrasound imaging systems, when augmented with lasers and other light sources to create photoacoustic imaging systems, light is transmitted and sound is received. This alternative approach provides optical absorption information, rather than the acoustic reflectivity information that is provided with traditional ultrasound imaging system, while retaining the four major benefits noted above. Ultrasound and photoacoustic images may also be interleaved to improve the overall clinical experience.

In both ultrasound and photoacoustic imaging systems, the sound received by an array of acoustic sensors is typically converted to an interpretable image through the beamforming process, which is often the first line of software defense against poor quality images. However, the beamforming process has historically suffered from recurring limitations that produce poor image quality in a subset of patients. In particular, traditional beamforming procedures rely on assumptions about wave propagation (e.g., sound speed, acoustic pathways) that are not true in the presence of significant inter- and intrapatient variations. For example, a single, direct path from an ultrasound or photoacoustic source to the acoustic receiver is often assumed, but this assumption does not consider the presence of multi-path acoustic scattering or reflections that occur in the presence of acoustic impedance differences.

Advanced techniques have been implemented to improve beamformers, image quality, and diagnostic interpretability, yet no historical advance combines multiple benefits in a single image formation or signal processing step [1]. As a result, potentially useful information provided by the same raw data is either absent or prolonged, considering that advanced methods may be time consuming to implement and are typically applied in succession, rather than in parallel. A deep learning approach has the benefit of learning from multiple training examples, rather than relying on flawed assumptions. This benefit opens the door for the delivery of clear, interpretable images that combine multiple benefits in parallel, based on a single input of raw sensor data [1]. An example of this approach is presented in **Figure 31.1**.

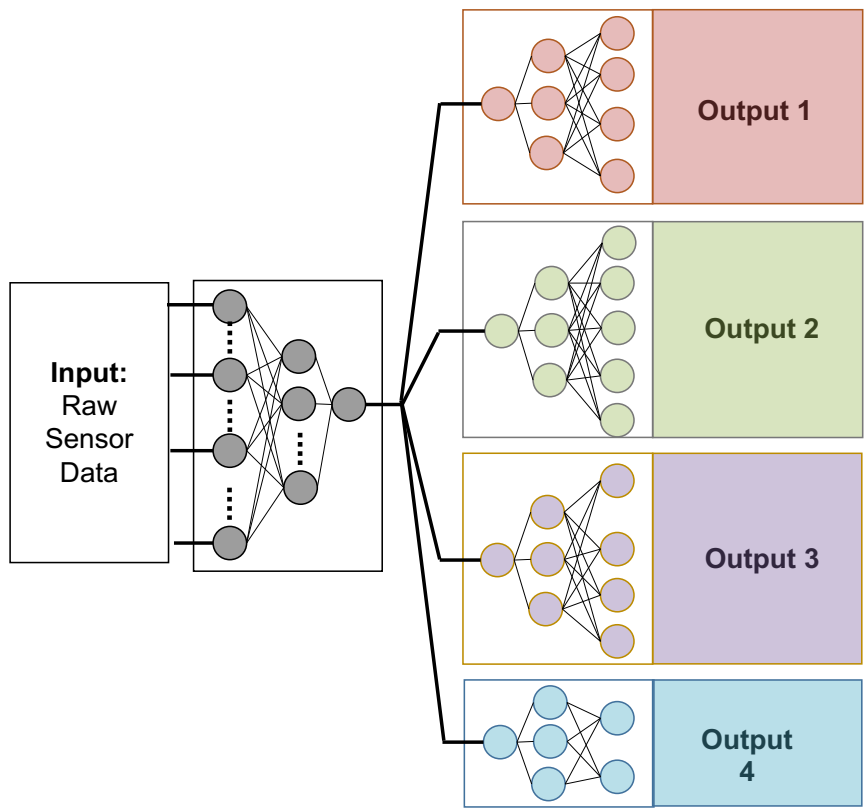

**Figure 31.1 Deep-learning-enhanced ultrasound imaging.** General concept of supplying a single input of raw ultrasound sensor data, from which multiple outputs are simultaneously produced in parallel with the assistance of deep neural networks (DNNs).

**Current and Future Challenges**

Three significant challenges surround the capability of deep learning to create ultrasound and photoacoustic images that combine multiple image formation and signal processing techniques in a single step. First, ultrasound and photoacoustic sources must be localized and detected in the presence of noise and artifacts. Based on the Hyugens-Fresnel principle [2], this challenge can be reduced to detection of point-like sources, which can represent single scatterers within tissue, needle or catheter tips, photoacoustic signals from an optical fiber tip, or individual microbubbles. When the acoustic response from a point source travels outward from the source to the transducer, the shape of the recorded wavefront is determined by the distance from the source to the acoustic receivers. Thus, sources that are closer to an array of acoustic receivers will have a different recorded wavefront shape than sources that are farther away. This unique shape-to-depth relationship can be learned by deep neural networks (DNNs) [3,4]. Advantages include the ability to spatially locate acoustic sources with high precision and accuracy in comparison to images created within the diffraction limits imposed by the beamforming process [4,5].

The second challenge is accurate segmentation of an imaging target or disease feature of interest. The underlying goal is to produce an image that emphasizes structures of interest for a particular application and deemphasizes surrounding structures. Segmentation is typically performed after image formation, but if the image quality is poor, then the segmentation will also suffer. This is additionally problematic when ultrasound or photoacoustic imaging systems are operated by less experienced users or when automated tasks rely on segmentations to deliver a diagnosis, treatment, or medical assessment.

The third challenge is the time required to create images with advanced beamforming methods, such as coherence-based images. For example, the short-lag spatial coherence (SLSC) [6] beamformer successfully reduces acoustic clutter in cases where traditional clutter reduction methods, such as harmonic imaging, fails. SLSC beamforming also has the ability to determine the fluid or solid content of suspicious masses [7]. Despite these benefits, SLSC and similarly advanced beamformers can be computationally intensive to implement, which has hindered integration into existing clinical systems.

Additional challenges that must be overcome to enable widespread future impact include the ability to understand, interpret, and predict expected network outcomes and failure points. This ability will enable systematic development of new deep learning approaches. In addition, providing the most optimal speed up of advanced beamforming approaches will expand existing potential, and smart integration with robotic approaches will promote future possibilities for fully automated procedures.

**Advances in Science and Technology to Meet Challenges**

DNNs were either applied or created to address the three significant challenges noted above. Point source detection was demonstrated with multiple networks, including AlexNet, Resnet, and VGG-16 [5]. Segmentation and feature detection from raw data was demonstrated with U-Nets [1]. CohereNet [8] was created to calculate coherence functions for the advanced SLSC beamformer. CystNet1 [9,10] and CystNet2 [10] each consist of one encoder and two decoders, and these networks were built to simultaneously image and segment cysts from raw ultrasound data in parallel, rather than perform the traditional sequential approach. Figure 2 shows the architecture of a selection of these DNNs.

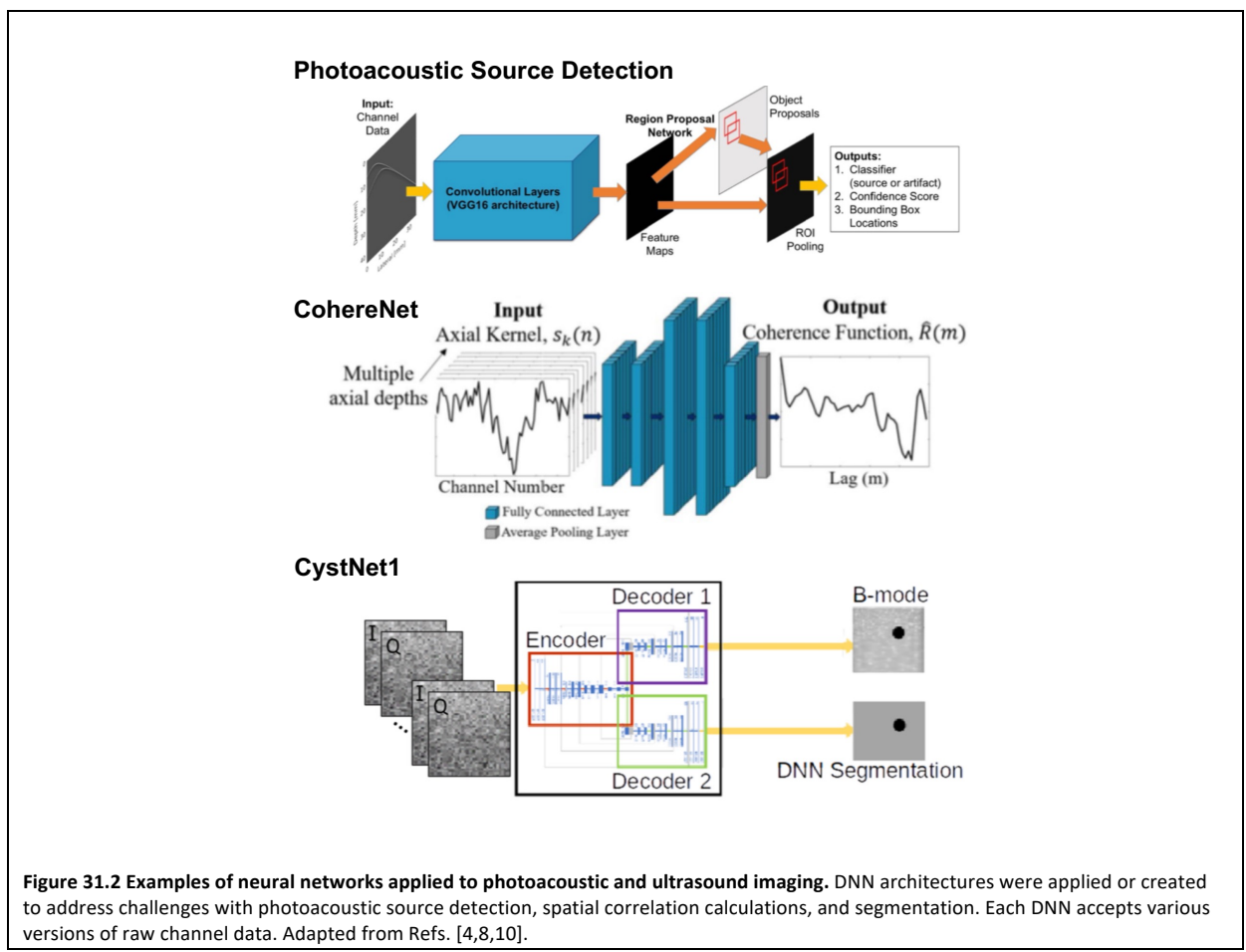

**Figure 31.2 Examples of neural networks applied to photoacoustic and ultrasound imaging.** DNN architectures were applied or created to address challenges with photoacoustic source detection, spatial correlation calculations, and segmentation. Each DNN accepts various versions of raw channel data. Adapted from Refs. [4,8,10].

These advances either independently or collectively demonstrate the feasibility of creating multiple outputs from a single input of raw channel data with DNNs. The collective demonstration exists because it is possible to concatenate multiple DNNs in parallel, thus providing an approach to input raw data to each concatenated network and achieve simultaneous outputs from each input. While many of the networks noted above were trained with simulated data that mimicked the physics of wave propagation [5,9,10] or *in vivo* breast data which contains variability arising from highly heterogeneous breast tissue [8], it is a critical advance that each DNN operated on sufficiently variable data sources relative to the training data. This generalizability highlights the success of the training process.

An apparent tradeoff between providing human-interpretable images and integrating advances with robotics has also emerged as a direct outcome of the science and technology implemented to meet existing challenges. For example, a DNN can be created to achieve a specific task for robotic integration, such as implementation of visual servoing to find and stay centered on a target of interest. The output of this DNN can be coordinates, rather than a human-interpretable image that is then used to extract coordinates [11]. On the other hand, a human-interpretable image is useful for supervising the automated procedure, intervening if necessary, and providing interpretable reports if there is a runtime error. Thus, the two seemingly competing approaches between robot and human data formatting have symbiotic advantages when operating in parallel [9].

The availability of resources is anticipated to further advance the field with regard to open-source implementations that lower entry barriers into an otherwise specialized field. One such resource was made possible through the Challenge on Ultrasound Beamforming with Deep Learning [12,13]. Outcomes of this challenge include freely available datasets, code, and trained network weights, which may collectively be employed to benchmark new approaches.

**Concluding Remarks**

Ultrasound and photoacoustic imaging are two technologies that use the same sensing hardware to make images. After raw sensor data is received by ultrasound and photoacoustic imaging systems, there are multiple signal processing and beamforming steps that can be implemented to address a variety of healthcare challenges across multiple organs, diagnoses, and procedures. Deep learning provides a viable pathway to implement multiple approaches in parallel, possibly in a single signal processing step. This pathway is promising to overcome previous barriers to producing high-quality images for all patients.

**Acknowledgements**

*Thanks to all of my past and present Photoacoustic and Ultrasonic Systems Engineering (PULSE) Lab researchers and clinical collaborators who contributed to this direction of research, as well as to all of my CUBDL co-organizers and data contributors who helped to make the open-source resources available. This work was supported in part by the NIH Trailblazer Award (Grant No. R21-EB025621), the NIH Trailblazer Award Supplement (Grant No. R21-EB025621-03S), the NSF CAREER Award (Grant No. ECCS 1751522), the NSF SCH Award (Grant No. NSF IIS-2014088), the Computational Sensing and Medical Robotics Research Experience for Undergraduates Program (Grant No. NSF EEC 1852155), and NIH R01-EB032960. More information is available at https://pulselab.jhu.edu and https://cubdl.jhu.edu.*

# 32 — Enabling equitable access to deep-learning solutions

Bruno M. Saraiva[1], Guillaume Jacquemet[2,3,4,5], Ricardo Henriques[1,6]

1. Instituto Gulbenkian de Ciência, Oeiras, Portugal
2. Turku Bioscience Centre, University of Turku and Åbo Akademi University, Turku, Finland
3. Faculty of Science and Engineering, Cell Biology, Åbo Akademi University, Turku, Finland
4. Turku Bioimaging, University of Turku and Åbo Akademi University, FI- 20520 Turku, Finland
5. InFLAMES Research Flagship Center, Åbo Akademi University, FI- 20520, Turku, Finland
6. MRC-Laboratory for Molecular Cell Biology, University College London, London, UK

**Status**

The machine learning field has been a critical ally in advancing and improving microscopy image analysis, with several machine learning algorithms automating common tasks such as image segmentation and classification. Recently there has been an incredibly rapid development of new microscopy image analysis approaches, thanks to the current boom in developing deep learning (DL) techniques. A DL algorithm will look at data, labelled in the case of supervised learning or unlabelled in the case of unsupervised learning, and through self-optimisation, will infer data features that it can then use to perform the desired task. When correctly implemented, it can provide its users access to expert-level performance at unprecedented speed. However, the implementation of DL approaches is highly dependent on the quality and quantity of data used for training and, typically, on the availability of high-performance computers. When used appropriately, DL techniques have shown incredible performance in several image analysis problems such as segmentation (U-Net [1]; StarDist [2]), classification (YOLOv2 [3]), denoising (Noise2Void [4]), restoration (CARE [5]), and super-resolution, either via super pixelation or improving reconstruction algorithms (DeepSTORM [6]).

Although there is no question about the high impact potential that DL has in image analysis, users are often required to either pay for closed-source tools or have programming knowledge to take advantage of these approaches poses a barrier to the widespread adoption of DL techniques in most research fields. In recent years, several efforts have been made to bring these techniques to the non-expert scientific community to tackle these issues. Notorious examples include DeepImageJ [7], which focuses on implementing pre-trained DL models as plugins for ImageJ/FIJI [8,9], the most widely used tool for image analysis (**Figure 32.1**). Another example is ZeroCostDL4Mic [10], which provides easy-to-use Google Colab notebooks that allow users to train and use DL models for their microscopy image analysis without programming knowledge (**Figure 32.1**). Considering ZeroCostDL4Mic runs on Google Colab, it provides users with a cloud-based solution that eliminates the need for dedicated hardware to use DL approaches for their image analysis tasks. Besides these two, many other projects make available tools using DL microscopy image analysis tools, such as Cellpose [11], CSBDeep [12], DeepMIB [13], and many others.

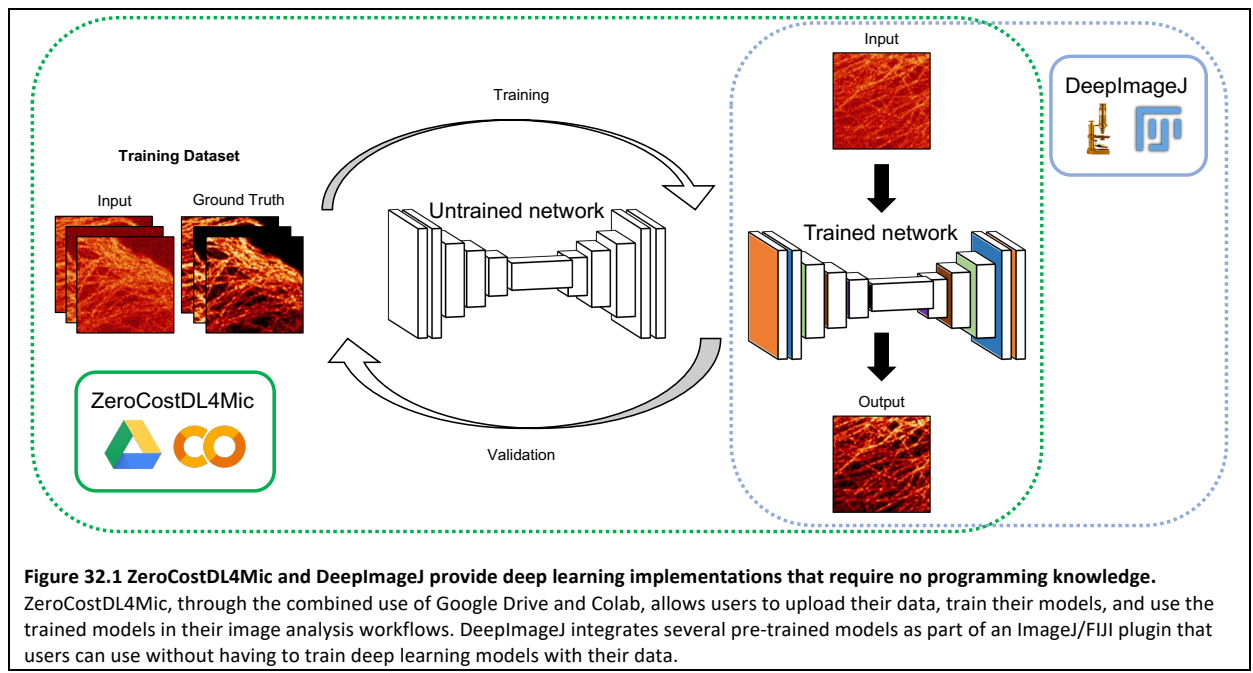

**Figure 32.1 ZeroCostDL4Mic and DeepImageJ provide deep learning implementations that require no programming knowledge.** ZeroCostDL4Mic, through the combined use of Google Drive and Colab, allows users to upload their data, train their models, and use the trained models in their image analysis workflows. DeepImageJ integrates several pre-trained models as part of an ImageJ/FIJI plugin that users can use without having to train deep learning models with their data.

**Current and Future Challenges**

While DL has been providing researchers with new approaches to enhance their image analysis capabilities, it is still a technology with its shortcomings that makes it not easily accessible to every researcher around the globe. As of October 2022, the number of publications containing "deep learning microscopy" as search keywords compared to publications with only "microscopy" as keyword shows that even with access to image acquisition equipment, access and adoption of DL approaches is still not widespread, with US and China leading the way (**Figure 32.2**).

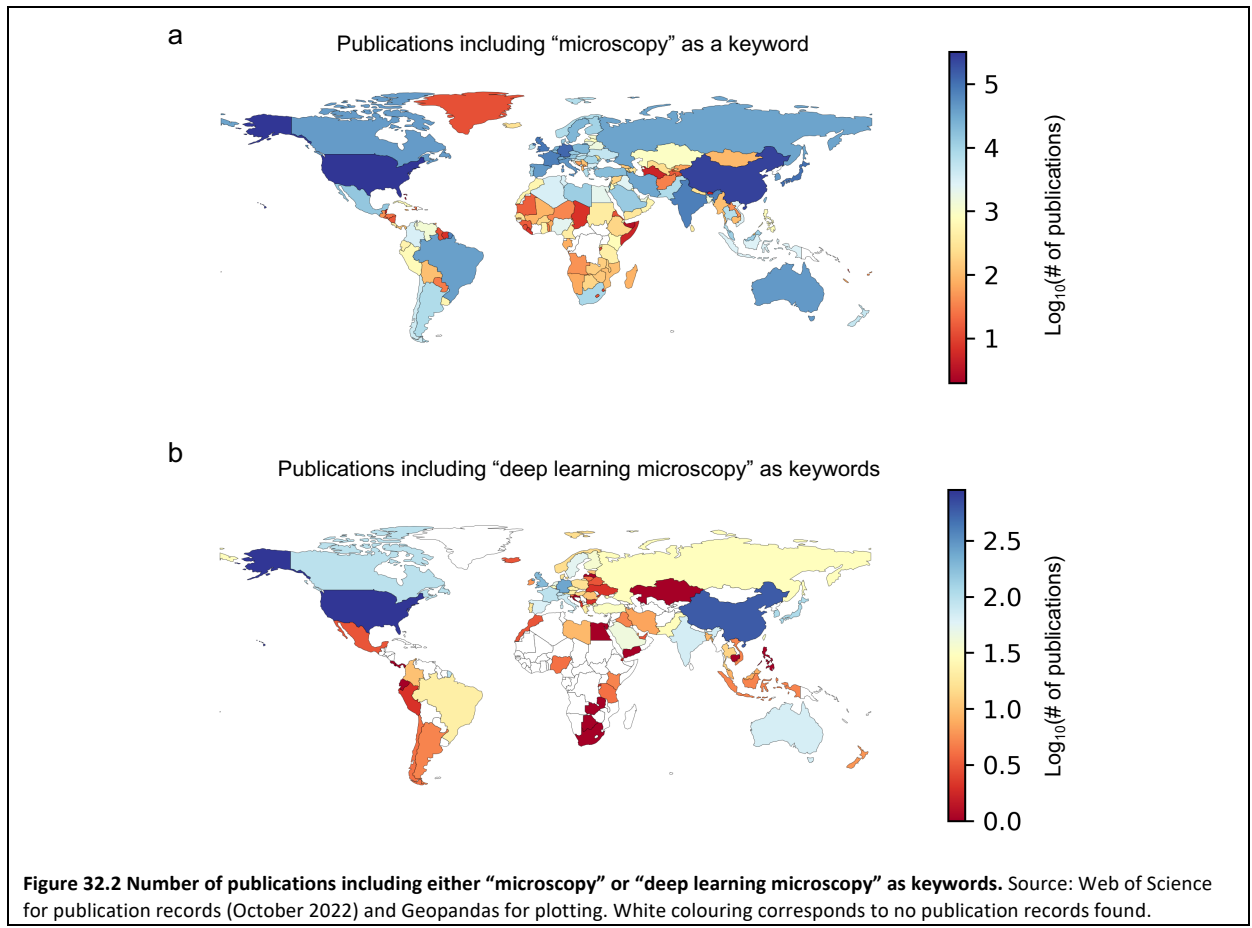

**Figure 32.2 Number of publications including either "microscopy" or "deep learning microscopy" as keywords.** Source: Web of Science for publication records (October 2022) and Geopandas for plotting. White colouring corresponds to no publication records found.

Deep learning algorithms require high computational power, which is expensive and poses a barrier for researchers with limited access to research funds. ZeroCostDL4Mic, which can be run entirely on Google Colab which currently has a free usage plan. However, this plan limits the amount of computational power and storage space that can be used, and there are no guarantees that this might not change or that DL approaches will not evolve to a point where they require more than what is currently provided for free. In addition to this cost, there is also the question of environmental impact of implementing and using DL approaches, as not only the energy consumption is considerably high, but there is also the issue of carbon emissions along with hardware production and distribution.

One of the major frameworks for developing DL approaches, TensorFlow, is mainly funded and developed by a private company. While currently it might operate with an open-source model, this might change in the future. This could create a first access problem and further increase the economic burden of implementing DL solutions. In addition, to use DL as a tool in their research, researchers must have technical knowledge of programming and image analysis. Although ZeroCostDL4MIC and DeepImageJ remove the need for users to know programming, users are still limited to using the included DL models. Recently, the BioImage Model Zoo [14] project has been developed to create a standard for deploying DL models for image analysis and a community-driven database of pre-trained models. Nonetheless, even with access to pretrained models and frameworks that require little to no programming knowledge, users still need to know when to use specific DL models, how to generate/access and pre-process the necessary data, how to analyse the output of these models, and how to validate the results [15]. Considering how DL approaches rely on proper datasets for training the models, generating, accessing, and storing data can also be an issue due to the associated costs. All of this can be especially challenging for researchers who are not yet versed in the DL field.

**Advances in Science and Technology to Meet Challenges**
Deep learning is a technology that, although very powerful, carries with it a considerable economic burden to its potential users. Due to this, it is not yet accessible to researchers worldwide, especially those where funds for scientific research are not easily obtained. Projects like DeepImageJ[7] help alleviate this burden by providing users with the means to use already established pretrained models. However, researchers must have their own solutions if they want to work or use new DL models. Google Colab, as used by ZeroCostDL4Mic [10], is an option. However, its free usage plan has limitations and is still entirely reliant on a service provided by a private company with its own financial interests. As such, creating publicly funded cloud-based solutions that can be commonly used by researchers worldwide will be key to making DL accessible to every researcher. The European project AI4Life is focused on bringing sustainable quality research infrastructure and services to enable life sciences researchers to access DL image analysis tools by creating a bridge between life and computer sciences. The Chan Zuckerberg Initiative is also contributing to bring DL to life sciences researchers by funding several projects that focus on implementing DL approaches in Napari [16], a Python based open-source image processing tool. As a key factor in DL, data access can also be a constraint in using DL in microscopy. Ensuring that the acquired data follow the FAIR principles (Findable, Accessible, Interoperable, Reusable) can promote data sharing to the whole scientific community and enable DL solutions to research groups that might not have the means to generate the required data. An example of how sharing data can have a real impact is the work developed by Abdurahman *et al.* [17], in which by using a publicly available dataset they were able to implement a DL strategy to detect malaria parasites in thick blood smear microscopic images. Sharing data and pre-trained models will also be key in reducing the carbon footprint inherent to the need for high-computational power required for DL.

In addition to the economic burden that implementing DL approaches entails, the knowledge required to take advantage of this technology hinders the adoption of DL for image analysis. The image.sc forum is an example of a community-driven knowledge network that can help and guide new users who want to deepen their knowledge in image analysis, including with DL implementations. Online courses and training sessions will also be fundamental to bringing DL to every researcher, as they inherently have fewer associated costs, making them more inclusive than courses requiring in-person attendance, and reducing the carbon footprint associated with travelling to in-person events.

**Concluding Remarks**
There is no question that deep learning revolutionised the field of microscopy image analysis. DL approaches have outperformed many classical image analysis tasks, providing researchers with state-of-the-art performance at unprecedented speeds. However, due to the need for specific knowledge and equipment to implement a DL approach, it is still a tool that is not easily accessible to all researchers. Several recent projects have contributed to make DL approaches more accessible by removing the need for programming knowledge to use DL for microscopy image analysis. Nevertheless, there are more challenges to be solved. Having access to quality data is a fundamental prerequisite for DL model training. However, accessing the data required for a DL approach is not always possible for all research groups. Understanding when and how to use DL and what pre-processing is needed can also be a limiting factor that could be solved by creating a community-driven knowledge network and training in using these approaches. As a scientific community, we need to join efforts to develop and implement strategies that can make DL equitable and available to the global scientific community.

**Acknowledgements**

*R.H. and B.S. are supported by Gulbenkian Foundation and received funding from the European Research Council (ERC) under the European Union's Horizon 2020 research and innovation programme (grant agreement No. 101001332), the e European Commission through the Horizon Europe program (AI4LIFE project, grant agreement 101057970-AI4LIFE), the European Molecular Biology Organization (EMBO) Installation Grant (EMBO-2020-IG4734), and the Chan Zuckerberg Initiative Visual Proteomics Grant (vpi-0000000044). G. J. is supported by the Academy of Finland (G.J. 338537), the Cancer Society of Finland (G.J.), Åbo Akademi University Research Foundation (G.J., CoE CellMech), the Solution for Health strategic funding to Åbo Akademi University (G.J.) and the InFLAMES Flagship Programme of the Academy of Finland (G.J., 337531).*

## 33 — Deployment of deep learning applications

Wei Ouyang[1], Trang Le[2]

1. Science for Life Laboratory, Department of Applied Physics, KTH - Royal Institute of Technology, 171 65 Stockholm, Sweden
2. Department of Bioengineering, Stanford University, Stanford, CA 94305, USA

**Status**

Due to their extraordinary performance, deep learning models have gained rapid popularity in microscopy image analysis. While deep neural networks are performant in a wide range of tasks, they pose unprecedented challenges in software design and deployment due to the amount of data and computation required for running them, especially in the training phase. Under the hood, most of the deep learning methods are implemented using one of the few popular deep learning frameworks including PyTorch [1] and Tensorflow [2], which are implemented in Python. Source code for these methods is wrapped into repositories and shared via online platforms such as GitHub under permissive licenses. The openness of the deep learning field greatly contributes to the wide spread of popular models and further development, it has become a common practice.

In addition to the raw python code, Jupyter notebooks [3], which split the source code into sections and are surrounded with explanatory text in an executable document, are often provided for demonstrating the usage of the interface functions in the companion code repository. Notebooks are widely used for educational purposes and provided in hands-on workshops and tutorials in microscopy image analysis. To improve the reproducibility, container-based cloud execution services such as Binder [4] and Google Colaboratory are provided for free and become valuable resources in the deployment of deep learning applications. Colaboratory in particular, provides free GPU which is ideally suited for distributing deep learning tools and it is in fact used by ZeroCost4Mic [5], which provides a collection of curated notebooks for running deep learning based microscopy image analysis configured via interactive user interface elements in the Colaboratory notebooks.

While it is flexible to use Python source code or Jupyter notebooks, users with little programming experience can easily adapt the tool according to instructions provided by the developers. For users without programming skills, it is still challenging to adapt the code or notebook to work with conventional software, e.g. ImageJ [6], in a more complete analysis workflow. Despite it being challenging to run deep learning in programming languages other than Python, Java solutions such as DeepImageJ [7] have been proposed. To address the need, napari8, which is a trending Python-based image analysis software supported by Chan Zuckerberg Initiative, is being developed and gaining traction in the community. The aim of the project is to provide a fast, interactive, multi-dimensional image viewer for browsing, annotating, and analyzing large multi-dimensional images.

Furthermore, due to the data- and computation-hungry nature of deep learning methods, web and cloud computing has become increasingly important for the further scaling of the applications to handle massive datasets with almost unlimited access to data storage and compute power. On that front, web based platforms such as ImJoy [9], CDeep3M [10], and DeepCell Kiosk [11] are developed for supporting deep learning applications running on the server side. Since the computations are carried out on remote servers that are maintained by IT experts or developers, users can use these tools with little or no setup, in most cases, using a web browser to access the user interface. This type of deployment approach is more scalable compared to the conventional desktop software, however, it requires transmission of potentially large amounts of data to remote servers which can be limited by the bandwidth of the internet connection, it poses challenges on server-side data confidentiality and privacy concerns. In practice, the user will also require more feature-rich software running fully in the web browser or in the cloud to avoid moving data between local vs remote in a more complete analysis workflow.

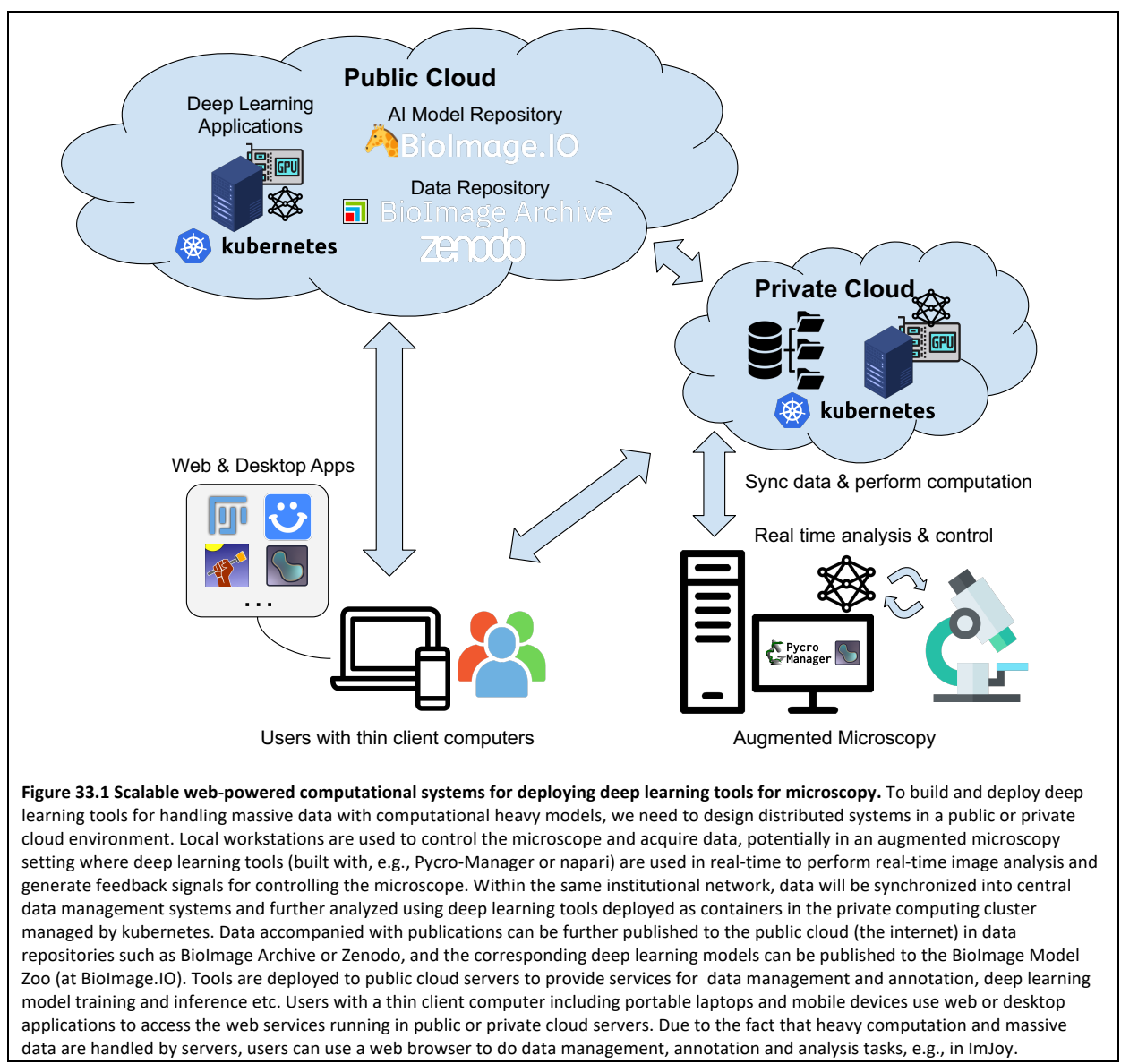

**Figure 33.1 Scalable web-powered computational systems for deploying deep learning tools for microscopy.** To build and deploy deep learning tools for handling massive data with computational heavy models, we need to design distributed systems in a public or private cloud environment. Local workstations are used to control the microscope and acquire data, potentially in an augmented microscopy setting where deep learning tools (built with, e.g., Pycro-Manager or napari) are used in real-time to perform real-time image analysis and generate feedback signals for controlling the microscope. Within the same institutional network, data will be synchronized into central data management systems and further analyzed using deep learning tools deployed as containers in the private computing cluster managed by kubernetes. Data accompanied with publications can be further published to the public cloud (the internet) in data repositories such as BioImage Archive or Zenodo, and the corresponding deep learning models can be published to the BioImage Model Zoo (at BioImage.IO). Tools are deployed to public cloud servers to provide services for data management and annotation, deep learning model training and inference etc. Users with a thin client computer including portable laptops and mobile devices use web or desktop applications to access the web services running in public or private cloud servers. Due to the fact that heavy computation and massive data are handled by servers, users can use a web browser to do data management, annotation and analysis tasks, e.g., in ImJoy.

**Current and future challenges**

Currently, it is a pressing need for the community to work together to solve the challenges in the deployment of deep learning tools. For desktop software, the challenges include figuring out how to ship the software packages, complex dependencies, making it easier to install, and reliably work under the several types of mainstream operating systems. Web and server-based deployment options are becoming more common. It alleviates the deployment issues by delegating the task to IT experts for setting up the complex software environment and accessing cloud storage and computational resources.

Future-proof AI systems for microscopy require a scalable human-compatible framework. To work with the ever-increasing amount of data, it is inevitable to utilize a centralized computing cluster hosted by an institutional IT department or in the cloud. In the meantime, users tend to use a "thin" client such as a laptop, tablet, or mobile devices for accessing the services. Different from conventional desktop software with user interface and the computational code coupled to the same software module, cloud-facing software requires a major change in the design pattern that separates the user interface and the compute parts. As shown in **Figure 33.1**, while the user interface parts run in the user's web browser or a desktop client, the main parts contain deep learning models and other heavy computation runs in one or multiple remote servers in a public or private computing cluster. The two parts need to be synchronized via communication over the internet, and often implemented in different programming languages, e.g., HTML/CSS/Javascript for the interface and Python for the compute part. To make the transition smoother for the next generation of tool developers for microscopy image analysis, we will likely need coordinated efforts in the community to create tools and platforms, and produce educational materials to simplify the process.

In addition, reducing the computational costs and making machine learning training more environmentally friendly is an important aspect to consider when distributing deep learning tools. Meanwhile, for patient related microscopy images, data confidentiality and privacy represent another dimension of challenges in the deployment of deep learning tools.

**Advances in science and technology to meet challenges**

To address the challenges in shipping desktop software with complex dependencies, conda-like virtual environments or container-based solutions are used as an alternative with the price of an increased package size. For example, to take advantage of the existing Java-based developers in the ImageJ community, there is an ongoing effort of building bridges between Python and Java in the pyimagej project to allow easy integration of deep learning models in ImageJ. Pycro-Manager is another software which connects python with micro-manager and further enables real-time deep learning powered image analysis and feedback control during image acquisition (**Figure 33.1**). In the meantime, since it contains binary compiled for a specific operation system, it does not guarantee that it works for all the cases. For cloud based solutions, Container orchestration platforms represented with Kubernetes are becoming the de-facto standard for institutional IT and academic cloud providers to provide a managed environment for developers to deploy their deep learning tools. AI model training and serving software such as KubeFlow and Nvidia Triton inference server makes it easier to serve AI models for production and executed remotely via HTTP-based interface. For data confidentiality, while private computing clusters (as shown in **Figure 33.1**) may address some of the challenges, federated learning [12] enables the training of powerful models while keeping data private in their own data lakes.

To facilitate the sharing of AI models, repositories such as the BioImage Model Zoo [13] (https://bioimage.io) are developed to facilitate the sharing of pre-trained models, and softwares is joined by a consortium to define common model formats to enable cross-compatibility. These efforts make it easier to distribute deep learning models and can be used in multiple software. Meanwhile, re-using existing models either as is or a warm start for training can greatly reduce efforts in producing models and further contribute to climate change. In addition, model compression techniques such as knowledge distillation are used to reduce the model size and accelerate the model execution in e.g. augmented microscopy to provide real-time feedback.

In the browser, building a rich and powerful user interface is becoming easier and more reliable. WebAssembly, which enables compiling and running foundational scientific software packages written in C/C++, Rust, etc. in the browser. It makes it possible to reuse Python libraries such as numpy, scipy, pandas and scikit-image for loading and processing images directly in the web browser and paves the way for creating powerful in-browser image analysis tools with easy-to-use user interface. ImJoy is a framework built for taking advantage of the web ecosystem and providing a remote procedure call layer to connect plugins running in the browser or in a remote server. On top of ImJoy, ImageJ.JS (https://ij.imjoy.io) is a tool we developed by comping ImageJ in java to javascript which is now being used by ~1000 users per day.

**Concluding remarks**

With massive natural language models such as OpenAI GPT-3, Codex [14] or ChatGPT, it allows generating executable source code in various programming languages such as Python and Javascript from plain English. This opens a new door for future deep learning tool developers to create simplified voice or chat-bot like interfaces for reducing the complexity of user interface design and making the tools more flexible and human-compatible. However, the sheer size and computation required to run massive models virtually rejects the access for low-budgeted research entities, and it makes tech giants become the "natural monopoly". The wider AI community will need to join efforts and explore the way forward.

Overall, the wide adoption of AI solutions in microscopy imaging is leading to the paradigm shift to a future of augmented microscopy powered by human-in-the-loop AI, and generating profound changes in the way we understand biology and contribute to precision medicine and healthcare.

**Acknowledgments**

*This work is funded by the Knut and Alice Wallenberg Foundation (grant no. 2016-0204 and 2018.0172).*

## 34 — DeepTrack 2

Benjamin Midtvedt[1], Jesus Pineda[1]

1. University of Gothenburg, Department of Physics, Sweden

**Status**

The ability to quantitatively analyze the microscopic world was enabled by two relatively recent advances [1,2]. First, digital video microscopy has allowed researchers to numerically represent visual data and record it for later analysis. Second, the explosive growth of computing power has made the expensive, high-dimensional analysis of video recordings possible.

Recently, we have seen a new wave of developments in the analysis of microscopy video data, thanks to the power of deep learning. Indeed, deep learning has shown remarkable performance on many common tasks in microscopy, such as cell counting, object detection, cell morphometry, trajectory reconstruction, super-resolution microscopy, diagnostics, object classification, and cross-modality transformations [3,4]. However, deep learning has not yet been broadly adopted as an analysis tool in digital microscopy, mainly because of the significant barrier-to-entry to the development of custom deep learning solutions for microscopy data.

Here, we present DeepTrack 2, an open-source Python library equipped with all the necessary tools to produce a full, end-to-end microscopy-analysis pipeline [1]. A common example of such a pipeline designed to extract mechanistic and spatio-temporal information from biological data entails the following steps:

1. The image is prepared for analysis.
2. Objects of interest are detected and measured to extract morphological and intensity information.
3. Detections in different video frames are connected into trajectories.
4. Time-resolved information is combined to gain both measures of both local and global properties.

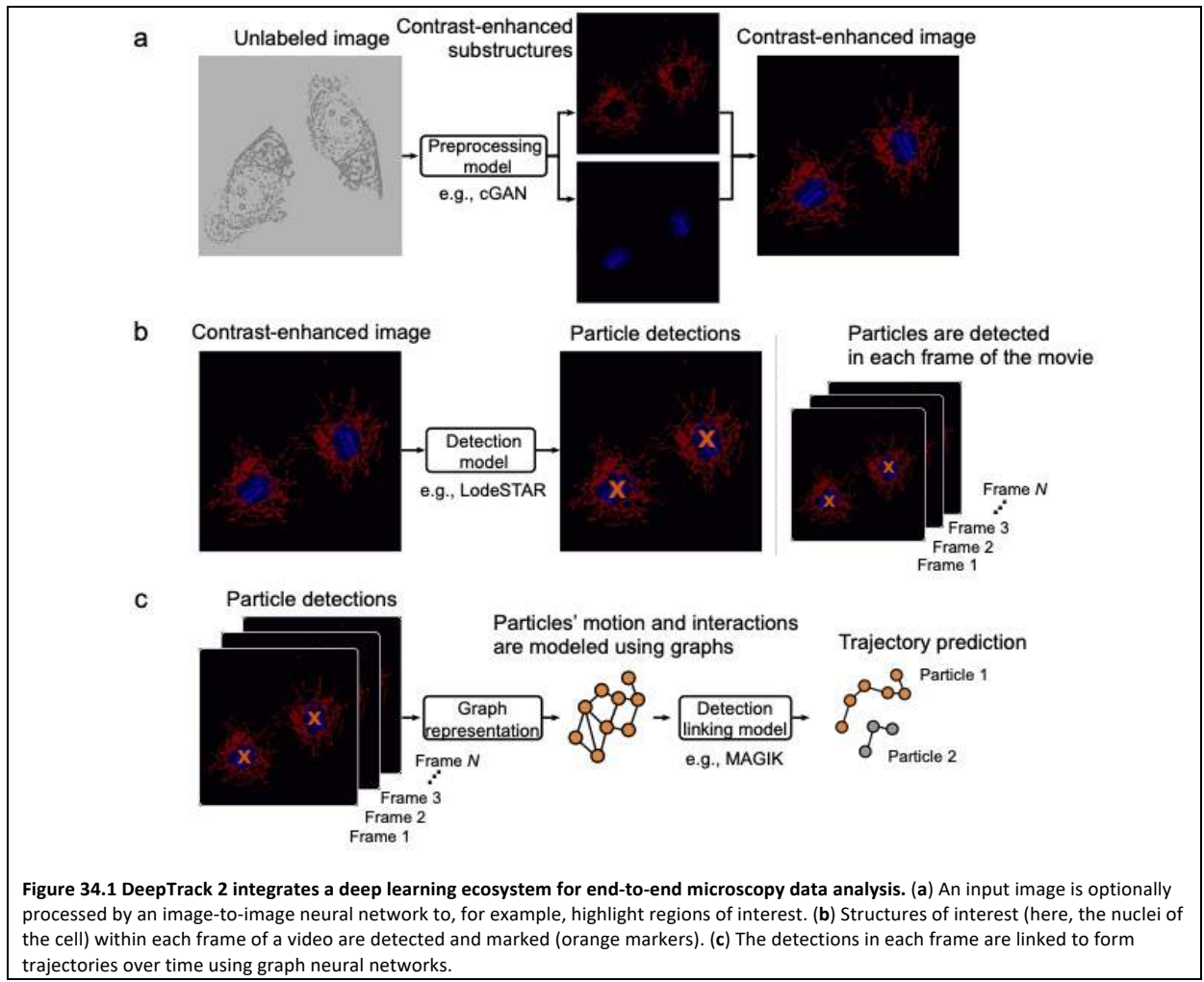

**Figure 34.1 DeepTrack 2 integrates a deep learning ecosystem for end-to-end microscopy data analysis.** (**a**) An input image is optionally processed by an image-to-image neural network to, for example, highlight regions of interest. (**b**) Structures of interest (here, the nuclei of the cell) within each frame of a video are detected and marked (orange markers). (**c**) The detections in each frame are linked to form trajectories over time using graph neural networks.

DeepTrack 2 provides deep-learning solutions for each step of this processing pipeline, allowing users with varying programming experience to use and optimize the available solutions to their data, goals, and challenges. For step 1., DeepTrack 2 integrates tools for data normalization, noise suppression, augmentations, and even more advanced methods such as virtual staining to produce high-contrast and high-specificity images (**Figure 34.1a**). For step 2., DeepTrack 2 ships with state-of-the-art models ready for training, such as LodeSTAR [5] for label-free particle detection, U-Net for semantic segmentation, and YOLO for simultaneous detection and classification (**Figure 34.1b**). For step 3. and 4., DeepTrack 2 uses MAGIK [6], a state-of-the-art graph-based network which both connects observations into trajectories and extracts information about the dynamics of the system from spatio-temporal data (**Figure 34.1c**).

Recent research has demonstrated the benefit of DeepTrack 2 to analyze microscopy data. For example, **Figure 34.2a** shows how virtual staining can unveil high-quality visualizations of complex biological systems from cheap-to-capture brightfield images using conditional generative adversarial neural networks (cGAN) [7]; **Figure 34.2b** demonstrates that close-to-perfect object detection can be achieved from just a single unannotated image using LodeSTAR [5]; and **Figure 34.2c** shows how graph neural networks such as MAGIK can be used to connect detections into traces even if the cells divide [6]. Furthermore, DeepTrack 2 has been used for the characterization of microplanktons from Holographic microscopy images [8] and the monitoring of active droploids, a new class of active matter systems [9].

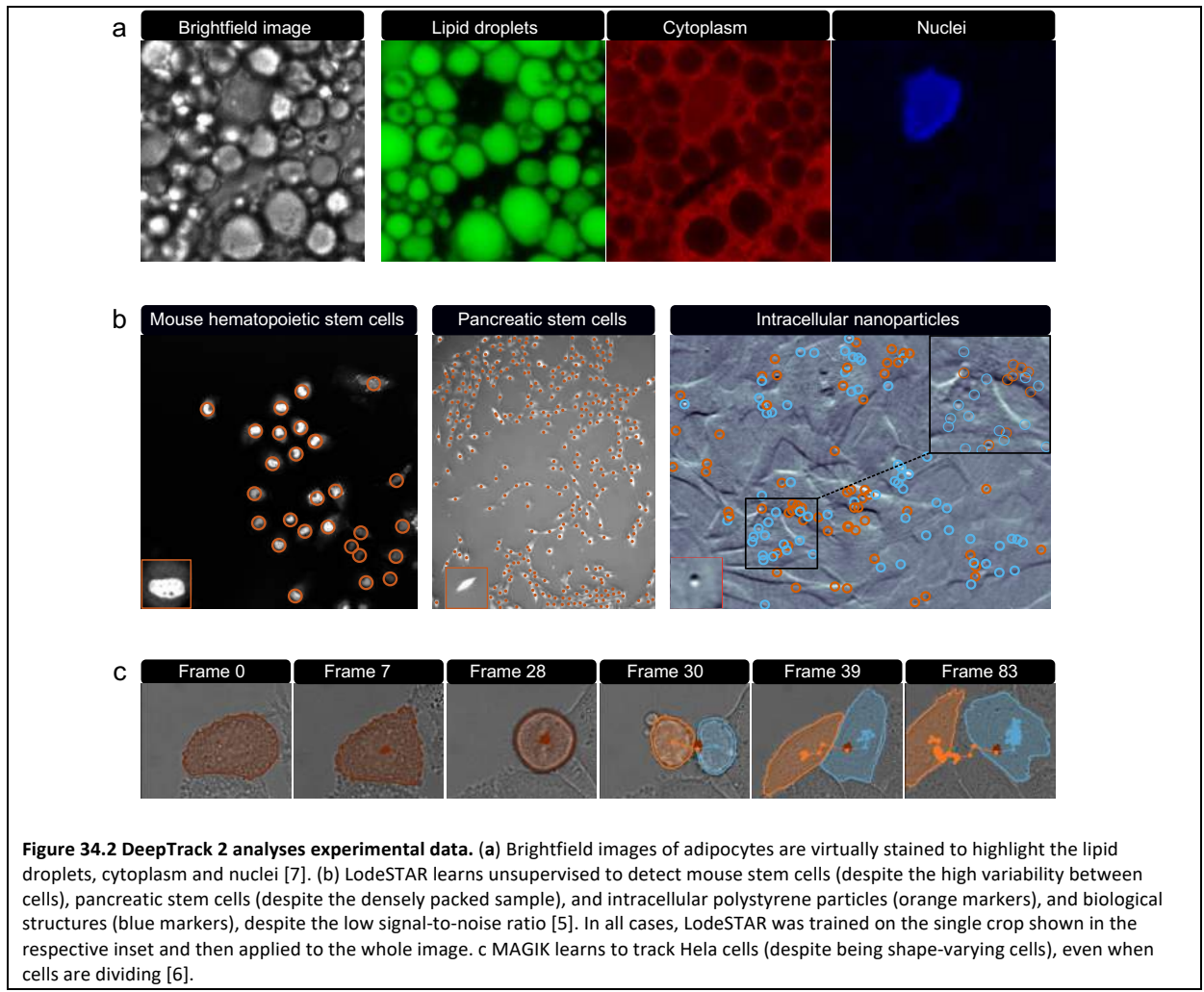

**Figure 34.2 DeepTrack 2 analyses experimental data.** (**a**) Brightfield images of adipocytes are virtually stained to highlight the lipid droplets, cytoplasm and nuclei [7]. (**b**) LodeSTAR learns unsupervised to detect mouse stem cells (despite the high variability between cells), pancreatic stem cells (despite the densely packed sample), and intracellular polystyrene particles (orange markers), and biological structures (blue markers), despite the low signal-to-noise ratio [5]. In all cases, LodeSTAR was trained on the single crop shown in the respective inset and then applied to the whole image. c MAGIK learns to track Hela cells (despite being shape-varying cells), even when cells are dividing [6].

**Current and Future Challenges**

The transition to deep-learning-enhanced analysis has not been without challenges. Firstly, the high variability between imaging modalities and object samples has made publicly available datasets an unreliable source of training data. The dataset would need to contain the same object of interest, imaged through a near-identical optical device, and annotated with the desired ground truth. The likelihood of these coinciding is slim-to-none.

Secondly, for many applications, the expected quality of analysis exceeds human precision. Examples include sub-pixel localization of objects for super-resolution microscopy and force calibrations, object detection in noisy images, 3d-microscopy and more. Consequently, even high-quality human annotations are insufficient for training these neural networks.

Thirdly, neural networks have deservedly gained a reputation as black-box functions, calling into question their reliability for clinical and industrial use. Consequently, deep learning has struggled to gain traction beyond a research tool. Fundamentally, this is because neural networks only yield answers with little explanation of how that answer was acquired.

Finally, the rapid development of methods and model architectures has resulted in a scattered field without a unified interface. Consequently, comparing methods is prohibitively difficult. Moreover, reusing deep learning methods for new data often requires re-implementing them from scratch, which is a daunting task for non-experts.

These four challenges combined have resulted in a slow adoption of new deep learning methods, despite their significant advantage in performance. DeepTrack 2 attempts to address these issues through a three-pronged approach:

1. The use of synthetic data for training neural networks to help reduce the reliance on annotated data while generating perfect ground truths [1,8].
2. The development of label-free, low-shot, and interpretable models to promote low-cost methods with less opaque neural networks [5,6].
3. The unification of methods into a consistent interface, by continuously implementing state-of-the-art methods and encouraging authors to contribute their methods [10].

**Advances in Science and Technology to Meet Challenges**

We identify three avenues of research essential to bring out the full potential of deep learning for quantitative microscopy:

1. Further development of unsupervised or self-supervised methods, particularly for tasks such as segmentation, trajectory reconstruction, and classification. While the scaffolding exists for these developments through methods such as self-distillation and contrastive learning, they have yet to be optimized for the specific challenges of microscopy.
2. Development of interpretable neural networks. The recently popularized attention-based neural networks have allowed much more transparent neural networks than previously possible. Attentive neural networks can highlight the parts of the data that lead to a particular conclusion, providing insight into the reason behind an answer. Currently, this is mainly used for classification tasks. However, we see no reason why attention-like mechanisms cannot be incorporated into neural networks designed for other tasks in microscopy, significantly increasing the interpretability and thereby the trustworthiness of the neural networks.
3. Design of advanced simulation methods for various optical devices. Unlike the impossibly complex macroscopic world, the physics of the microscopic world can feasibly be fully simulated. As such, the need for annotated data can be done away with entirely by developing faster and more accurate simulation techniques.

**Concluding Remarks**

Deep learning is undeniably a powerful tool for quantitative microscopy. Nonetheless, it has remained a research tool instead of reaching the hands of clinicians. We identify four key challenges that need to be overcome to reach the full potential of deep learning for quantitative microscopy. These are: a reliable source of training data that is general enough to match the specific problem and experimental device of the end user, a method of training neural networks beyond the limit of human accuracy for annotation, the development of less opaque models, and a unified interface for using these models.

We also consider the field in a good state to tackle these problems. Self-supervised methods are on the rise, and interpretable layers such as attention layers are taking over the field. Targeted development of these two approaches for microscopy may lead to the widespread adoption of reliable deep learning methods, revealing physical and biological insights encoded in the data in an unsupervised manner.

## 35 —DeepImageJ

Estibaliz Gómez-de-Mariscal[1], Daniel Sage[2], Arrate Muñoz-Barrutia[3,4]

1. Instituto Gulbenkian de Ciência, Oeiras, Portugal
2. Biomedical Imaging Group and EPFL Center for Imaging, Ecole Polytechnique Fédérale de Lausanne (EPFL) Lausanne, Switzerland
3. Bioengineering Department, Universidad Carlos III de Madrid, Leganés, Spain
4. Instituto de Investigación Sanitaria Gregorio Marañón, Madrid, Spain

**Status**

As in other scientific fields, deep learning (DL) has achieved outstanding performance for many microscopy image processing tasks [1,2]. Its potential to reveal visual features in complex images has changed the game in microscopy image analysis; this is especially the case when the visual features are unexplainable by humans, like for image restoration (e.g., CARE [3]), for dense nuclei detection (e.g., StarDist [4]), for super-resolution localization microscopy (e.g., DECODE [5]), or for correlative microscopy imaging. Although recent contributions translating DL to microscopy imaging have empowered researchers with the capacity to build powerful pipelines, the IT barrier is still too high for the large majority of end-users. It requires technical competencies and programming expertise to fully exploit this new technology. Life scientists are usually uncomfortable with Python, the preferred language for DL frameworks (TensorFlow, PyTorch). Instead, they vastly prefer the friendly user interface of Java-based software like ImageJ [6], Icy [7], or QuPath [8]. ImageJ is currently the most used software in cell imaging, as it provides a rich palette of bioimage analysis tools. Moreover, it can interoperate with many other platforms in a unique ecosystem, including drivers for microscopy or data analysis packages. Nevertheless, the hardware restrictions of Java have prevented its use as the reference tool for bioimage analysis with machine learning. Hence, the users of ImageJ have been refrained from deploying DL technology in their image analysis pipelines.

In the last few years, several initiatives were already taken to integrate DL features in ImageJ. Several teams (DeepClass4Bio [9], CSBDeep [3], deepImageJ [10]) are working to provide access to pre-trained DL models through ImageJ plugins or ImageJ macros. In the context of ImageJ, deepImageJ is the generic consumer of pre-trained DL models. On the other hand, well-established ImageJ plugins have also integrated DL in their pipelines. For instance, TrackMate [11] has included cell segmentation with CellPose or StarDist to extend the versatility of the tracking pipeline. Finally, the BioImage Model Zoo [12] is a community-based initiative that gives effective access to DL technology. It proposes a standardized format tailored for life scientists to share and deploy trained DL models across open bioimage analysis software (e.g., ImageJ, Ilastik [13], ImJoy [14], ZeroCostDL4Mic [15], QuPath).

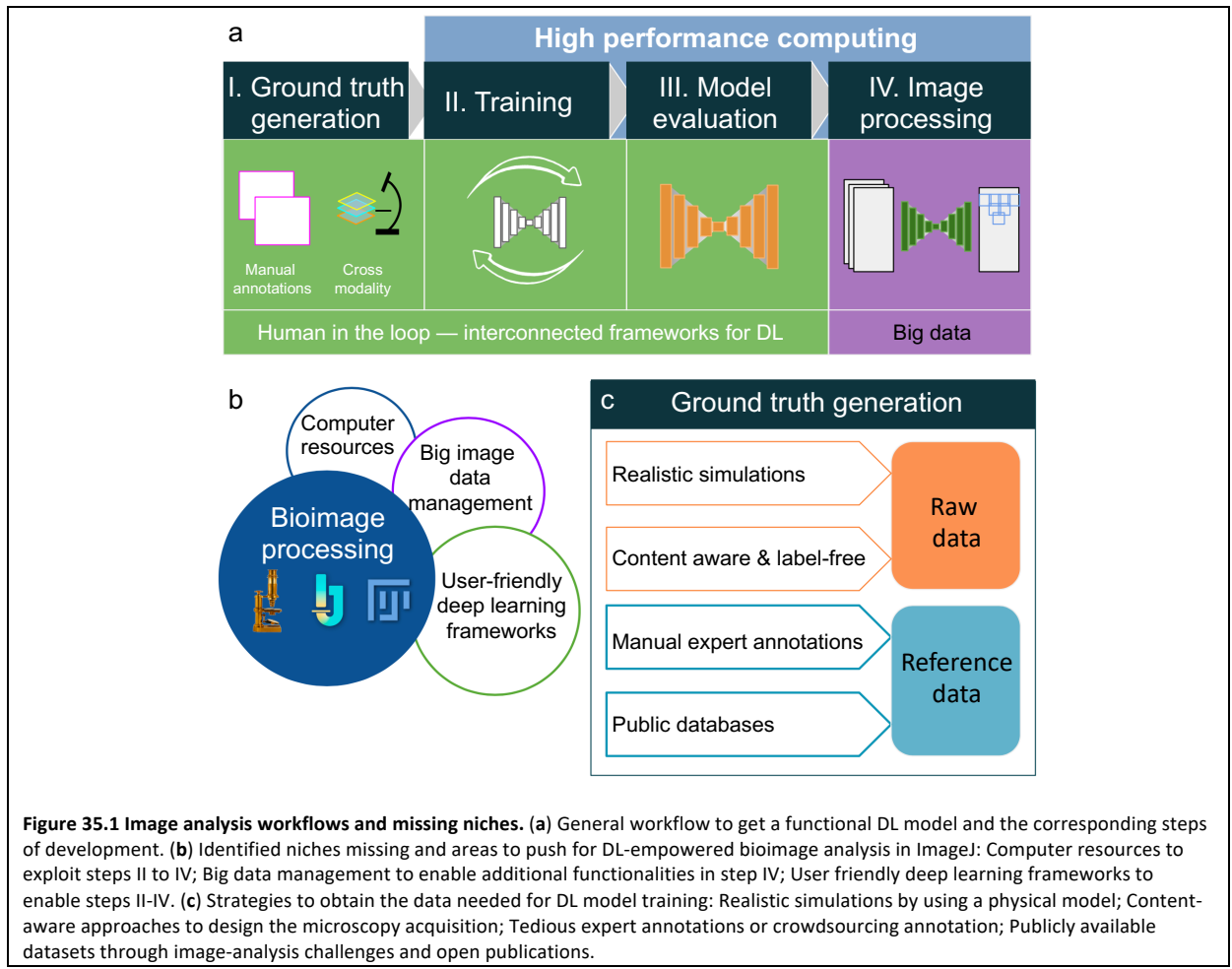

**Figure 35.1 Image analysis workflows and missing niches.** (**a**) General workflow to get a functional DL model and the corresponding steps of development. (**b**) Identified niches missing and areas to push for DL-empowered bioimage analysis in ImageJ: Computer resources to exploit steps II to IV; Big data management to enable additional functionalities in step IV; User friendly deep learning frameworks to enable steps II-IV. (**c**) Strategies to obtain the data needed for DL model training: Realistic simulations by using a physical model; Content-aware approaches to design the microscopy acquisition; Tedious expert annotations or crowdsourcing annotation; Publicly available datasets through image-analysis challenges and open publications.

**Current and Future Challenges**

We identify challenges at two levels for the full integration of DL in microscopy image analysis. Current DL approaches have the capacity to boost the limits of microscopy image acquisition [3,5]. Still, their performance relies on the quality and the quantity of data, both acquired raw images and annotated or reference images (**Figure 35.1c**). Usually, scientific images are acquired to answer specific biological questions. Unfortunately, researchers look for a suitable method to analyze once all the images are acquired. We envision that integrating the data analysis at the initial steps of the project could significantly improve the imaging workflow's robustness (**Figure 35.1a**). This is particularly demanding for data-centric approaches when the images are the heart of the image-analysis task. The acquisition protocol should be appropriately designed for this purpose, including a large variability of examples in order to reduce the time and effort needed to obtain scientific results.

The deployment of DL methods routinely is not straightforward. Important technical niches in the ImageJ ecosystem (**Figure 35.1b**) are 1) the lack of user-guided training, 2) the easy use of DL models on large images, 3) the connection to high-performance computing resources (GPU), and 4) the interactive insertion of DL models in complex bioimage analysis pipelines. Likewise, a more optimal integration of DL into bioimage analysis pipelines relies mainly on the education of the practitioners in machine learning and image analysis. In practice, it is essential to train the researchers on the appropriate use cases and warn them about the limitations and the risks of the DL technology. The pandemic significantly fomented the participation in recorded online courses (Neubias Academy, EMBO and EMBL) and the edition of user-guides of good practices. The community needs to keep the momentum for transferring this knowledge. Therefore, this is a crucial challenge to tailor the current DL knowledge and technology to users in terms of data acquisition and usable computational tools for analysis.

As part of Artificial Intelligence (AI) technology, DL is a very active field. Preparedness for the integration of upcoming new methods in ImageJ is still a bottleneck. Some still missing techniques are 1) human-in-the-loop, 2) auto machine learning (AutoML), 3) self-supervised or weakly supervised training, and 4) tracking of biological particles. Ultimately, the next generation of DL approaches seeks smart workflows capable of embedding the prior knowledge to a guided analysis or some sparse information to run weakly supervised training. All these developments require a deep understanding of both DL and bioimaging, and therefore, the close collaboration between developers and end-users. Thus, this is a great challenge to push DL to its full potential for smart imaging in biomedical research.

**Advances in Science and Technology to Meet Challenges**

DL is data-hungry, and its reusability can only be accomplished through model training or fine-tuning. Aware of this, the focus on acquiring findable, accessible, interoperable, and reusable (FAIR) data is notoriously growing in academia. The number of publicly available databases tailored for DL model training is increasing more quickly in computer vision than in microscopy (Data Science Bowl [2], MONAI (https://monai.io)). Data generators and simulators are another source of microscopy images for training; they are specifically suitable when the physical laws of the image formation model are well identified.

The performance of the neural networks depends on the amount of data available or on the accessibility to validated trained models. For example, the resolution and signal-to-noise ratio needed for a specific measurement will vary according to the structures to analyze. Some works provide guidelines about these features so researchers can optimize their image acquisition [3-5]. Furthermore, recent approaches focus on adjusting the content in the image — content aware approaches [3]. The aim is to push the limits of data acquisition and virtually recover biologically relevant information from simpler or more sample-friendly acquisitions [3, 5]. Ultimately, the ability of DL algorithms to encode a large amount of information from the images in latent spaces enables the discovery of different biological behaviors.

Existing user-friendly software for DL usually targets a specific step of a general DL-based image analysis workflow. For example, ZeroCostDL4Mic has democratized the training procedure and the assessment of several neural network architectures. It allows most of the steps in Figure 1A using the free cloud computing Google Collaboratory. ImJoy [18] proposes a framework to interact with virtual content beyond DL directly in the browser without the need of any technical installation steps. This is a big step towards connecting with cloud computing. Ilastik is the reference tool for image segmentation using machine learning and now, it can also run DL trained models. For ImageJ we found third-party plugins, mainly for prediction, DeepClass4Bio [9], CSBDeep [3], StarDist [4] and deepImageJ [10]. The latter has enabled the generic use of trained models of the Bioimage Model Zoo.

As the need for more powerful computational hardware increases with new AI-based approaches, cloud computing or network-embedded infrastructures are essential to ensure translational technology in microscopy imaging. Common infrastructure is maybe the key to reduce the carbon footprint of the energy-intensive training. The new European infrastructure AI4Life

(https://www.eu-openscreen.eu/projects/ai4life.html) aims to provide sustainable, intuitive, and highest quality research services and infrastructures that will enable all life scientists to exploit machine learning to improve the utility and interpretability of image data.

**Concluding Remarks**

Integrating DL in ImageJ as the initial steps provided by DeepImageJ is crucial for the computational microscopy imaging area. Friendly access (up to zero-code) and accessibility to DL tools are required for democratized access to powerful DL solutions. By open access to pre-trained models and the related data, the users can test them and understand their potential and the limitations of this new technology. Community-driven developments such as those provided by the Bioimage Model Zoo are boosting the reach of DL integration in open-source software devoted to analyzing microscopy images. The initiative behind Bioimage Model Zoo is pushing forward the resolution of the challenges presented in the previous section. Namely, the standardization of the model format facilitates the cross-compatibility between different platforms and will contribute to the dissemination of the robust models. The availability of advanced DL tools enables the application of holistic approaches for life-sciences research. In particular, technological advances precede life-science research advances and facilitate the realization of breakthrough discoveries. So, in conclusion, we are at the gates of knowing the next 'AlphaFold' for microscopy imaging.

**Acknowledgements**

*We acknowledge the support of Ministerio de Ciencia e Innovación, Agencia Estatal de Investigación, under grant PID2019-109820RB-I00, MCIN/AEI/10.13039/501100011033/, co-financed by European Regional Development Fund (ERDF), 'A way of making Europe' and the European Commission through the Horizon Europe program (AI4LIFE project, grant agreement 101057970-AI4LIFE). E.G.M. was supported by the Gulbenkian Foundation and the European Molecular Biology Organization (EMBO) Installation Grant (EMBO-2020-IG-4734) (granted to the Optical Cell Biology laboratory at Instituto Gulbenkian de Ciência) and Postdoctoral Fellowship (EMBO ALTF 174-2022).*

## 36 — Hackathons to spur innovation

Ebba Josefson Lindqvist[1], Johanna Bergman[1]

1. AI Sweden, Gothenburg, Sweden

**Status**

Research and innovation do not happen in isolation. Existing and acting within an ecosystem is becoming increasingly important for organisations as AI is becoming a key technology in almost all industries [1]. Few organisations have all the competence, data, and infrastructure needed to fully apply machine learning at scale. To address real world research and innovation challenges, both extensive domain knowledge and machine learning expertise is needed. Hackathons are one way of addressing this issue and facilitate collaborations.

The machine learning community has a long tradition of hackathons and competitions [2]. There are several platforms, the most famous being Kaggle and Codalab [3,4]. Hackathons are traditionally focused on either a specific problem within machine learning or a problem formulation around a dataset. AI Sweden, the Swedish national centre for applied artificial intelligence has hosted several hackathons focusing on industry data and problem formulations from industry partners [5]. The key success factors have been, top management commitment from industry, a clear problem formulation, close collaboration with domain experts, accessible data, well defined evaluation metrics, and access to the computational resources needed.

One recent example within microscopy was the Adipocyte Cell Imaging Challenge, hosted together with AstraZeneca. The problem formulation and data was provided by domain experts at AstraZeneca. AI Sweden invited researchers and AI experts, provided a collaborative platform, access to data and computational infrastructure. The task was to utilise machine learning to predict the content of fluorescence images from the corresponding bright field images [6] (see figure 1 for example images). Results from this specific hackathon resulted in new industry-academia collaborations, publications [7], and most importantly solutions that could be directly applied by AstraZeneca benefitting their research.

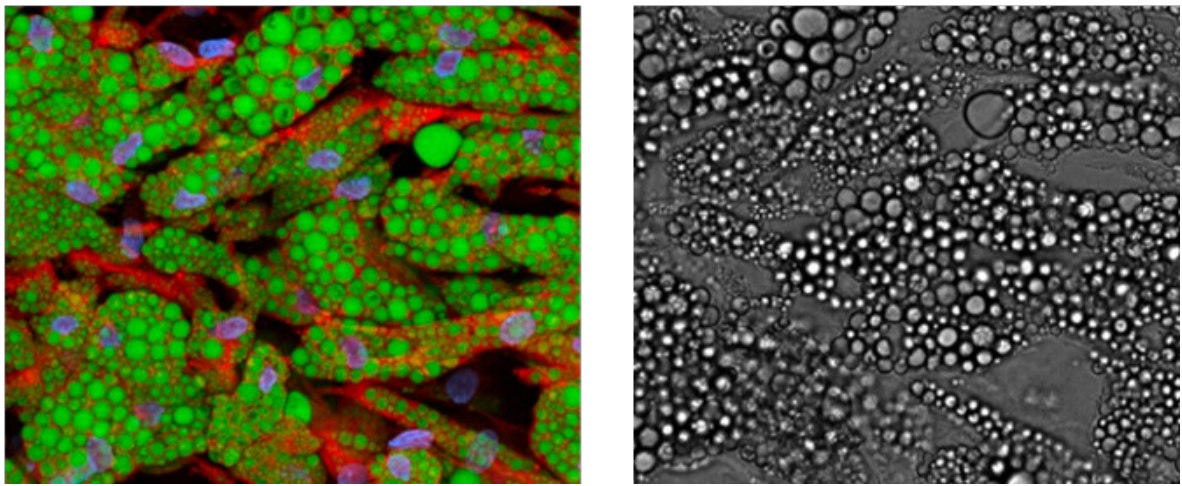

**Figure 36.1 Adipocyte Cell Imaging Challenge.** Example images from the Adipocyte Cell Imaging Challenge hosted by AI Sweden and AstraZeneca in 2021.

**Current and Future Challenges**

There are several challenges that need to be addressed in order to make open hackathons more beneficial for the application of machine learning to microscopy. Namely:

*Data access*

Access to data is key for developing machine learning applications for microscopy. In many cases the data needed is collected within industry. Such data is rarely accessible for academic researchers, startups or other companies. At the same time, these organisations typically hold a lot of the machine learning competence needed but still may not be able to use all the resources needed to make the full potential use of the data collected.

The main reasons for industry data to not be shared are e.g. the traditional way of thinking of the data as proprietary and too valuable to share openly as, most of the time, considerable investments have been made in collecting the data. Sharing data also opens up for questions concerning legal issues, uncertainty regarding business models for data, and how to know the value of your data. Legal considerations may include questions regarding Intellectual Property, GDPR (General Data Protection Regulation), and what is the appropriate copyright licence to be used. In general the more openly a dataset may be used the more it opens up for new advances in research. To open up for access to a dataset may also be crucial to enable publications in ML and microscopy to enable peer-reviews. This requires an open and new mindset for many industry actors.

*Domain knowledge and cross-disciplinary research*

To understand the data, problem formulation, and interpret the results, extensive domain knowledge is needed. During a hackathon, access to domain knowledge and possibility to work in cross-functional teams is of critical importance. From the Adipocyte Cell Imaging Challenge, we could see that teams with cross-disciplinary expertise performed better than teams coming solely from the machine learning field. This despite the fact that all teams had the opportunity to have daily contact with the domain experts from AstraZeneca.

Traditionally pharmaceutical companies have a relatively long-time perspective for research and focus a lot on their internal resources and established academic research collaborations.

The trends in applied machine learning research are diverging in two directions. One direction towards being as open as possible with strong community efforts, collaborations, and grass-root research initiatives where data, code, and models are typically shared openly and where open source is the default option. On the other hand many advancements in the field are seen within big companies with a lot of resources, in this case models and data are typically not shared. Common for both directions is the speed and amount of resources needed (both in terms of researchers, data, and computational resources).

Naturally, research that also has a large component of data collection or even physical experiments (for example microscopy) have a longer time frame. As machine learning is becoming an important part of other research fields with the potential to lead to disruptive advances, the different traditions, trends, and time perspectives need to be addressed.

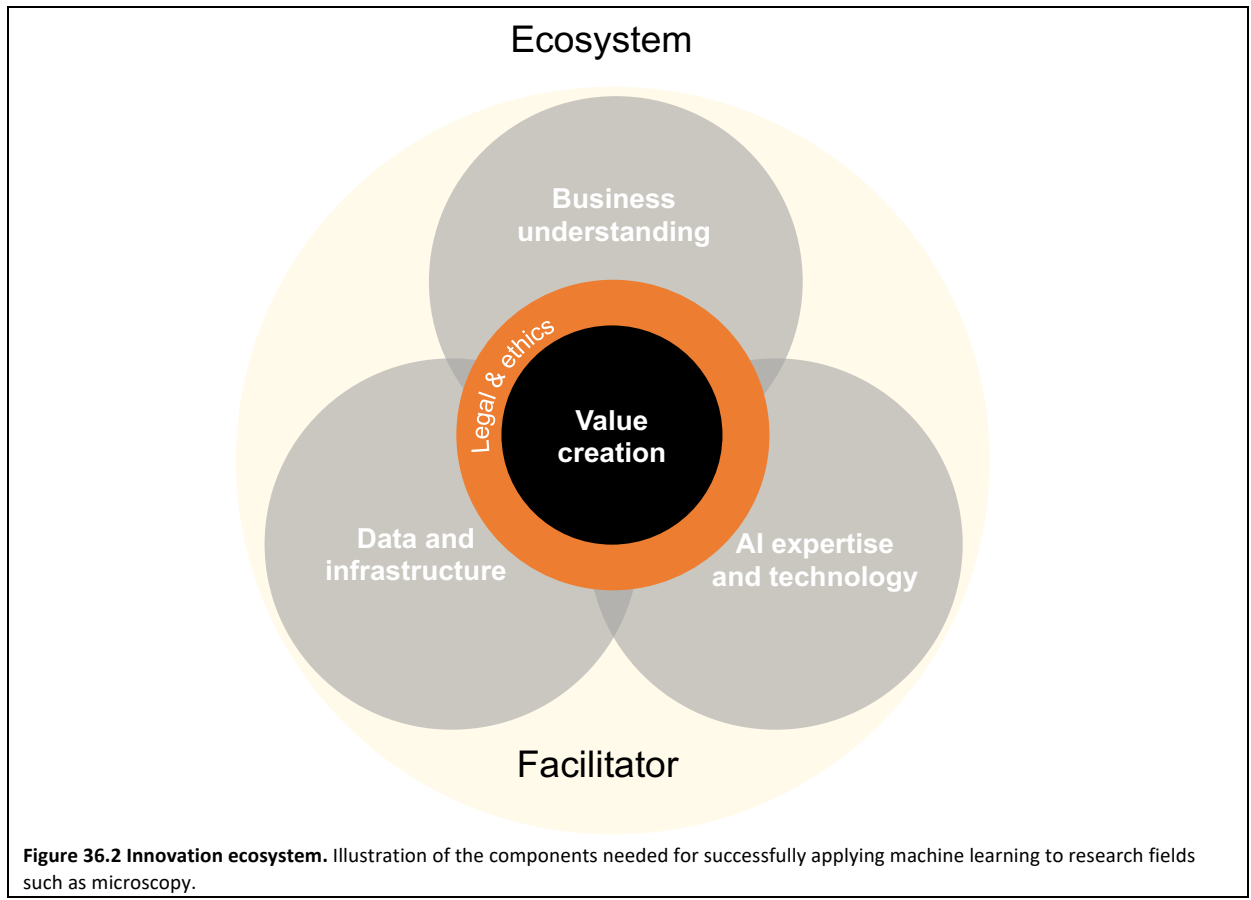

**Figure 36.2 Innovation ecosystem.** Illustration of the components needed for successfully applying machine learning to research fields such as microscopy.

**Advances in Science and Technology to Meet Challenges**

We see two main topics for advancements to further benefit from hackathons and competitions in the intersection of machine learning and microscopy research, especially around facilitating collaboration between industry, startups, and academia, and the broader machine learning community.

*Data sharing & data business models*

Industry data sharing is challenging for several reasons. Our experience is that the organisation of the hackathon itself forces the participating parties to come up with pragmatic solutions for data access, from both a technical and a legal perspective, a process that is often lengthy in other projects. Taking a different and more open approach in such a process potentially accelerates the development of new best practices for data sharing.

In addition to data access we see a strong need for developing business models and legal frameworks to share data and code to create lasting effects of hackathons and competitions. It is equally important to change the mindset of the data being proprietary and create an understanding of the value of sharing data more openly.

*Community building and cross-disciplinary research*

To succeed with hackathons, building a strong community and platform for collaboration is key. Building a strong community of researchers from different disciplines and organisations will benefit the research field and development of machine learning applications for microscopy. Organising hackathons could be one way of building such a community.

## Concluding Remarks

To conclude, we see that hackathons have potential to further integrate the microscopy research field with machine learning as one way of exploring new collaborations, methods, and ways of working. In addition, we see an opportunity to build a strong research community around microscopy and machine learning. To enable such a community, collaborative platforms, data access, legal frameworks and business models around data are needed. Furthermore, cross-disciplinary collaboration and strong domain expertise is key to success when machine learning is applied to research questions in other fields.

Shorter challenge driven competitions or hackathons should be seen as a way of exploring and initiating new (sometimes unexpected) research collaborations that have the potential to lead to continued long term collaborations. This could also be one way of addressing and challenging the different time perspectives and research traditions of industry and academia, accelerate data sharing, and invite a broader community to take part in applied research.

Collaborative research will be increasingly important as machine learning becomes an important tool for microscopy research. Based on the experience from the Adipocyte Cell Imaging Challenge, hackathons could be one enabler for accelerating access to industry data developing the collaborations needed within machine learning research for microscopy. Neutral organisations, such as AI Sweden, or dedicated challenge platforms (e.g., Kaggle, Codalab), can play a facilitating role and reach a broader community.

## Acknowledgements

*We want to direct a special thank you to AstraZeneca for giving us the opportunity to co-host the Adipocyte Cell Imaging Challenge as well as all the participating teams proving that hackathons are a powerful tool for building new collaborations and accelerating applied research.*